\documentclass[11pt,a4paper]{article}
\usepackage{epsfig}
\usepackage[T1]{fontenc}    
\usepackage{graphics}
\usepackage{graphicx}
\usepackage{pstricks,pst-coil,pst-fill,pst-plot}
\usepackage[fleqn]{amsmath}    
\usepackage{amssymb}    
\usepackage{amsfonts}   
\usepackage{verbatim}   
\usepackage{mathrsfs}   
\usepackage{dsfont}
\usepackage{euscript}
\usepackage{yfonts}
\usepackage{enumerate}     
\usepackage{amsthm}         
\usepackage{txfonts}
\usepackage{marvosym}
\usepackage{vmargin}        
\usepackage{wasysym}		

\setmarginsrb{2.5cm}{2cm}{2.5cm}{2cm}{1cm}{1cm}{1cm}{1.6cm}
 \makeatletter
\@addtoreset{equation}{section}
 \makeatother

\providecommand{\bysame}{\leavevmode\hbox to3em{\hrulefill}\thinspace}
\providecommand{\MR}{\relax\ifhmode\unskip\space\fi MR }

\providecommand{\href}[2]{#2}

\newcommand{\bra}[1]{\big< \,#1\,|}
\newcommand{\ket}[1]{|\,#1\, \big>}
\newcommand{\braket}[2]{\big<\,#1 \, \big|\,  #2  \big>}

\let\ua=\uparrow

\let\tend=\rightarrow

\long\def\symbolfootnote[#1]#2{\begingroup%
\def\thefootnote{\fnsymbol{footnote}}\footnote[#1]{#2}\endgroup}

\newtheorem{theorem}{Theorem}[section]
\newtheorem{prop}[theorem]{Proposition}
\newtheorem*{theorem*}{Theorem}
\newtheorem{cor}[theorem]{Corollary}

\newtheorem{lemme}[theorem]{Lemma}
\newtheorem{hypothesis}[theorem]{Hypothesis}

\def\Proof{\medskip\noindent {\it Proof --- \ }}

\def\qed{\hfill\rule{2mm}{2mm}}

\newcommand\beq{\begin{equation}}
\newcommand\enq{\end{equation}}
\newcommand\bem{\begin{multline}}
\newcommand\enm{\end{multline}}

\def\beqa{\begin{eqnarray}}
\def\eeqa{\end{eqnarray}}
\def\ba{\begin{array}}
\def\ea{\end{array}}
\def\det{\operatorname{det}}

\newcommand{\f}[2]{{\ensuremath{%
    \mathchoice%
    {\dfrac{#1}{#2}}
    {\dfrac{#1}{#2}}
    {\frac{#1}{#2}}
    {\frac{#1}{#2}}
}}}
\newcommand{\tf}[2]{\ensuremath{#1/#2}}
\newcommand{\pa}[1]{\ensuremath{\left(#1\right)}}

\def\a{\alpha}

\def\be{\beta}
\def\ga{\gamma}
\def\Ga{\Gamma}

\def\de{\delta}

\def\De{\Delta}
\def\eps{\epsilon}
\def\veps{\varepsilon}
\def\la{\lambda}
\def\La{\Lambda}

\def\sg{\sigma}
\def\vsg{\varsigma}

\def\Ups{\Upsilon}
\def\ups{\upsilon}
\def\th{\theta}

\def\vth{\vartheta}

\def\Om{\Omega}
\def\om{\omega}
\def\vp{\varphi}

\newcommand{\mc}[1]{\ensuremath{\mathcal{#1}}}
\newcommand{\mf}[1]{\ensuremath{\mathfrak{#1}}}
\newcommand{\msc}[1]{\ensuremath{\mathscr{#1}}}

\newcommand{\bs}[1]{\ensuremath{\boldsymbol{#1}}}

\DeclareFontFamily{OT1}{pzc}{}
\DeclareFontShape{OT1}{pzc}{m}{it}{<-> s * [1.10] pzcmi7t}{}
\DeclareMathAlphabet{\mathpzc}{OT1}{pzc}{m}{it}

\def \i{ \mathrm i}

\newcommand{\ov}[1]{\ensuremath{\overline{#1}}}
\newcommand{\wt}[1]{\ensuremath{\widetilde{#1}}}
\newcommand{\wh}[1]{\ensuremath{\widehat{#1}}}

\newcommand{\Int}[2]{\ensuremath{\int\limits_{#1}^{#2}}}
\newcommand{\Oint}[2]{\ensuremath{\oint\limits_{#1}^{#2}}}

\newcommand{\sul}[2]{\ensuremath{\sum\limits_{#1}^{#2}}}
\newcommand{\pl}[2]{\ensuremath{\prod\limits_{#1}^{#2}}}

\newcommand{\R}{\ensuremath{\mathbb{R}}}
\newcommand{\Cx}{\ensuremath{\mathbb{C}}}

\newcommand{\Dp}[1]{\ensuremath{\partial_{#1}}}

\newcommand{\ex}[1]{\ensuremath{\e{e}^{#1}}}

\def\Res{\operatorname{Res}}

\newcommand{\op}[1]{ \boldsymbol{ \texttt{#1} } }

\newcommand{\Norm}[1]{\ensuremath{\big| \big| #1 \big|\big| }}
\newcommand{\norm}[1]{\ensuremath{  || #1 || }}

\newcommand{\dd}{\mathrm{d}}
\newcommand{\e}[1]{\ensuremath{\mathrm{#1}}}

\newcommand{\intff}[2]{\ensuremath{ [  #1 \,; #2 ] }}

\newcommand{\intoo}[2]{\ensuremath{ ]  #1 \,; #2 [ }}

\newcommand{\intn}[2]{\ensuremath{[\![ \, #1 \,;\, #2 \,]\!]}}

\begin{document}

\begin{center}
\begin{LARGE}
{\bf Form factors of bound states in the XXZ chain}
\end{LARGE}

\vspace{1cm}

\vspace{4mm}
{\large Karol K. Kozlowski \footnote{e-mail: karol.kozlowski@ens-lyon.fr}}%
\\[1ex]
Univ Lyon, Ens de Lyon, Univ Claude Bernard, CNRS, Laboratoire de Physique, F-69342 Lyon, France. \\[2.5ex]

\par 

\end{center}

\vspace{40pt}

\centerline{\bf Abstract} \vspace{1cm}
\parbox{12cm}{\small}

This work focuses on the calculation of the large-volume behaviour of form factors of local operators in the XXZ spin-$1/2$ chain taken between the ground state and an 
excited state containing bound states. 
The analysis is rigorous and builds on various fine properties of the string solutions to the Bethe equations and certain technical hypotheses. These technical 
hypotheses are satisfied for a generic excited state. 
The results obtained in this work pave the way for extracting, starting from the first principles, the large-distance and long-time asymptotic behaviour of the XXZ chain's two-point functions 
just as the so-called edge singularities of their Fourier transforms.

\vspace{40pt}

\section*{Introduction}

Form factors of local operators constitute the elementary objects encoding the dynamics of a model. 
Although form factors cannot be computed in closed form for a general model, the state of the art is much more satisfactory in the case of quantum integrable models. 
First explicit computations of form factors in such models go back to the works \cite{KarowskiWeiszFormFactorsFromSymetryAndSMatrices,KirillovSmirnovFirstCompleteSetBootstrapAxiomsForQIFT,SmirnovFormFactors}. 
There the form factor of various 1+1 dimensional integrable quantum field theories have been obtained by means of the 
boostrap axioms. The analysis of form factors in lattice quantum integrable models has been pioneered 
by Jimbo and Miwa \cite{JimboMiwaFormFactorsInMassiveXXZ} on the example of the XXZ spin-1/2 chain at anisotropy greater than $1$. All the results mentioned so far were obtained for massive models directly in the infinite volume,
this owing to the existence of a mathematically well-posed description in such a setting. Indeed, Eigenstates of infinite volume massive integrable models can be labelled
by using continuous rapidities and well-defined isolated variables. This property allows one to characterise the  infinite volume form factors in terms of a sequence of densities in the continuous rapidities. 
The situation becomes much more involved in models having a massless spectrum as the very presence of massless modes renders such a description impossible, 
see \textit{e.g.} \cite{KozMailletMicroOriginOfc=1CFTUniversality}. For this reason, one can only achieve a meaningful description of the 
form factors in massless models by first keeping the volume finite and then extracting their large-volume asymptotic behaviour.  
However, obtaining finite volume expression for the form factors turned out to be a rather challenging task, even in the case of quantum integrable models. 
In the case of interacting models, it could only be achieved recently thanks to the invention of the algebraic Bethe Ansatz \cite{FaddeevSklyaninTakhtajanSineGordonFieldModel}
followed by the calculation of norms \cite{KorepinNormBetheStates6-Vertex} and scalar products \cite{SlavnovScalarProductsXXZ} of Bethe vectors and, finally, 
the resolution \cite{KMTFormfactorsperiodicXXZ} of the quantum inverse scattering problem. 
All these ingredients put together led \cite{KMTFormfactorsperiodicXXZ} to determinant based representations for the form factors of local operators
of the finite volume XXZ spin-1/2 chain. In the massless antiferromagnetic phase of the model, the size of the determinants describing the ground to excited states
form factors grows linearly with the model's volume. Thus, the analysis of the large-volume behaviour of the form factors demands to extract the
large-size behaviour of the underlying class of determinants. Such a problem was first investigated by N. Slavnov in \cite{SlavnovFormFactorsNLSE} who 
studied the large-size asymptotics of determinants describing the form factors of the current operator in the non-linear Schrödinger model. This analysis
was then improved and extended in the works \cite{KozKitMailSlaTerEffectiveFormFactorsForXXZ,KozKitMailSlaTerThermoLimPartHoleFormFactorsForXXZ}. There, 
the authors considered the massless regime of the XXZ chain at finite magnetic field and extracted the large-volume behaviour of the form factors 
taken between the ground state and a class of excited states of particle-hole type.  This explicit control on the large-volume behaviour of the particle-hole form factors allowed to derive, 
on the level of heuristic but physically quite plausible arguments, 
the large-distance asymptotic behaviour of two \cite{KozKitMailSlaTerRestrictedSums} and multi-point \cite{KozKitMailTerMultiRestrictedSums} correlation functions in the XXZ chain. 
Furthermore, such a large-volume asymptotic behaviour permitted to identify the amplitudes arising in the long-distance asymptotic expansion as the thermodynamic limit of properly normalised in the volume
form factors of local operators \cite{KozKitMailSlaTerXXZsgZsgZAsymptotics,KozKitMailSlaTerEffectiveFormFactorsForXXZ}. 
It also led to establishing \cite{KozMailletMicroOriginOfc=1CFTUniversality} a correspondence between a weak limit of operators in the lattice XXZ chain and certain vertex operators arising in the free boson model.

In fact, for massless models having a pure particle-hole spectrum such as the non-linear Schrödinger model, the control on the large-volume asymptotic behaviour of form factors
was enough so as to derive \cite{KozKitMailSlaTerRestrictedSumsEdgeAndLongTime} and even prove \cite{KozReducedDensityMatrixAsymptNLSE},  
under technical hypotheses of convergence of auxiliary series, the asymptotic behaviour of dynamical correlation functions in this model. The work \cite{KozKitMailSlaTerRestrictedSumsEdgeAndLongTime} also confirmed the predictions for the 
value of the so-called edge exponents that were predicted earlier on by means of the non-linear Luttinger liquid approach
\cite{GlazmanImambekovComputationEdgeExpExact1DBose,GlazmanImambekovDvPMTCompletTheoryNNLL,GlazmanKamenevKhodasPustilnikNLLLTheoryAndSpectralFunctionsFremionsFirstAnalysis}. 

The above stresses the prominent role that the large-volume behaviour of form factors in massless models plays in the analysis of various properties of a model's correlation functions. 
However,  the study of dynamical properties of more complex models such as the spin-$\tf{1}{2}$
XXZ chain demands more work. Indeed, on top of particle-hole excitations, this model also exhibits bound states. These were argued, on many instances, see   \textit{e.g} 
\cite{AffleckPereiraWhiteEdgeSingInSpin1-2, AffleckPereiraWhiteSpectralFunctionsfor1DLatticeFermionsBoundStatesContributions,KohnoDynamDominantStringExcitationXXZChain}, to play a role in the 
dynamical properties of the chain. As a consequence, in order to deal appropriately with the time and space dependent properties of the chain one first needs to access to 
the large-volume asymptotic behaviour of the form factors of local operators involving, on top of particle-hole excitations, the bound states as well. 

One should stress that more effort is needed to extend the techniques of the large-volume analysis of form factors to the case of form factors involving bound states. 
This stems, in particular, from the way the bound states are characterised within the Bethe Ansatz; these are described in terms of certain complex valued solutions
to the Bethe equations which agglomerate, in the thermodynamic limit, into complexes called strings. A given string can be characterised in terms of its length and
its central root, the so-called string centre. The first investigation of the large-volume $L$ behaviour of quantities building up the
XXZ chain's form factors between Eigenstates containing bound states concerned the norm of a Bethe Eigenvector.
The authors of  \cite{KirillovKorepinNormsStateswithStrings} took a formal large-$L$ limit of the determinant representing the norm of a Bethe Eigenvector containing
bound states and managed to reduce the dependence on the parameters of the bound state to solely one on the string centres. More recently, the authors of \cite{CauxHagemansMailletDynamicalCorrFunctXXZinFieldPlots}
generalised the former result to the \textit{per se} case of form factors involving bound states. 
The rewriting of the form factor determinants they obtained was enough so as to allow them a numerical calculation of the form factors. However, the mentioned handlings do not appear to allow for a 
rigorous analysis, mainly due to the difficulties associated with controlling the remainders within such an approach. 
In fact, owing to the exponential in the volume divergencies that appear in the entries of the form factor determinants due to the presence of bound states,
the rigorous analysis of the large-volume behaviour of such form factors demands
to have a very precise characterisation of the large-$L$ behaviour of string solutions. 
The first full analysis of the large-volume behaviour of form factors involving bound states has been carried out in \cite{KozDugaveGohmannSuzukiLargeVolumeBehaviourFFMassiveXXZ} for the massive regime of the XXZ chain. 
The analysis relied on certain technical hypotheses and on the results of \cite{BabelondeVegaVialletStringHypothesisWrongXXZ,DestriLowensteinFirstIntroHKBAEAndArgumentForStringIsWrong,VirosztekWoynarovichStudyofExcitedStatesinXXZHigherLevelBAECalculations}
where higher level Bethe equations characterising the complex solutions to the Bethe equations above the ground state in this sector have been obtained. 
In fact, due to the importance played by the bound states in the XXZ chain, starting from he pioneering work of Bethe \cite{BetheSolutionToXXX},
a rather extensive literature has been devoted to the analysis of complex solutions to the Bethe equations. 
Until recently, a rigorous description of bound states could only be achieved for the ferromagnetic regime of the XXZ chain 
\cite{BabbittThomasPlancherelFormulaInfiniteXXX,ThomasGSRepForFerromagneticXXX}. 
In \cite{KozProofOfStringSolutionsBetheeqnsXXZ}, the author managed to characterise, on rigorous grounds, the structure of a wide class of complex solutions in the massless regime of the antiferromagnetic chain
in the presence of a finite magnetic field. I refer to that paper for a review of the history of complex  solutions to the XXZ chain Bethe Ansatz equations. 
The  results of \cite{KozProofOfStringSolutionsBetheeqnsXXZ} opened the way to a rigorous analysis of the large-$L$ behaviour of bound state form factors in the XXZ chain that is developed in this work. 

\vspace{2mm} 
The goal of the present paper is to determine, on rigorous grounds, the large-volume asymptotic behaviour of the form factors of local operators in the XXZ spin-$1/2$ chain, this 
while keeping a  uniform in respect to the excited states, control on the remainder.  These large-volume asymptotics will then be used, in a subsequent publication, 
so as to demonstrate, under mild assumptions, the non-linear Luttinger liquid model-based predictions for the edge exponents characterising the singular behaviour of Fourier transforms of two-point
functions in the chain and also extract the long-time and large-distance asymptotic behaviour
of two-point functions in the model.

 The paper is organised as follows. Section \ref{Section modele et resultats} introduces the model and states, without giving too much details on the building blocks, 
 the overall form taken by the large-$L$ asymptotic behaviour of form factors of local operators that is obtained in the present work. This section also contains a list of various notations that will be employed in the core of the paper. 
 Section \ref{Section fonctions speciales} is of technical nature and discusses all the solutions to the linear integral equations that drive the thermodynamics of the chain. 
Section \ref{Section Counting functions asymptotics} gathers various auxiliary results that are necessary for the analysis of the large-volume behaviour of form factors. 
Section \ref{Section FF local ops introduction} presents the starting, determinant-based, expressions for the form factors of local operators in the model. 
It also contains the precise statement, with all building blocks given explicitly, of the main theorem proven in this work. 
Sections \ref{Section large-L analysis of Dbk et Dex}, \ref{Section Analyse de Areg} and \ref{Section analyse de Asing} focus on the extraction of the large-volume behaviour of the various sub-constituents of the form factors. 
The paper contains various appendices where several technical results of use to the analysis are established. 
Appendix \ref{Appendix Asymptotics auxiliary integrals} is devoted to the asymptotic analysis of the integral transforms that arise in the course of the analysis. 
Appendix \ref{Appendix auxiliary bounds} establishes a certain amount of bounds that appear helpful for the analysis developed in the core of the paper. 
Finally, Appendix \ref{Appendix integrales auxiliaires} list certain identities involving the special functions that are used in the paper as well as an evaluation of 
certain auxiliary integrals that appear in the course of the analysis.

\section{The model and main results}
\label{Section modele et resultats}

\subsection{The model}

The XXZ spin-$\tf{1}{2}$ chain refers to a system of interacting spins in one dimension described by the Hamiltonian  
\beq
\op{H}_{\De} \, = \, J \sum_{a=1}^{L} \Big\{ \sigma^x_a \,\sigma^x_{a+1} +
  \sigma^y_a\,\sigma^y_{a+1} + \De  \,\sigma^z_a\,\sigma^z_{a+1}\Big\} \; . 
\label{ecriture hamiltonien XXZ}
\enq
$\op{H}_{\De}$ is an operator on the Hilbert space $\mf{h}_{XXZ}=\otimes_{a=1}^{L}\mf{h}_a$ with $\mf{h}_a \simeq \Cx^2$. 
The matrices $\sg^{w}$, $w=x,y,z$ are the Pauli matrices and $\sg_a^{w}$ stands for the operator on $\mf{h}_{XXZ}$ which acts as the Pauli matrix $\sg^{w}$
on $\mf{h}_a$ and as the identity on all the other spaces appearing in the tensor product defining $\mf{h}_{XXZ}$. 
The Hamiltonian depends on two coupling constants: $J>0$ which represents the so-called exchange interaction and $\De$
which parametrises the anisotropy in the coupling between the spins in the longitudinal and transverse directions. 
In this paper I shall focus on the range of anisotropy $-1 < \De < 1$ which corresponds to the so-called massless anti-ferromagnetic regime.
In the following, I shall adopt the parametrisation 
\beq
\De= \cos(\zeta) \qquad  \e{with}  \qquad \zeta \in \intoo{0}{\pi}  \; .
\label{ecriture parametrisation anisotropie}
\enq

The XXZ Hamiltonian commutes with the total spin operator $\op{S}^z=\sum_{a=1}^{L} \sg_a^z$. 
It can thus be diagonalised in each sub-space $\mf{h}^{(N)}_{XXZ}$ corresponding to a fixed Eigenvalue of $\op{S}^z$:
\beq
\mf{h}_{XXZ}^{(N)}\; = \; \Big\{ \ket{v} \in \mf{h}_{XXZ}  \; :  \; \op{S}^z \ket{v} \, = \, (L-2N) \ket{v} \Big\}\qquad \e{so}\; \e{that} \qquad 
\mf{h}_{XXZ} \; = \; \bigoplus\limits_{N=0}^{L}\mf{h}_{XXZ}^{(N)} \;. 
\enq
As a consequence, one can embed $\op{H}_{\De}$ in an external longitudinal magnetic field $h$ and rather focus on the Hamiltonian  $\op{H}_{\De,h}=\op{H}_{\De} \, - \,  \tf{h  \op{S}^{z} }{2}$, this without altering
the diagonalisation problem. The effect of the magnetic field will be to change the value of the integer $N$ labelling the subspace $\mf{h}_{XXZ}^{(N)}$ which contains the model's ground state.

When $-1< \De < 1$ and for magnetic fields  $h\geq h_{\e{c}}=8J \cos^{2}\big( \tf{\zeta}{2} \big)$, the Hamiltonian $\op{H}_{\De,h}$ is in its ferromagnetic regime and the ground state belongs to the 
$\mf{h}^{(0)}_{XXZ}$ subspace. When the magnetic field is below the critical value $h_{\e{c}}$, $0\leq h < h_{\e{c}}$, the model is an antiferromagnet. 
Then, the ground state belongs to the subspace $\mf{h}_{XXZ}^{(N)}$ with $N$ such that $\tf{N}{L}\tend D \in \intoo{0}{\tf{1}{2}}$. As will be discussed later on, the value of $D$
is fixed by $h$. In the following, I will focus on the regime $0< h < h_{\e{c}}$. The reason is that, on the one hand, when $h=0$ and $L\tend +\infty$, the structure of the complex solutions to the Bethe equations has been advocated 
\cite{BabelondeVegaVialletStringHypothesisWrongXXZ,DestriLowensteinFirstIntroHKBAEAndArgumentForStringIsWrong,WoynaorwiczHLBAEMAsslessXXZ0Delta1} to change drastically 
in respect to the string picture argued by Bethe. A rigorous description of the $L\tend + \infty$ behaviour of the complex solutions when $h=0$ is still an open problem. Furthermore, owing to the 
unboundedness in $L$ of the solutions to the ground state Bethe equations, \textit{c.f.} \cite{KozProofOfDensityOfBetheRoots}, various brand new features will appear in the $L\tend+\infty$
analysis of the form factors. On the other hand, for $h\geq h_{\e{c}}$, the problem is trivial. 

\subsection{The main result}

As observed by Bethe \cite{BetheSolutionToXXX} for $\De=1$ and generalised to any $\De$ by Orbach \cite{OrbachXXZCBASolution}, Eigenstates $\ket{\Ups}$ of $\op{H}_{\De,h}$ 
can be constructed by the so-called Bethe Ansatz. Within this approach, the Eigenstates $\ket{\Ups}$ are constructed as certain combinatorial sums  depending on a set of auxiliary parameters $\Ups=\{\mu_a\}_1^{|\Ups|}$
which solve a system of transcendental equations, the so-called Bethe equations. The completeness of the Bethe Eigenstates for the XXZ chain, as built from solutions to the Bethe equations, is a complicated issue
which can however be settled for a certain inhomogeneous deformation of the model \cite{TarasovVarchenkoCompletenessBAXXZ} or by adopting a slightly more general point of view, as it has been done
for the XXX Heiseiberg chain \cite{MukhinTarasovVarchenkoCompletenessandSimplicitySpectBA}. I will not dwell on such issues in the present work, and simply consider Eigenstates $\ket{\Ups}$ which can
be build from the Bethe Ansatz. 

In the following, I will consider Eigenstates $\ket{\Ups}$ which have, in the thermodynamic limit,  a finite excitation energy above the ground state. In this limit, the Eigenstate can be described in terms of

\begin{itemize}

\item massless excitations characterised by the integers $\{p_a^{\ups} \}_{a=1}^{ n^{(p)}_{\ups} } \cup \{h_a^{\ups} \}_{a=1}^{ n^{(h)}_{\ups} }  $ with $\ups\in \{L,R\}$;

\item massive excitations characterised by the parameters $\mf{C} =  \big\{ \{ c_{p}^{(a)} \}_{a=1}^{n_p^{(z)}} \big\}_{p=1}^{p_{\e{max}}} $ and $\Ups^{(h)}_{\e{off}}=\{ \mu^{(h)}_{a} \}_{a=1}^{ n^{(h)}_{\e{off}} } $

\item Umklapp integers $\ell_{\ups}$ and centred Umklapp integers $\ell^{\varkappa}_{\ups}$ \;. 

\end{itemize}

\noindent All these quantities are defined precisely in the core of the paper, \textit{c.f.} Sections \ref{Sous Section definition des racines etat excite}-\ref{Sous Section excited state Ctg Fct} and 
Section \ref{Sous Section parametrisation limite thermo}. 

\vspace{2mm}

The form factors of local operators correspond to the expectation values of the local operators $\sg^{\ga}_1$ taken between two Eigenstates of $\op{H}_{\De,h}$
\beq
\msc{F}^{(\ga)}_{\Ups;\Om} \; = \; \f{ \bra{\Ups} \sg_1^{\ga} \ket{\Om} }{ \norm{\Ups} \cdot \norm{\Om} } \;. 
\enq

The main result of this paper is summarised in the theorem below.

\begin{theorem}
 
 Let $\La$ be the set of Bethe roots describing the ground state at magnetic field $0<h< h_{\e{c}}$ and $\Ups$ be the set of Bethe roots  described above and such that the roots of $\Ups$
 are spaced at least as described in Hypothesis  \ref{Hypothesis espacement des cordes}.

\noindent Assume that the integers  $p_a^{\ups}$ and $h_a^{\ups}$ and the parameters forming $\mf{C}\cup\Ups^{(h)}_{\e{off}}$ are finite in number. Assume that the integers  
$p_a^{\ups}$ and $h_a^{\ups}$ and the parameters forming $\mf{C}\cup\Ups^{(h)}_{\e{off}}$ are bounded in $L$. 
Given $\ga=z$ or $\ga=+$, then there exists $0<r< \tf{1}{4}$ such that  the below asymptotic expansion holds 
\bem
 \Big|  \msc{F}^{(\ga)}_{\Ups;\La}\Big|^2  \; = \; 
\Oint{ \Dp{}\mc{D}_{0,r} }{} \pl{ \ups \in \{L,R\} }{}\Bigg\{ \f{ G^2\big(1-\sg_{\ups} \op{f}_{\ups}^{\,(\varrho)}   \big) }{ G^2\big(1-\sg_{\ups}[\op{f}_{\ups}^{\,(\varrho)}  -\sg_{\ups}\ell_{\ups}] \big) }
\f{ \mc{R}_{n^{\ups}_{p}, n^{\ups}_{h} }\big( \{p_a^{\ups}\};\{h_a^{\ups} \} \mid-\sg_{\ups} \op{f}_{\ups}^{\,(\varrho)}  \big) } { \big( \tf{ L }{ 2 \pi }\big)^{ ( \op{f}_{\ups}^{\,(\varrho)} - \sg_{\ups} \ell_{\ups})^2 } } \Bigg\} \\
\times \f{ \mc{F}^{(\ga)}\Big( \Ups^{(h)}_{\e{off}}; \mf{C}  ; \{ \ell_{\ups}^{\varkappa} \}  \mid \varrho \Big)   \cdot  \Big( 1\, + \, \e{O}\Big( \tf{\ln L}{L} \Big) \Big) }
{ \prod_{ \mu \in \Ups^{(h)}_{\e{off}} }^{} \big\{  L p^{\prime}_1(\mu)   \big\} \cdot \prod_{ r=1  }^{ p_{\e{max}} } \prod_{a=1}^{n_r^{(z)}}  \big\{  L p^{\prime}_r\big( c_a^{(r)} \big)   \big\}   }
 \cdot \f{\dd \varrho}{2\i\pi \varrho }  \;. 
\label{ecriture DA FF theorem ppl expliquant forme resultat}
\end{multline}

\end{theorem}

The structure of the answer is the following. The contour integral over the auxiliary variable $\varrho$ plays the role of a regularisation.  

The leading asymptotics can be split in two contributions. 
The one appearing on the first line of \eqref{ecriture DA FF theorem ppl expliquant forme resultat} corresponds to the massless modes. 
The pre-factor containing the Barnes-$G$ functions is a normalisation constant. The function $\mc{R}_{n^{\ups}_{p}, n^{\ups}_{h} }$ should be though of as 
the form factor density squared associated with the massless modes of the model. The contribution is weighted by the non-integer power 
of the volume $ L  ^{ -( \op{f}_{\ups}^{\,(\varrho)} - \sg_{\ups} \ell_{\ups})^2 } $ which should be thought of as the fundamental spectral "volume"
carried by the massless excitation. Here $\sg_{L}=-1$ and $\sg_{R}=1$, $\ell^{\varkappa}_{\ups} \in \mathbb{Z}$ are the Umklapp integers associated with the excited state $\Ups$ and 
$\op{f}_{L/R}^{\,(\varrho)}$ is related to the value taken by the shift function associated with the  excited state $\Ups$ on the left/right end of the Fermi zone. 

The contribution appearing in the second line of \eqref{ecriture DA FF theorem ppl expliquant forme resultat} corresponds to the massive modes. 
 The function $\mc{F}^{(\ga)}$ should be though of as 
the form factor density squared associated with the massive modes of the model. $\mc{F}^{(\ga)}$ is a smooth function of the variables $ \Ups^{(h)}_{\e{off}}$ and  $\mf{C}$.  The contribution is weighted by an integer power
of the volume: exactly one power for each massive parameter present in $\mf{C}\cup \Ups^{(h)}_{\e{off}}$. The functions $\tf{p^{\prime}_r}{2\pi}$,
represent the density at which each of the variables condense in the thermodynamic limit.

The expressions for the various constituents of the leading asymptotics are a bit bulky and their definition demands a certain amount of auxiliary objects.
Such details are thus postponed to Section \ref{Sous Section principal theorem} where, also, a slightly more general version of the result is presented in Theorem \ref{Theorem principal du papier}.

\subsection{Main notations}

\begin{itemize}

 \item Given a set $S$, $|S|$ stands for its cardinal, $\bs{1}_S$ for its indicator function. 
 
 \item $I_{\a}$ stands for the interval $I_{\a}=\intff{-\a}{\a}$.
 
 \item Given a set $S$, $\mc{S}_{\de}(S)$ stands for a $\de$-neighbourhood of $S$, namely $\mc{S}_{\de}(S)= \big\{ z \in \Cx \; : \;  |d(z,S) |< \de  \big\}$ where $d(z,S)$ is the distance
of the point $z$ to the set $S$ that is  induced by the canonic distance on $\Cx$. For instance, $\mc{S}_{\de}(\R)$ is the strip of width $2\de$ centred on $\R$. 
\item $\mc{D}_{z,\de}$ stands for the open disk of radius $\de$ centred at $z$.

 \item Given two sets $A,B$, I adopt the shorthand notation for products and sums involving
\beq
\pl{ \substack{\la,\mu \in \\ A\setminus B } }{} f(\la) \; \equiv \; \f{ \pl{\la\in A }{} f(\la)  }{   \pl{\la\in B }{} f(\la)   }
\qquad \e{and} \qquad 
\sul{ \substack{\la,\mu \in \\ A\setminus B } }{} f(\la) \; \equiv \; \sul{\la\in A }{} f(\la)  \, - \,  \sul{\la\in B }{} f(\la)   \;. 
\label{ecriture convention produit et somme difference ensemble}
\enq
In the case when $B \subset A$, this convention reproduces the value of a product or sum over elements of the set $A\setminus B$.

\item The function $\ln$ refers to the principal branch of the logarithm. Unless stated otherwise, it is this branch that will be used in the formulae. 

\item Given $x\in \R$, $\lfloor x \rfloor$ is the integer part of $x$, namely the closest integer lower or equal to $x$.  

\item Given a closed non-intersecting curve $\msc{C}$ in $\Cx\setminus \{\i\pi \mathbb{Z} \}$, $\e{Ext}(\msc{C})$ and $\e{Int}(\msc{C})$ denote, respectively, the interior and exterior 
of $\msc{C}$ in $\Cx\setminus \{\i\pi \mathbb{Z}  \}$. 

\item Given an open subset $O\subset \Cx$, $\Dp{}O$ denotes its canonically oriented boundary. 

\item In general $C,c, \wt{C}, C^{\prime},...$ denote positive constants appearing in the various  bounds. The value of these constants may change from one line of an equation to another
without further specifications.

 \item $W_{k}^{\infty}(\R)$ stands for the $L^{\infty}$-based Sobolev space of order $k \in \mathbb{N}$, namely 
 \beq
W_{k}^{\infty}(\R) \; = \; \Big\{ f   \, : \, f^{(p)}\in L^{\infty}(\R) \quad 0 \leq p \leq k \Big\} \quad \e{with}\, \e{norm} \qquad 
\norm{ f }_{ W_{k}^{\infty}(\R) } \, = \, \max_{0\leq p \leq k} \norm{ f^{(k)} }_{L^{\infty}(\R)} \;. 
\enq

\item Let $\Ga$ and $G$ be the Euler Gamma and the Barnes-$G$ functions. I adopt the hypergeometric-like notation for products and ratios of such functions
\beq
\Ga\left( \ba{c} a_1,\dots, a_k \\ b_1,\dots, b_{\ell} \ea \right) \; = \; \f{ \pl{s=1}{k} \Ga(a_s) }{  \pl{s=1}{s} \Ga(b_s)   } \qquad \e{and} \qquad
G\left( \ba{c} a_1,\dots, a_k \\ b_1,\dots, b_{\ell} \ea \right) \; = \; \f{ \pl{s=1}{k} G(a_s) }{  \pl{s=1}{s} G(b_s)   } \;. 
\enq

\end{itemize}

\section{The special functions at play}
\label{Section fonctions speciales}

The observables associated with the thermodynamic limit of integrable models are described through a collection of solutions to linear 
integral equations. This section provides the description of all the functions of this type that arise in the context of the present analysis. I first discuss the bare quantities which appear as driving terms in the linear
integral equations and then introduce the solutions to these equation, the so-called dressed quantities. 

\subsection{The bare quantities}
\subsubsection{Integral kernels}

The function
\beq
K(\la\mid \eta ) \; =  \;      \f{1}{2\i\pi} \bigg\{ \coth(\la-\i\eta) - \coth(\la+\i\eta) \bigg\} \, = \,   \f{ \sin(2\eta)   }{ 2\pi \sinh(\la + \i \eta)  \sinh(\la  - \i \eta)  }  
\label{ecriture fonction K de lambda et eta}
\enq
plays an important role in the analysis. When $\eta$ is specialised to $\zeta$ introduced in \eqref{ecriture parametrisation anisotropie}, $K(\la \! \mid \! \zeta)$ corresponds to the integral kernel of the operator that drives the linear integral
equations describing the thermodynamic observables in the model. In fact, below, whenever the auxiliary argument will dropped, 
$K(\la)$   will stand precisely for this function:
\beq
K(\la) \; \equiv \; K(\la\mid \zeta) \; . 
\enq
Moreover, sums of $\i \zeta$-shifted kernels $K$ arise in the description of the quantities associated with the bound states: 
\beq
K_{r,s}(\om) \,  \equiv  \, \sul{\ell=1}{r} \sul{k=1}{s} K\Big( \om + \i\f{\zeta}{2}\big(r-s-2(\ell-k)  \big) \Big)  \;.
\label{definition noyau K r s}
\enq
These sums are symmetric in $r,s$ and can be recast in terms of a reduced number of functions $K(\la\mid \eta)$ evaluated at different values of $\eta$ as 
\bem
K_{r,s}(\om) \, = \, 
K\Big(\om \mid \tfrac{1}{2} \zeta(r+s) \Big) \, + \, \big(2-\de_{1,s} -\de_{1,r} \big) \cdot K\Big(\om \mid \tfrac{1}{2} \zeta(r+s-2) \Big)  \\
\, + \hspace{-2mm} \sul{ p = \big[ \f{r-s+2}{2} \big] }{r-2} \hspace{-4mm} \big(w_{p-1}^{(r,s)}-w_{p+1}^{(r,s)} \big) \cdot  K\Big(\om \mid \tfrac{1}{2} \zeta(r-s-2p) \Big) 
\label{expression explicite noyau K r s}
\end{multline}
where $\de_{a,b}$ is the Kronecker symbol and 
\beq
w_p^{(r,s)}\; \equiv \; \min(r,s+p)-\max(0,p) \, = \, w_{r-s-p}^{(r,s)} \;. 
\enq
%
%

Note that, as a particular case, one has
\beq
K_{1,r}(\om) \, = \, 
K\Big(\om \mid \tfrac{1}{2} \zeta(r+1) \Big) \, + \, \big(1-\de_{1,r} \big) \cdot K\Big(\om \mid \tfrac{1}{2} \zeta(r-1) \Big) \;. 
\enq
One has analogous identities to \eqref{expression explicite noyau K r s} relatively to the double product
\beq
\Phi_{r,p}\big( \la \big) \, = \, \Phi_{p,r}\big( \la \big) \, = \, 
\pl{k=1}{p} \pl{s=1}{r} \Bigg\{ \f{  \sinh\Big(\la +\tfrac{\i}{2}\zeta\big[ p-r-2(k-s) \big] \Big)    }{ \sinh\Big(  \la + \tfrac{\i}{2}\zeta\big[ p-r-2(k-s+1) \big] \Big)   }  \Bigg\} \;. 
\label{definition produite double sur racine corde}
\enq
For general $r$ and $p$, it can be recast as  
\beq
\Phi_{r,p}\big( \la \big) \, = \, 
\pl{\ell = \big[ \tfrac{r-p+1}{2} \big] }{r-1} \Bigg\{ \f{  \sinh\Big(\la +\i\zeta\big[ \tfrac{p-r}{2}+\ell \big] \Big)    }{ \sinh\Big(  \la - \i\zeta \big[\tfrac{p-r}{2} +\ell+1\big] \Big)   }     \Bigg\}^{ w_{\ell}^{(r,p)}- w_{\ell+1}^{(r,p)}} \;. 
\label{formule explicite pour produit double Phi r p sur racines corde}
\enq
This expression slightly simplifies when one of the integers is set to one:
\beq
\Phi_{1,p}\big( \la \big) \, = \, \Phi_{p,1}\big( \la \big) \, = \, \f{  \sinh\Big(\la +\i\zeta \tfrac{p-1}{2}  \Big)    }{ \sinh\Big(  \la - \i\zeta \tfrac{p+1}{2} \Big)   }    \; . 
\enq
The representation \eqref{formule explicite pour produit double Phi r p sur racines corde} allows one to deduce the structure of the zeroes and poles of $\Phi_{r,p}$ close to $\R$. 

\noindent Let $0 \leq r,\ell \leq p_{\e{max}}$. Then there exists $\de>0$ such that 
\begin{itemize}
 \item when $r\not=\ell$ $\Phi_{r,\ell}$ has no poles or zeroes in $\mc{S}_{\de}(\R)$;
 
 \item  $\Phi_{r,r}$ has no poles  in $\mc{S}_{\de}(\R)$ and a unique zero at $0$ which has multiplicity one. 

\end{itemize}

\subsubsection{The bare phases}

\begin{figure}[ht]
\begin{center}

\begin{pspicture}(7,3)

\psline[linestyle=dashed, dash=3pt 2pt]{->}(0,0.5)(3,0.5)

\rput(3,0.2){$\R$}
\rput(0.5,0.2){$0$}
\psdots(0.5,2.5)
\rput(0,2.2){$\i\eta$}
\psline[linestyle=dashed, dash=3pt 2pt](0,2.5)(3,2.5)

\rput(3,2.2){$\R+\i\eta$}

\psline{-}(0.5,0.5)(0.5,2)
\psline{-}(0.5,2)(2,2)
\psline{->}(0.5,1.5)(0.5,1.6)
\psline{->}(1.5,2)(1.6,2)
\psdots(2,2)
\rput(2.1,1.7){$\la$}

\psline[linestyle=dashed, dash=3pt 2pt]{->}(4,0.5)(7,0.5)

\rput(7,0.2){$\R$}
\rput(4.5,0.2){$0$}
\psdots(4.5,2.5)
\rput(4.8,2.2){$\i\eta$}
\psline[linestyle=dashed, dash=3pt 2pt](4,2.5)(7,2.5)

\rput(7,2.2){$\R+\i\eta$}

\psline{-}(4.5,0.5)(4.5,2.3)
\pscurve{-}(4.5,2.3)(4.3,2.5)(4.5,2.7)
\psline{-}(4.5,2.7)(4.5,3)
\psline{-}(4.5,3)(6.5,3)
\psline{->}(4.5,1.5)(4.5,1.6)
\psline{->}(5.5,3)(5.6,3)
\psdots(6.5,3)
\rput(6.6,2.7){$\la$}

\end{pspicture}
\caption{Prescription for the contour $\Ga_{\la}$ plotted for two values of $\la$ having imaginary part below and above $\eta$ in the case where $0<\eta<\tf{\pi}{2}$ and $0<\Im(\la)<\tf{\pi}{2}$. }
\end{center}
 \label{Figure systeme de poles a eviter}
\end{figure}
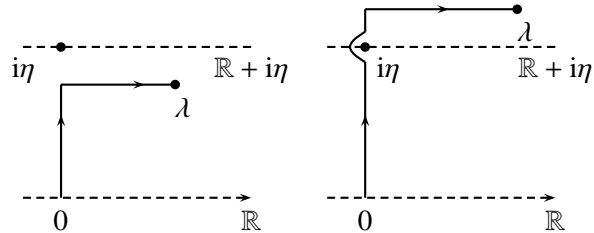

The bare phase $\vth(\la\mid \eta)$ is defined as the below ante-derivative of $2\pi K(\la\mid \eta)$:
\beq
\vth(\la\mid \eta) \, = \, 2\pi \Int{ \Ga_{\la}  }{}  K(\mu-0^+\mid \eta ) \cdot \dd \mu  \qquad \e{with} \qquad \Ga_{\la} \; = \; \intff{ 0 }{ \i \Im(\la) }\cup \intff{\i \Im(\la) }{ \la } \;. 
\label{ecriture rep int bare phase}
\enq
The $-0^+$ prescription indicates that the poles of the integrand at $\pm \i \eta +\i \pi \mathbb{Z}$ should be avoided from the left, \textit{c.f.} Fig.~\ref{Figure systeme de poles a eviter}.   
Throughout the paper, I agree upon  $\th(\la)=\vth(\la\mid \zeta)$. 

It will be convenient to introduce $r,p$ analogues of the bare phase built up from the kernel $K_{r,p}$:
\beq
\th_{r,p} \big( \la  \big) \, = \, 2 \pi \Int{ \Ga_{\la} }{} K_{r,p}(\mu-0^+) \cdot \dd\mu \;. 
\label{definition phase nue typre r p}
\enq
Just as in \eqref{ecriture rep int bare phase}, the contour $\Ga_{\la}$ avoids from the left the poles of the integrand lying on $\i\R$. Also, later on, I shall use the notation $\th(\la)=\th_{11}(\la)$.

The $r,p$ bare phases arise as a result of summation of bare phases evaluated at so-called string configurations. One has, for any $\om$ where the sum makes sense, 
\beq
\sul{k=1}{p} \th\Big(\om + \i \tfrac{\zeta}{2}\big(p+1-2k\big) \Big) \; = \; \th_{p,1}(\om)+ \pi m_p(\zeta) 
\quad \e{with} \quad 
m_p(\zeta)  \; = \;   \Big(2-p-\de_{p,1} + 2 \lfloor \zeta \tfrac{p-1}{2 \pi} \rfloor + 2 \lfloor \zeta \tfrac{p+1}{2 \pi} \rfloor \Big) \;. 
\label{ecriture somme partielle sur corde pĥase nue et definition entier mr de zeta}
\enq
This identity can be established by decomposing the sum as
\beq
\sul{k=1}{p} \th\Big(\om + \i \tfrac{\zeta}{2}\big(p+1-2k\big) \Big) \; = \; \lim_{\eps \tend 0^+} \Big\{ \mf{s}_1(\eps)+\mf{s}_2(\eps) \Big\}
\enq
with 
\beq
 \mf{s}_1(\eps)\, = \, \sul{k=1}{p} \Big\{ \th\Big(\om +\i\eps + \i \tfrac{\zeta}{2}\big(p+1-2k\big) \Big) \, - \,\th\Big( \i\eps + \i \tfrac{\zeta}{2}\big(p+1-2k\big) \Big) \Big\}
\, = \, 2\pi \Int{\Ga_{\om} }{} K_{p,1}\big(\mu-0^++\i\eps \big) 
\enq
and  
\beq
 \mf{s}_2(\eps)\, = \, \sul{k=1}{p}   \,\th\Big( \i\eps  + \i \tfrac{\zeta}{2}\big(p+1-2k\big) \Big)  \;. 
\enq
The second term can be evaluated as a telescopic sum and by using the definition of the principal value integral. One eventually gets $\mf{s}_2(\eps)_{\mid \eps=0^+} \; = \; \pi m_p(\zeta)$
while the $\eps \tend 0^+$ limit of the first sum is easily taken.

\subsection{The solutions to the Lieb equation}

It is well known since the work of H\'{u}lthen \cite{HultenGSandEnergyForXXX} that  solutions to linear integral equation of the type $\big(\e{id}+\op{K}_{I_Q}\big)\big[f]=g$
describe the thermodynamic properties of the XXZ chain. Here $\op{K}_{I_Q}$ is the integral operator on 
$L^2\big( I_Q\big)$ acting as
\beq
\op{K}_{I_Q}\big[f](\la) \; = \; \Int{-Q}{Q} K(\la-\mu) f(\mu) \dd \mu \;. 
\enq
It is a classical fact, see \textit{e.g.}  \cite{KozDugaveGohmannThermoFunctionsZeroTXXZMassless,Yang-YangXXZStructureofGS},  
that the operator $\e{id}+\op{K}_{I_Q}$ is invertible for any $Q\in \R^+$. Its inverse takes the form $\e{id}-\op{R}_{I_Q}$ where the resolvent operator $\op{R}_{I_Q}$
is characterised by its integral kernel $R_{I_Q}(\la,\mu)$.

\vspace{2mm}

Given some parameter $Q>0$, the dressed momentum is defined as the solution to the linear integral equation
\beqa
p(\la\mid Q) & = &    \vth\big( \la \mid \tf{\zeta}{2} \big)   \, - \, \Int{-Q}{Q} \th(\la-\mu) p^{\prime}(\mu\mid Q) \cdot \f{\dd \mu}{2\pi} \label{ecriture equation moment habille avec derivee}  \\
& = & 
\vth\big( \la \mid \tf{\zeta}{2} \big)  \, -\, \f{ p(Q\mid Q) }{2\pi} \Big( \th(\la-Q)+\th(\la-Q) \Big)   \, - \, \Int{-Q}{Q} K(\la-\mu) p (\mu\mid Q) \cdot  \dd \mu \;. 
\label{ecriture equation moment habille}
\eeqa
Given $D\in \intff{0}{\tf{1}{2}}$, there exists a unique  \cite{KozDugaveGohmannThermoFunctionsZeroTXXZMassless,Yang-YangXXZStructureofGS} 
$Q_D \in \intff{0}{+\infty}$ such that $p\big(Q_D \mid Q_D \big)=\pi D$. The dressed momentum in strictly increasing on $\R$ and strictly decreasing on $\R+\i\tf{\pi}{2}$:
\beq
p^{\prime}(\la\mid Q) >0 \qquad \e{and} \qquad p^{\prime}(\la+\i\tf{\pi}{2}\mid Q) < 0 \qquad \e{for} \;\; \la \in \R\;.  
\enq
The dressed energy is defined as the solution to the linear integral equation
\beq
\Big( \e{id} + \op{K}_{I_Q} \Big)\big[ \veps(*\mid Q) \big](\la) \; = \; \mf{e}(\la) \qquad \e{with} \qquad  \mf{e}(\la)\, = \, h - 2 J \sin(\zeta) \vth^{\prime}\big( \la \mid \tfrac{1}{2}\zeta \big) \;. 
\label{definition energie habille et energie nue}
\enq
The $r$-string dressed momentum and energy are defined, respectively, by 
\beq
  p_{r}\big( \om \mid Q \big) \; = \; \sul{\ell=1}{r} p\Big( \om +\i\f{\zeta}{2}\big( r + 1 - 2\ell) \mid Q    \Big)  
\label{defintion moment habille r-corde}
\enq
and
\beq
\veps_{r}\big( \om \mid Q \big) \; = \; \sul{\ell=1}{r} \veps\Big( \om +\i\f{\zeta}{2}\big( r + 1 - 2\ell \big)  \mid Q \Big)  \;. 
\label{defintion energie habille r-corde}
\enq
I refer to Proposition \ref{Proposition DA energie et impulsion} for an explanation of the origin of such a denomination. 
It follows from \eqref{ecriture equation moment habille avec derivee} that $p^{\prime}_r$ admits the integral representation
\beq
p_r^{\prime}(\om \mid Q ) \, = \, 2\pi K\big(\om   \mid r \tfrac{\zeta}{2}\big)
\, - \, \Int{ - Q }{ Q }  K_{r,1}\big(  \om  - s \big) \cdot p^{\prime}\big( s  \mid Q \big) \cdot \dd s \;. 
\label{ecriture eqn lin pour densite r-corde}
\enq
One can show \cite{KozProofOfStringSolutionsBetheeqnsXXZ} that the dressed momenta of $r$-string excitations are non-vanishing 
\beq
 | p^{\prime}_{r}(\la) | > 0 \qquad \e{for} \qquad \la \in \R\cup \big\{ \mathbb{\R}+\i\tf{\pi}{2} \big\} 
\label{ecriture positivite densite r corde}
\enq
provided\symbolfootnote[2]{It is conjectured in \cite{KozProofOfStringSolutionsBetheeqnsXXZ} that, in fact, the property holds true throughout the regime $\zeta \in \intoo{ \tf{\pi}{2} }{ \pi }$} that
\begin{itemize}

 \item[$i)$] $\zeta \in \intoo{0}{\tf{\pi}{2}}$ ;
\item[$ii)$] $\zeta \in \intoo{\tf{\pi}{2}}{\pi}$ and the additional conditions hold  
$\left\{ \ba{cc } \sin\big( \tfrac{r-1}{2}\zeta \big)\sin\big( \tfrac{r+1}{2}\zeta \big)<0 & \e{ if } \;   \la\in \R \vspace{2mm} \\
\cos\big( \tfrac{r-1}{2}\zeta \big)\cos\big( \tfrac{r+1}{2}\zeta \big)<0  & \e{if} \;  \la\in \R+\i\tfrac{\pi}{2} \ea \right.$. 
\end{itemize}

\noindent The dressed phase is defined as the solution to the linear integral equation
\beq
\phi(\la,\mu\mid Q) \, = \, \f{ 1 }{ 2 \pi }  \th\big( \la -\mu  \big)  \, - \, \Int{-Q}{Q} K(\la-\nu) \phi(\nu, \mu \mid Q) \cdot \dd \nu \;. 
\enq
In their turn, its $r$-sum generalisations are defined as 
\beq
\phi_{r,1}(\la,\mu\mid Q) \, = \, \f{ 1  }{ 2 \pi }  \th_{r,1}\big( \la -\mu  \big) \, - \, \Int{-Q}{Q} K(\la-\nu) \phi_{r,1}(\nu, \mu \mid Q) \cdot \dd \nu \;. 
\enq
In fact, $\phi_{r,1}$ is related to $r$ sums of the dressed phase similarly to \eqref{ecriture somme partielle sur corde pĥase nue et definition entier mr de zeta}.

Finally, there is yet another solution of importance to the thermodynamics of the XXZ spin-$1/2$ chain the so-called dressed charge $Z(\la\mid Q)$ solving 
\beq
\Big( \e{id} + \op{K}_{I_Q} \Big)\big[ Z(*\mid Q)  \big](\la)  \; = \; 1 \;. 
\enq
The dressed charge is closely related to the dressed phase for the below identities hold \cite{KorepinSlavnovNonlinearIdentityScattPhase}:
\beq
\phi(\la,Q\mid Q) \, - \,   \phi(\la,-Q\mid Q) \, + \, 1 \; = \; Z(\la\mid Q) \quad \e{and} \quad 1+\phi(Q,Q\mid Q) - \phi(-Q,Q\mid Q) \, = \, \f{1}{ Z(Q\mid Q) } 
\label{ecriture identites entre phase et charge habilles} 
\enq

It is clear from the form taken by the linear integral equations that their solutions are holomorphic in some open strip $\mc{S}_{\de}(\R)$ centred around $\R$.

\section{The counting functions}
\label{Section Counting functions asymptotics}

\subsection{The ground state roots}

Throughout the paper,  $\La=\{ \la_a\}_1^{|\La|} $ will denote the set build up from the Bethe roots describing the model's ground state in the spin $L-2|\La|$ sector. 
It was shown in \cite{Yang-YangXXZproofofBetheHypothesis}  that the ground state Bethe roots 
are real valued and correspond to a solution  to the below logarithmic Bethe equation
\beq
 \f{ \vth\big( \la_a \mid \tf{\zeta}{2} \big) }{ 2\pi } \, -\, \f{1}{2 \pi L } \sul{ b=1 }{|\La|} \vth\big(\la_a-\la_b \mid \zeta \big) \, + \, \f{|\La|+1}{2L} \; = \; \f{a}{L} \qquad a=1,\dots,|\La|\,.  
\label{ecriture eqns de log Bethe GS roots} 
\enq
It was shown in \cite{KozProofOfDensityOfBetheRoots} that, for $L$-large enough and irrespectively of the value of $\zeta$, there exist a unique real valued solution $\La$
to \eqref{ecriture eqns de log Bethe GS roots}. This $\La$ forms a dense distribution, when $L\tend +\infty$ with $\tf{|\La|}{L}\tend D$,  on the interval $\intff{-Q_D}{Q_D}$, the Fermi zone
of the model.

The value of $q=Q_{D_{gs}}$ of the endpoint of the Fermi zone corresponding to the overall ground state, \textit{viz}. the ground state of $\op{H}_{\De,h}$ on $\mf{h}_{XXZ}$, 
is fixed uniquely in terms of the magnetic field $h$: $q$ corresponds to the unique positive solution, \textit{c.f.}  \cite{KozDugaveGohmannThermoFunctionsZeroTXXZMassless} for more details, to the equation
$\veps(q \!\mid\! q)=0$ where $\veps(\la \!\mid \!Q)$ has been defined in \eqref{definition energie habille et energie nue}.
In particular, when $h_{\e{c}}>h>0$, the parameter $q$ runs through $\intoo{0}{+\infty}$ and the ground state is anti-ferromagnetic. 
 The value of $q$ being fixed by the magnetic field, the thermodynamic limit of the \textit{per} site
magnetisation in the ground state is given by $1-2D_{gs}$ where $D_{gs}$ is expressed in terms of $q$ as $\pi D_{gs}=p(q\mid q)$. Note that the allowed range for $q$ implies, \textit{c.f.}  \cite{KozDugaveGohmannThermoFunctionsZeroTXXZMassless}, 
that $D_{gs} \in \intoo{0}{\tf{1}{2}}$. When specialising the endpoint $Q$ of integration to $q$ in \eqref{definition energie habille et energie nue}, the dressed energy is such that 
\cite{KozDugaveGohmannThermoFunctionsZeroTXXZMassless} 
\beq
\veps(\la\mid q) <0 \quad \e{on} \; \intoo{-q}{q} \quad \e{and}  \quad 
\veps(\la \mid q) > 0 \quad \e{on} \quad \Big\{ \ov{\R} \setminus \intoo{-q}{q} \Big\} \cup \Big\{ \ov{\mathbb{\R}}+\i\tf{\pi}{2} \Big\} \;. 
\label{ecriture ppte sgn energie habille}
\enq
The dressed energies $r$-string excitations with $r\geq 2$ are all strictly positive \cite{KozProofOfStringSolutionsBetheeqnsXXZ}, \textit{i.e.} for any $r\geq 2$, there exists $c_r>0$ such that 
\beq
\veps_{r}(\la \mid q) \geq c_r>0    \qquad \e{on} \qquad \R\cup \big\{ \mathbb{\R}+\i\tf{\pi}{2} \big\}\; 
\label{ecriture positivite energie r corde}
\enq
this provided\symbolfootnote[3]{Again, it is conjectured that, in fact, \eqref{ecriture positivite energie r corde} remains true for $\zeta \in \intoo{ \tf{\pi}{2} }{ \pi }$ irrespectively of
the auxiliary conditions.} that 
\begin{itemize}

 \item[$i)$] $\zeta \in \intoo{0}{\tf{\pi}{2}}$ ;
\item[$ii)$] $\zeta \in \intoo{\tf{\pi}{2}}{\pi}$ and  
$\left\{ \ba{ccc} \sin\big( \tfrac{r-1}{2}\zeta \big)\sin\big( \tfrac{r+1}{2}\zeta \big)<0 \;, \; \e{and}\;  \sin(r\zeta)<0  & \e{ if }  \;  \la\in \R \vspace{2mm} \\
\cos\big( \tfrac{r-1}{2}\zeta \big)\cos\big( \tfrac{r+1}{2}\zeta \big)<0 \;, \; \e{and}\;  \sin(r\zeta)>0 & \e{if} \; \la\in \R+\i\tfrac{\pi}{2} \ea \right.$. 
\end{itemize}

When $L$ is finite, the ground state $\ket{\La}$ belongs to the sector with $|\La|$ spins down, where the integer $|\La|$ grows with $L$ in such a way that $\tf{|\La|}{L}\tend D_{gs}$. 
In order to avoid technical complications\symbolfootnote[2]{In principle one could have that $\big|D_{gs}-\tf{|\La|}{L}\big|=c/L$ for some $c>0$. In such a case, there would arise additional terms in the various intermediate expansions
obtained in the core of the paper. These are not hard to deal with but would make numerous expressions much bulkier without bringing anything new to the physics. 
See \cite{EckleTruongWoynarovichNonAnalyticCorrectionsForXXZInFiniteMagFieldANdCFTSpectrum} for a more extensive discussion of this issue.} in the analysis that will follow, I will assume that 
$\big|D_{gs}-\tf{|\La|}{L}\big|=\e{O}(L^{-2})$.

\vspace{3mm}
Throughout the paper, all functions solving a linear integral equation driven by  $\e{id}+\op{K}_{I_{q}}$ will be denoted with the auxiliary argument omitted, \textit{e.g.} $p(\la)\equiv p(\la\mid q)$, $\phi(\la,\mu) \equiv \phi(\la,\mu\mid q)$. 
Likewise, the integral operator, resp. its resolvent operator and the associated integral, will be denoted as $\e{id}+\op{K}$, resp. $\e{id}-\op{R}$ and $R(\la,\mu)$. 
 
\subsection{The ground state counting function}

In order to study the large-$L$ behaviour of various quantities expressed in terms of the Bethe roots, it appears convenient to introduce their counting 
function following \cite{DeVegaWoynarowichFiniteSizeCorrections6VertexNLIEmethod,KlumperBatchelorNLIEApproachFiniteSizeCorSpin1XXZIntroMethod}:
\beq
\wh{\xi}_{\La}(\om) \,  = \, \f{ 1 }{ 2\pi } \vth\big( \om\mid \tf{\zeta}{2} \big) \, -\, \f{1}{2 \pi L } \sul{ \la \in \La }{}\vth\big(\om-\la\mid \zeta \big) \, + \, \f{|\La|+1 }{2L} \;. 
\enq
The latter satisfies, by construction, $\wh{\xi}_{\La}(\la_a) = \tf{a}{L}$. 
The ground state counting function was rigorously characterised in \cite{KozProofOfDensityOfBetheRoots}.  
\begin{theorem}
 \label{Theorem pte pples gd L ctg fct}

There exists $\de>0$ such that, for $L$ is large enough, 

\begin{itemize}

\item the function $\wh{\xi}_{\La}$ is a biholomorphism from  $\mc{S}_{\de}(\R)$ onto its image;

\item the restriction of $\wh{\xi}_{\La}$ to $\R$ is strictly increasing;
\item  for any $\eps, C>0$, there exists $c>0$ such that  $ \pm \Im\Big(\, \wh{\xi}^{\prime}_{\La}(\om) \Big) > c $ 
for $|\Re(\om)|\leq C$ and $\eps \leq  \pm \Im(\om) \leq \de $; 

\item $\wh{\xi}_{\La}(\la) \, = \,  \tfrac{ p(\la) }{2\pi}  \big( 1+\e{O}(L^{-2}) \big) + \tfrac{|\La|+1}{2L} $ uniformly on $\mc{S}_{\de}(\R)$. 
 
\end{itemize}

\end{theorem}

These properties ensure that, given two parameters $\tau_{L}, \tau_{R} \in \intoo{-\tf{1}{8} }{ \tf{1}{8} }$, one can define unambiguously  two roots $\wh{q}_{L}$ and $\wh{q}_R$  
as the unique solution on $\R$ to the equations
\beq
\wh{\xi}_{\La}\big( \,\wh{q}_{L} \big) \; = \; \f{ \tau_L + \tf{1}{2} }{ L } \qquad \e{and} \qquad 
\wh{\xi}_{\La}\big( \, \wh{q}_{R} \big) \; = \; \f{|\La|+\tau_R+\tf{1}{2}}{ L } \,.  
\label{definition parametres hat q L et R}
\enq
The properties enjoyed by $\wh{\xi}_{\La}$ ensure that, for some bounded in $L$ constants $C_{\ups}$, $\ups\in \{L,R\}$, 
\beq
  \wh{q}_{R}-q    \,  =\, \f{ C_R \tau_R }{L} \, + \,  \e{O}\Big( \f{\tau_R^2}{L^2}  \Big) \qquad \e{and}  \qquad    \wh{q}_L + q     \, = \,  \f{ C_L \tau_L }{L} \, + \,  \e{O}\Big( \f{\tau_L^2}{L^2}  \Big)   \; .
\label{ecriture DA racine q har R et L}
\enq
The interval $\intff{\, \wh{q}_L}{\wh{q}_R}$ can be though of as a finite volume version of the Fermi zone. The parameters $\tau_{\ups}$, $\ups \in \{L,R\} $, 
appearing in the definition \eqref{definition parametres hat q L et R} of $\wh{q}_{\ups}$ will play the role of regularisation parameters as will be explained in 
Section \ref{SousSection fct shift} to come.

\begin{figure}[ht]
\begin{center}

\begin{pspicture}(7,7)

\psline[linestyle=dashed, dash=3pt 2pt]{->}(1,4)(6.7,4)
\psdots(1.5,4)(6,4) 
\rput(0.7,3.5){ $\f{ \tau_L + \tf{1}{2} }{ L } $ }
\rput(7.2,3.4){  $\f{|\La| + \tau_R + \tf{1}{2} }{ L } $  }

\rput(0.7,6.2){$\f{ \tau_L + \tf{1}{2} }{ L }  + \i\ga$}
\rput(6.8,6.1){$\f{ |\La| + \tau_R + \tf{1}{2} }{ L } + \i\ga$}

\rput(0.5,1.8){$\f{ \tau_L + \tf{1}{2} }{ L }   - \i\ga$}
\rput(6.8,1.5){$\f{ |\La| + \tau_R + \tf{1}{2} }{ L } - \i\ga$}

\psline{-}(1.5,2.5)(1.5,5.5)
\pscurve{-}(1.5,5.5)(1.65,5.9)(2,6)

\psline{-}(2,6)(5.5,6)
\pscurve{-}(5.5,6)(5.85,5.9)(6,5.5)

\psline{-}(6,5.5)(6,2.5)
\pscurve{-}(6,2.5)(5.85,2.1)(5.5,2)

\psline{-}(5.5,2)(2,2)
\pscurve{-}(2,2)(1.65,2.1)(1.5,2.5)




\rput(3.5,6.5){ $ \wh{\Ga}^{(+)} $ }
\rput(4,1.5){ $ \wh{\Ga}^{(-)} $ }

\psline[linewidth=2pt]{->}(5,2)(5.1,2)
\psline[linewidth=2pt]{->}(4,6)(3.9,6)

\end{pspicture}
\caption{ Contour  $\wh{\Ga}=\wh{\Ga}^{(+)} \cup \wh{\Ga}^{ (-) }$, with $\wh{\Ga}^{  (\pm) }  \, = \, \wh{\Ga} \cap \mathbb{H}^{\pm}$. $\ga>0$ is a sufficiently small but otherwise fixed constant.
The contour $\msc{C}$ is defined by  $ \msc{C}  = \wh{\xi}_{\La}^{-1}\big(\, \wh{\Ga} \, \big) $.  }
\end{center}
\label{Figure contour Gamma}
\end{figure}
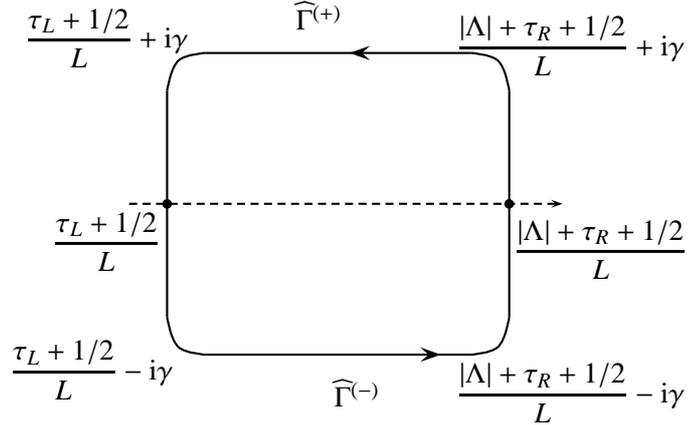

The ground state counting function allows one to introduce an auxiliary contour $\msc{C}$ which passes through $\wh{q}_{L}$ and $\wh{q}_{R}$. This contour  will play an important role
in the analysis. In order to construct $\msc{C}$, one first defines a contour $\wh{\Ga}$  according to Fig.~\ref{Figure contour Gamma} and then sets $ \msc{C}  = \wh{\xi}_{\La}^{-1}\big(\, \wh{\Ga} \, \big) $. 
The parameter $\ga>0$ appearing in the definition of $\wh{\Ga}$ is taken small enough, in particular such that $\wh{\Ga} \subset \wh{\xi}_{\La}\big( \mc{S}_{\de}(\R) \big)$, but otherwise $L$-independent.

\subsection{ $\a$-twist and $\mf{b}$-deformation of the ground-state roots}

In the following , I will consider the set of $\a$-twisted, $\a\in \R$,  solution to the ground state Bethe Ansatz equations, \textit{viz}. the real valued solution $\La^{(\a)} \, = \, \{ \la_a^{(\a)} \}_1^{|\La|} $ to 
 $\wh{\xi}_{\La^{(\a)}}\big( \la_a^{(\a)} \big)=\tf{a}{L}$ with $a=1,\dots,|\La|$, where  $\wh{\xi}_{\La^{(\a)}}$ is the associated counting function 

\beq
\wh{\xi}_{\La^{(\a)}}(\om) \,  = \, \f{ 1 }{ 2\pi } \vth\big( \om\mid \tf{\zeta}{2} \big) \, -\, \f{1}{2 \pi L } \sum_{ \la \in \La^{(\a)} }\vth\big(\om-\la\mid \zeta \big) \, + \, \f{|\La|+1 -2\a_{\La} }{2L}\,.  
\label{ecriture eqns de log Bethe GS roots} 
\enq
Here, I assume that $\a_{\La} \in \R$. It can be proven within the techniques of \cite{KozProofOfDensityOfBetheRoots} that $\wh{\xi}_{\La^{(\a)}}$ enjoys the conclusion of Theorem \ref{Theorem pte pples gd L ctg fct}
with the minor difference that 
\beq
\wh{\xi}_{\La^{\!(\a)}}(\la) \, = \,  \tfrac{ p(\la) }{2\pi}  \big( 1+\e{O}(L^{-2}) \big) + \tfrac{|\La|+1}{2L}  - \f{ \a_{\La} }{ L} Z(\la)  \qquad \e{uniformly}\,\e{on} \quad \mc{S}_{\de}(\R) \; . 
\enq
In the following, the parameter $\a_{\La}$ will be assumed to be small enough in L, 
\beq
\a_{\La} \; = \; \e{O}\big( L^{-k_{\La}} \big) 
\label{ecriture estimation alpha La}
\enq
with $k_{\La}$ large enough. 
It will also appear convenient to introduce the $\mf{b}$-deformed counting function of the $\a$-twisted ground state roots. 
\beq
\wh{\xi}^{\,(\mf{b})}_{\La^{(\a)}} \; = \; \wh{\xi}_{\La^{(\a)}}\, + \, \f{\mf{b}}{L} 
\enq
and denote by $\La_{\mf{b}}^{\! (\a)}$ the collection of $\mf{b}$-deformed ground state roots 
\beq
\La_{\mf{b}}^{\! (\a)}=\Big\{ \la_k^{(\a)}\!(\mf{b}) \Big\}_{k=1}^{|\La|} \qquad \e{with} \qquad \la_k^{(\a)}\!(\mf{b}) \; = \; \wh{\xi}_{ \La^{(\a)} }^{\, -1}\Big( \f{k-\mf{b} }{ L} \Big) \quad k=1,\dots, |\La|\;. 
\label{ecriture definition ensemle Lab et ses racines}
\enq
The properties of $\wh{\xi}_{ \La^{(\a)} }$ ensure that, for $L$ large enough, the set $\La_{\mf{b}}^{\! (\a)}$ is well defined and that all 
the maps $u\mapsto \la_k^{(\a)}\!(u)$ are holomorphic on $\mc{D}_{0,1}$. 

\begin{prop}
\label{Proposition beta deformation des racines Lambda}
 
Let  $\{\mu_a^{(s)}\}_1^{n_{\e{sg}}}$ and $\Ups^{(\e{in})}$ be two given sets of real valued parameters.
 For any $r  \leq \tf{1}{4}$,  there exists $\de r $ bounded in $L$  and a $L$-independent constant $C>0$ such that, 
 for $L$ large enough and for any $\mf{b}\in \Cx$ satisfying $|\mf{b}| \, = \,  r + \tf{\de r}{L}$, it holds  
\beq
d\big( \La_{\mf{b}}^{\! (\a)} , \big\{ \mu_a^{(s)} \big\}_1^{n_{\e{sg}}} \big) \,  > \, \f{ C }{ L^{2} } 
\label{condition distance racine mua sing au reseau Lambda}
\enq
for any $\a_{\La}$ as in \eqref{ecriture estimation alpha La}. If, in addition, one also has $\mf{b}\not\in \R$, then it holds 
\beq
\La_{\mf{b}}^{\! (\a)} \; \cap \;   \Big\{  \Ups^{(\e{in})}\cup  \big\{ \mu_a^{(s)} \big\}_1^{n_{\e{sg}}} \Big\} \; = \; \emptyset \;  . 
\label{condition separation ensemble Lambda b et Ups in}
\enq

\end{prop}

In the following, this proposition will be applied to the sets $\Ups^{(\e{in})}$ and parameters $\{\mu_a^{(s)}\}_1^{n_{\e{sg}}}$ as defined in Sections \ref{Sous Section definition des racines etat excite}-\ref{Sous Section excited state Ctg Fct},
equations \eqref{definition ensemble Ups in}-\eqref{definition des parametres mu sing}, to come.

\Proof 

Since $\wh{\xi}_{\La}$ preserves the sign of the imaginary part, so does its inverse. 
Clearly, the same does hold for  $\wh{\xi}_{ \La^{(\a)} }$. Since the set $ \Ups^{(\e{in})}\cup  \big\{ \mu_a^{(s)} \big\}_1^{n_{\e{sg}}} $ is real valued,
\eqref{condition separation ensemble Lambda b et Ups in} clearly holds, this irrespectively of the choice of $\de r$ provided that the latter is small enough. 

Let $G$ be some relatively compact open neighbourhood of $\intff{-q}{q}$ containing all the roots $\La_{u}^{(0)}$ when $u$ runs through the closed unit disc. 
It follows from the definition \eqref{ecriture definition ensemle Lab et ses racines} of the roots that for any $u \in \mc{D}_{0,1}$, these admit the large-$L$ expansion 
\beq
\la_a^{(0)}(u) \, = \, \la_a^{(0)}(0) \, - \,  \f{u}{L \, \wh{\xi}^{\prime}_{\La}\big( \la_a^{(0)}(0) \big) }  \, + \,  \f{u^2}{2L^2  } \Big( \, \wh{\xi}^{-1}_{\La}\Big)^{\prime\prime}\big(\tfrac{a}{L} \big)
\;+\; \e{O}\big( L^{-3} \big) \;. 
\label{ecriture DA roots}
\enq
Thus, for $L$ is large enough,  if $u=|u|\ex{\i\vp}$ with $|\sin \vp|>\tf{1}{\sqrt{2}}$, one has 
\beq
d\big( \la_a^{(0)}(u), \R)  \, > \,  \f{ |u|  }{ 2 \sqrt{2} L c_{\La}  } \qquad \e{with} \quad  c_{\La} \, = \, \e{inf}_{G} \big| \, \wh{\xi}^{\, \prime}_{\La}\big| >0 \;. 
\enq

Therefore, owing to the real-valuedness of the parameters $\{\mu_a^{(s)}\}_1^{n_{\e{sg}}}$, the lower bound \eqref{condition distance racine mua sing au reseau Lambda} automatically 
holds for such $\mf{b}$'s. 

Now suppose that  $|\cos \vp|>\tf{1}{\sqrt{2}}$ and set $\varrho=r\ex{\i\vp}$. For each $k$ pick $a_k$ such that $\big| \Re\big( \la^{(0)}_{a_k}(\varrho)-\mu_k^{(s)} \big) \big|$ is minimal. Then owing to the asymptotic expansion 
\eqref{ecriture DA roots} and $|\la_{a}^{(0)}(0)-\la_{a+1}^{(0)}(0)|\geq \big\{ c_{\La} L \big\}^{-1}$, it holds
\beq
\big| \Re\big( \la^{(0)}_{a }(\varrho)-\mu_k^{(s)} \big) \big| \geq \f{1}{2 L c_{\La} } \qquad \e{for} \; \e{any} \; a\not= a_k \;, 
\label{ecriture espacement racines lambda b et mua s}
\enq
with $c_{\La}$ as given above. For $L$ large enough, given $\ell \leq n_{\e{sg}}+1$ and 
\beq
\mf{b}_{\ell} \, = \, \Big( r + \f{ \ell}{L c_{\La} } \Big) \cdot \ex{\i\vp}  \qquad \e{one} \,\e{has} \qquad
\big| \Re\big( \la^{(0)}_{a }(\varrho) -\la^{(0)}_{a } (\mf{b}_{\ell}) \big) \big| \leq \f{n_{\e{sg}}+2}{ L^2 c_{\La}^2 }
\;. 
\label{ecriture espacement racines lambda b et lambda beta}
\enq
The asymptotic expansion \eqref{ecriture DA roots} ensures that if $\ell\not=p$,  
\beq
\big| \Re\big( \la^{(0)}_{a }(\mf{b}_{\ell})-\la^{(0)}_{a }(\mf{b}_{p}) \big) \big| > \f{ \cos(\vp) }{ 2 c_{\La}^2 L^2  } \qquad \e{so} \, \e{that} \qquad 
\big| \Re\big( \la^{(0)}_{a_k }(\mf{b}_{\ell_k}) - \mu_k^{(s)} \big) \big| < \f{ 1 }{ 4 \sqrt{2} c_{\La}^2 L^2  }
\enq
for at most one $\ell_{k} \in \intn{1}{n_{\e{sg}}+1}$. 
In other words, for any $\ell$ in the non empty set $\intn{1}{n_{\e{sg}}+1}\setminus \{ \ell_k \}_1^{n_{\e{sg}}}$, it holds
$4 \sqrt{2} c_{\La}^2 L^2  \big| \Re\big( \la^{(0)}_{a_k }(\mf{b}_{\ell_k})- \mu_k^{(s)} \big) \big| > 1 $. 
Owing to \eqref{ecriture espacement racines lambda b et mua s} and \eqref{ecriture espacement racines lambda b et lambda beta}, this last lower bound
holds, with possibly a different constant,  not only for $a_k$, but for any $a=1,\dots |\La|$, provided that $L$ is large enough. 
Finally, since $\a_{\La}=\e{O}(L^{-k_{\La}})$ with $k_{\La}$ large enough, one has $\la^{(0)}_{a }(\mf{b})-\la^{(\a)}_{a }(\mf{b})=\e{O}(L^{-k_{\La}})$ so that 
\eqref{condition distance racine mua sing au reseau Lambda} holds for $\La^{(\a)}_{\mf{b}}$.

\qed

\subsection{The structure of an excited state's Bethe roots} 

\label{Sous Section definition des racines etat excite}

As already stated, within the Bethe Ansatz, one constructs excited states of $\op{H}_{\De,h}$ as vectors $\ket{\Ups}$ parametrised by sets $\Ups = \{\mu_a\}_1^{|\Ups|}$ build up from solutions to the 
logarithmic Bethe Ansatz equations
\beq
 \f{ \vth\big( \mu_a \mid \tf{\zeta}{2} \big) }{ 2\pi } \, -\, \f{1}{2 \pi L } \sul{ b=1 }{|\Ups|} \vth\big(\mu_a-\mu_b \mid \zeta \big) \, + \, \f{  |\Ups| +1-2\a_{\Ups} }{2L} \; = \; \f{\ell_a}{L} 
 \qquad \e{with} \qquad  a=1,\dots, |\Ups|  
\label{ecriture alpha twised log BAE}
\enq
where $\ell_a\in \mathbb{Z}$ are some integers and where one should set $\a_{\Ups}=0$. The roots $\mu_a$ can be real or complex valued \cite{BetheSolutionToXXX} but always appear in complex conjugated pairs \cite{VladimirovProofConjInvSolBetheEqns} so that 
$\Ups^*=\Ups$. This property remains true as long as $\a_{\Ups}$ is real. 
For a generic state with $|\Ups|$ and $L$ arbitrary, the characterisation of the Bethe roots seems extremely difficult -see \cite{CauxHagemansDeformedStringsInXXX} for a numerical investigation at small $|\Ups|$ and $L$-. 
However, for excited states close in structure to those of the ground state, namely when $|\Ups|/L\tend D_{gs}$ and when $\Ups$ mainly consist of real roots forming a dense distribution on $\intff{-q}{q}$, one can characterise the Bethe roots completely. 
First reasonings of the sort go back to Bethe \cite{BetheSolutionToXXX}. The arguments raised by Bethe were improved and sophisticated in the works 
\cite{BabbittThomasPlancherelFormulaInfiniteXXX,DestriLowensteinFirstIntroHKBAEAndArgumentForStringIsWrong,KorepinAnalysisofBoundStateConditionMassiveThirring,TakahashiSuzukiFiniteTXXZandStrings,ThomasGSRepForFerromagneticXXX}. 
However, most of the arguments and especially those that were rigorous \cite{BabbittThomasPlancherelFormulaInfiniteXXX,ThomasGSRepForFerromagneticXXX}, were concentrated on the sector where $|\Ups|$ is fixed and $L\tend +\infty$. 
The rigorous description of the complex valued solutions in the case described above has only been achieved recently in \cite{KozProofOfStringSolutionsBetheeqnsXXZ}
by the author, this when $0<D_{gs} <\tf{1}{2}$.  I refer to \cite{KozProofOfStringSolutionsBetheeqnsXXZ} for a thorough discussion of the history of the subject. 

In the following,  the set $\Ups=\{\mu_a\}_1^{|\Ups|} $ will be built out from a solution to the $\a_{\Ups}$-twisted logarithmic Bethe equations \eqref{ecriture alpha twised log BAE}  with real generic $\a_{\Ups}$
satisfying to similar bounds as $\a_{\Ups}$ \textit{viz}. $\a_{\Ups}=\e{O}\big( L^{-k_{\Ups} }\big)$. 
 $\Ups$ will describe an excitation over the ground state  at finite magnetic field $h_{\e{c}}> h >0$.
By this I mean that $\big| |\La|-|\Ups| \big|$ is fixed and finite in $L$ and that the set $\{\ell_a\}$ differs from $\{1,\dots, |\La|\}$ only by a finite in $L$ number of integers. 
The conclusion of \cite{KozProofOfStringSolutionsBetheeqnsXXZ} is that the set $\Ups$
can be partitioned as
\beq
\Ups \; = \; \Big\{ \Ups^{(\e{in})}\setminus \Ups^{(h)}  \Big\} \cup \Ups^{(p)} \cup \Ups^{(z)} \;. 
\label{ecriture decomposition racines Ups}
\enq
The sets building up this partition are characterised as follows 
\begin{itemize}
\item $\Ups^{(\e{in})}\setminus \Ups^{(h)}$ is build out of the real roots contained in $\Ups $ that are located inside of the interval $\intff{\, \wh{q}_L }{ \,\wh{q}_R}$ and that do not form part of a string of complex roots. 
In the thermodynamic limit, the elements of  $\Ups^{(\e{in})}$ 
form  a dense distribution on the Fermi zone $\intff{-q}{q}$. $\Ups^{(h)}$ is built out of certain roots which form "holes" in this dense distribution. Such roots are called hole roots. 
\item $\Ups^{(p)}$ contains roots belonging to $\big\{ \R \setminus \intff{\, \wh{q}_L }{ \,\wh{q}_R} \big\} \cup \big\{ \R+\i\tf{\pi}{2} \big\} $. Such roots are called particle roots. To avoid complications, 
the set $\Ups^{(p)}$ will be taken to be bounded in $L$. 
\item $\Ups^{(z)}$ contains the genuinely complex roots. These organise into complexes called strings. Two contiguous elements of a string are separated, up to exponentially small corrections in $L$, by $\i\zeta$.
A given string is  centred either around  $\R$ or around $\R+\i\tf{\pi}{2}$. Depending on the parity of its length -odd or even-, a given string may or may not contain an $\R\cup \big\{ \R+\i\tf{\pi}{2} \big\}$ valued root.
However, since the complex roots form strings that should be considered as a whole, such $\R\cup \{\R+\i \tf{\pi}{2} \}$ valued roots ought to be included in the set $\Ups^{(z)}$ rather than be considered
as a particle root belonging to $\Ups^{(p)}$ or some of the $\Ups^{(\e{in})}$ roots. It is convenient to parametrise  the set $\Ups^{(z)}$ as 
\beq
\Ups^{(z)} \; = \; \bigg\{ \Big\{ \big\{ c_a^{(r)} +\i\f{\zeta}{2}\big( r + 1 - 2k \big) \, + \, \de_{a,k}^{(r)} \big\}_{k=1}^{r} \Big\}_{a=1}^{n_r^{(z)}} \bigg\}_{r=2}^{ p_{\e{max}} } \;. 
\label{definition ensemble Ups z}
\enq
Within this parametrisation, the centres $c_a^{(r)}$ of the strings belong to $\R\cup \big\{ \R+\i\tf{\pi}{2} \big\}$. The complex roots form strings of length $r$, with $r=2,\dots, p_{\e{max}}$. 
There are $n_r^{(z)}$ different strings of length $r$, each characterised by the centre $c_{a}^{(r)}$ with $a=1,\dots, n_{r}^{(z)}$.  
Finally, the parameters $\de_{a,k}^{(r)}$ represent the so-called string deviations and are exponentially small in $L$.
Without making any explicit emphasis on the decay rate, we shall simply estimate all such corrections as $\e{O}\big( L^{-\infty} \big)$, \textit{viz}. $\de_{a,k}^{(r)}=\e{O}\big( L^{-\infty} \big)$.  

\end{itemize}

In the following, $n_{\e{tot}}$ will refer to the total number of roots differing from the bulk, namely, 
\beq
n_{\e{tot}} \; = \; \big| \Ups^{(p)} \big| + \big| \Ups^{(h)} \big| + \sul{ r=2 }{ p_{\e{max}} } r n_r^{(z)} \qquad \e{and} \qquad 
n_{\e{tot}}^{(z)} \; = \;   \sul{ r=2 }{ p_{\e{max}} } r n_r^{(z)}  \;. 
\label{definition nombre total excitations}
\enq
I will always assume in the following that $n_{\e{tot}}=\e{O}\big( \sqrt{L} \big)$ \;.

\subsection{The excited state counting function}
\label{Sous Section excited state Ctg Fct}
The various sets arising in the decomposition \eqref{ecriture decomposition racines Ups} can be characterised more precisely with the help of the counting function 
subordinate to $\Ups$. As opposed to the case of the ground state roots, some care is needed in defining the latter
since the presence of strings of odd length $p\geq 3$ generates a singular behaviour around some points belonging to a shrinking neighbourhood of $\R$. 
For such a reason, it is convenient to decompose the counting function into its regular and singular parts
\beq
\wh{\xi}_{\Ups}(\om) \, = \,  \wh{\xi}_{\Ups_{\e{reg}}}(\om) \, + \, \wh{\xi}_{\Ups_{\e{sing}}}(\om) \;. 
\enq
The singular part if built out of the complex roots which collapse to $\R\pm\i\zeta$. These roots will be called singular. 
Since the set $\Ups$ is invariant under complex conjugation \cite{VladimirovProofConjInvSolBetheEqns}, it is enough to focus on the roots
collapsing to $\R+\i\zeta$. These are gathered into the set 
\beq
Z^{(s)}=\big\{ z \in \Ups \, : \, \Im(z)\tend \i\zeta \big\} \;. 
\enq
 It is these roots that give rise to the singular part of the counting function:
\beq
 \wh{\xi}_{ \Ups_{\e{sing}} }(\om) \, = \, \f{1}{L} \sul{    \be +\i\zeta  \in Z^{(s)}   }{} \Int{\be^{*}}{ \be } \coth(s-\om) \cdot \f{ \dd s }{2\i \pi } 
 \, = \, \f{1}{2\i\pi L } \sul{    \be +\i\zeta\in Z^{(s)}   }{} \ln \bigg( \f{ \sinh(\be -\om) }{ \sinh(\be^{*} -\om) } \bigg) \;. 
\label{definition partie sing fct cptge}
\enq
The singular part of the counting function has cuts along the segments $\intff{\be}{\be^{*}}+\i\pi \mathbb{Z}$, $\be \in Z^{(s)}$ and its derivative 
\beq
 \wh{\xi}_{\Ups_{\e{sing}}}^{\, \prime}(\om) \, = \, \f{1}{L} \sul{   \be +\i\zeta \in Z^{(s)}   }{} K\big( \om - \Re(\be) \mid \Im(\be) \big) 
\enq
may change sign on $\R$ depending on the values of $\Im(\be)$. Due to the exponential smallness of the string deviation, one has that $\Im(\be)=\e{O}(L^{-\infty})$ 
and thus $ \wh{\xi}_{\Ups_{\e{sing}}}^{\, \prime}$ will be exponentially large in a small, $\e{O}(L^{-\infty})$ open neighbourhood of $\Re(\be)$ and exponentially small 
on $\mc{S}_{\de}(\R)$ provided that one is uniformly away from the set $Z^{(s)}-\i\zeta$. 

In its turn, the regular part of the counting function is a holomorphic function in some open, $L$-independent, strip $\mc{S}_{\de}(\R)$, $\de>0$, around $\R$. It is defined as  
\bem
 \wh{\xi}_{\Ups_{\e{reg}}}(\om) \, = \, \frac{1}{2\pi} \vth \big( \om \mid \tfrac{\zeta}{2} \big) \, - \, \f{1}{2\pi L }\sul{ \mu \in \Ups \setminus \Ups^{(z)}   }{  } \th(\om-\mu)  \\
\, - \, \f{ 1 }{ 2\pi L} \sul{r=2}{ p_{\e{max}} } \sul{ a=1 }{ n_r^{(z)} } \vth^{(\e{reg})}_{r;a}\big(\om-c_a^{(r)} \mid \{ \de_{a,k}^{(r)} \} \big)
\, + \, \f{ |\Ups|+1-2 \a_{\Ups} }{2L} \;. 
\end{multline}
The function $ \vth^{(\e{reg})}_{r;a}$ is a regularisation of the total bare phase associated to a given string:
\bem
 \vth^{(\e{reg})}_{r;a}\big(\om \mid \{ \de_{a,k}^{(r)} \} \big)  \; = \; 2\pi \Int{ \Ga_{\om} }{} \bigg\{ \sul{k=1}{r} K\Big(\mu-0^+-\i\f{\zeta}{2}\big[ r + 1 - 2k \big] - \de^{(r)}_{a,k} \Big)  \\
\, - \, \tfrac{1}{2\i\pi} \cdot \bs{1}_{Z}\big( c_{a}^{(r)} \big)  \Big[ \coth\Big( \mu - 0^+ - \de^{(r)}_{a,\frac{r-1}{2}} \Big)  \, -  \,  \coth\Big(\mu-0^+ - \de^{(r)}_{a,\frac{r+3}{2}} \Big) \Big]  \Bigg\}\cdot \dd \mu \;. 
\label{definition fct theta sommee des racines cordes}
\end{multline}
One has that $\bs{1}_{Z}\big( c_{a}^{(r)} \big)  =1$ if $c_a^{(r)}\in \R$ and $r\geq 3$ is odd and  $\bs{1}_{Z}\big( c_{a}^{(r)} \big)  = 0$ otherwise. 
The counter term present in the second line of \eqref{definition fct theta sommee des racines cordes} is only there if one deals with a string that contains singular roots. 
One can check that \eqref{definition fct theta sommee des racines cordes} does define a  holomorphic function in a neighbourhood of $\R$.

The set $\Ups^{(\e{in})}$ can be defined in terms of the $\Ups$-counting function as
\beq
\Ups^{(\e{in})}  \; = \; \Big\{ \mu \in \e{Int}\big( \msc{C} \big) \; : \; \ex{2\i\pi L \wh{\xi}_{\Ups}(\mu)} -1\, = \, 0 \Big\} \setminus \Ups^{(z)} \;. 
\label{definition ensemble Ups in}
\enq
 In other words, $\Ups^{(\e{in})}$ contains all the zeroes of $\ex{ 2\i\pi L \wh{\xi}_{\Ups} } -1$ lying inside of the contour $ \msc{C} $ defined in Fig.~\ref{Figure systeme de poles a eviter},  
with the exception of those zeroes that correspond to an element of a string
which squeezes down to $\intff{ \wh{q}_L }{ \wh{q}_R }$ in the $L\tend +\infty$ limit. The set $ \Ups^{(\e{in})} $ contains most of the Bethe roots building up the set $\Ups$. 
It also contains a certain amount of extra roots, the holes; these are roots of $\ex{ 2\i\pi L \wh{\xi}_{\Ups} } -1$ located inside of the interval $\intff{\,\wh{q}_L}{\, \wh{q}_R}$ 
which do not coincide with a root solving the logarithmic Bethe equations \eqref{ecriture alpha twised log BAE}, hence leading to the definition 
\beq
\Ups^{(h)}\, = \, \Ups^{(\e{in})} \setminus \Ups \; = \; \Big\{ \mu \in \e{Int}\big( \msc{C} \big) \; : \; \ex{2\i\pi L \wh{\xi}_{\Ups}(\mu)} -1\, = \, 0 \;\;\e{but} \;\; \mu \not\in \Ups \Big\} \;. 
\enq
Finally, the set containing the particle roots can be constructed as 
\beq
\Ups^{(p)}\; = \; \Ups\setminus \Big\{ \Ups^{(z)} \cup\Ups^{(\e{in})} \Big\} \;.  
\enq
One can check that the above definitions are indeed consistent with the decomposition \eqref{ecriture decomposition racines Ups}.

It will appear useful, for further handling, to single out a sub-class of singular roots, namely those whose real part is inside of $\intff{\, \wh{q}_L }{ \wh{q}_R }$: 
\beq
\Big\{ \be_a^{(s)} \Big\}_{ 1 }^{ n_{\e{sg}} } \, = \, \Big\{  z -\i\zeta\, : \, z  \in \Ups^{(z)} \, \e{satisfying} \,  \Im(z) \tend \zeta \quad \e{and} \quad \Re(z) \in \intff{ \, \wh{q}_L }{ \wh{q}_R } \Big\}   \;. 
\label{definition des parametres beta sing}
\enq
Necessarily, a root $\be_a^{(s)}+\i\zeta$ belongs to a string of odd length centred on $\intff{ \, \wh{q}_L }{ \wh{q}_R }$. The central root of such a string will be denoted by $\mu_{a}^{(s)}$. 

The $n_{\e{sg}}$ roots $\mu_{a}^{(s)}$ are then collected into the set 
\beq
\Big\{ \mu_a^{(s)} \Big\}_{ 1 }^{  n_{\e{sg}} } \, = \, \Big\{  z \in \Ups^{(z)} \, : \,  \Im(z) \tend 0 \; \e{and} \; \Re(z) \in \intff{ \wh{q}_L }{ \wh{q}_R } \Big\}  \;. 
\label{definition des parametres mu sing}
\enq
It is also advantageous to introduce the difference set 
\beq
\daleth \, = \, \big\{ \be_a^{(s)} \big\}_{ 1 }^{  n_{\e{sg}} } \setminus \big\{ \mu_a^{(s)} \big\}_{ 1 }^{  n_{\e{sg}} } 
\enq
where the parameters $\be_a^{(s)}$ and $\mu_a^{(s)}$ are defined, respectively, in 
\eqref{definition des parametres beta sing} and \eqref{definition des parametres mu sing}. 
Then according to the convention \eqref{ecriture convention produit et somme difference ensemble} one has 
\beq
\sul{\a \in \daleth}{} f(\a) \, = \, \sul{a=1}{  n_{\e{sg}} } \Big\{ f\big(  \be_a^{(s)} \big) \, - \, f\big(  \mu_a^{(s)} \big) \Big\} \qquad \e{and} \qquad
\pl{\a \in \daleth}{} f(\a) \, = \, \pl{a=1}{  n_{\e{sg}} } \f{  f\big(  \be_a^{(s)} \big) }{  f\big(  \mu_a^{(s)} \big)  } \;. 
\label{ecriture convention somme et prod racines sing}
\enq

To close this section, I state a technical hypothesis that will be used in the analysis of the large-$L$ behaviour. This hypothesis concerns lower bounds in $L$ 
on the separation of the string centres from the sets $\Ups^{(\e{in})}\cup\Ups^{(p)}$ as well as on the separation between string centres of equal parity strings 
and the separation away from zero of the string centres of even length. I will also assume that the set $\Ups^{(z)}$ does not contain so-called exact strings, \textit{viz}. strings 
where some of its constituents are \textit{exactly} spaced by $\i\zeta \; \e{mod}[\i\pi]$. In particular, this means that there are no strings of length $r$
 such that $(r-1)\zeta/\pi$ is an integer. Finally, I will  assume that the set $\Ups$ does not contain repeating elements.

\begin{hypothesis}
\label{Hypothesis espacement des cordes}

 There exists a constants $C>0$ and $0 \leq \ups  <  \tf{1}{2}$ such that 
\beq
d\Big(  \Ups^{(\e{in})}\cup \Ups^{(p)} ,  c_a^{(r)} + \de^{(r)}_{a,\frac{r+1}{2}}  \Big) \,  >  \, C\cdot \Big( \Im\big(\de^{(r)}_{a,\frac{r-1}{2}} \big) \Big)^{ \ups } 
\label{propriete espacement ctre corde et particule trou}
\enq
for  any  choice of the $2$-uple $(a,r)$, $ 2\leq r \leq p_{\e{max}}$  and $1 \leq a \leq n_r^{(z)}$, with $r$ being odd. 

There exists $C>0$ and $\kappa>0$ such that 
\beq
\big| c_a^{(p)}-c_b^{(r)} \big|  > C\cdot L^{-\kappa}  \quad for   \; any \;\; (a,p)\not=(b,r)
\label{propriete espacement ctres cordes}
\enq
such that $p$ and $r$ have the same parity.

The elements of even length strings are not too close of the points $\pm \tf{\i\zeta}{2}$, namely there exists $C>0$ such that 
\beq
\big| c_{a}^{(r)} +\de^{(r)}_{a;\frac{r}{2}}  \big| \, > \,   C \Big| \de^{(r)}_{a;\frac{r}{2}} - \de^{(r)}_{a;\frac{r}{2}+1} \Big|^{1/2}  \;, 
\label{hypothese sur borne inf sur espacement des diverses racines}
\enq
 for  any choice of the $2$-uple $(a,r)$, $ 2\leq r \leq p_{\e{max}}$  and $1 \leq a \leq n_r^{(z)}$, with $r$ being odd.

There are no string constituents exactly spaced by $\i\zeta$, namely $\de_{a,k}^{(r)}-\de_{a,k+1}^{(r)}$, $k=1,\dots,r-1$ for any $a$ and $r$.  
 In particular, if $\tf{\zeta}{\pi}\in \mathbb{Q}$ then there are no strings of length $r$ such that $(r-1)\zeta \in \pi \mathbb{Z}$ 

The set $\Ups$ does not contain repeating elements. 

\end{hypothesis}

According to the classification that will be established in \cite{KozProofOfStringSolutionsBetheeqnsXXZ}, these hypotheses are satisfied hold for most $\Ups$ of interest. 
In fact, by playing with $\a_{\Ups}$ small enough and inhomogeneously deforming the model with inhomogeneities $\xi_k\,=\, \e{O}\big( L^{-k_{\xi}}\big)$ and $k_{\xi}$ large enough, 
one can have these assumptions to hold for any state $\Ups$. I will however not discuss these issues further in the present work. 

Furthermore, some of the results obtained in the core of the paper are independent of Hypothesis \ref{Hypothesis espacement des cordes}. 
It will be  made explicit in the statement of a proposition or a theorem whenever this or that assumption will be used in a proof. 
The lack of exact strings and repeating elements will however be used tacitly in the following.

Using Rouch\'{e}'s theorem, it is an easy corollary of the above hypothesis that, for some $C>0$ large enough, 
\beq
\big| \mu_a^{(s)}- \be_a^{(s)} \big| \, \leq \,  C |\Im\big( \be_a^{(s)} \big)|^{1-\ups}\, .
\label{ecriture deviation des mua au bea}
\enq

 \subsection{The shift function}
\label{SousSection fct shift}

It will appear convenient in the following to introduce the shift function $\wh{F}$ associated with the roots $\La^{ \!(\a) }$ and $\Ups$
\beq
\wh{F}(\om) \, = \, L\, \Big(\, \wh{\xi}_{\La^{ \!(\a) }}(\om) \, - \,  \wh{\xi}_{\Ups}(\om) \Big)\;. 
\enq
The shift function is such that 
\beq
\ex{2\i\pi \wh{F}(\om) } \; = \; \ex{2\i\pi \a} \pl{ \la \in \La^{ \!(\a) } }{} \bigg\{ \f{ \sinh\big( \om-\la+\i\zeta \big)  }{  \sinh\big( \om-\la-\i\zeta \big) } \bigg\}
\cdot \pl{ \mu \in \Ups }{} \bigg\{ \f{  \sinh\big( \om - \mu - \i\zeta \big) }{ \sinh\big( \om - \mu + \i\zeta \big)  } \bigg\} \; = \;  
\ex{2\i\pi \a} \f{  V_{\Ups;\La^{ \!(\a) }}(\om-\i\zeta) } {  V_{\Ups;\La^{ \!(\a) }}(\om+\i\zeta) }
\label{ecriture explicite produit shift function}
\enq
where 
\beq
 V_{\Ups;\La^{ \!(\a) }}(\om) \; = \; \f{ \pl{\mu \in \Ups}{}  \sinh(\om-\mu) }{ \pl{\la \in \La^{ \!(\a) }}{}  \sinh(\om-\la) }  \qquad \e{and} \qquad \a=\a_{\Ups}-\a_{\La} \,. 
\label{definition fct V Ups et La}
\enq

It is time to explain the appropriate choice of the parameters $\tau_{\ups} \in \intoo{-\tf{1}{8} }{ \tf{1}{8} }$, $\ups\in \{L,R\}$. Let $\varkappa_{\ups}$ be 
the integers such that $-\tf{1}{2} \, \leq  \, \wh{F}\big( \,\wh{q}_{\ups} \big)   - \varkappa_{\ups} \, < \, \tf{1}{2}$. The 
 parameters $\tau_{\ups}$ should be chosen in such a way that

 \begin{itemize}
  \item there exists an $L$-independent constant $\eps_{\Ups}>0$ such that 
\beq
-\f{1}{2}+\eps_{\Ups}\, < \, \wh{F}\big( \,  \wh{q}_{\ups} \big) -\tau_{\ups} - \varkappa_{\ups} \, < \, \f{1}{2}-\eps_{\Ups}\; , \e{with} \;\; \ups\in \{L,R\} \, ; 
\label{bornes sur fct shift en q hat left right}
\enq

\item  there exists a constant $C>0$ such that
\beq
\underset{ \substack{ \be+\i\zeta \\ \in Z^{(s)} } }{\e{min}} \Big\{ d\big(\be,\msc{C}\big), d\big(\be^{*},\msc{C}\big) \Big\} \; > \; \f{C}{L} \;. 
\label{ecriture propriete espacement contour aux racines singulieres}
\enq
 \end{itemize}

 It is clear that there are numerous ways to satisfy the first constraint \eqref{bornes sur fct shift en q hat left right}.  
 To convince oneself that the second constraint \eqref{ecriture propriete espacement contour aux racines singulieres} can also be fulfilled, 
one should observe that the large-$L$ behaviour of $\wh{\xi}_{\La}$ ensures that changing $\tau_{\ups} \hookrightarrow \tau_{\ups}^{\prime}$ shifts $\wh{q}_{\ups}$ by a factor 
$\tf{ \big(\tau_{\ups}-\tau^{\prime}_{\ups}\big) }{ \{ L\, \wh{\xi}^{\prime}_{\La}(\, \wh{q}_{\ups} ) \} }$ plus some higher order corrections in $L^{-1}$.  
Since the singular roots $\be$ in \eqref{ecriture propriete espacement contour aux racines singulieres} squeeze with exponential speed on $\R$, the resulting
change in $\msc{C}$ is indeed enough so as to fulfil \eqref{ecriture propriete espacement contour aux racines singulieres}.

 Since $\a_{\La}=\e{O}\big( L^{ - k_{\La} } \big)$, the property \eqref{bornes sur fct shift en q hat left right} allows one to be in a situation where all three functions 
\beq
1 - \exp\Big\{ 2\i\pi L \wh{\xi}_{\La} \Big\} \quad ,  \quad 1 - \exp\Big\{ 2\i\pi L \wh{\xi}_{ \La^{\!(\a)} }^{\,(\mf{b}) } \Big\}   \quad \e{and} \quad 1 - \exp\Big\{ 2\i\pi L \wh{\xi}_{\Ups}  \Big\}  
\nonumber
\enq
are uniformly away from zero at $\wh{q}_{\ups}$. This non-vanishing is a crucial property for the analysis to come. 
In particular, it entails that the function $1-\ex{2\i\pi L \wh{\xi}_{ \Ups}} $ is non-vanishing on $\msc{C}$, and thus that $\Ups^{(\e{in})}$ is well defined. 
\eqref{ecriture propriete espacement contour aux racines singulieres} allows one to avoid the case when the singular roots will be approaching too close to $\wh{q}_{\ups}$, what would generate additional problems in the analysis.

It will also appear useful, at some later stage, to introduce the regular part of the shift function, 
\beq
\wh{F}_{\e{reg}}(\tau) \, = \, L \, \Big(\, \wh{\xi}_{ \La^{\!(\a)} } (\tau) \, - \, \wh{\xi}_{ \Ups_{\e{reg}} }(\tau) \, \Big) \;.
\label{definition partie reg fct shift}
\enq

\subsection{The asymptotic expansion of the $\Ups$-counting function}

Below, I establish the large-$L$ expansion of the counting function $\wh{\xi}_{\Ups}$ 
on the basis of a certain properties it enjoys. These properties are established in \cite{KozProofOfStringSolutionsBetheeqnsXXZ} and this demands a separate kind of analysis. 
Thus, for the purpose of the present paper, they can just be though of as a set of hypothesis under which the conclusion of the analysis does hold. 

\begin{itemize}

\item The restriction of $\wh{\xi}_{\Ups}$ to compact subsets of $\R$ is strictly increasing uniformly in $L$.

\item  For any $\eps,C>0$, there exists $c>0$ such that  $ \pm \Im\Big(\, \wh{\xi}^{\,\prime}_{\Ups}(\om) \Big) > c $ for $|\Re(\om)|\leq C$ and $\eps \leq  \pm \Im(\om) \leq \de $. 

\item There exists $C^{\prime}>0$ and $\de>0$ such that $\norm{ \wh{F}_{\e{reg}} }_{ \mc{S}_{\de} (\R) } < C^{\prime} $, with $ \wh{F}_{\e{reg}}$ as in \eqref{definition partie reg fct shift} . 

\end{itemize}

The strategy to obtain the asymptotic expansion of $\wh{\xi}_{\Ups}$ consists in writing down a non-linear integral equation (NLIE) satisfied by it 
\cite{DestriDeVegaAsymptoticAnalysisCountingFunctionAndFiniteSizeCorrectionsinTBAFirstpaper,KlumperBatchelorNLIEApproachFiniteSizeCorSpin1XXZIntroMethod,KlumperBatchelorPearceCentralChargesfor6And19VertexModelsNLIE}. 
This NLIE can be written precisely because the above properties does hold. Its very form does allow one for an easy calculation of the asymptotic
expansion of the counting function. The asymptotic expansion obtained below slightly differs, in structure, from the one obtained in \cite{KozProofOfStringSolutionsBetheeqnsXXZ} owing to a different choice, more appropriate for the present analysis, 
of the contour $\msc{C}$ in the NLIE satisfied by $\wh{\xi}_{\Ups}$. More precisely, by changing the definition of such a contour, one changes the definition
of  "particle" and   "hole" root, \textit{viz}. which roots are called "particle"  and which ones are called "hole" . For two given contours $\msc{C}$ and $\msc{C}^{\prime}$, one can always construct a bijection between the 
sets $\Ups^{(p/h)}_{\msc{C}}$ and $\Ups^{(p/h)}_{\msc{C}^{\prime}}$, but the latter can quickly become rather complicated. 

Prior to stating the result, I introduce a convenient auxiliary function that arises in the intermediate analysis 
\beq
\wh{u}_{\Ups}(\om) \, = \, \left\{ \ba{cc} -2\i\pi L \cdot \big[ \,  \wh{\xi}_{\Ups_{\e{reg}}}(\om) + \wh{\xi}_{\Ups_{\e{sing}}}(\om) \big]  \, + \, \wh{u}_{\Ups}^{\, (+)}(\om)  & \om \in \mathbb{H}^+ \vspace{1mm} \\ 
					    \wh{u}_{\Ups}^{\, (-)}(\om)  & \om \in \mathbb{H}^-  \ea \right.  \quad \e{with} \quad  \wh{u}_{\Ups}^{\, (\eps)}(\om) \, = \, \ln \Big[ 1\, - \, \ex{2\i\pi \eps L \wh{\xi}_{\Ups}(\om)} \Big] \;. 
\label{definition hat u Ups}
\enq
Although its $\La$ counterpart will not be used immediately, I already mention the analogous auxiliary function built from the counting function for the $\mf{b}$-deformed ground state roots $\La_{\mf{b}}^{\! (\a)}$
\beq
\wh{u}_{ \La_{\mf{b}}^{\! (\a)} }(\om) \, = \, \left\{ \ba{cc} -2\i\pi L \cdot \wh{\xi}_{ \La^{\! (\a)}  }^{\, (\mf{b})}(\om) \, + \, \wh{u}_{ \La_{\mf{b}}^{\! (\a)} }^{\, (+)}(\om) & \om \in \mathbb{H}^+   \vspace{1mm} \\ 
					    \wh{u}_{ \La_{\mf{b}}^{\! (\a)} }^{\, (-)}(\om)  & \om \in \mathbb{H}^-  \ea \right.  
	      \quad \e{with} \quad  \wh{u}_{ \La_{\mf{b}}^{\! (\a)} }^{\, (\eps)}(\om) \, = \, \ln \bigg[ 1\, - \, \ex{2\i\pi \eps L  \wh{\xi}_{ \La^{\! (\a)}  }^{\, (\mf{b})} (\om)} \bigg]\;. 
\label{definition hat u La}
\enq

Also, for further convenience, I agree upon 
\beq
\sg_{R}=1 \qquad \e{and} \qquad \sg_{L}=-1 \; . 
\label{definition parametres sg L et R}
\enq

\begin{prop}
 \label{Proposition DA ctg fct}
 The regular part of the counting function $\wh{\xi}_{\Ups}$ can be recast as 
\beq
\wh{\xi}_{\Ups_{\e{reg}}}(\om) \, = \, \frac{1}{2\pi} p(\om  ) \, - \, \f{1}{L} \Big( F (\om) \, + \, \a_{\La} Z(\om)  \Big) \, + \, \mf{R}_N\big[\, \wh{\xi}_{\Ups}\big](\om)
\label{ecriture DA fct de cptage}
\enq
 where, setting $\a=\a_{\Ups} \, - \, \a_{\La}$, 
\beq
F (\om) \; = \; \bigg( \a + \f{|\La|-|\Ups|}{2} \bigg)\,  Z(\om \big) \;+\; \sul{ r=2 }{ p_{\e{max}} } \sul{ a=1 }{ n_r^{(z)} } \phi_{r;1}(\om,c_a^{(r)} )
\; + \hspace{-3mm} \sul{ \mu \in \Ups^{(p)}\setminus \Ups^{(h)} }{} \hspace{-2mm} \phi( \om, \mu ) \; - \hspace{-1mm} \sul{ \ups \in \{L,R\} }{}\varkappa_{\ups} \sg_{\ups} \phi(\om, \sg_{\ups} q \, )  \;. 
\label{definition fct shit semi limite thermo}
\enq
The remainder term takes the form 
\bem
\mf{R}_N\big[\, \wh{\xi}_{\Ups}\big](\om) \; = \; \f{1}{L} \sul{a=1}{ n_{\e{sg}}} \Big\{ \phi\big( \om, \mu_a^{(s)} \big) \, - \,  \phi\big( \om, \be_a^{(s)} \big) \Big\} 
\, - \, \sul{\eps=\pm 1 }{} \Int{ \msc{C}^{(\eps)} }{} R\big(\om , s \big) \cdot \bigg\{  \f{ \wh{u}^{\,(\eps)}_{\Ups}(s) }{ 2\i\pi L }\,-  \, \de_{\eps,+}  \, \wh{\xi}_{\Ups_{\e{sing}}}(s) \bigg\} \cdot \dd s \\
+ \sul{ \ups\in \{R,L\}  }{} \sg_{\ups}\Bigg\{ \f{-1}{L} \Big[ L \, \wh{\xi}_{\Ups_{\e{sing}}}(\, \wh{q}_{\ups})  \, + \, \wh{F}(\, \wh{q}_{\ups})_{\mid _{\a_{\La}=0} } -\varkappa_{\ups}-\tau_{\ups}    \Big]
\cdot \Big[ \phi(\om, \wh{q}_{\ups}) \,- \, \phi(\om, \sg_{\ups}q \, )   \Big] \\
\, + \,  \Int{ \sg_{\ups} q }{ \wh{q}_{\ups} } R\big(\om , s \big) \cdot \Big[ \wh{\xi}_{ \Ups_{\e{reg}} }(s)- \wh{\xi}_{ \Ups_{\e{reg}} }(\, \wh{q}_{\ups}) \Big] \cdot \dd s \Bigg\}  
\, + \; \f{1}{2} \Big(D-\f{|\La|}{L} \Big) \cdot \Big[\phi(\om,q)+\phi(\om,-q) \Big] \\
\; - \; \f{1}{2\pi L } \sul{p=2}{p_{\e{max}} } \sul{ a = 1 }{ n_p^{(z)} } \Big( \e{id} \, - \, \op{R} \Big) \bigg[ \vartheta^{(\e{reg})}_{p;a}\big( *-c_{a}^{(p)}\mid \{\de_{a,k}^{(p)} \} \big) \, - \, \th_{p,1}\big(*-c_{a}^{(p)}\big)  \bigg]
%
%
\end{multline}
where $\sg_{\ups}$ has been defined in \eqref{definition parametres sg L et R}, $\e{id}-\op{R}$ is the inverse to $\e{id}+\op{K}$ and $R(\la,\mu)$ stands for the integral kernel of the resolvent. 
Given  $\de>0$ small enough and for any $k$, one has the bounds
\beq
\Norm{ \mf{R}_N\big[\, \wh{\xi}_{\Ups}\big] }_{ W_{k}^{\infty}\big( \mc{S}_{\de}(\R) \big) } \; = \; \e{O}\big( L^{-2} \big)  \qquad and \qquad 
  \Norm{  \, \wh{F}_{\e{reg}}  - F \, }_{ W_{k}^{\infty}\big( \mc{S}_{\de}(\R) \big) } \; = \; \e{O}\big( L^{-1} \big)   \;. 
\label{estimee reste NLIe et deviation Freg a F}
\enq

\end{prop}

Below, I only sketch the proof and solely insist on the most important points since these will play a role in the various other estimates carried out in the paper. 
I refer to \cite{KozProofOfStringSolutionsBetheeqnsXXZ} for more details.

\Proof 

To start with, observe that  the function 
\beq
\wh{u}_{\Ups}^{\,\prime}(s) \; = \; \f{ 2\i\pi L \, \wh{\xi}_{\Ups}^{\prime}(s) }{ \ex{2\i\pi L  \wh{\xi}_{\Ups}(s)} - 1   } \quad \e{has}\, \e{simple}\, \e{poles} \,\e{at} \qquad   
\left\{ \ba{cl} \Ups^{(\e{in})}\cup\big\{ \mu_a^{(s)} \big\}_{ 1 }^{  n_{\e{sg}} }  & \e{with} \; \e{ residue } \; 1  \vspace{2mm} \\
	\big\{ \be_a^{(s)} \big\}_{ 1 }^{  n_{\e{sg}} }   &  \e{with} \;  \e{residue} \;  -1 \ea \right. 
\label{ecriture systeme poles et zeros u ups prime}
\enq
and that these are its only poles inside of $\msc{C}$. The regular part of the counting function can thus be recast as
\beq
 \wh{\xi}_{\Ups_{\e{reg}}}(\om) \, = \, \f{ \vth \big( \om \mid \tf{\zeta}{2} \big) }{2\pi} \, - \, \Oint{ \msc{C} }{}  \th(\om-s) \wh{u}^{\, \prime}_{\Ups}(s)\f{ \dd s }{ 4\i \pi^2 L  }
\, + \, \mf{R}^{(1)}(\om)
\, + \, \f{1}{L} \Theta\Big(\om\mid \Ups^{(h)}; \Ups_{\e{tot}}^{(z)} \Big)
\, + \, \f{ |\Ups|+1-2 \a_{\Ups} }{2L} 
\enq
where 
\beq
\Theta\Big(\om\mid \Ups^{(h)}; \Ups_{\e{tot}}^{(z)} \Big) \; = \; \f{1}{2\pi  }\hspace{-3mm} \sul{ \mu \in \Ups^{(h)}\setminus \Ups^{(p)}  }{  } \hspace{-2mm} \th(\om-\mu) 
\, - \, \f{ 1 }{ 2\pi } \sul{p=2}{ p_{\e{max}} } \sul{ a=1 }{ n_p^{(z)} } \th_{p,1}\big(\om -c_a^{(p)} \big)
\enq
and, using the notation \eqref{ecriture convention somme et prod racines sing}, the remainder takes the form 
\beq
\mf{R}^{(1)}(\om) \, = \,  - \f{1 }{2\pi L }  \sul{ \mu  \in \daleth }{  }  \th\big( \om - \mu\big) 
\, - \, \f{ 1 }{ 2\pi L } \sul{p=2}{ p_{\e{max}} } \sul{ a=1 }{ n_p^{(z)} } \Big\{ \vth^{(\e{reg})}_{p;a}\big(\om-c_a^{(p)} \mid \{ \de_{a,k}^{(p)} \} \big)  \, - \,  \th_{p,1}\big(\om -c_a^{(p)} \big)  \Big\}\;. 
\enq
Upon integrations by part, the contour integral can be recast as 
\bem
\, - \, \Oint{ \msc{C} }{}  \th(\om-s) \wh{u}^{\, \prime}_{\Ups}(s)\f{ \dd s }{ 4\i \pi^2 L  } \, = \, -\Int{ \wh{q}_L }{ \wh{q}_R }K(\om-s) \cdot \wh{\xi}_{ \Ups_{\e{reg}} }^{\, (\e{sym})}(s) \cdot \dd s
-\mf{d}\big[ \,\wh{u}_{\Ups}^{\, (\e{sym})} \big] \big(\, \wh{q}_R \big) \cdot \f{ \th(\om-\wh{q}_R) }{ 2\pi  L }\\
 \, + \,  \mf{d}\big[ \,\wh{u}_{\Ups}^{\, (\e{sym})} \big] \big( \,  \wh{q}_L \big) \cdot \f{ \th(\om-\wh{q}_L) }{ 2\pi  L } 
 \, -\; \sul{\eps= \pm }{} \Int{ \msc{C}^{(\eps)} }{}  K(\om-s) \Big[\, \wh{u}^{\,(\eps)}_{\Ups}(s) \,- \, 2\i\pi L \de_{\eps;+}\wh{\xi}_{\Ups_{\e{sing}}}(s)  \Big] \cdot \f{ \dd s }{ 2\i\pi L } \;. 
\end{multline}
Here, I agree upon
\beq
\wh{\xi}_{ \Ups_{\e{reg}} }^{\; (\e{sym})} \, = \,  \wh{\xi}_{ \Ups_{\e{reg}} }-\tf{(|\La|+1)}{2L} \; , \qquad   \wh{u}_{\Ups}^{\, (\e{sym})} \, = \,  \wh{u}_{\Ups} +\i\pi (|\La|+1) \bs{1}_{\mathbb{H^+}} 
\enq
and that,  given a  piecewise continuous function $f$ on $\msc{C}$ and continuous on $\msc{C}^{(\pm)}=\msc{C}\cap \mathbb{H}^{\pm}$, the operator $\mf{d}[f]$ is defined by
\beq
\mf{d}[f]\big( s \big) \,  = \, \f{-1}{2\i\pi} \lim_{\eps\tend 0^+} \Big\{ f(s+\i\eps)-f(s-\i\eps)  \Big\} \;. 
\label{definition symbole de f}
\enq
By using that, for $\sg \in \intff{ - \tf{1}{2} }{ \tf{1}{2} }$, one has  
\beq
\ln \big[ 1+\ex{2\i\pi \sg} \big]\, - \, \ln \big[ 1+\ex{-2\i\pi \sg} \big] \, = \; 2\i\pi \sg\; , 
\label{condition saut log neper}
\enq
the variations of $\wh{u}_{\Ups}^{\, (\e{sym})}$ can be recast as 
\beq
 \mf{d}\big[\,\wh{u}_{\Ups}^{\, (\e{sym})}\big]( \wh{q}_{\ups} \big) \; = \; L \, \wh{\xi}_{\Ups}^{\; (\e{sym})}\big( \, \wh{q}_{\ups} \big) \,+\,  L \, \wh{\xi}_{\Ups_{ \e{sing} } } \big( \, \wh{q}_{\ups} \big)  
 \, + \wh{F}\big( \, \wh{q}_{\ups} \big)_{\mid _{\a_{\La}=0} } - \varkappa_{\ups}-\tau_{\ups} \;, 
\qquad  \ups \in \{L, R\}\;. 
\enq
After some additional algebra, one gets that 
\beq
\, - \, \Oint{ \msc{C} }{}  \th(\om-s) \wh{u}^{\, \prime}_{\Ups}(s)\f{ \dd s }{ 4\i \pi^2 L  } \, = \, - \op{K}\Big[   \wh{\xi}_{ \Ups_{\e{reg}} }^{\, (\e{sym})} \Big](\om) 
\, - \, \Big( \f{D}{2} -\f{\varkappa_{R}}{L}  \Big) \f{ \th\big(\om-q \, \big) }{ 2\pi } \, - \, \Big( \f{D}{2} + \f{\varkappa_{L}}{L}  \Big) \f{ \th\big(\om+q \, \big) }{ 2\pi }
\, + \; \mf{R}^{(2)}\Big[ \wh{\xi}_{ \Ups } \Big](\om)
\nonumber
\enq
where the remainder term takes the form 
\bem
\mf{R}^{(2)}\big[ \wh{\xi}_{ \Ups } \big](\om) \; = \; - \sul{\eps=\pm 1 }{} \Int{ \msc{C}^{(\eps)} }{} K\big(\om - s \big) \cdot \bigg\{  \f{ \wh{u}^{\,(\eps)}_{\Ups}(s) }{ 2\i\pi L }\, - \, \de_{\eps,+}  
\, \wh{\xi}_{\Ups_{\e{sing}}}(s) \bigg\} \cdot \dd s   \, + \, \Big(D-\f{|\La|}{L}\Big) \f{ \th\big(\om-q \, \big) +\th\big(\om + q \, \big)   }{ 4\pi }   \\
+ \sul{ \ups\in \{R,L\}  }{} \sg_{\ups}\Bigg\{ \f{-1}{2\pi L} \Big[ L \wh{\xi}_{ \Ups_{\e{sing}} }(\, \wh{q}_{\ups})  \, + \, \wh{F}(\, \wh{q}_{\ups})_{\mid _{\a_{\La}=0} } -\varkappa_{\ups}-\tau_{\ups}    \Big] 
\cdot \Big[ \th(\om- \wh{q}_{\ups}) \,- \, \th(\om- \sg_{\ups} q \, )   \Big] \\
\, + \,  \Int{ \sg_{\ups} q }{ \wh{q}_{\ups} } K\big(\om - s \big) \cdot \Big[ \wh{\xi}_{ \Ups_{\e{reg}} }(s)- \wh{\xi}_{ \Ups_{\e{reg}} }(\, \wh{q}_{\ups}) \Big] \cdot \dd s \Bigg\} \;. 
\nonumber
\end{multline}
Thus, the regular part of the counting function satisfies to the non-linear integral equation
\bem
\Big( \e{id}+\op{K} \Big) \Big[    \wh{\xi}_{ \Ups_{\e{reg}} }^{\, (\e{sym})}  \Big](\om) \; = \; 
\f{ \vth \big( \om \mid \tf{\zeta}{2} \big) }{2\pi} \, - \, \f{D}{4\pi L} \Big( \th\big(\om-q\,\big) + \th\big(\om+q\,\big)  \Big) \, + \, \f{1}{L} \Theta\Big(\om\mid \Ups^{(h)}; \Ups_{\e{tot}}^{(z)} \Big)  \\
+\f{1}{2\pi L}  \Big( \varkappa_{R} \th\big(\om-q\, \big)  \, - \, \varkappa_{L} \th\big(\om + q \, \big)  \Big) 
\, + \, \f{ |\Ups|-|\La|-2 \a_{\Ups} }{2L} \, + \, \mf{r}^{(1)}(\om) \, + \,  \mf{r}^{(2)}\big[ \wh{\xi}_{ \Ups } \big](\om)  \;. 
\end{multline}
The representation \eqref{ecriture DA fct de cptage} follows upon inverting the operator  $\e{id}+\op{K}$. The bounds on the remainder $\mf{R}_{N}\big[ \, \wh{\xi}_{\Ups} \big]$ result
from the estimates \eqref{ecriture DA racine q har R et L} and  \eqref{ecriture estimee hat u Omega reg sur C en L1} as well as the fact that, for $p\geq 3$,  
\beq
 \vth^{(s)}_{p;a}\big(\om \mid \{ \de_{a,k}^{(p)} \} \big)  \; = \; \th_{p,1}\big(\om  \big)  \,  +  \, \e{O}\big( \,  \wt{\de}  \, \big)   \qquad \e{with} \qquad \wt{\de} \, = \,  \e{max}\big| \de^{(p)}_{a,k} \big| 
\enq
where $\th_{p,1}$ has been defined in \eqref{definition phase nue typre r p}. This allows one to conclude on the negligibility of the remainder. 
\qed

\hspace{3mm}

\subsection{Asymptotic expansion of the excitation energy and momentum}

The  momentum  and energy of an Eigenstate $\ket{\Ups}$ are given by the expressions \cite{OrbachXXZCBASolution}
\beq
\wh{\mc{P}}_{\Ups} \; = \; \sul{\mu \in \Ups}{}  \vth\big(\mu \mid \tf{\zeta}{2} \big)  \qquad \e{and} \qquad 
\wh{\mc{E}}_{\Ups} \; = \; \Big( J \De \, - \,  \f{h}{2} \Big)L \; + \; \sul{\mu \in \Ups}{} \mf{e}(\mu) 
\enq
where the bare energy $\mf{e}$ is as given in \eqref{definition energie habille et energie nue}. The relative excitation energy $\wh{\mc{E}}_{  \Ups\setminus \La  } $ 
and momentum $\wh{\mc{P}}_{\Ups\setminus \La  } $ of an excited state $\ket{\Ups}$ in respect to the  ground state $\ket{\La}$ are defined as the differences
\beq
\wh{\mc{E}}_{\Ups\setminus \La} = \wh{\mc{E}}_{\Ups} - \wh{\mc{E}}_{\La}  \qquad \e{and} \qquad 
\wh{\mc{P}}_{\Ups\setminus \La} = \wh{\mc{P}}_{\Ups} - \wh{\mc{P}}_{\La}  \; .
\label{ecriture energie en impulsion excitation}
\enq

Below, I establish the large-$L$  expansion of the slightly more general quantities 
$\wh{\mc{E}}_{ \Ups \setminus \La_{\mf{b}}^{\!(\a)} } $ and $\wh{\mc{P}}_{ \Ups \setminus \La_{\mf{b}}^{\!(\a)}  } $. 
The latter is a rather direct consequence of the two lemmata below.

\begin{lemme}
\label{Lemme DA somme relative Ups vs La}

Let $f$ be holomorphic in an neighbourhood of $I_{q}$, regular on $\Ups^{(z)}_{\e{tot}}$ and assume that there exists $\de_{f}>0$ such that  
\beq
f^{\prime} \, \in  \, L^{\infty}\bigg(  \mc{S}_{\de_{f}} \Big(\Big\{\R+\i k \tfrac{\zeta}{2}+\i s \tfrac{\pi}{2}, k\in \intn{ -2(p_{\e{max}}-1) }{ 2(p_{\e{max}}-1) } , s \in \{0,1\} \Big\}  \Big)  
\setminus \mc{D}_{\pm \i \frac{\zeta}{2},\frac{1}{2}\de_{f}} \bigg)
\label{conditions de dominantion de f pour reecrire sommes finies}
\enq
and
\beq
\big| f^{\prime}(z) \big| \, \leq \, \f{ C}{ d^{\ell}\big(z, \pm \i \tf{\zeta}{2} \big) }   \qquad on \qquad \mc{D}_{\pm \i \frac{\zeta}{2},\de_{f} } \setminus\{ \pm \i \tf{\zeta}{2}  \} \;, 
\label{condions domination locale f}
\enq
for  some $C>0$ and $\ell \in \mathbb{N}$. Further, assume that \eqref{hypothese sur borne inf sur espacement des diverses racines} holds. 

Let 
\beq
\mc{S}_{\Ups\setminus  \La_{\mf{b}}^{\!(\a)} }[f]  \; = \; \sul{ \mu \in \Ups}{} f(\mu) \, - \, \sul{ \la \in \La_{\mf{b}}^{\!(\a)}  }{} f(\la) \;. 
\enq
Then, it holds
\bem
\mc{S}_{\Ups\setminus \La_{\mf{b}}^{\!(\a)}  }[f] \; = \; \Big( \a \,+ \, \tfrac{|\La|-|\Ups|}{2} \Big) \Int{-q}{q} Z(s) f^{\prime}(s) \dd s \; + \;
\sul{r=2}{p_{\e{max}} } \sul{a=1}{n_r^{(z)} } \op{v}_{r}[f]\big( c_a^{(r)}\big) \; + \; \mf{b} \hspace{-2mm} \sul{\ups \in \{L,R\} }{}  \hspace{-2mm} \sg_{\ups} f\big(\sg_{\ups} q \big) \\
\, + \hspace{-3mm} \sul{  \mu \in    \Ups^{(p)}\setminus \Ups^{(h)}  }{} \hspace{-2mm} \op{v}_1[f](\mu)  \, - \, \sul{\ups\in \{L,R\}}{} \sg_{\ups} \varkappa_{\ups} \op{v}_1[f](\sg_{\ups} q )  \; + \; \e{O}\Big( L^{-1} \Big) 
\label{definition somme relative Ups vs La}
\end{multline}
where $\a$ is as in \eqref{definition fct V Ups et La}, 
\beq
\op{v}_{r}[f](\om) \; = \; f_{r}(\om) \,-\, \Int{-q}{q} f(s) \Dp{s} \phi_{r,1}\big(s,\om\big) \cdot \dd s 
\, + \sul{ \ups \in \{L,R\}}{} \hspace{-2mm} \sg_{\ups} f\big( \sg_{\ups}q\big) \phi_{r,1}\big(\sg_{\ups}q , \om \big) 
\enq
with
\beq
f_r(\om) \, = \, \lim_{\eps\tend 0^+} \sul{k=1}{r} f\Big( \om + \i\tfrac{\zeta}{2} \big(r+1-2k \big) +\i\eps \Big)  \;. 
\enq
\end{lemme}
\Proof 
Given $\Om \in \{\Ups, \La_{\mf{b}}^{\!(\a)} \}$, I agree upon 
\beq
 \left\{ \ba{cc} \wh{\xi}_{\Om_{\e{reg}} } \; = \; \wh{\xi}_{\Ups_{\e{reg}} }   \\ 
					    \wh{\xi}_{\Om_{\e{sing}} } \; = \; \wh{\xi}_{\Ups_{\e{sing}} }   \ea \right. \;\;  \e{for} \quad \Om=\Ups \; \quad  \e{and} \quad
 \left\{ \ba{cc} \wh{\xi}_{\Om_{\e{reg}} } \; = \; \wh{\xi}_{\La^{\!(\a)} }^{\, (\mf{b})}   \\ 
	 \wh{\xi}_{\Om_{\e{sing}} } \; = \; 0   \ea \right.   \e{for} \quad \Om=\La^{\!(\a)}_{\mf{b}}   \;. 
\label{definition Uomega sing et reg}
\enq
Then, for such an $\Omega$, by repeating the handlings outlined in the proof of Propositon \ref{Proposition DA ctg fct} and using that 
\beq
\wh{u}_{\Om}^{(+)}(\, \wh{q}_{\ups} ) -  \wh{u}_{\Om}^{(-)}(\, \wh{q}_{\ups} )  \,  =  \, 2\i\pi \tau_{\ups} +2\i\pi \de_{\Om;\Ups} \Big( \varkappa_{\ups}-\wh{F}(\,\wh{q}_{\ups}) \Big)_{\mid_{ \a_{\La}=0 } }
\, + \, 2\i\pi \mf{f}_{\ups} \de_{\Om;\La^{\!(\a)}_{\mf{b}}} \; , 
\enq
where $\mf{f}_{\ups} = L\Big(\,  \wh{\xi}^{\,(\mf{b})}_{ \La^{\!(\a)} }(\,\wh{q}_{\ups}) - \wh{\xi}_{ \La }(\,\wh{q}_{\ups}) \Big)$, 
one gets 
\bem
\sul{\mu \in \Om}{} f(\mu) \, = \,  L \Int{ \wh{q}_L }{ \wh{q}_R } \wh{\xi}_{\Om_{\e{reg}}}^{\, \prime}(s) f(s) \cdot \dd s \, - 
\sul{ \ups \in \{L,R\}}{} \hspace{-2mm} \sg_{\ups} \Big\{ \tau_{\ups} +\de_{\Om;\Ups} \Big( \varkappa_{\ups}-\wh{F}(\,\wh{q}_{\ups}) \Big)_{\mid_{ \a_{\La}=0 } }\, + \, \mf{f}_{\ups} \de_{\Om;\La^{\!(\a)}_{\mf{b}}} \Big\} f\big(\, \wh{q}_{\ups} \big)\\
\, + \, \de_{\Om;\Ups}\hspace{-3mm} \sul{ \substack{ \mu \in \Ups^{(z)}\cup \Ups^{(p)} \\ \cup \daleth \setminus \Ups^{(h)} }      }{} \hspace{-3mm}  f(\mu)
-\,\sul{\eps=\pm}{} \Int{ \msc{C}^{(\eps)} }{} \Big\{ \, \wh{u}^{\, (\eps)}_{\Om}(s) \, + \, 2\i\pi L \, \wh{\xi}_{\Om_{\e{sing} }}^{\, \prime} (s) \Big\} f(s) \cdot \f{ \dd s }{ 2 \i \pi } \;. 
\end{multline}
Here, I agree upon 
\beq
\de_{\Om;\Ups}=1 \quad \e{if} \quad  \Om=\Ups \quad \e{and} \quad  \de_{\Om;\Ups}=0 \quad \e{if} \quad \Om=\La^{\!(\a)}_{\mf{b}} \; . 
\label{definition kronecker of a set}
\enq

The last integral can be estimated, by means of Lemmata \ref{Lemme borne sur magnitude fct cptge sing} and \ref{Lemme estimation propriete generales ctg fct}, to be a $\e{O}(L^{-1})$. Likewise, the bound 
\eqref{ecriture deviation des mua au bea} ensures that $\sum_{ \mu \in   \daleth       }^{}   f(\mu) \, = \, \e{O}(L^{-\infty})$ and \eqref{ecriture bornes ctg fct sing sur C} yields 
$\wh{F}(\,\wh{q}_{\ups})\,= \, \wh{F}_{\e{reg}}(\,\wh{q}_{\ups})\, + \, \e{O}\big( n_{\e{sg}}L^{-\infty} \big)$. Thus, one gets that 
\beq
\mc{S}_{\Ups\setminus \La^{\!(\a)}_{\mf{b}} }[f] \; = \; -\Int{ \wh{q}_{L} }{ \wh{q}_{R}  } f(s) \wh{F}_{\e{reg}}^{\, \prime}(s) \dd s 
\, - \, \sul{ \ups \in \{L,R\} }{} \sg_{\ups} \cdot   \Big( \varkappa_{\ups}-\wh{F}_{\e{reg}}(\,\wh{q}_{\ups}) -\mf{b} \Big)f(\,\wh{q}_{\ups}) \; + \hspace{-3mm}
\sul{ \mu \in \Ups^{(z)}\cup \Ups^{(p)} \setminus \Ups^{(h)} }{} \hspace{-3mm} f(\mu) \; + \; \e{O}\Big( L^{-1} \Big) \;. 
\enq
It then remains to recall  the form \eqref{ecriture DA fct de cptage} of the asymptotic
expansion of $\wh{\xi}_{\Ups_{\e{reg}}}$, the bounds \eqref{ecriture DA racine q har R et L} on the deviations of the endpoints $\wh{q}_{\ups}$
from $\sg_{\ups} q$ and to neglect the exponentially small string deviations
what can be achieved by using either the $L^{\infty}$ property in \eqref{conditions de dominantion de f pour reecrire sommes finies}
or, in the case of strings of even length, the bounds in \eqref{condions domination locale f} and hypothesis \eqref{hypothese sur borne inf sur espacement des diverses racines}. 
Doing so produces  $\e{O}\big( n^{(z)}_{\e{tot}}\cdot L^{-\infty}\big)$ corrections.  \qed

\begin{lemme}
\label{Lemme reecriture operateur vp}
Let $\mf{t}[f]$ be the unique solution to the linear integral equation 
\beq
\Big(\e{id}\, + \, \op{K} \Big)\big[\mf{t}[f] \big] (\om) \, = \, f(\om) - \tfrac{1}{2\pi} \sul{\ups \in \{L,R\} }{} \sg_{\ups} \mf{t}[f](\sg_{\ups} q)  \th(\om-\sg_{\ups} q) 
\enq
and let $ \mf{t}_r[f](\om) \, = \, \lim_{\eps \tend 0^+}\sum_{k=1}^{r}  \mf{t}[f]\Big( \om + \i\tfrac{\zeta}{2} \big(r+1-2k \big) +\i\eps \Big)$. 

Then, for $\om \in \R\cup \{\R+\i\tf{\pi}{2}\}$, it holds
\beq
\op{v}_r[f](\om) \, = \, \mf{t}_r[f](\om) \, + \, \tfrac{ \wt{m}_r(\zeta , \om) }{ 2 } \cdot \sul{\ups \in \{L,R\} }{} \sg_{\ups} \mf{t}[f](\sg_{\ups} q)  \;, 
\label{ecriture relation vr de f comme taur de f}
\enq
with
\beq
\wt{m}_r\big(\zeta , \om\big) \; = \; m_r(\zeta) \, - \, 2 \bs{1}_{\R+\i\frac{\pi}{2}}(\om) 
\bigg\{  \sul{\eps=\pm}{} \e{sgn}\bigg( 1 \, +\,  2\lfloor \f{r+\eps}{2\pi}\zeta \rfloor \, - \, \f{ r+\eps }{ \pi }\zeta\bigg)   \; - \; \de_{r,1} \bigg\}
\label{definition fct mr de zeta et omega}
\enq
$m_r(\zeta)$ as defined in \eqref{ecriture somme partielle sur corde pĥase nue et definition entier mr de zeta}. One also has
\beq
\Int{-q}{q} Z(s) f^{\prime}(s) \cdot \dd s  \;  =\; \sul{ \ups \in \{L,R\} }{} \sg_{\ups} \mf{t}[f](\sg_{\ups} q) 
\label{ecriture integrale f contre Z}
\enq
\end{lemme}

\Proof 

By taking derivatives and integrating by parts, one gets that $\Dp{s}\phi_{r,1}(s,\mu)$ solves the linear integral equation
\beq
\Big( \e{id}+\op{K} \Big)\big[  \Dp{*}\phi_{r,1}(*,\mu) \big](\om) \; = \; K_{r,1}(\om-\mu) \, +  \hspace{-2mm} \sul{\ups \in \{L,R\} }{}\hspace{-2mm} \sg_{\ups} K\big(\om-\sg_{\ups}q\big)\,  \phi_{r,1}\big(\sg_{\ups}q,\mu\big) \;. 
\enq
Likewise, owing to \eqref{ecriture somme partielle sur corde pĥase nue et definition entier mr de zeta},  one readily infers that $\mf{t}_r[f]$ can be expressed as
\beq
\mf{t}_r[f](\om) \; = \; f_r(\om) \, -\,  \Int{-q}{q} K_{r,1}(\om-s) \mf{t}[f](s) \dd s \, -  \sul{ \ups \in \{L , R \} }{} \hspace{-2mm}  \sg_{\ups} \mf{t}[f](\sg_{\ups} q) \cdot 
\Big\{\tfrac{1}{2\pi}\th_{r,1}\big( \om - \sg_{\ups} q \big) + \tfrac{m_r(\zeta)}{2}   \Big\}
\enq
The first identity in \eqref{ecriture relation vr de f comme taur de f} then follows upon substituting $f$ and $f_r$ in terms of $\mf{t}[f]$ and $\mf{t}_r[f]$ into the definition of $\op{v}_r[f]$ 
and using that $\th_{r,1}$ is odd on $\R$ while
\beq
\th_{r,1}(u)+\th_{r,1}(-u) - 2\pi \de_{r,1} \, = \, -2\pi \sul{\eps=\pm}{} \e{sgn}\bigg( 1 \, +\,  2\lfloor \f{r+\eps}{2\pi}\zeta \rfloor \, - \, \f{ r+\eps }{ \pi }\zeta\bigg) \qquad \e{when} \quad u \in \R+\i\f{\pi}{2} \;.
\enq
 Finally, the last identity \eqref{ecriture integrale f contre Z} is a consequence of $Z=\big(\e{id}+\op{K} \big)^{-1}[1]$ and the fact that 
$\big(\e{id}+\op{K} \big)\big[  \big( \mf{t}[f] \big)^{\prime} \big]=f^{\prime}$. \qed

One in now in position to establish the large-$L$ expansion of $\wh{\mc{E}}_{\Ups\setminus \La_{\mf{b}}^{\!(\a)} }$ and $\wh{\mc{P}}_{ \Ups\setminus \La_{\mf{b}}^{\!(\a)} }$
\begin{prop}
\label{Proposition DA energie et impulsion}
Let $\mf{b}=\varrho+\e{O}(L^{-1})$. One has the large-$L$ expansion 
\beq
\wh{\mc{E}}_{\Ups\setminus \La_{\mf{b}}^{\!(\a)} } \, = \, \sul{ \mu \in \Ups^{(p)} }{} \veps(\mu)  \, - \, \sul{ \mu \in \Ups^{(h)} }{} \veps(\mu)   \, + \, 
					      \sul{r=2}{p_{\e{max}} } \sul{a=1}{n_r^{(z)} } \veps_{r}\big( c_a^{(r)}\big)   \; + \; \e{O}\Big( L^{-1} \Big) 
\enq
and, with $\a$ as in \eqref{definition fct V Ups et La}, 
\bem
\wh{\mc{P}}_{\Ups\setminus \La_{\mf{b}}^{\!(\a)} } \, = \,  \sul{r=2}{p_{\e{max}} } \sul{a=1}{n_r^{(z)} } p_{r}\big( c_a^{(r)}\big) \, + \hspace{-3mm} \sul{ \mu \in \Ups^{(p)}\setminus \Ups^{(h)}   }{} \hspace{-3mm} p(\mu)  
\, + \, 2\varrho \th\big(q\mid \tfrac{\zeta}{2} \big) \\
\; + \; \Big( 2\a + |\La|-|\Ups| + \sul{p=2}{ p_{\e{max}} } \sul{a=1}{n_p^{(z)}}\wt{m}_p\big(\zeta,c_a^{(p)} \big) \, + \, \sul{\mu \in \Ups^{(p)} }{} \wt{m}_1\big(\zeta,\mu \big)  \, - \, \varkappa_L \, - \, \varkappa_R \Big) p(q) 
\; + \; \e{O}\Big( L^{-1} \Big) 
\end{multline}
where $\wt{m}_p(\zeta, \om ) $ is as defined by \eqref{definition fct mr de zeta et omega}. 

\end{prop}
\Proof 

First observe that $\wh{\mc{E}}_{ \Ups \setminus \La_{\mf{b}}^{\!(\a)} } \, =  \, \mc{S}_{ \Ups \setminus \La_{\mf{b}}^{\!(\a)} } [ \mf{e}  ]$ and   $\wh{ \mc{P} }_{ \Ups \setminus \La_{\mf{b}}^{\!(\a)}}  \, = \, 
\mc{S}_{ \Ups \setminus \La_{\mf{b}}^{\!(\a)}}\big[ \vth\big( * \mid \tf{\zeta}{2} \big) \big]$
where $\mc{S}_{\Ups\setminus \La}$ is as defined in \eqref{definition somme relative Ups vs La}. 
Since both $\vth\big( * \mid \tf{\zeta}{2} \big)$ and  $\mf{e}$ satisfy to the property \eqref{conditions de dominantion de f pour reecrire sommes finies}, 
Lemmata \ref{Lemme DA somme relative Ups vs La} and \ref{Lemme reecriture operateur vp} guarantee that one has 
\bem
\mc{S}_{\Ups\setminus \La_{\mf{b}}^{\!(\a)}}[f]  \; =   \hspace{-2mm}  \sul{\ups\in \{L,R\}}{} \hspace{-2mm} \sg_{\ups} \Big( \a \,+ \, \tfrac{|\La|-|\Ups|}{2} 
\, + \, \tfrac{1}{2} \sul{ r=2 }{ p_{\e{max}} }\sul{a=1}{n_r^{(z)}}\wt{m}_r\big(\zeta,c_a^{(r)} \big)  \, + \,  \tfrac{1}{2} \sul{\mu \in \Ups^{(p)} }{} \wt{m}_1\big(\zeta,\mu \big) 
\, - \, \varkappa_{\ups}  \Big)   \cdot  \mf{t} [f](\sg_{\ups} q )    \\
\, + \, \mf{b} \sul{\ups\in \{L,R\}}{} \hspace{-2mm} \sg_{\ups}   f\big( \sg_{\ups} q  \big) 
\; + \;\sul{r=2}{p_{\e{max}} } \sul{a=1}{n_r^{(z)} } \mf{t}_{r}[f]\big( c_a^{(r)}\big) \, + \, \sul{ \mu \in \Ups^{(p)}\setminus \Ups^{(h)} }{} \mf{t}[f](\mu)  \; + \; \e{O}\Big( L^{-1} \Big) \;. 
\end{multline}
It then only remains to observe that 
\beq
\mf{t}[ \vth\big( * \mid \tf{\zeta}{2} \big)](\mu) \; = \;  p(\mu) \qquad \e{and} \qquad 
\mf{t}[ \mf{e}](\mu) \, = \, \veps(\mu)  
\enq
so as to conclude. \qed 



\subsection{Parametrisation in the thermodynamic limit}
\label{Sous Section parametrisation limite thermo}

Proposition \ref{Proposition DA energie et impulsion} shows that the particle, hole and string roots correspond to different kinds of excitations in the model.
Owing to \eqref{ecriture positivite energie r corde} one concludes that the strings correspond to various kinds of massive excitations.
Since $\veps$ changes sign at $\pm q$, \textit{c.f.} \eqref{ecriture ppte sgn energie habille}, and that $|\veps(\la)|> c >0$ uniformly away from $\pm q$ on $\R \cup \{ \R+\i\tf{\pi}{2}\}$, 
one infers that the particle and hole excitations can give
rise to massive and massless excitations depending on the proximity of their parameters to the endpoints of the Fermi zone. If a particle or hole 
root collapses, when $L\tend +\infty$, on the Fermi boundary then it will produce a vanishing, in the thermodynamic limit, contribution to the excitation 
energy of that state. However, if it stays at finite distance from the Fermi zone, then owing to \eqref{ecriture ppte sgn energie habille}, it will generate a 
finite positive contribution and hence correspond to a massive excitation. 

Massive and massless excitation contribute rather differently to the large-$L$ behaviour of a form factor. Thus, it is convenient to
distinguish between such excitations and decompose the particle and hole sets $\Ups^{(p)}$ and $\Ups^{(h)}$ as
\beq
\Ups^{(p)} \; = \; \Ups^{(p)}_{\e{off}} \, \cup \,  \Ups^{(p)}_{R} \,  \cup \, \Ups^{(p)}_{L} \qquad \e{and} \qquad 
\Ups^{(h)} \; = \; \Ups^{(h)}_{\e{off}} \, \cup \,  \Ups^{(h)}_{R} \,  \cup \, \Ups^{(h)}_{L} \;. 
\label{ecriture partition ens part trou spectre massif et massless}
\enq

The roots contained in $ \Ups^{(p)}_{R/L}$ and  $ \Ups^{(h)}_{R/L}$ generate the massless excitations in the model. 
These roots correspond to the solutions to $\wh{\xi}_{\Ups}\big( \mu \big) \, = \,  \tf{ m }{ L }$ where 
\beq
m \in \left\{ \ba{cc}  \lfloor L \, \wh{\xi}_{\Ups}\big( \, \wh{q}_{\ups} \big)  \rfloor + \sg_{\ups} \big\{ 1, 2,\dots  \big\}  &\e{if}  \; \mu \in \Ups^{(p)}_{\ups} \vspace{1mm} \\ 
  \lfloor L \, \wh{\xi}_{\Ups}\big( \,  \wh{q}_{\ups} \big)  \rfloor - \sg_{\ups} \big\{ 0, 1,\dots  \big\}  &\e{if}  \; \mu \in \Ups^{(h)}_{\ups} \ea \right.   \;\; , \quad \ups \in \{L,R\}\; . 
\enq
In other words, the roots giving rise to the massless excitations can be parametrised by integers $p_a^{\ups}, h_a^{\ups} \in \mathbb{N}$, $\ups \in \{L,R\}$, so that they solve
\beq
\wh{\xi}_{\Ups}\big( \mu \big)  \; = \; \f{1}{L} \big( |\La| - \varkappa_{R} + p_a^{R} + 1 \big)  \quad \e{for} \quad \mu \in \Ups^{(p)}_{R} \; , \qquad 
\wh{\xi}_{\Ups}\big( \mu \big)  \; = \; \f{1}{L} \big( |\La| - \varkappa_{R} - h_a^{R} \big)  \quad \e{for} \quad  \mu \in \Ups^{(h)}_{R} 
\label{definition entiers particule et trou bord droit}
\enq
and 
\beq
\wh{\xi}_{\Ups}\big( \mu \big)  \; = \; \f{1}{L} \big( -\varkappa_{L}-p_a^{L}\big)  \quad \e{for} \quad  \mu \in \Ups^{(p)}_{L} \; ,  \qquad 
\wh{\xi}_{\Ups}\big( \mu \big)  \; = \; \f{1}{L} \big( -\varkappa_{L}+h_a^{L}+1\big)  \quad \e{for} \quad  \mu \in \Ups^{(h)}_{L}  \;. 
\label{definition entiers particule et trou bord gauche}
\enq
Here, the integers $\varkappa_L, \varkappa_{R}$ are precisely those arising in \eqref{bornes sur fct shift en q hat left right}. 
This representation implies that, for any $\mu \in \Ups^{(p/h)}_{\ups}$, 
\beq
\big| \mu-\sg_{\ups} q \big| \; = \;\e{O}\Big( \f{ |k_a^{\ups}|  }{ L }  \Big)
\label{definition estimee m}
\enq
where $k_a^{\ups}$ is the integer arising in the definition \eqref{definition entiers particule et trou bord droit} or \eqref{definition entiers particule et trou bord gauche} of the root $\mu$. 
Note that $\varkappa_{\ups}$ does not arise in the remainder \eqref{definition estimee m} owing to the condition \eqref{bornes sur fct shift en q hat left right} and hypothesis \eqref{propriete espacement ctre corde et particule trou}
which ensures that for any $\mu \in \Ups^{(p/h)}_{R/L}$ one has $\wh{F}=\wh{F}_{\e{reg}} \, + \, \e{O}\big( n_{\e{sg}} L^{-\infty} \big)$.  

The off-boundary roots $\Ups^{(p/h)}_{\e{off}}$ generate massive particle-hole excitations. They solve the equation $\wh{\xi}_{\Ups}\big( \mu \big)  \; = \; \tf{m}{L} $  were $m$ is 
such that $\lim (\tf{m}{L})  \not\in \big\{0,D\}$. 
Thus, such roots are located uniformly away from $\intff{-q}{q}$. 

For further purposes,  it will appear convenient to introduce the shorthand notations for the number and for the  discrepancies between the number of particles and holes in the left, right and off-boundary collections of roots: 
\beq
n_{\ups}^{(p/h)} \, = \, \big| \Ups^{(p/h)}_{ \ups } \big| \qquad \e{and} \; \e{set} \qquad 
\ell_{\ups} \; = \; \big| \Ups^{(p)}_{ \ups } \big|  \, - \,  \big| \Ups^{(h)}_{ \ups } \big| \; \quad \e{for} \quad \;  \ups \in \{L, R, \e{off} \} \;. 
\label{definition notation compactes pour modes massless massif et leur differences sur zone Fermi}
\enq

It will also appear useful, in the following, to introduce a parametrisation allowing one to interpret the off-critical particle roots as one-strings having no-string deviations: 
\beq
\Ups^{(p)}_{\e{off}}\; = \; \Big\{  c_a^{(1)} \Big\}_{a=1}^{n_z^{(1)}} \qquad \e{with} \quad  n_z^{(1)}=\big| \Ups^{(p)} \big| \qquad \e{and} \qquad   \de_{a,1}^{(1)}=0 \quad \e{for} \, \e{any} \, a \; . 
\label{definition centre cordes pour Ups p off}
\enq
This allows to collect all the string centres associated with all possible lengths $r=1,\dots, p_{\e{max}}$ into a single set 
\beq
\mf{C}\;= \;  \Big\{  \big\{ c_a^{(p)} \big\}_{a=1}^{n_p^{(z)}} \Big\}_{p=1}^{ p_{\e{max}} } \;. 
\label{definition ensemble centres cordes}
\enq
I stress that $\mf{C}$ does not contain the particle roots which collapse on either of the two edges of the Fermi zone.

When specialising to excited states having the decomposition \eqref{ecriture partition ens part trou spectre massif et massless}, the regular part of the counting function admits the asymptotic expansion
\beq
\wh{\xi}_{\Ups_{\e{reg}}}(\om) \, = \, \frac{1}{2\pi} p(\om  ) \, - \, \f{1}{L} F_{\infty} (\om) \, + \, \e{O}\Big( \f{ \descnode }{L^2}\Big)
\enq
in which
\beq
F_{\infty} (\om) \; = \; \Big( \a + \f{|\La|-|\Ups|}{2} \Big) Z(\om \big) \;+\; \sul{ p=2 }{ p_{\e{max}} } \sul{ a=1 }{ n_p^{(z)} } \phi_{p;1}(\om,c_a^{(p)} )
\; + \hspace{-3mm} \sul{ \mu \in \Ups^{(p)}_{\e{off}}\setminus \Ups^{(h)}_{\e{off}} }{} \hspace{-2mm} \phi( \om, \mu ) \; + \hspace{-1mm} \sul{ \ups \in \{L,R\} }{}\ell^{\varkappa}_{\ups}  \phi(\om, \sg_{\ups} q \, )  
\label{ecriture limite thermo de la fct de cptge}
\enq
represents the thermodynamic limit of the shift function in presence of such excitations while 
the remainder is uniform and holomorphic in some fixed strip around $\R$. The control term in the remainder is given by 
\beq
 \descnode \; = \; 
  \sul{ \ups\in \{L,R\} }{} \hspace{-2mm}  \, \bigg\{  \sul{ a = 1 }{ n_{\ups}^{(p)} } \big( p_a^{\ups} + \tf{1}{2} \big) \, + \,\sul{ a = 1 }{ n_{\ups}^{(h)} } \big( h_a^{\ups} + \tf{1}{2} \big) \bigg\} \;. 
\label{definition control descnode}
\enq
Here and in the following, it will be tacitly assumed that $\descnode/L\leq 1$. 

The thermodynamic shift function $F_{\infty}$ is written, among other things, in terms of the Fermi boundary Umklapp integers
\beq
\ell_{\ups}^{\varkappa} \; = \; \ell_{\ups}-\sg_{\ups} \varkappa_{\ups}  
\label{definition ell ups varkappa}
\enq
that are expressed  in terms of the integers $\ell_{\ups}$ and $\varkappa_{\ups}$ introduced, respectively, in \eqref{definition notation compactes pour modes massless massif et leur differences sur zone Fermi}, 
\eqref{bornes sur fct shift en q hat left right}.

Note that \eqref{ecriture limite thermo de la fct de cptge} entails the bound 
\beq
\norm{ F^{(\varrho)} }_{W_k\big(\mc{S}_{\de}(\R) \big)}\leq C n_{\e{tot}}^{(\e{msv})}  \qquad \e{so}\;\e{that} \quad 
|\varkappa_{\ups}|\leq C n_{\e{tot}}^{(\e{msv})},
\label{borne sur entier varkappa et sur F rho}
\enq
where 
\beq
n_{\e{tot}}^{(\e{msv})} \; = \; \big| \Ups^{(p)}_{\e{off}} \big| \, + \, \big| \Ups^{(h)}_{\e{off}} \big| \, + \, n_{\e{tot}}^{(z)} \, + \, |\ell^{\varkappa}_{L}|\, + \, |\ell^{\varkappa}_{R}| \, + \, 1 \;. 
\label{definition n tot msv}
\enq
In the following, the number $n_{\e{tot}}^{(\e{msv})}$ will be taken to be bounded in $L$.

\noindent The excitation energy and momentum of such excited states takes the form 
%
%
\beqa
\wh{\mc{P}}_{\Ups\setminus\La_{\mf{b}}^{\!(\a)}} & = & \mc{P}_{\e{ex}}^{(\varrho)} \; + \; \f{2\pi  }{L} \hspace{-2mm} \sul{ \ups\in \{L,R\} }{} \hspace{-2mm} \sg_{\ups} \, \bigg\{  \sul{ a = 1 }{ n_{\ups}^{(p)} } p_a^{\ups} 
\, + \,\sul{ a = 1 }{ n_{\ups}^{(h)} } \big( h_a^{\ups} + 1 \big) \bigg\}
\; + \; \e{O}\Big( \f{1}{L}, \f{  \descnode^2}{L^2} \Big)  \label{ecriture DA excitation momentum} \\
\wh{\mc{E}}_{\Ups\setminus\La_{\mf{b}}^{\!(\a)}} & = & \mc{E}_{\e{ex}} \; + \; \f { 2\pi   v_F }{ L } \hspace{-2mm} \sul{ \ups\in \{L,R\} }{} \hspace{-1mm}  \bigg\{  \sul{ a = 1 }{ n_{\ups}^{(p)} } p_a^{\ups} 
\, + \,\sul{ a = 1 }{ n_{\ups}^{(h)} } \big( h_a^{\ups} + 1 \big) \bigg\}
\; + \; \e{O}\Big( \f{1}{L}, \f{  \descnode^2}{L^2} \Big)
\label{ecriture DA excitation energy}
\eeqa
where $ v_F  = \eps^{\prime}(q)/p^{\prime}(q)$ is the Fermi velocity,  
\bem
 \mc{P}_{\e{ex}}^{(\varrho)} \; = \;  \sul{r=2}{p_{\e{max}} } \sul{a=1}{n_r^{(z)} } p_{r}\big( c_a^{(r)}\big) \, + \hspace{-3mm} \sul{ \mu \in \Ups^{(p)}_{\e{off}} \setminus \Ups^{(h)}_{\e{off}}   }{} \hspace{-3mm} p(\mu) 
+2 \varrho  \th\big( q \mid \tfrac{\zeta}{2} \big) \\
\;+ \; \Big( \sul{ r=2 }{ p_{\e{max}} } \sul{a=1}{n_r^{(z)}} \wt{m}_r\big(\zeta,c_a^{(r)} \big)  \; +  \sul{\mu \in \Ups^{(p)}_{\e{off}} }{} \wt{m}_1\big(\zeta,\mu \big) \, + \, |\La| \, - \, |\Ups| \, + \, 2\a \, + \,  \ell^{\varkappa}_R-\ell^{\varkappa}_L \Big) \, p(q)
\end{multline}
and
\beq
 \mc{E}_{\e{ex}} \; = \;  \sul{r=2}{p_{\e{max}} } \sul{a=1}{n_r^{(z)} } \veps_{r}\big( c_a^{(r)}\big) \; +  \hspace{-3mm} \sul{ \mu \in \Ups^{(p)}_{\e{off}}\setminus \Ups^{(h)}_{\e{off}}   }{} \hspace{-3mm} \veps(\mu) \;. 
\label{definition P ex et E ex}
\enq
Above, the $\e{O}(1/L, \descnode^2/L^2)$ symbol means that the corrections are either of order $\e{O}(1/L)$ and do not depend on the integers $p_{a}^{\ups}$ and $h_{a}^{\ups}$ or are of order 
$\e{O}(1/L^2)$ but with at most a quadratic dependence on the integers. \textit{Per se} the statement is  not a direct consequence of Proposition \ref{Proposition DA energie et impulsion} in that one should
push the asymptotic expansion of $\wh{\mc{P}}_{ \Ups\setminus\La_{\mf{b}}^{\!(\a)} }$ and  $\wh{\mc{E}}_{ \Ups\setminus\La_{\mf{b}}^{\!(\a)} }$ obtained there one order further, namely up to $\e{O}(L^{-2})$. 
This is technically involved but does not present any major conceptual difficulty. We thus state the result without sketching its proof.

 \subsection{The $\varrho$-regularised shift function}

 For technical purposes, it appears convenient to introduce the $\mf{b}$-regularised shift function:
\beq
\wh{F}^{\,(\mf{b})} \, = \, L\, \Big(\,  \wh{\xi}_{\La^{\!(\a)}}^{\,(\mf{b})} - \wh{\xi}_{\Ups} \Big)\; = \; \wh{F}+\mf{b}
\qquad \e{and} \qquad \wh{F}^{\,(\mf{b})}_{\e{reg}} \; = \; \wh{F}_{\e{reg}}+\mf{b} \;. 
\enq
Quite analogously, one defines $\varrho$-regularisations $F^{(\varrho)}$ and $F^{(\varrho)}_{\infty}$ of the functions $F$ \eqref{definition fct shit semi limite thermo}
and $F_{\infty}$ \eqref{ecriture limite thermo de la fct de cptge} introduced earlier on
\beq
F^{(\varrho)} \; = \; F\, + \, \varrho  \qquad \e{and} \qquad  F^{(\varrho)}_{\infty} \; = \; F_{\infty} \, + \, \varrho \;. 
\label{definition fct shift semi thermo et thermo beta reg}
\enq
The main advantage of the $\varrho$-deformation is that the zeroes of the functions
\beq
1-\ex{2\i\pi F^{(\varrho)}_{\infty} } \; , \quad 1-\ex{2\i\pi F^{(\varrho)} } \quad \e{and} \quad 1-\ex{2\i\pi \wh{F}^{\,(\mf{b})} } \;. 
\enq
have nicer properties. 
Characterising these will be the aim of the lemma below.

\begin{lemme}
\label{Lemme structure zeros F}

Assume that hypothesis \eqref{propriete espacement ctres cordes} holds. Let $F_{\infty}\!\mid_{\a=0}$ be the thermodynamic limit of the shift function associated with 
the sets of parameters $\big(\mf{C},\Ups^{(p)}_{\e{off}}, \ell_{\ups}^{(\varkappa)} \big)$. 
There exist 
\begin{itemize}
 
 \item an open neighbourhood $\msc{V}_{F}$ of $\intff{-q}{q}$ in $\Cx$ 
 
 \item  parameters $\wt{\eps}>0$ and $1/4>r_1>r_2>0$ 

 \end{itemize}
such that, 
\begin{itemize}
 \item for any shift function satisfying $\norm{ F^{(\varrho)} -  F_{\infty}\!\mid_{\a=0} }_{ L^{\infty}(\msc{V}_F) } < \wt{\eps}$
 
 \item for any $\varrho, \mf{b} \in \Cx$ such that $|\varrho|=r$, for some $r\in \intff{r_2}{r_1}$ and   $\varrho-\mf{b}=\e{O}(L^{-1})$
\end{itemize}
one has 

\begin{itemize}
\label{Lemme distributions zeros fct comptage}

\item [i)] $d\big( \Dp{}\msc{V}_{F} ,  Z^{(s)}-\i\zeta \big)>C^{\prime}$ for some $C^{\prime}>0$, $ \msc{V}_{F} \cap \big\{ Z^{(s)}-\i\zeta \big\} \, = \, \big\{ \be_{a}^{(s)} \big\}_1^{ n_{\e{sg}}^{\prime} }$ with 
\beq
 \be_{a}^{(s)} \in \e{Int}(\msc{C}) \quad for\;\;  a=1,\dots, n_{\e{sg}} \qquad  and  \qquad 
 \be_{a}^{(s)} \in \e{Ext}(\msc{C})\quad  for  \; \; a=n_{\e{sg}}+1,\dots, n_{\e{sg}}^{\prime} \; .
\enq

\item[ii)] The functions $t \mapsto \ex{2\i\pi F^{(\varrho)}_{\infty}(t)}\mid_{\a=0}$ and $t \mapsto \ex{2\i\pi F^{(\varrho)}(t)}$ are both holomorphic and non-zero on $\msc{V}_F$. 
The function $t \mapsto \ex{2\i\pi \wh{F}^{\,(\mf{b})}(t)}$ is meromorphic on $\msc{V}_F$. Its only zeroes and poles inside on $\msc{V}_F$ are 
the simple poles at $\be_{a}^{(s)}$ and the simple zeroes at $\big(\be_{a}^{(s)}\big)^*$ with $a=1,\dots,  n_{\e{sg}}^{\prime}$. 

\item[iii)]  The function $t\mapsto 1-\ex{ 2\i\pi \wh{F}^{\,(\mf{b})} (t)}$ admits $ n_{\e{sg}}^{\prime}+ \ell_F$ zeroes in  $\msc{V}_{F}$ with $\ell_{F} \, = \,  \e{O}\Big( n_{\e{tot}}^{(\e{msv})} \Big) $
and $ n_{\e{tot}}^{(\e{msv})}$ as given by  \eqref{definition n tot msv}. The zeroes partition as
\beq
Z \, = \, \{ \mf{z}_{a}^{(s)} \}_{ 1 }^{  n_{\e{sg}}^{\prime} }   \qquad  and  \qquad W \, = \, \big\{  \mf{w}_{a }   \big\}_{ a=1 }^{ \ell_{F}  }  \;. 
\label{ecriture zeros fct shift exponentiee}
\enq
The zeroes $\mf{w}_{a}$ are simple and well separated:  there exists $\eps>0$ such that  
\beq
\big| \mf{w}_a - \mf{w}_{b}\big|>2\eps \;\; if\;\; a\not=b  \qquad  and  \qquad 
d\Big(W, \big\{Z^{(s)}-\i\zeta\big\}\cup \Ups^{(h)}_{\e{off}}\cup \{\pm q\} \Big) \, > \, 2\eps \;. 
\label{ecriture borne zero 1 moins fct shift exponent}
\enq
The zeroes $\mf{z}_{a}^{(s)}$ are all simple and there exists $C>0$ such that 
\beq
C^{-1}  |\Im\big( \be^{(s)}_a \big)| \, \leq \, \big|\,  \mf{z}_{a}^{(s)} - \be_a^{(s)} \big| \, \leq  \, C    |\Im\big( \be^{(s)}_a \big)|  \; \qquad and \qquad 
C^{-1}  \, \leq \, \bigg| \f{  \mf{z}_{a}^{(s)} - \big( \be_a^{(s)} \big)^* }{  \mf{z}_{a}^{(s)} - \be_a^{(s)}  } \bigg| \, \leq  \, C \;. 
\label{ecriture bornes zeros za proche de beas}
\enq

 \item[iv)] There exists $\de>0$ and  $c>0$ such that 
\beq
 \e{min}_{t\in \msc{C} } \Big\{ \big|1-\ex{2\i\pi F^{(\varrho)}(t)} \big| \,, \,  \big|1-\ex{2\i\pi\wh{F}^{\,(\mf{b})}(t)} \big| \Big\} \geq c  \qquad and \qquad 
\norm{ \wh{F}^{\,(\mf{b})}-F^{(\varrho)} }_{W^{\infty}_k(\msc{C}) }=\e{O}(L^{-1}) \; .
\label{bornes sur fct shift upper and lower}
\enq
The contour $\msc{C}$ depends on $\de$ according to \eqref{definition contours locaux C ups et C ups pm}. 
Furthermore, given $\eps$ as in point ii), for each $w\in W$, there exists an $L$-independent $z_{w}$ such that $w$ is the only zero contained in  $\mc{D}_{z_w,\eps}$ and 
\beq
 \e{min}_{t\in  \Dp{}\msc{V}_F  \cup_{w\in W} \Dp{}\mc{D}_{z_w,\eps} } \Big\{ \big|1-\ex{2\i\pi F^{(\varrho)}(t)} \big| \,, \,  \big|1-\ex{2\i\pi\wh{F}^{\,(\mf{b})}(t)} \big| \Big\} \geq c    \; .
\label{bornes sur fct shift upper and lower}
\enq

\end{itemize}

\end{lemme}

Note that the roots $\be^{(s)}_a$, for  $ a=1,\dots, n_{\e{sg}} $ are precisely the roots appearing in \eqref{definition des parametres beta sing}. The roots 
$ \be_{a}^{(s)} $ with $a=n_{\e{sg}}+1,\dots, n_{\e{sg}}^{\prime}$ are singular roots whose real part lies outside of $\intff{\wh{q}_L}{\wh{q}_R}$ and which are contained in $\msc{V}_F$.

In fact, the lemma can be applied to shift-functions $\wh{F}^{(b)}$ and $F^{(\varrho)}$ that are not necessarily built from 
configurations of roots $\big( \Ups^{(z)}, \Ups^{(p)}, \Ups^{(h)} \big)$ of an excited state which approach the 
configuration $\big(\mf{C},\Ups^{(p)}_{\e{off}}, \ell_{\ups}^{(\varkappa)} \big)$. It is enough that the associated shift function 
is not too far away from a given thermodynamic limit $F_{\infty}^{(\varrho)}\mid_{\a=0}$.

\Proof

The properties of the dressed charge and phases ensure that $ F_{\infty}^{(\varrho)} \!\mid_{\a=0}$ given by \eqref{ecriture limite thermo de la fct de cptge}, and hence 
$1-\ex{2\i\pi F^{(0)}_{\infty}}\!\mid_{\a=0}$, is holomorphic on a sufficiently small neighbourhood of $\intff{-q}{q}$. 
Therefore, $1-\ex{2\i\pi F^{(0)}_{\infty}}\!\mid_{\a=0}$ will only have isolated zeroes there. One can thus always pick an open neighbourhood
$\msc{V}_F$ that is relatively compact, such that the function does not vanish on its boundary and such that the only zeroes contained in 
$\msc{V}_F$ are real and located on $\intff{-q}{q}$. It is also clear that one can choose $\msc{V}_F$ such that 
$d\big(\Dp{}\msc{V}_F, Z^{(s)}-\i\zeta \big)>C^{\prime}$ for some constant $C^{\prime}$. 

The statements of point $ii)$ are evident.

To establish $iii)$, let  $u_a$ with $a=1,\dots,\ell_F^{(\infty)}$ be the zeroes of $ 1 - \ex{ 2 \i \pi F^{(0)}_{\infty} } \!\mid_{\a=0}$ in $\msc{V}_F$ and let $m_a$ denote the multiplicity of $u_a$. 
In virtue of the local behaviour of holomorphic functions, there exist $\vsg, \vsg^{\prime} >0 $, an integer $n_k$ and a biholomorphism $g_k$  
\beq
g_k \, : \, \left\{ \ba{ccc}  \mc{D}_{u_k, \vsg} & \tend  &  g_k \big( \mc{D}_{u_k, \vsg} \big) \supset \mc{D}_{0, \vsg^{\prime}}  \\ 
			  u_k & \mapsto & 0    \ea  \right. 
\qquad \e{such} \, \e{that} \; \; 
 F^{(0)}_{\infty} \! \mid_{\a=0} \, = \,  n_k \, + \, \tfrac{1}{2\i\pi} \big( g_k \big)^{m_k}  
\enq
on $ \mc{D}_{u_k, \vsg}$. It thus follows that, for any $\varrho $ small enough,   $1-\ex{2\i\pi F^{(\varrho)}_{\infty}}\!\mid_{\a=0}$ admits $m_k$ simple zeroes
$u_{k,r}$, $r=1,\dots, m_k$, on $\mc{D}_{u_k,\vsg}$ which satisfy  $|u_{k,r}-u_{k,r^{\prime}}|\, > \, C\, |\varrho|^{\tf{1}{m_k}}$ for some constant $C$ if $r\not= r^{\prime}$. 

\vspace{2mm}

For an arbitrary choice of $|\varrho|$, the roots $u_{k,r}$ may come arbitrarily close to the set $\big\{ Z^{(s)}-\i\zeta\big\}\cup \Ups^{(h)}_{\e{off}}\cup \{\pm q\} $
provided that the latter has a  non-zero intersection with $ \mc{D}_{w_k,\vsg}$. In such a case, denote $\{ v_{k,r} \}_{r=1}^{d_a}$ the points of intersection:
\beq
\Big( \big\{ Z^{(s)}-\i\zeta\big\}\cup \Ups^{(h)}_{\e{off}}\cup \{\pm q\} \Big) \cap \mc{D}_{u_k,\vsg} \, = \, \{ v_{k,r} \}_{r=1}^{d_a}
\enq
The roots $u_{k,r}$ will be uniformly away from the set $\big\{ Z^{(s)}-\i\zeta\big\}\cup \Ups^{(h)}_{\e{off}}\cup \{\pm q\} $ as long as
$|\varrho|$ is at finite distance from the set $\{ |g_k(v_{k,r})|^{m_k} \, , \,1\leq k \leq  \ell^{(\infty)}_{F} \; \e{and} \;  1\leq r \leq m_k \}$. 
This can be always done since one deals with a finite collection of points.  

As a consequence, there exists $\eps>0$ and $|\varrho|>0$ small enough such that all the zeroes of $1-\ex{2\i\pi F^{(\varrho)}_{\infty}}\mid_{\a=0}$ in $\msc{V}_{F}$ are simple, at least distant by $4\eps$ from each other, and such that 
any zero $w$ satisfies
\beq
d\big( w, \big\{ Z^{(s)}-\i\zeta\big\}\cup \Ups^{(h)}_{\e{off}}\cup \{\pm q\} \big) \,  > \,  3\eps \;. 
\label{ecriture distance zero F infty alpha zero beta a ensemble problematiques}
\enq
One can even pick $r_1 , r_2$ such that this property does hold uniformly in $r_1> |\rho|>r_2$. 

Let $Z_F^{(\infty)}=\{ u_{a,k} \}$ be the collection of these simple zeroes and set $\ell_F=|Z_F^{(\infty)}|$. Lemma \ref{Lemme borne sup zeros fct holomorphe}
applied to  the function $1-\ex{2\i\pi F^{(\varrho)}_{\infty}}\mid_{\a=0}$ on the compact set $\ov{\msc{V}}_{F}$ and some larger, fixed simply connected domain $U$ containing $\ov{\msc{V}}_{F}$ such that 
this function is holomorphic on $\ov{U}$ and has no zeroes on $\Dp{}U$ ensures that 
\beq
\ell_{F} \; \leq \; C \f{  \ln \Norm{ 1-\ex{2\i\pi F^{(\varrho)}_{\infty}}\mid_{\a=0}  }_{L^{\infty}\big( \msc{V}_{F} \big) }  }{  \ln \Norm{ 1-\ex{2\i\pi F^{(\varrho)}_{\infty}}\mid_{\a=0} }_{L^{\infty}\big( U \big) } }
\; \leq \; C^{\prime} n_{\e{tot}}^{(\e{msv})}\;. 
\enq
The last bound follows from \eqref{borne sur entier varkappa et sur F rho}.

The maximum principle applied to 
$\Big\{ 1-\ex{2\i\pi F^{(\varrho)}_{\infty}} \Big\}^{-1}_{\mid \a=0}$ on $ \msc{V}_{F}\setminus \cup_{z \in Z_F^{(\infty)} } \mc{D}_{z, \eps }$
ensures that there exist $c>0$ such that 
\beq
\Big| 1-\ex{2\i\pi F^{(\varrho)}_{\infty}}\!\mid_{\a=0} \Big| > 2c  \quad \e{on} \quad \Dp{}\msc{V}_{F}\setminus \cup_{ z \in Z_F^{(\infty)} } \Dp{} \mc{D}_{z,\eps} \;. 
\label{ecriture minoration 1 moins thermo ctg fct}
\enq
There exists $\wt{\eps}>0$ small enough, such that for any $F^{(\varrho)}$ satisfying $\norm{ F^{(\varrho)} - F_{\infty}^{(\varrho)}\!\mid_{\a=0} }_{ L^{\infty}(\msc{V}_F) } < \wt{\eps}$
it holds 
\beq
\Big| \ex{2\i\pi F^{(\varrho)}}-\ex{2\i\pi F^{(\varrho)}_{\infty}}\mid_{\a=0}  \Big| < \f{c}{2}  \quad \e{on} \quad \Dp{}\msc{V}_{F}\setminus  \cup_{ z \in Z_F^{(\infty)} } \Dp{} \mc{D}_{z,\eps}  \;. 
\label{ecriture distance exposants F thermo et F semi thermo}
\enq

By applying Rouch\'{e}'s theorem first on  $\msc{V}_F$ and then on the discs $\mc{D}_{z,\eps}$ to the function  $1-\ex{2\i\pi F^{(\varrho)}_{\infty}}$
and $1-\ex{2\i\pi F^{(\varrho)}_{\infty}}\!\mid_{\a=0} $ one obtains that  $1-\ex{2\i\pi F^{(\varrho)}_{\infty}}$ will have $\ell_F=|Z_F^{(\infty)}|$ simple zeroes on $\msc{V}_F$
and that any such zero will belong to a unique disk  $ \mc{D}_{z,\eps}$ for some $z \in Z_F^{(\infty)} $. Furthermore, 
by applying the maximum principle to the inverse function on $\msc{V}_F\setminus \cup_{z\in Z^{(\infty)}_F } \mc{D}_{z,\eps}$ and using 
\eqref{ecriture minoration 1 moins thermo ctg fct}-\eqref{ecriture distance exposants F thermo et F semi thermo} one gets the lower bound 
\beq
\Big| 1-\ex{2\i\pi F^{(\varrho)} }  \Big| > \f{3c}{2}  \quad \e{on} \quad  \msc{V}_{F}\setminus \cup_{ z \in Z_F^{(\infty)} }  \mc{D}_{z,\eps}  \;. 
\label{ecriture borne inf de exp Fbeta moins un}
\enq

\vspace{2mm}

It remains to focus on $\wh{F}^{(\varrho)}$.  
It follows from $d\big(\Dp{}\msc{V}_F, Z^{(s)}-\i\zeta \big)>C^{\prime}$, equation \eqref{ecriture distance zero F infty alpha zero beta a ensemble problematiques}, the fact that $|\mf{b}-\varrho|=\e{O}( L^{-1} )$
and the estimate \eqref{ecriture estimee norme Wk ctg fct sing loin des sings} on $\wh{\xi}_{ \Ups_{\e{sing}} }$ that  
\beq
\norm{ \wh{F}^{\,(\mf{b})} \,-\, \wh{F}^{\,(\mf{b})}_{\e{reg}} }_{L^{\infty}\big(\Dp{}\msc{V}_{F}\setminus \cup_{ z \in Z_F^{(\infty)} } \Dp{} \mc{D}_{z,\eps} \big) } \; = \;  \e{O}\Big( L^{-\infty} \Big) 
\enq
and thus
\beq
\norm{ \wh{F}^{\,(\mf{b})} \,-\, F^{\,(\varrho)} }_{L^{\infty}\big(\Dp{}\msc{V}_{F}\setminus \cup_{ z \in Z_F^{(\infty)} } \Dp{} \mc{D}_{z,\eps} \big) } \; = \; \e{O}\big( L^{-1} \big)
\enq
owing to \eqref{estimee reste NLIe et deviation Freg a F}. Thus, for $L$ large enough, 
\beq
\Big| \ex{2\i\pi F^{(\varrho)}}-\ex{2\i\pi\wh{F}^{\,(\mf{b})} }   \Big| < \f{c}{2}  \quad \e{on} \quad \Dp{}\msc{V}_{F}\setminus  \cup_{ z \in Z_F^{(\infty)} } \Dp{} \mc{D}_{z,\eps} \;. 
\enq

The function $1-\ex{2\i\pi \wh{F}^{\,(\mf{b}) } }$ is meromorphic in $\msc{V}_F$ with simple poles at $\{ \be_a^{(s)} \}_{ 1 }^{ n_{\e{sg}}^{\prime} }=\big\{Z^{(s)}-\i\zeta \big\}\cap \msc{V}_F$. 
Applying the meromorphic generalisation of Rouch\'{e} theorem to the functions $1-\ex{2\i\pi F^{(\varrho)} }$ and  $1-\ex{2\i\pi \wh{F}^{\,(\mf{b})}}$
on $\msc{V}_F$ ensures that $1-\ex{2\i\pi \wh{F}^{\,(\mf{b})}}$ has $n_{\e{sg}}^{\prime}+| Z^{(\infty)}_{F}|$ zeroes in $\msc{V}_F$. 
Then, by focusing on $\mc{D}_{z,\eps}$ with $z\in Z_F^{(\infty)}$, since $1-\ex{2\i\pi \wh{F}^{\,(\mf{b}) }}$ is holomorphic there owing to the lower bound 
\eqref{ecriture distance zero F infty alpha zero beta a ensemble problematiques}, one gets that $1-\ex{2\i\pi \wh{F}^{\,(\mf{b})}}$ admits a unique zero $\mf{w}_a\in \mc{D}_{z,\eps}$ that is simple. 
The spacing properties of the $z\in Z^{(\infty)}_{F}$ then ensure that \eqref{ecriture borne zero 1 moins fct shift exponent} holds.
Also, the lower bound \eqref{ecriture minoration 1 moins thermo ctg fct} holds with $2c$ replaced by $c$ and $F^{(\varrho)}_{\infty}\!\mid_{\a=0}$ replaced by $\wh{F}^{\,(\mf{b})}$. 

\vspace{2mm}

It remains to focus on the neighbourhoods $\mc{D}_a \equiv \mc{D}_{\be_a^{(s)}, R |\Im(\be_a^{(s)})|}$ of the $\be_a^{(s)}$'s. Here $R>0$ will be assumed large enough. 
Let 
\beq
\wh{F}_a^{\,(\mf{b})}=\wh{F}_{\e{reg}}^{\,(\mf{b})} \, - \, L\, \wh{\xi}_{ \Ups_{\e{sing}}^{\, (a)}  } \quad \e{with} \quad 
\wh{\xi}_{\Ups_{\e{sing}}^{\, (a)}  }(\om) \,= \, \f{1}{2\i\pi L} \sul{ \substack{ \be \in Z^{(s)}-\i\zeta \\ \be\not=\be_a^{(s)}  }  }{}  \ln \bigg( \f{ \sinh(\be-\om) }{ \sinh(\be^*-\om)  } \bigg) \;. 
\label{definition xi Ups sing local regulier en a}
\enq
Then, by virtue of \eqref{propriete espacement ctres cordes}, 
the estimate \eqref{ecriture estimee norme Wk ctg fct sing regularisee en a} ensures that, on $ \mc{D}_{a}$,
\beq
\wh{F}_a^{\,(\mf{b})} \,= \, \wh{F}_{\e{reg}}^{\,(\mf{b})}  \, + \,  \e{O}\Big( L^{ -\infty} \Big) \qquad \e{leading}\; \e{to} \qquad 
\norm{\wh{F}_a^{\,(\mf{b})} - F^{(\varrho)}  }_{ L^{\infty}\big(  \mc{D}_{a} \big)   } \, \leq \, \f{ C^{\prime} }{ L } \;. 
\enq
Thus, it follows from \eqref{ecriture borne inf de exp Fbeta moins un} that  one has the lower bound $\big| 1 -\ex{2\i\pi \wh{F}_a^{\,(\mf{b})} }  \big| \, \geq \, c$  on  $\ov{\mc{D}}_{a}$. 
Since, on $\Dp{}\mc{D}_{a}  $, 
\beq
\big| \ex{2\i\pi \wh{F}_a^{\,(\mf{b})} }  \big|  \; \leq \;  \ex{ C^{\prime} n_{\e{tot}}^{(\e{msv})} |\Im(\be^{(s)}_{a} )| R  }  \; \leq \; C^{\prime}
\enq
one has the bound 
\beq
\big| \ex{2\i\pi \wh{F}^{\,(\mf{b})}  }  -\ex{2\i\pi \wh{F}_a^{\,(\mf{b})} }  \big| \, \leq \,  C 
  \bigg| \f{ \sinh\big[  \sqrt{(R^2+2)}\, |\Im(\be_a^{(s)})| \big] }{   \sinh\big[ R \, |\Im(\be_a^{(s)})| \big]  } -1\bigg|
\leq \f{c}{2} 
\enq
provided that $R$ is large enough. 

One is in position to apply the meromorphic generalisation of Rouch\'{e}'s theorem to the functions $1 -\ex{2\i\pi \wh{F}_a^{\,(\mf{b}) } }  $ and $1 -\ex{2\i\pi \wh{F}^{\,(\mf{b}) } }  $. 
 Since $1 -\ex{2\i\pi \wh{F}_a^{\,(\mf{b}) } }  $ is holomorphic on $\mc{D}_{a} $ while 
$1 -\ex{2\i\pi \wh{F}^{\,(\mf{b}) }  }  $ has a simple pole at $\be^{(s)}_a$, it follows that it has also a simple zero $\mf{z}_a^{(s)}$ on $\mc{D}_{a} $.
The upper bound appearing in the \textit{lhs} of \eqref{ecriture bornes zeros za proche de beas} follows from the fact that $\mf{z}_a^{(s)} \in \mc{D}_{a}$. Going back to the very definition of $\mf{z}_a^{(s)}$ and 
using that $\big| \ex{2\i\pi \wh{F}_a^{\,(\mf{b})} }  \big| $ is bounded on $\mc{D}_{a} $, one gets the upper and lower bound in the \textit{rhs} of \eqref{ecriture bornes zeros za proche de beas}. 
Adopting the parametrisation $\mf{z}_a^{(s)}\, = \, \be_a^{(s)}+\tau$  one gets from  the \textit{rhs} of \eqref{ecriture bornes zeros za proche de beas} 
\beq
\Big|  \f{ \mf{z}_a^{(s)}\, - \, \big(\be_a^{(s)}\big)^* }{ \mf{z}_a^{(s)}\, - \, \be_a^{(s)} }   \Big| \, = \, \Big| 1 + 2\i \f{\Im(\be_a^{(s)}) }{ \tau } \Big| \leq C
\qquad \e{so}\; \e{that} \quad 
\f{2 \big| \Im(\be_a^{(s)}) \big|  }{ \tau }  \leq C +1 \;. 
\enq

\vspace{3mm}

It thus only remains to establish the lower bounds \eqref{bornes sur fct shift upper and lower}-\eqref{bornes sur fct shift upper and lower} stated in point $iii)$. 
Clearly there exists $\de>0$ defining the contours $\msc{C}$ such that $\msc{C}^  \subset  \msc{V}_F\setminus  \cup_{ z \in Z_F^{(\infty)} } \mc{D}_{z,\eps}  $. 
There, \eqref{ecriture borne inf de exp Fbeta moins un} implies that, in particular, one has  $ \big| 1-\ex{2\i\pi F^{(\varrho)} }\big| > \tf{3c}{2}$ on $\msc{C}$. 
The bounds \eqref{ecriture bornes ctg fct sing sur C}, \eqref{estimee reste NLIe et deviation Freg a F} and $| \varrho-\mf{b}|=\e{O}\big(L^{-1} \big)$ ensure that
$\norm{\wh{F}^{\, (\mf{b})} - F^{(\varrho)} }_{W_k(\msc{C}) } = \e{O}(L^{-1})$. On that account, the lower bounds \eqref{bornes sur fct shift upper and lower}-\eqref{bornes sur fct shift upper and lower} follow  
provided that  $L$ large enough.  \qed

\section{The form factors of local operators}
\label{Section FF local ops introduction}

Recall that $\ket{\Ups}$ stands for the Eigenvector associated with the set $\Ups$ of Bethe roots solving the $\a_{\Ups}$-twised Bethe Ansatz equation \eqref{ecriture alpha twised log BAE}. 
In its turn, the vector $\ket{\La^{\!(\a)} }$ stands for the $\a_{\La}$-twisted ground state, namely the Bethe vector built out of the $\a_{\La}$ deformation of the ground state Bethe equations \eqref{ecriture eqns de log Bethe GS roots}.

\subsection{A regular representation for the form factors}

The longitudinal and transverse form factors admit determinant representations \cite{KMTFormfactorsperiodicXXZ}. 
A rewriting of these expressions that is more adapted for the further handlings of this paper was obtained in \cite{KozKitMailSlaTerThermoLimPartHoleFormFactorsForXXZ}. 
\textit{Per se}, this rewriting is only valid if $\Ups^{(\e{in})}\cap\La^{\!(\a)}=\emptyset$ as, otherwise, one should understand the formulae as limits of coinciding parameters. 
Dealing with such limits would introduce various technical complications to the large-volume analysis. One can bypass this problem by deforming 
the $\a_{\La}$-twisted ground state roots $\La^{\!(\a)}=\La_{0}^{\!(\a)} \hookrightarrow \La_{\mf{b}}^{\!(\a)}$, \textit{c.f.} \eqref{ecriture definition ensemle Lab et ses racines}, 
entering in the expression of the form factor so that one has always $\La_{\mf{b}}^{\!(\a)} \cap \Ups=\emptyset$ provided that $\mf{b}$ belongs to some close loop $\msc{L}$ around the origin.
If the deformation of the form factor is holomorphic $\mf{b}$ belonging to the interior of $\msc{L}$, then one can build on the calculation of residues so as to reconstruct the original expression
from an integration along $\mf{b}\in\msc{L}$.

Prior to stating the main result of this sub-section, namely a contour integral based determinant representations for the form factors, I need to introduce a few shorthand notations. 

The set functions $\mc{D}$ and $\mc{W}$ are defined by the double products
\beq
 \mc{W}\big(\La;  \Ups \big) \; = \;  \f{ \pl{\la \in \La}{} \pl{ \mu \in \Ups}{} \Big\{ \sinh(\la-\mu-\i\zeta)  \sinh(\mu-\la-\i\zeta) \Big\} } 
 {  \pl{  \la,\la^{\prime}   \in \La  }{}   \sinh(\la-\la^{\prime}-\i\zeta)  \pl{  \mu,\mu^{\prime}    \in \Ups  }{}   \sinh(\mu-\mu^{\prime}-\i\zeta)    }
\enq
and
\beq
 \mc{D}\big(\La;  \Ups \big) \; = \;  \f {  \pl{   \la \not= \la^{\prime}    \in \La  }{}   \sinh(\la-\la^{\prime})  \pl{  \mu \not= \mu^{\prime}   \in \Ups }{}   \sinh(\mu-\mu^{\prime})    }
 { \pl{\la \in \La}{} \pl{ \mu \in \Ups}{} \Big\{ \sinh(\la-\mu)  \sinh(\mu-\la) \Big\} } \;. 
\enq

$\Xi_{\Om}$ corresponds to a  set-dependent matrix associated with a set of Bethe roots $\Om$ whose entries are given by 
\beq
\big[ \Xi_{\Om} \big]_{ab} \; = \; \de_{ab} \, + \, \f{ K(\nu_a-\nu_b) }{ L\, \wh{\xi}_{\Om}^{\, \prime}(\nu_b) }\qquad \e{upon}\;\e{taking} \; \e{the} \; \e{parametrisation} \qquad \Om \, = \, \{ \nu_a \}_{ 1 }^{ |\Om| }  \;. 
\label{definition matrice Xi Omega}
\enq 
Above, $\wh{\xi}_{\Om}$ is the counting functions associated with the roots $\Om$. As shown in \cite{KorepinNormBetheStates6-Vertex}, 
determinants of these matrices are the main building block of the norm of a Bethe vector.

Finally, I need to introduce the coefficients $\wh{C}^{\,(\ga)}[f](\La;\Ups)$ with $\ga = z$ or $\ga=+$. The parameter $\ga$ distinguishes between the case of form factors
associated with longitudinal ($\ga=z$) or transverse ($\ga=+$) operators. The longitudinal coefficient takes the form 
\beq
\wh{C}^{\,(z)}[f](\La;\Ups) \, = \, \f{2}{\pi^2} \sin^2\bigg( \tfrac{1 }{2} \big( \, \wh{\mc{P}}_{\Ups \setminus \La } -\pi \a\big) \bigg)  \cdot \f{ \sin^2\big[ \pi \a \big] }{  \sin^2\big[ \pi f(\th) \big]  }\cdot 
   \f{ \Big(  \det_{ \Ga(\La) }\! \Big[ \e{id}+\wh{\op{U}}_{\a;\th}^{\, (z)}[f] \, \Big] \Big)^2 }{ \pl{\eps=\pm}{} \Big\{  V_{\Ups;\La}\big( \th+ \eps \i\zeta \big) \Big\} } \;. 
\label{definition coefficient initial hat C z}
\enq
The definition of $\wh{C}^{\,(z)}[f](\La;\Ups)$ contains an arbitrary parameter $\th$. The fact that the ratio defining $\wh{C}^{\,(z)}[f](\La;\Ups)$ does not depend on $\th$ has been established in \cite{KozKitMailSlaTerXXZsgZsgZAsymptotics}. 
In its turn, the transverse coefficient reads
\beq
\wh{C}^{\, (+)}[f](\La;\Ups) \, = \,    \sin^2 \big(\zeta \big)    \cdot 
   \f{ \Big(  \det_{ \Ga(\La) }\!\Big[ \e{id}+\wh{\op{U}}^{\,(+)}_{\a;\th}[f] \Big] \Big)^2 }{ \pl{\eps=\pm}{} \Big\{  V_{\Ups;\La} \big( \tf{\eps \i\zeta }{ 2} \big) \Big\} }  \;. 
\label{definition coefficient initial hat C +}
\enq
The expression for $\wh{C}^{\,(\ga)}[f](\La;\Ups)$ involves  Fredholm determinants of the integral operators $\wh{\op{U}}^{\,(\ga)}_{\a;\th}[f]$ acting on $L^2\big(\Ga(\La)\big)$, where 
the contour $\Ga(\La)$ is a small counterclockwise  loop around the set $\La$ which avoids and does not surround any  other singularities of the integral kernel. 

I stress that, \textit{a priori}, the expression for $\wh{C}^{\,(\ga)}[f](\La;\Ups)$ is only well defined when $\La\cap\Ups=\emptyset$, 
the function $f$ is holomorphic on some small neighbourhood of $\La$, $f(\th)\not\in \mathbb{Z}$  and the function $1-\ex{2\i\pi f}$ does not vanish on $\La$. 

The integral kernels of the operators $\wh{\op{U}}^{\,(\ga)}_{\a;\th}[f]$ take the form
\beq
\wh{U}^{(\ga)}_{\a; \th}[f]\big( \om,\om^{\prime} \big) \; = \;     V_{\Ups;\La}^{-1}\big(\om^{\prime}+\i\zeta)\cdot V_{\Ups;\La}\big(\om^{\prime})
\cdot \f{ \mc{K}^{(\ga)}_{\a;\th} \big( \om , \om^{\prime} \big) }{ 1\, - \, \ex{2\i\pi f(\om^{\prime}) }   }
\enq
where $V_{\Ups;\La}$ has been defined in \eqref{definition fct V Ups et La}, 
\beq
\mc{K}^{(z)}_{\a;\th} \big( \om , \om^{\prime} \big) \; = \; K_{\a}\big( \om - \om^{\prime} \big) \, - \,  K_{\a}\big( \th - \om^{\prime} \big) \qquad \e{with} \quad
K_{\a}(\om) \, = \, \f{1}{2\i\pi} \Big\{ \ex{2\i\pi \a} \coth(\om-\i\zeta) \, - \, \coth\big( \om + \i\zeta \big) \Big\} 
\enq
and
\beq
\mc{K}^{(+)}_{\a;\th} \big( \om , \om^{\prime} \big) \; = \; \f{1}{2\i\pi} 
\bigg\{ \f{ \sinh\big(\om^{\prime}+\tf{3\i\zeta}{2} \big) }{ \sinh\big(\om^{\prime}-\tf{\i\zeta}{2} \big) \sinh\big(\om - \om^{\prime}-\i \zeta \big)  } 
\, - \,  \f{ \ex{2\i\pi \a} \sinh\big(\om^{\prime}-\tf{3\i\zeta}{2}  \big) }{ \sinh\big(\om^{\prime}+\tf{\i\zeta}{2}  \big) \sinh\big(\om - \om^{\prime}+\i \zeta \big)  }   \bigg\}  \;. 
\enq
Note that only the longitudinal kernel $\mc{K}^{(z)}_{\a;\th}$ does exhibit a dependence on $\th$.

\begin{prop}
 \label{Proposition rep reguliere pour les FF}

The form factors of local operators admit the representation 
\beqa
 \f{  \bra{ \La^{ \! (\a) } } \sg_1^{z} \ket{  \Ups } \bra{ \Ups } \sg_{m+1}^{z} \ket{ \La^{ \! (\a) }  }  }{ \braket{ \La^{ \! (\a) } }{ \La^{ \! (\a) } }  \cdot\braket{ \Ups }{  \Ups  }     }  \Big|_{  \a_{\La}=\a_{\Ups} }
 &=  &
 \ex{\i m \wh{\mc{P}}_{\Ups \setminus \La^{ \! (\a) } }}    \cdot \Oint{  \substack{ \Dp{}\mc{D}_{0,r_L} \\  \setminus \{\pm r_L \}  } }{}  \f{\Dp{}^2}{\Dp{}\a^2}\wh{\mc{S}}^{\, (z)}\big(  \La^{ \! (\a) }_{\mf{b}}  ;  \Ups \big)_{  | \a=0 }   \f{\dd \mf{b} }{2\i\pi \mf{b}} \vspace{2mm} \; , 
 \label{representation FF SgZ b regularisee} \\ 
\f{  \bra{ \La^{ \! (\a) } } \sg_1^{-} \ket{  \Ups } \bra{ \Ups } \sg_{m+1}^{+} \ket{ \La^{ \! (\a) }  }  }{ \braket{ \La^{ \! (\a) } }{ \La^{ \! (\a) } }  \cdot\braket{ \Ups }{  \Ups  }     } \Big|_{ \a_{\La}=\a_{\Ups}  }
& = & (-1)^m \cdot 
\ex{\i m \wh{\mc{P}}_{\Ups \setminus \La^{ \! (\a) } } } \cdot  \Oint{ \substack{ \Dp{}\mc{D}_{0,r_L} \\  \setminus \{\pm r_L \}  } }{} \wh{\mc{S}}^{\, (+)}\big(  \La^{ \! (\a) }_{\mf{b}}  ; \Ups \big) _{  | \a=0 }  \f{\dd \mf{b} }{2\i\pi \mf{b} } \; , 
\label{representation FF Sg+ b regularisee} 
\eeqa
where $\wh{\mc{P}}_{ \Ups \setminus \La^{ \! (\a) } } $ has been defined in \eqref{ecriture energie en impulsion excitation} and $\a=\a_{\Ups}-\a_{\La}$. Further, I have set 
\beq
\wh{\mc{S}}^{\,(\ga)}\big(  \La^{ \! (\a) }_{\mf{b}}  ; \Ups \big) \; = \; 
\pl{ \la \in \La^{ \! (\a) }_{\mf{b}} }{}  \bigg\{ \f{\big( \ex{2\i\pi \wh{F}^{\,(\mf{b})}(\la)}-1 \big) \cdot  \big( \ex{-2\i\pi \wh{F}^{\, (\mf{b})}(\la)}-1 \big)  }
				  { 2\i\pi L \, \wh{\xi}^{\prime}_{\La^{ \! (\a) }} (\la) } \bigg\}
\cdot
 \f{ \mc{D}\big(\La^{ \! (\a) }_{\mf{b}};  \Ups \big) \cdot  \mc{W}\big(\La^{ \! (\a) }_{\mf{b}};  \Ups \big) \cdot \wh{C}^{(\ga)}\big[ \wh{F}^{\, (\mf{b})} \big](\La^{ \! (\a) }_{\mf{b}};\Ups) }
 { \pl{ \mu \in \Ups}{} \Big\{  2\i\pi L \,  \wh{\xi}^{\, \prime}_{\Ups} (\mu)  \Big\} \cdot  \det \big[ \Xi_{\Ups} \big] \cdot \det \big[ \Xi_{ \La^{ \! (\a) }_{\mf{b}} } \big]  } \;. 
\label{ecriture expression initiale hat S gamma}
\enq
Finally, the integration in \eqref{representation FF SgZ b regularisee}-\eqref{representation FF Sg+ b regularisee} runs through a disc of radius $r_{L}= r + \tf{\de r}{L}$ when $r$ is 
such that the conclusions of Lemma \ref{Lemme structure zeros F} hold  while $\de r$ as given in Proposition \ref{Proposition beta deformation des racines Lambda}. 

Here, according to \eqref{definition matrice Xi Omega}, one has 
$\big( \Xi_{\La^{ \! (\a) }_{\mf{b}}}\big)_{ab}=\de_{ab}+K\Big( \la_a^{ (\a) }(\mf{b})-\la_a^{ (\a) }(\mf{b}) \Big) \cdot \Big\{ L \, \wh{\xi}^{\, \prime}_{\La^{ \! (\a) }}\big(\la_a^{ (\a) }(\mf{b}) \big) \Big\}^{-1}$.

\end{prop}

\Proof 

The starting point is the determinant representation obtained in \cite{KozKitMailSlaTerThermoLimPartHoleFormFactorsForXXZ}. It represents the form factors of the local operators 
as in \eqref{representation FF SgZ b regularisee}-\eqref{representation FF Sg+ b regularisee} but with the functions $\wh{\mc{S}}^{\,(\ga)}\big(  \La^{ \! (\a) }_{\mf{b}}  ; \Ups \big) $
being replaced by 
\beq
\wh{\mc{S}}^{\,(\ga)}_{\e{BA}}\big(  \La^{ \! (\a) }  ; \Ups \big) \; = \; 
\pl{ \la \in \La^{ \! (\a) }  }{}  \bigg\{ \f{ \big( \ex{-2\i\pi L \wh{\xi} _{\Ups}(\la)}-1 \big) \cdot  \big( \ex{2\i\pi L \wh{\xi} _{\Ups} (\la)}-1 \big)  }{ 2\i\pi L \, \wh{\xi}^{\prime}_{\La^{ \! (\a) }} (\la) } \bigg\}
\cdot
 \f{ \mc{D}\big(\La^{ \! (\a) };  \Ups \big) \cdot  \mc{W}\big(\La^{ \! (\a) };  \Ups \big) \cdot \wh{C}^{(\ga)}_{\e{BA}}\big[ \wh{F} \big](\La^{ \! (\a) };\Ups) }
 { \pl{ \mu \in \Ups}{} \Big\{  2\i\pi L \,  \wh{\xi}^{\, \prime}_{\Ups} (\mu)  \Big\} \cdot  \det \big[ \Xi_{\Ups} \big] \cdot \det \big[ \Xi_{\La^{ \! (\a) }} \big]  } \;. 
\label{ecriture mc S BA}
\enq
Since, in such a situation, the integrand does not depend on $\mf{b}$, the contour integral can be taken and simply gives $1$. 

The coefficient $\wh{C}^{(\ga)}_{\e{BA}}$ is as defined in \eqref{definition coefficient initial hat C z}-\eqref{definition coefficient initial hat C +}
with the sole difference that one should replace the Fredholm determinant arising in its definition by the determinant of the finite matrix 
\beq
\det_{|\La|}\Big[ \e{id} \, + \,  \wh{U}^{(\ga)}_{ \mf{b} } \Big]_{\mid \mf{b}=0 } \qquad \e{with} \qquad 
\Big( \wh{U}^{(\ga)}_{\mf{b}} \Big)_{k\ell} \, = \,  
\f{ V_{\Ups;\La^{ \! (\a) }_{\mf{b}}}^{-1}\Big( \la_{\ell}^{ (\a) }(\mf{b})+\i\zeta \Big) }{  \Big(V_{\Ups;\La^{ \! (\a) }_{\mf{b}}}^{-1}\Big)^{\prime}\Big( \la_{\ell}^{ (\a) }(\mf{b}) \Big) }
\cdot \f{ \mc{K}^{(\ga)}_{\a;\th} \Big( \la_{k}^{ (\a) }(\mf{b})  , \la_{\ell}^{ (\a) }(\mf{b}) \Big) }{ 1\, - \, \ex{-2\i\pi L \wh{\xi}_{\Ups}\big( \la_{\ell}^{ (\a) }(\mf{b}) \big) }   } \;. 
\label{ecriture determinant de replacement au Fredholm}
\enq

The main issue is that the expression \eqref{ecriture mc S BA} has an apparent $0/0$ indeterminacy if $\Ups^{(\e{in})}\cap \La^{ \! (\a) }  = \emptyset$, which, however, can be resolved. 
The more general case of $\wh{\mc{S}}^{\,(\ga)}_{\e{BA}}\big(  \La^{ \! (\a) } _{\mf{b}}  ; \Ups \big) $ which reduces to \eqref{ecriture determinant de replacement au Fredholm} when $\mf{b}=0$.
Lemma \ref{Lemme DA det Ups} ensures that $\det \big[ \Xi_{\La^{ \! (\a) } _{\mf{b}}} \big]\not=0$ for any $|\mf{b}|\leq 1$, 
thus no singularity can issue from the norm determinant. Furthermore, since $\wh{\xi}_{\La^{ \! (\a) } }$ is a biholomorphism on $\mc{S}_{\de}(\R)$, $\big\{\wh{\xi}^{\prime}_{\La^{ \! (\a) } }(\La^{ \! (\a) } _b)\big\}\cap\{0\}=\emptyset $ 
for any $|\mf{b}|\leq 1$, provided that $L$ is large enough. 

 There are three possible origins of poles:
\begin{itemize}
 \item[i)]  If $\Ups\cap \La^{ \! (\a) } _{\mf{b}} = \Om \not= \emptyset $ then $ 1\, - \, \ex{-2\i\pi L \wh{\xi}_{\Ups}\big( \la_{\ell}^{ (\a) }(\mf{b}) \big) }=0$ if $\la_{\ell}^{ (\a) }(\mf{b})\in \Om$. First assume that it is a first order zero. 
Then, the pole appearing in the lines $\ell$ such that $\la_{\ell}^{ (\a) }(\mf{b}) \in \Om$, is compensated by the zero appearing in 
$\Big\{ \Big(V_{\Ups;\La^{ \! (\a) } _{\mf{b}}}^{-1}\Big)^{\prime}\big( \la_{\ell}^{ (\a) }(\mf{b}) \big) \Big\}^{-1}$. 
In their turn, the double poles appearing in $\mc{D}\big(\La^{ \! (\a) } _{\mf{b}};\Ups\big)$ are cancelled by the double zeroes of the prefactors $\ex{\pm 2\i\pi L \wh{\xi} _{\Ups}(\la)}-1$. 
Finally, if $\la_{\ell}^{ (\a) }(\mf{b})\in \Om$ is a higher order zero, say $ 1\, - \, \exp\big\{-2\i\pi L \wh{\xi}_{\Ups}(\la) \big\}  \, =  \, (\la-\la_{\ell}^{ (\a) }(\mf{b}))^k g(\la)$, then it is enough to distribute the factors going to zero 
partly so as to compensate the double poles appearing in $\mc{D}\big(\La^{ \! (\a) } _{\mf{b}};\Ups\big)$ and partly inside of the lines of the determinant that diverge. 

\item[ii)]  If $\Ups^{(h)}\cap \La^{ \! (\a) } _{\mf{b}} = \Om^{(h)} \not= \emptyset $, then the pole appearing in the lines $\ell$ such that $\la_{\ell}^{ (\a) }(\mf{b})\in \Om^{(h)}$
 will be cancelled by the zeroes of the prefactors $\ex{\pm 2\i\pi L \wh{\xi} _{\Ups}(\la)}-1$. 

\item[iii)] The expression is regular when some of the elements of $\La^{ \! (\a) }_{\mf{b}}$ coincide since  
the poles   appearing in the concerned lines of $ \wh{U}^{(\ga)}_{\mf{b}} $ owing to the vanishing of $\Big(V_{\Ups;\La^{ \! (\a) }_{\mf{b}}}^{-1}\Big)^{\prime}\big(\la_{\ell}^{ (\a) }(\mf{b}) )$ 
are compensated by the associated zeroes of the Vandermonde's determinants present in $\mc{D}\big(\La^{ \! (\a) }_{\mf{b}}; \Ups \big)$. 
 
\end{itemize}

 The above  pieces of information then allow one to convince oneself that the function 
\beq
G: \mf{b} \mapsto G(\mf{b}) \; \equiv \; 
\pl{ \la \in \La^{ \! (\a) }_{\mf{b}}  }{}  \bigg\{ \f{ \big( \ex{-2\i\pi L \wh{\xi} _{\Ups}(\la)}-1 \big) \cdot  \big( \ex{2\i\pi L \wh{\xi} _{\Ups} (\la)}-1 \big)  }{ 2\i\pi L \, \wh{\xi}^{\prime}_{\La^{ \! (\a) }} (\la) } \bigg\}
\cdot
 \f{ \mc{D}\big(\La^{ \! (\a) }_{\mf{b}};  \Ups \big) \cdot  \mc{W}\big(\La^{ \! (\a) }_{\mf{b}};  \Ups \big) \cdot \wh{C}^{(\ga)}_{\e{BA}}\big[ \wh{F}^{\,(\mf{b})} \big](\La^{ \! (\a) }_{\mf{b}};\Ups) }
 { \pl{ \mu \in \Ups}{} \Big\{  2\i\pi L \,  \wh{\xi}^{\, \prime}_{\Ups} (\mu)  \Big\} \cdot  \det \big[ \Xi_{\Ups} \big] \cdot \det \big[ \Xi_{\La^{ \! (\a) }_{\mf{b}}} \big]  }  
\nonumber
\enq
is holomorphic on an open neighbourhood of $\mc{D}_{0,r_L}$, with $r_L$ as in the statement of the Proposition. 
Hence
\beq
\wh{\mc{S}}^{\,(\ga)}_{\e{BA}}\big(  \La^{ \! (\a) }  ; \Ups \big) \; = \; G(0) \; = \; \Oint{ \Dp{}\mc{D}_{0,r_L} }{} G(\mf{b}) \cdot \f{ \dd \mf{b} }{2\i\pi \mf{b} } 
 \; = \; \Oint{ \Dp{}\mc{D}_{0,r_L} \setminus \{\pm r_L \} }{} \hspace{-4mm} G(\mf{b}) \cdot \f{ \dd \mf{b} }{2\i\pi \mf{b} } \;, 
\enq
where the last equality follows from neglecting the zero measure set $\{ \pm r_L \}$.  

For any $\mf{b} \in \big\{ \Dp{}\mc{D}_{0,r_L} \setminus \{\pm r_L \} \big\}$ 
one has $\La^{ \! (\a) }_{\mf{b}}\cap \Big\{ \{ \mu_a^{(s)} \}_1^{ n_{\e{sg}} } \cup \Ups^{(\e{in})} \Big\}=\emptyset$ owing to Proposition \ref{Proposition beta deformation des racines Lambda}. 
 It then remains to observe that  $\ex{ \mp 2\i\pi L \wh{\xi} _{\Ups}(\la)} = \ex{ \pm 2\i\pi \wh{F}^{(\mf{b})} }$ for any $\la \in \La^{ \! (\a) }_{\mf{b}}$ and that, by using residue calculations, 
the discrete determinant can be recast in terms of a Fredholm determinant acting on the contour $\Ga(\La^{ \! (\a) }_{\mf{b}})$, this precisely because the other
singularities of the integral kernel are disjoint from $\La^{ \! (\a) }_{\mf{b}}$, as stipulated in Lemma \ref{Lemme structure zeros F}.  \qed

\subsection{The large-volume behaviour of the form factors}
\label{Sous Section principal theorem}

I have now introduced enough notation to state, in detail, the main result of the paper which was already stated, in a weaked form and without making the building blocks explicit, 
in the introduction. 
The large-$L$ behaviour is described by a certain amount of auxiliary functions that I first need to discuss. Almost all such
functions involve the thermodynamic limit of the shift function $F$ as given in \eqref{ecriture limite thermo de la fct de cptge} and of the centred Fermi boundary Umklapp integers 
$\ell_{\ups}^{\varkappa}$ introduced in \eqref{definition ell ups varkappa}. I first focus on the non-universal, \textit{i.e.} operator dependent, part. 

To start with, consider the function
\beq
G\big(\Ups^{(h)}_{\e{off}};\mf{C}; \{ \ell_{\ups}^{\varkappa} \} \mid \tau \big) \, = \, 
\f{ \pl{ p=1 }{ p_{\e{max}} } \pl{ a=1 }{ n_p^{(z)} } \Phi_{1,p}\big( c_{a}^{(p)} - \tau  \big)    }{ \pl{ \mu \in \Ups^{(h)}_{\e{off}} }{ }  \Phi_{1,1}\big( \mu - \tau \big) }
\cdot \pl{ \ups \in \{L,R\} }{} 
\bigg( \f{ \sinh(\sg_{\ups} q-\tau) }{ \sinh(\sg_{\ups}q-\tau-\i\zeta) } \bigg)^{   \ell_{\ups}^{\varkappa} } 
\enq
and the integral transform $\op{C}_{I_q}[f](\om)=\int_{I_q}{} \coth(s-\om) f(s) \cdot \tf{\dd s}{(2\i\pi)}$
which arise as the building block of the integral kernel
\beq
U_{\a;\th}^{(\ga)}[f]\big( \om , \tau \big) \; = \; \ex{ 2\i\pi \big[ \op{C}_{I_q}[f ](\tau)   -  \op{C}_{I_q}[f ](\tau + \i \zeta) \big]  }
\cdot G\big(\Ups^{(h)}_{\e{off}};\mf{C}; \{ \ell_{\ups}^{\varkappa} \} \mid \tau \big) \cdot \f{  \mc{K}_{\a;\th}^{(\ga)}\big( \om , \tau \big) }{  1-\ex{2\i\pi f (\tau) } } \;. 
\label{definition noyau integral U limite thermo}
\enq
Let $\varrho \in \Dp{}\mc{D}_{0,r}$ with $r$ such that the conclusion of Lemme \ref{Lemme structure zeros F} hold. The integral kernel $U_{\a;\th}^{(\ga)}[F^{(\varrho)}]\big( \om , \tau \big) $
defines a trace class\symbolfootnote[2]{ The trace class follows from the fact that the integral kernel is smooth and that the operator acts on functions supported on a finite number 
of compact curves, \textit{c.f.} \cite{DudleyGonzalesBarriosMetricConditionForOpToBeTraceClass} } operator 
$\op{U}_{\a;\th}^{(\ga)}[F^{(\varrho)}]$ on $L^{2}\big( \Ga_{q}  \big)$. 
Here $\Ga_q$ is a small counter-clockwise loop around $I_q$ which avoids the zeroes of $1-\exp\big\{ 2\i\pi  F^{(\varrho)} (\tau) \big\}$ but encircles 
$\pm q$ as well as the elements of the set $\Ups^{(h)}_{\e{off}}$. The very existence of the contour is guaranteed by Lemma \ref{Lemme structure zeros F}.

The Fredholm determinants of the operators $\op{U}_{\a;\th}^{(\ga)}[F^{(\varrho)}]$ arise as the main building blocks of the non-universal, operator dependent, part of the form factor's asymptotics
\bem
C^{(z)}\big( \Ups^{(h)}_{\e{off}};\mf{C}; \{ \ell_{\ups}^{\varkappa} \} \mid \varrho \big) \; = \; \f{2}{\pi^2 } \cdot  \sin^2\!\Big( \tfrac{1}{2} \big( \mc{P}_{\e{ex}}^{(\varrho)}-\pi \a  \big) \Big)     
\cdot \bigg( \f{ \sin[\a \pi] }{ \sin[\pi F^{(\varrho)} (\th) ] } \bigg)^2  \cdot \Big( \det\Big[ \e{id} \, + \, \op{U}_{\a;\th}^{(z)}[F^{(\varrho)}] \,\Big] \Big)^2
  \\
\times \pl{\eps=\pm}{} \Bigg\{  \pl{\ups \in \{L,R\}}{}  \hspace{-2mm} \Big\{ \sinh[\th+\i \eps \zeta-\sg_{\ups} q ] \Big\}^{-\ell_{\ups}^{\varkappa}}  
\f{  \ex{ -2\i\pi \op{C}_{I_q}[F^{(\varrho)} ](\th+\i\eps \zeta) } \prod_{ \mu \in \Ups^{(h)}_{\e{off}} }{ }  \sinh(\th+\i\eps \zeta-\mu) }
{  \pl{ p=1 }{ p_{\e{max}} } \pl{ a=1 }{ n_p^{(z)} } \pl{k=1}{p} \sinh\Big(\th+\i\eps \zeta-c_{a}^{(p)}-\i\tfrac{\zeta}{2}(p+1-2k) \Big)     }
\Bigg\}   
\label{definition cste Cz non univ}
\end{multline}
and
\bem
C^{(+)}\big( \Ups^{(h)}_{\e{off}};\mf{C}; \{ \ell_{\ups}^{\varkappa} \} \mid \varrho \big)\; = \;   \sin^2\big(  \zeta \big)  \cdot \Big( \det\Big[ \e{id} \, + \, \op{U}_{\a;\th}^{(+)}[F^{(\varrho)}]\, \Big] \Big)^2   \\
\times \pl{\eps=\pm}{} \Bigg\{  
\f{  \ex{ -2\i\pi \op{C}_{I_q}[F^{(\varrho)} ](\i\eps \tf{\zeta}{2}) } \prod_{ \mu \in \Ups^{(h)}_{\e{off}} }{ }  \sinh( \i\eps \tf{\zeta}{2} - \mu ) }
{ \pl{ p=1 }{ p_{\e{max}} } \pl{ a=1 }{ n_p^{(z)} } \pl{k=1}{p} \sinh\Big(\i\tfrac{\zeta}{2}(2k+\eps-1-p)   -c_{a}^{(p)} \Big)     }
\pl{ \ups \in \{L,R\} }{} \Big\{ \sinh[ \i \eps \tf{\zeta}{2} -\sg_{\ups}q ]\Big\}^{ -\ell^{\varkappa}_{\ups} }
\Bigg\}\; . 
\label{definition cste C+ non univ}
\end{multline}
The function $\mc{P}_{\e{ex}}^{\!(\varrho)}$ appearing  above is as defined in \eqref{definition P ex et E ex}.  

\vspace{2mm}

The next building block of the form factor's asymptotics is the function $\mf{X}_{\e{tot}}\big( \Ups^{(h)}_{\e{off}}; \mf{C}  ; \{ \ell_{\ups}^{\varkappa} \}  \mid \varrho \big)$. 
Its definition involves the auxiliary integral transforms
\beq
\aleph^{(r)}[f](\mu) \, = \, \Int{-q}{q} f(s) \sul{\eps=\pm}{} \Big\{ \coth\big[\mu-s+\i \eps \tfrac{\zeta}{2}(r+1) \big] - \coth\big[\mu-s+\i \eps \tfrac{\zeta}{2}(r-1) \big]  \Big\}\cdot \dd s  
\label{definition transformee aleph r}
\enq
and
\beq
\aleph^{(\e{bd})}[f](\mu) \, = \, 2\Int{-q}{q} \f{ f(s)-f(\mu) }{ \tanh(s-\mu) } \cdot \dd s \, - \, \sul{\eps=\pm}{} \Int{-q}{q} f(s)\coth(s-\mu+\i\eps \zeta) \cdot \dd s \;. 
\label{definition transformee aleph bd}
\enq
The function itself takes the form 
\bem
\mf{X}_{\e{tot}}\Big( \Ups^{(h)}_{\e{off}}; \mf{C}  ; \{ \ell_{\ups}^{\varkappa} \} \mid \varrho  \Big) \; = \; \bigg|  \f{ \sinh(2q) }{ \sinh(2q+\i\zeta) }  \bigg|^{ 2 \ell_R^{\varkappa}\ell_L^{\varkappa} }
\cdot \f{ \mf{X}\Big( \Ups^{(h)}_{\e{off}}; \mf{C}  ; \{ \ell_{\ups}^{\varkappa} \}  \Big) }{ \big[ \sin(\zeta) \big]^{ n_{\e{off}}^{(p)} + n_{\e{off}}^{(h)}  }  } 
\cdot \f{ \pl{r=1}{ p_{\e{max}} } \pl{ a=1 }{ n_r^{(z)} } \exp\Big\{ \aleph^{(r)}[F^{(\varrho)}](c_a^{(r)}) \Big\}   }{   \pl{ \mu \in \Ups^{(h)}_{\e{off}} }{  } \exp\Big\{ \aleph^{(1)}_-[F^{(\varrho)}](\mu) \Big\}  }
\\
\times \pl{ \ups \in \{L,R\} }{}  \Bigg\{  \f{  \ex{ \ell_{\ups}^{\varkappa} \, \aleph^{(\e{bd})}[F^{(\varrho)}](\sg_{\ups}q)  }  }{  (2\pi)^{\ell_{\ups}^{\varkappa}}  }   \bigg( \f{ \sinh(2q) }{ \sin(\zeta) } \bigg)^{ (\ell_{\ups}^{\varkappa})^2 }
 \Bigg\} 
\label{definition mfX} 
\end{multline}
where, $ \aleph^{(1)}_-$  corresponds to the $-$ boundary value of the function and I agree upon 
\bem
\mf{X}\Big( \Ups^{(h)}_{\e{off}}; \mf{C}  ; \{ \ell_{\ups}^{\varkappa} \}  \Big) \; = \; \pl{r=2}{  p_{\e{max}} }  \bigg\{ \f{1  }{   \sin( r \zeta) } \bigg\}^{ n_r^{(z)} }
\cdot \pl{ \ups \in \{L,R\}}{}  \f{ \pl{ r=1 }{ p_{\e{max}} } \pl{ b = 1 }{ n_{r}^{(z)} }   \Big\{ \Phi_{1,r}\big(\sg_{\ups} q -c_b^{(r)}\big)  \Phi_{1,r}\big(c_b^{(r)}-\sg_{\ups} q \big) \Big\}^{\ell_{\ups}^{\varkappa} }   }
{  \pl{ \la \in \Ups^{(h)}_{\e{off}} }{} \Big\{ \Phi_{1,p}\big(\sg_{\ups} q -\la \big)  \Phi_{1,p}\big(\la-\sg_{\ups} q \big) \Big\}^{\ell_{\ups}^{\varkappa} }     }  \\
\times \f{  \pl{   p,r=1    }{ p_{\e{max}} }    \underset{  (p,a)\not= (r,b) } { \pl{ a=1 }{ n_p^{(z)} }   \pl{ b=1 }{ n_r^{(z)} } }\Phi_{r,p}\big(c_a^{(p)}-c_b^{(r)}\big)  
	      \cdot \pl{ \substack{ \la\not=\la^{\prime} \\  \la,\la^{\prime} \in \Ups^{(h)}_{\e{off}}  } }{}    \Phi_{1,1}\big(\la-\la^{\prime}\big)  }
{ \pl{\la \in \Ups^{(h)}_{\e{off}} }{} \pl{p=1}{ p_{\e{max}} } \pl{ a=1 }{ n_p^{(z)} }  \Phi_{1,p}\big(\la-c_a^{(p)}\big)  \Phi_{1,p}\big(c_a^{(p)}-\la\big) } \;. 
\label{definition fonction G cal}
\end{multline}

Finally, I introduce the two functionals 
\beq
 \op{H}_{0}[ f ] \; = \; \Int{-q}{q} \f{ f^{\prime}(s)f(t)- f(s)f^{\prime}(t) }{ 2\tanh(s-t)} \dd t \dd s \; +  \hspace{-2mm} \sul{ \ups \in \{L,R\} }{}
\sg_{\ups} f(\sg_{\ups}q) \Int{-q}{q} \f{ f(s)-f(\sg_{\ups}q) }{ \tanh(s-\sg_{\ups} q) } \cdot \dd s  
\label{definition de la fnelle H0}
\enq
%
%
and 
\beq
\op{H}_1\big[ f \big] \; = \;  -\Int{-q}{q} \f{ f(t) f(s)  }{ \sinh^2(t-s-\i\zeta)  }   \cdot \dd s \dd t \;. 
\label{definition de la fnelle H1}
\enq

\begin{theorem}
\label{Theorem principal du papier}
 
Assume that the $\a_{\La}$-twisted Bethe roots $\Ups$ for an excited state satisfy Hypothesis \ref{Hypothesis espacement des cordes}. 
There exist a radius of integration $r\in \intoo{0}{1/4}$, fixed by $\Ups^{(h)}_{\e{off}}, \mf{C}$ and $\ell^{\varkappa}_{\ups}$, and such that,
 uniformly in $|\a_{\La}| \leq  L^{ -3 }  $,  one has the large-$L$ asymptotic expansion
\bem
 \bigg|  \f{  \bra{ \Ups } \sg_{ 1 }^{\ga} \ket{ \La^{\!(\a)}  }   }{ \norm{ \La^{\!(\a)} }  \cdot \norm{ \Ups }     } \bigg|_{\a=0}^2  \; = \; 
\Oint{ \Dp{}\mc{D}_{0,r} }{} \pl{ \ups \in \{L,R\} }{}\Bigg\{ \f{ G^2\big(1-\sg_{\ups} \op{f}_{\ups}^{\,(\varrho)}   \big) }{ G^2\big(1-\sg_{\ups}[\op{f}_{\ups}^{\,(\varrho)}  -\sg_{\ups}\ell_{\ups}] \big) }
\f{ \mc{R}_{n^{\ups}_{p}, n^{\ups}_{h} }\big( \{p_a^{\ups}\};\{h_a^{\ups} \} \mid-\sg_{\ups} \op{f}_{\ups}^{\,(\varrho)}  \big) } { \big( \tf{ L }{ 2 \pi }\big)^{ ( \op{f}_{\ups}^{\,(\varrho)} - \sg_{\ups} \ell_{\ups})^2 } } \Bigg\} \\
\times \f{ \mc{F}^{(\ga)}\Big( \Ups^{(h)}_{\e{off}}; \mf{C}  ; \{ \ell_{\ups}^{\varkappa} \}  \mid \varrho \Big)  }
{ \prod_{ \mu \in \Ups^{(h)}_{\e{off}} }^{} \big\{  L p^{\prime}(\mu)   \big\} \cdot \prod_{ r=1  }^{ p_{\e{max}} } \prod_{a=1}^{n_r^{(z)}}  \big\{  L p^{\prime}_r\big( c_a^{(r)} \big)   \big\}   }
\cdot  \bigg\{ 1\, + \,  \mf{R}_{L} \bigg\}   
\f{\dd \varrho}{2\i\pi \varrho } 
\;. 
\nonumber
\end{multline}
$\op{f}_{\ups}^{\,(\varrho)}$ is a constant built from the value of the shift function on the endpoints of the Fermi zone 
\beq
\op{f}_{\ups}^{\,(\varrho)} \, = \, \varkappa_{\ups}-F^{(\varrho)}(\sg_{\ups} q)  
\label{definition coeff f ups}
\enq
and $\mc{F}^{(\ga)} $ represents the form factor density squared associated with the massive modes:  
\bem
 \mc{F}^{(\ga)}\Big( \Ups^{(h)}_{\e{off}}; \mf{C}  ; \{ \ell_{\ups}^{\varkappa} \} \mid \varrho \Big) \; = \; (-1)^{|\La|-|\Ups^{(\e{in})}|+ |\Ups^{(h)}_{\e{off}} | } \cdot  \ex{  ( \op{H}_0+ \op{H}_1)[ F^{(\varrho)} ]  } 
\cdot \f{ \big(\mf{X}_{\e{tot}}\cdot \wt{C}^{(\ga)}\big)\big( \Ups^{(h)}_{\e{off}}; \mf{C}  ; \{ \ell_{\ups}^{\varkappa} \}  \mid \varrho \big)  }{ \det^2\big[ \e{id}+\op{K} \big] } \\
\times \pl{ \mu \in \Ups^{(h)}_{\e{off}} }{}  \Big(1\, - \, \ex{-2\i\pi F^{(\varrho)}(\mu) } \Big)^2   \cdot 
\pl{ \ups \in \{L,R\} }{}  \bigg\{  \f{ G^2\big(1-\sg_{\ups}[\op{f}_{\ups}^{\,(\varrho)} -\sg_{\ups}\ell_{\ups}] \big) }{  \big( \sinh(2q) p^{\prime}(q) \big)^{ ( \op{f}_{\ups}^{\,(\varrho)} - \sg_{\ups} \ell_{\ups})^2 }  } 
\cdot  (2\pi)^{-\sg_{\ups} F^{(\varrho)}(\sg_{\ups} q) }   \bigg\}  
\label{definition densite FF continue}
\end{multline}
where $G$ is the Barnes function, 
\beq
 \wt{C}^{(+)} \big( \Ups^{(h)}_{\e{off}}; \mf{C}  ; \{ \ell_{\ups}^{\varkappa} \} \mid \varrho \big) \, = \,  C^{(+)} \big( \Ups^{(h)}_{\e{off}}; \mf{C}  ; \{ \ell_{\ups}^{\varkappa} \}  \mid \varrho \big)_{\mid \a=0}
\nonumber
\enq
and
\beq
 \wt{C}^{(z)} \big( \Ups^{(h)}_{\e{off}}; \mf{C}  ; \{ \ell_{\ups}^{\varkappa} \} \mid \varrho \big) \, = \,  \Dp{\a}^2C^{(z)} \big( \Ups^{(h)}_{\e{off}}; \mf{C}  ; \{ \ell_{\ups}^{\varkappa} \} \mid \varrho \big)_{\mid \a=0} \;. 
\nonumber
\enq
 $\mc{R}_{n^{(p)}_{\ups}, n^{(h)}_{\ups} }\big( \{p_a\};\{h_a \} \mid \nu  \big)$ represents the form factor density squared associated with the massless modes
\bem
\mc{R}_{n_p, n_h }\big( \{p_a\};\{h_a \} \mid \nu  \big)  \; = \; 
\bigg( \f{\sin(\pi \nu) }{  \pi } \bigg)^{2 n_h} \f{ \pl{a<b}{n_h} (h_a-h_b)^2 \cdot  \pl{a<b}{n_p} (p_a-p_b)^2 }{  \pl{a=1}{n_h} \pl{b=1}{n_p} (h_a+p_b+1)^2   } \\
\times \pl{ a=1 }{n_p} \f{ \Ga^2 ( 1+p_a+\nu ) }{\Ga^2(1+p_a) } \cdot \pl{ a=1 }{n_h} \f{ \Ga^2 ( 1+h_a-\nu ) }{\Ga^2(1+h_a) } \;. 
\label{definition densite R discrete}
\end{multline}
Finally, $\mf{R}_{L}  $ is the remainder term that is controlled as 
\beq
\mf{R}_{L} =  \e{O}\Big(      \f{ \ln L+\descnode + \descnode_{\ln} }{L}  \Big)    \;. 
\enq
 $\descnode$ appearing in the control on the remainder has been introduced in  \eqref{definition control descnode} 
given the collection of particle/hole integers parametrising the massless modes associated with the excited state Bethe roots $\Ups$, one has 
\beq
\descnode_{\ln} \;= \; \sul{ x \in \{p_a^{\ups} , h_a^{\ups} \} }{}  \big( x+\tf{1}{2} \big) \Big| \ln \Big( \f{  x+\tf{1}{2} }{ L } \Big) \Big|  \;. 
\label{definition descnode log}
\enq
\end{theorem}

Note that the asymptotic expansion provided by the above theorem does hold, to the very same order of the remainder, if one replaces $F$ with $F_{\infty}$ in all expressions. 

The theorem follows from the integral representation provided by Proposition \ref{Proposition rep reguliere pour les FF} above and from the large-$L$ asymptotics obtained in Section 
\ref{Section large-L analysis of Dbk et Dex} Propositions \ref{Proposition DA Dbk} and \ref{Proposition asymptotiques Dex}, Section \ref{Section Analyse de Areg}{
Proposition \ref{Proposition asymptotiques Areg} and Section \ref{Section analyse de Asing}, Proposition \ref{Proposition asymptotiques Asing}. 
One should also note that in order to write the result in the stated form relatively to the integration contour, one should represent the integration variable $\de \varrho$
in $\mf{b}=\varrho + \tf{\de \varrho}{L} \in \mc{D}_{0,r_L}\setminus \{ \pm r_L\}$ as given in Proposition \ref{Proposition rep reguliere pour les FF}, as $\de \varrho= \tf{\de r \cdot \varrho }{r}$.

\subsection{A rewriting appropriate for the thermodynamic limit}

As shown in Proposition \ref{Proposition rep reguliere pour les FF}, in order to access to the large-$L$ asymptotics of the form factors, 
it is enough to obtain the asymptotic expansion of $\wh{\mc{S}}^{\,(\ga)}\big(\La^{\!(\a)}_{\mf{b}} ; \Ups\big) $, \textit{c.f.} \eqref{ecriture expression initiale hat S gamma},  uniformly in $\mf{b}\in \Dp{}\mc{D}_{0,r_L}\setminus \{\pm r_L\}$. 
For this purpose, one should first recast $\wh{\mc{S}}^{\,(\ga)}\big(\La^{\!(\a)}_{\mf{b}} ; \Ups\big) $ in a form more suited for taking the large-$L$ limit. 
After some algebra, one gets the decomposition 
\beq
\wh{\mc{S}}^{\,(\ga)}\big(\La^{\!(\a)}_{\mf{b}} ; \Ups\big) \, = \, \wh{\mc{D}}_{\e{bk}}\big(\La^{\!(\a)}_{\mf{b}} ; \Ups^{(\e{in})}\big) \cdot \wh{\mc{D}}_{\e{ex}}\big(\La^{\!(\a)}_{\mf{b}} ; \Ups \big) \cdot 
\wh{\mc{A}}^{\,(\ga)}_{\e{reg}}\big(\La^{\!(\a)}_{\mf{b}} ; \Ups \big) \cdot \wh{\mc{A}}_{\e{sing}}\big( \Ups \big) \;. 
\enq
The factor 
\beq
\wh{\mc{D}}_{\e{bk}}\big(\La^{\!(\a)}_{\mf{b}} ;  \Ups^{(\e{in})}\big) \; = \;  \pl{ \la \in \La^{\!(\a)}_{\mf{b}}}{}  \bigg\{ \f{   \ex{2\i\pi \wh{F}^{\, (\mf{b})}(\la)}-1      }{ 2\i\pi L \, \wh{\xi}^{\, \prime}_{\La^{\!(\a)}} (\la) } \bigg\} 
\cdot \pl{ \mu \in \Ups^{(\e{in})} }{}  \bigg\{ \f{   \ex{-2\i\pi \wh{F}^{\,(\mf{b})}(\mu)}-1    }{ 2\i\pi L \, \wh{\xi}^{\, \prime}_{ \Ups } (\mu) }  \bigg\} \cdot  \mc{D}\big(\La^{\!(\a)}_{\mf{b}};  \Ups^{(\e{in})} \big) 
\enq
gathers the contribution of the roots contained inside of  $\msc{C}$. The coefficient
\bem
\wh{\mc{D}}_{\e{ex}}\big(\La^{\!(\a)}_{\mf{b}} ;  \Ups \big) \; = \;  (-1)^{|\Ups^{(h)}|} (-\i)^{|\Ups^{(z)}|}  \cdot  \mc{D}\big(\Ups^{(h)};  \Ups^{(z)}_{\e{tot}} \big)  \cdot   \mc{W}\big( \Ups^{(h)} ; \Ups^{(z)}_{\e{tot}} \big) 
 \\ 
\times   \pl{p=2}{p_{\e{max}} } \pl{a=1}{n_p^{(z)}} \pl{k=2}{p} \Big\{ -\i \sinh\big( \de_{a,k-1}^{(p)} \, - \,  \de_{a,k}^{(p)}\big) \Big\} \cdot \f{ \pl{\mu \in \Ups^{(z)}_{\e{tot}} }{  }  V^{2}(\mu) }{   \pl{ \mu \in \Ups^{(h)} }{} V_{\mu}^2(\mu) } 
\cdot \pl{ \eps = \pm 1 }{} \Bigg\{ \f{   \pl{ \mu \in \Ups^{(h)} }{} V(\mu+\i\eps \zeta) }{ \pl{\mu \in \Ups^{(z)}_{\e{tot}} }{  }  V(\mu+\i\eps \zeta) } \Bigg\}
\cdot\f{ \pl{ \mu \in \Ups^{(h)} }{}  \Big\{ 2\i\pi L \wh{\xi}^{\prime}_{ \Ups } (\mu)   \Big\} }{ \pl{ \mu \in \Ups^{(p)} }{}  \Big\{ 2\i\pi L \wh{\xi}^{\prime}_{ \Ups } (\mu)   \Big\} }  
\nonumber
\end{multline}
takes into account all the "regular" prefactors depending on the particle and complex valued and roots 
\beq
\Ups^{(z)}_{\e{tot}} \; = \; \Ups^{(p)}\cup \Ups^{(z)} 
\label{definition Ups z tot}
\enq
just as on the hole roots $\Ups^{(h)}$. This factor does not contain any exponentially large or small behaviour in $L$.  $\wh{\mc{D}}_{\e{ex}}\big(\La^{\!(\a)}_{\mf{b}} ;  \Ups \big)$ is defined in terms of the auxiliary functions 
\beq
V(\om) \, \equiv \, V_{\Ups^{(\e{in})};\La^{\!(\a)}_{\mf{b}}}(\om) \; = \; \f{  \pl{\mu \in  \Ups^{(\e{in})} }{} \sinh(\om-\mu)  }{  \pl{ \la \in  \La^{\!(\a)}_{\mf{b}} }{} \sinh(\om-\la) } \qquad \e{and} \qquad
V_{\nu}(\om) \, = \, \f{  \pl{ \substack{\mu \in  \Ups^{(\e{in})}  \\ \not= \nu} }{} \sinh(\om-\mu)  }{  \pl{ \la \in  \La^{\!(\a)}_{\mf{b}} }{} \sinh(\om-\la) }
\enq
for any $\nu \in \Ups^{(\e{in})}$. The coefficient $\wh{\mc{A}}^{\,(\ga)}_{\e{reg}}\big(\La^{\!(\a)}_{\mf{b}} ; \Ups \big)$ contains the operator-dependent contributions to the form factor:
\beq
\wh{\mc{A}}^{\, (\ga)}_{\e{reg}}\big(\La^{\!(\a)}_{\mf{b}} ; \Ups \big) \; = \;  \mc{W}\big( \La^{\!(\a)}_{\mf{b}} ; \Ups^{(\e{in})} \big) \cdot 
\f{ \wh{C}^{\, (\ga)}\big[\wh{F}^{\,(\mf{b})} \big]\big(\La^{\!(\a)}_{\mf{b}};\Ups)  }{ \det\big[ \Xi_{\La^{\!(\a)}_{\mf{b}}} \big] \cdot \det \big[ \Xi_{\Ups^{(\e{in})}} \big] } 
\cdot \f{ \pl{ \la \in \La^{\!(\a)}_{\mf{b}} }{} \big( \ex{-2\i\pi \wh{F}^{\,(\mf{b})}(\la) } -1 \big)  }{ \pl{ \mu \in \Ups^{(\e{in})} }{} \big( \ex{-2\i\pi \wh{F}^{\,(\mf{b})}(\mu)} -1 \big)  } \;. 
\enq
Finally, $\wh{\mc{A}}_{\e{sing}}\big(\Ups \big) $ contains the different terms present in \eqref{ecriture expression initiale hat S gamma} that, when taken individually,
generate an exponentially large or small contributions in $L$:
\beq
\wh{\mc{A}}_{\e{sing}}\big( \Ups \big) \; = \; \pl{\mu \in \Ups^{(z)} }{} \bigg\{ \f{1}{ 2\pi L \wh{\xi}^{\prime}_{\Ups}(\mu) } \bigg\} 
\pl{p=2}{p_{\e{max}} } \pl{a=1}{n_p^{(z)}} \pl{k=2}{p}\Bigg\{ \f{1}{ -\i \sinh\big( \de_{a,k-1}^{(p)} \, - \,  \de_{a,k}^{(p)}\big) } \Bigg\} \cdot 
\f{ \det \big[ \Xi_{  \Ups^{(\e{in})}} \big]  }{ \det  \big[ \Xi_{\Ups} \big] } \;. 
\enq
%
%
%

%
%
%
%
%
%
%
%
%
%
%
%
%
%

\section{Analysis of $\wh{\mc{D}}_{\e{bk}}\big( \La^{\!(\a)}_{\mf{b}} ; \Ups^{(\e{in})} \big)$ and $\wh{\mc{D}}_{\e{ex}}\big( \La^{\!(\a)}_{\mf{b}} ; \Ups \big)$}
\label{Section large-L analysis of Dbk et Dex}

\subsection{Integral representations at finite $L$}

The $\i\pi$-periodic Cauchy transform subordinate to the contour $\msc{C}$ refers to the below integral transform 
\beq
\op{C}_{\msc{C}}\big[ f \big](\om) \; = \;  \Int{ \msc{C} }{} f(s) \coth(s-\om) \cdot \f{\dd s}{ 2 \i \pi }  \;. 
\enq
$\op{C}_{\msc{C}}$ plays an important role in the analysis to come. 
Its $\pm$-boundary values on $\msc{C}$ will be denoted by $\op{C}_{\msc{C};\pm}\big[ f \big]$. Given $f$  piecewise continuous on $\msc{C}$ 
and continuous on $\msc{C}^{(\pm)}=\msc{C}\cap \mathbb{H}^{\pm}$, the Cauchy transforms allow one to define two auxiliary transforms
\beq
\mc{L}_{\msc{C}}\big[ f \big](\om) = -\mc{C}_{\msc{C}}\big[ f \big](\om)  \, + \, \mf{d}[f]\big( \, \wh{q}_R \big) \ln \big[\sinh\big(\om-\wh{q}_R\big) \big] 
\, - \, \mf{d}[f]\big( \, \wh{q}_L \big) \ln \big[\sinh\big(\om-\wh{q}_L\big) \big]  
\enq
and
\beq
\wt{\mc{L}}_{\msc{C}}\big[ f \big](\om) = -\mc{C}_{\msc{C}}\big[ f \big](\om)  \, + \, \mf{d}[f]\big( \, \wh{q}_R \big) \ln \big[\sinh\big(\, \wh{q}_R-\om\big) \big] 
\, - \, \mf{d}[f]\big( \, \wh{q}_L \big) \ln \big[\sinh\big(\, \wh{q}_L-\om\big) \big]  
\enq
where the jump operator $\mf{d}[f]$ is as defined in \eqref{definition symbole de f}. 
As already stated earlier on, $\ln$ appearing above refers to the principal branch of the logarithm. Note that one has to recourse to $\pm$-boundary values to define the $\mc{L}_{\msc{C}}$
transforms on $\R$. The $\pm$ boundary values of these transforms on $\msc{C}$ will be denoted as $\mc{L}_{\msc{C};\pm}$ and $\wt{\mc{L}}_{\msc{C};\pm}$.

Finally, the double integral transform
\beq
\mc{A}_{\eta}[f,g]\; = \;   \Int{ \msc{C} }{} \f{ \dd s }{ 2\i\pi } \Int{ \msc{C}^{\prime} \subset \msc{C} }{} \f{ \dd t  }{ 2\i\pi}  \f{f^{\prime}(t) \,  g(s) }{ \tanh(t-s-\i\eta)  } 
\label{definition transfor integrale A eta}
\enq
will also be of use at some later stage. Here, the contour $ \msc{C}^{\prime} $ is a contour contained in $\msc{C}$ but infinitesimally close to it so that the poles at $t=s+\i\eta$ are located outside of $\msc{C}^{\prime}$. 
Note that this prescription is only necessary if $\eta=0^+$. 

\begin{lemme}
 
 Let $\chi \in \big\{ \La^{\!(\a)} _{\mf{b}}, \Ups\big\}$ and let $\eta \in \R$ be generic and small enough. Define the sums 
\beq
f_{\eta}\big(\om \mid \chi  \big)\, = \, \sul{ \a \in \chi^{(\e{in})} }{} \ln \big[ \sinh\big(\om-\a-\i\eta \big)  \big] 
\qquad and \qquad 
\wt{f}_{\eta}\big(\om \mid \chi  \big)\, = \, \sul{ \a \in \chi^{(\e{in})} }{} \ln \big[ \sinh\big(\a-\om-\i\eta \big)  \big] 
\enq
where 
\beq
\chi^{(\e{in})}=\La^{\!(\a)} _{\mf{b}} \quad  if \quad  \chi=\La^{\!(\a)} _{\mf{b}} \quad and  \quad \chi^{(\e{in})}=\Ups^{(\e{in})} \quad if  \quad \chi=\Ups \; .
\label{definition chi in pour La et Ups}
\enq
Then, the function $f_{\eta}$ and $\wt{f}_{\eta}$ can be recast as 
\beqa
f_{\eta}\big(\om \mid \chi \big) & = & \mc{L}_{\msc{C}}\big[ \, \wh{u}_{ \chi } \big](\om-\i\eta) \, + \, \bs{1}_{ \e{Int}( \msc{C} )  }(\om-\i\eta) \cdot \wh{u}_{ \chi } (\om-\i\eta)
\, + \, \de_{\chi;\Ups} \sul{ \a \in \daleth }{  } \ln \big[ \sinh\big(\om-\a-\i\eta \big) \big]     \\ 
\wt{f}_{\eta}\big(\om \mid \chi \big) & = & \wt{\mc{L}}_{\msc{C}}\big[ \, \wh{u}_{ \chi } \big](\om + \i\eta) \, + \, \bs{1}_{ \e{Int}( \msc{C} )  }(\om+\i\eta) \cdot \wh{u}_{\chi  } (\om+\i\eta)
\, + \, \de_{\chi;\Ups} \sul{ \a \in \daleth }{  } \ln \big[ \sinh\big(\a-\om -\i\eta \big) \big]  
\eeqa
where $\de_{\chi;\Ups}$ is as defined in \eqref{definition kronecker of a set}. Also, the arguments of the indicator functions $\e{Int}( \msc{C} ) $ should be understood in 
$\Cx /\{ \i\pi\mathbb{Z} \}$.

Likewise, given $\Om, \chi \in \big\{ \La^{\!(\a)} _{\mf{b}}, \Ups\big\}$ the double sum
\beq
\msc{S}_{\eta}\big( \Om ; \chi \big)\, = \, \sul{ \a \in \Om^{(\e{in})} }{}  \sul{ \be \in \chi^{(\e{in})} }{} \ln \big[ \sinh\big(\a-\be-\i\eta \big)  \big] 
\enq
can be re-expressed in the form 
\bem
\msc{S}_{\eta}\big( \Om ; \chi \big)\, = \,  \mc{A}_{\eta}\big[ \, \wh{u}_{\Om},\wh{u}_{\chi} \big]
\, + \,  \sul{ \a \in \Om^{(\e{in})} }{}  \bs{1}_{\e{Int}(\msc{C})}(\a-\i\eta) \cdot \wh{u}_{\chi}(\a-\i\eta) \, + \, \de_{\Om;\Ups}\sul{ \a \in \daleth}{} \mc{L}_{\msc{C}}\big[\, \wh{u}_{\chi} \big]\big( \a - \i\eta \big) \\
\, + \, \sul{\ups \in \{L,R\} }{} \sg_{\ups} \mf{d}\big[\,\wh{u}_{\chi}\big]\big(\,  \wh{q}_{\ups} \big) \wt{\mc{L}}_{\msc{C}}\big[\, \wh{u}_{\Om} \big]\big( \, \wh{q}_{\ups} + \i\eta \big)
 \, + \, \de_{\chi;\Ups} \sul{ \a \in \daleth}{} \wt{\mc{L}}_{\msc{C}}\big[\, \wh{u}_{\Om} \big]\big( \a + \i\eta \big)  \\
\, + \, \de_{\chi;\Ups}  \sul{ \a \in \daleth }{}  \bs{1}_{\e{Int}(\msc{C})}(\a+\i\eta) \cdot \wh{u}_{\Om}(\a+\i\eta) 
\, - \, \de_{\Om;\Ups} \de_{\chi;\Ups} \ln \mc{W}_{\eta}\Big(  \big\{ \be_a^{(s)} \big\}_{ 1 }^{  n_{\e{sg}} } ;  \big\{ \mu_a^{(s)} \big\}_{ 1 }^{  n_{\e{sg}} } \Big) \;. 
\end{multline}
Above, it is understood that $\wh{q}_{L/R} +\i\eta \in \e{Ext}(\msc{C})/\{ \i\pi\mathbb{Z} \}$. Also, given two sets $A,B$, $ \ln \mc{W}_{\eta}$ is defined as
\bem
\ln \mc{W}_{\eta} \big(  A ;  B \big) \, = \, \sul{ \substack{ \la\in A  \\ \mu \in B} }{} \Big\{ \ln \big[ \sinh\big(\la-\mu-\i\eta \big) \big] \, + \, \ln \big[ \sinh\big(\mu-\la-\i\eta \big) \big]\Big\} \\ 
\, - \, \sul{ \la, \la^{\prime} \in A  }{} \ln \big[ \sinh\big(\la-\la^{\prime}-\i\eta \big) \big]
\, - \, \sul{ \mu, \mu^{\prime} \in B  }{} \ln \big[ \sinh\big(\mu-\mu^{\prime}-\i\eta \big) \big] \;. 
\end{multline}

\end{lemme}

\Proof 

By definition of the set $\chi^{(\e{in})}$, one gets that 
\beq
f_{\eta}\big(\om \mid \chi \big)\, = \,
\Int{ \msc{C}\setminus \{ \om- \i \eta \} }{}  \wh{u}_{\chi}^{\,\prime}(s) \ln \big[ \sinh\big( \om - s -\i\eta \big) \big]  \cdot \f{ \dd s }{ 2 \i \pi }
\, + \, \de_{\chi;\Ups} \sul{ \a \in \daleth }{} \ln \big[ \sinh\big( \om - \a -\i\eta \big) \big] \;. 
\enq
Above,  $ \msc{C}\setminus \{ \om- \i \eta \}$ corresponds to the contour $\msc{C}$ where one has removed a tiny loop around the line 
$\om- \i \eta+\R^{+}$ so as to avoid the cut of the logarithm. Note that this regularisation of the contour $\msc{C}$ is only necessary when $\om-\i\eta$ lies inside of $\msc{C}$. 
The second term is only present when $\chi  \, = \, \Ups$ owing to \eqref{ecriture systeme poles et zeros u ups prime}.  The claim then results upon an integration by parts followed by 
a straightening of the contour up to $\msc{C}$ and picking up the pole at $s=\om-\i\eta$. 

The handlings are similar in what concerns the re-writing of $\wt{f}_{\eta}\big(\om \mid \chi \big)$. 
Finally, by using the expression for $f_{\eta}\big(\om \mid \chi \big)$ one recasts $\msc{S}_{\eta}\big( \Om ; \chi \big)$ as
\bem
\msc{S}_{\eta}\big( \Om ; \chi  \big) \, = \,  
\mc{A}_{\eta}\big[ \, \wh{u}_{\Om},\wh{u}_{ \chi } \big] 
\, + \, \mf{d}\big[\,\wh{u}_{\chi}\big] \big( \wh{q}_R\big) \wt{f}_{\eta}\big(\wh{q}_R \mid \Om \big) 
\, - \,\mf{d}\big[\,\wh{u}_{\chi}\big]\big( \wh{q}_L\big) \wt{f}_{\eta}\big(\wh{q}_L \mid \Om \big) \\
\, + \,   \sul{ \a \in \daleth }{}  \Big\{ \de_{\chi;\Ups} \cdot  \wt{f}_{\eta}\big( \a \mid \Om \big)  \, - \,
\de_{\Om;\Ups} \cdot  C_{\msc{C}}\big[ \wh{u}_{\chi}  \big](\a-\i\eta) \Big\}
\, + \,  \sul{ \a \in \Om^{(\e{in})} }{}  \bs{1}_{\e{Int}(\msc{C})}(\a-\i\eta) \cdot \wh{u}_{\chi}(\a-\i\eta)   \;. 
\end{multline}
Upon inserting the obtained  expression for $\wt{f}_{\eta}$ and after some algebra, one gets the claim. \qed

The above lemma allows one to recast various simple and double products appearing in the expression for the form factors. In fact, 
all the double and single products arising in the intermediate expressions will be recast as certain integral transformations of the function 
\beq
\wh{z}(\om) \, = \,  \wh{u}_{\Ups}(\om) \, - \, \wh{u}_{\La^{\!(\a)} _{\mf{b}}}(\om) \quad \e{which} \, \e{has} \, \e{jumps} \quad \mf{d}\big[\, \wh{z} \, \big] \big( \, \wh{q}_{\ups} \big) \, = \, - \varkappa_{\ups} 
\qquad \e{for}\quad  \ups \in \{L,R\} \;. 
\enq

To start with, I provide expressions for the function $V_{ \Ups ; \La^{\!(\a)} _{\mf{b}} }$ defined in \eqref{definition fct V Ups et La}.

\begin{cor}
\label{Corollaire DA V Ups La}
 Let $\om$ be such that, either, $\om \in \e{Ext}\big( \msc{C} \big) / \{ \i\pi\mathbb{Z} \}$ or  $\om \in \e{Int}\big( \msc{C} \big) / \{ \i\pi\mathbb{Z} \}$ and $\ex{2\i\pi \wh{F}^{\,(\mf{b})}(\om)}=1$. 
Then, it holds
\beq
V_{ \Ups ; \La^{\!(\a)} _{\mf{b}} }(\om) \, = \, \ex{ \mc{L}_{\msc{C}}[ \, \wh{z} \, ](\om) }   \pl{\mu \in \Ups^{(z)}_{\e{tot}} \setminus \Ups^{(h)} }{} \sinh(\om-\mu)   \cdot  \pl{\a \in \daleth}{} \sinh(\om-\a) 
\enq
and $\Ups^{(z)}_{\e{tot}}$  is as defined in \eqref{definition Ups z tot}. 

Furthermore, any such $\om$ that is also uniformly away from $\pm q$ and satisfies $d(\om,\Ups^{(z)})>cL^{-\kappa}$, one has the large-$L$ expansion 
\beq
V_{\Ups;\La^{\!(\a)} _{\mf{b}}}(\om) \, = \,  \ex{2\i\pi \op{C}_{ I_q^{\ua} }[F^{(\varrho)}](\om) }
\cdot \f{ \pl{r=1}{ p_{\e{max}} }  \pl{ a = 1 }{ n_r^{(z)} } \pl{k=1}{r} \sinh\Big(\om-c_a^{(r)}-\i \tfrac{\zeta}{2}\big(r+1-2k\big) \Big)  }
{  \pl{\ups \in \{L,R \} }{} \big\{ \sinh(\om-\sg_{\ups} q) \big\}^{-\ell_{\ups}^{\varkappa}} \cdot  \pl{\mu \in \Ups^{(h)}_{\e{off}} }{} \sinh(\om-\mu)     }
\cdot \Big( 1\, + \, \e{O}\Big( \f{\descnode }{L} \Big) \Big)   
\enq
where $\ell_{\ups}^{\kappa}$ is as defined in \eqref{definition ell ups varkappa} and $F^{(\varrho)}$ as defined in \eqref{definition fct shit semi limite thermo}. The expansion holds, to the same degree of precision
in the remainder, with $F^{(\varrho)}$ replaced by $F_{\infty}^{(\varrho)}$ defined in \eqref{ecriture limite thermo de la fct de cptge}. Finally, $I_q^{\ua}$ is a small deformation of $I_{q}$
which avoids $\om$ from above, in the case when $\om\in \msc{C}$.  
\end{cor}
Prior to stating the re-writing of $\mc{D}_{\e{bk}}$ and $\mc{W}$, one should introduce a convenient parametrisations of the local behaviour of the counting function $\wh{\xi}_{\Ups}$ in a neighbourhood of $\be_a^{(s)}$:
\beq
\ex{-2\i\pi L \wh{\xi}_{\Ups}( \om ) }  \, = \, \ex{-2\i\pi L \wh{\xi}_{\Ups_{\e{reg}}^{\, (a)}}( \om ) } \cdot 
\f{ \sinh\big[ \om- \big( \be_a^{(s)}\big)^*  \big] }{  \sinh\big[\om-  \be_a^{(s)}  \big] } \;. 
\label{ecriture decomposition fct cptge partie reg et dvgte vers racine sing}
\enq
The function $\wh{\xi}_{\Ups_{\e{reg}}^{\, (a)}}$ is called the locally regular counting function. Provided that hypothesis \eqref{propriete espacement ctres cordes} holds and $L$ is large enough,
 $\exp\Big\{ -2\i\pi L \wh{\xi}_{\Ups_{\e{reg}}^{\, (a)}}( \om ) \Big\}$ is non-vanishing for any $\om$  such that $|\om - \be^{(s)}_a| \leq  |\Im\big( \be^{(s)}_a \big)|^{ \tf{1}{4} }$.

\begin{prop}
\label{Proposition reecriture integrale D bulk}

 One has the integral representations
\beq
\wh{\mc{D}}_{\e{bk}}\big(\La^{\!(\a)} _{\mf{b}};\Ups^{(\e{in})}\big) \, = \, (-1)^{|\La|-|\Ups^{(\e{in})}|}\exp\Big\{  \mc{A}_{0}\big[ \, \wh{z}, \, \wh{z} \,\big] 
\, +   \sul{ \ups \in \{L,R\} }{} \sg_{\ups} \mf{d}\big[\, \wh{z}\,\big]\big( \, \wh{q}_{\ups} \big)  \wt{\mc{L}}_{\msc{C};-}\big[ \, \wh{z} \,  \big]\big( \, \wh{q}_{\ups} \big)  
%
%
\Big\} \cdot   \mc{R}_{\e{bk}}
\enq
where $\sg_{R}=+$, $\sg_{L}=-$, and, recalling \eqref{ecriture decomposition fct cptge partie reg et dvgte vers racine sing}, 
\beq
 \mc{R}_{\e{bk}}   \; = \; 
\mc{D}\big(   \big\{ \be_a^{(s)} \big\}_{ 1 }^{  n_{\e{sg}} }   ;    \big\{ \mu_a^{(s)} \big\}_{ 1 }^{  n_{\e{sg}} }   \big)  \cdot 
\pl{\a \in \daleth }{} \Bigg\{ \f{ \ex{ \wt{\mc{L}}_{\msc{C}} [\,  \wh{z} \, ]( \a )   + \mc{L}_{\msc{C}}[\, \wh{z} \, ]( \a )   }  }
{  1-\ex{-2\i\pi L \wh{\xi}_{\La^{\!(\a)} }^{\,(\mf{b})} (\a) }    }  \Bigg\} \cdot 
\pl{ a = 1 }{  n_{\e{sg}}  } \Bigg\{   \f{ -\ex{-2\i\pi L \wh{\xi}_{\Ups_{\e{reg}}^{\, (a)}}\big( \be_a^{(s)} \big) }  }
{ 2\i\pi L \wh{\xi}_{\Ups}^{\prime}\big( \mu_a^{(s)} \big) }  \cdot \sinh\big[ \be_a^{(s)}  -  \big( \be_a^{(s)} \big)^* \big] \Bigg\}  \;. 
\enq
Likewise, it holds 
\beq
\mc{W}\big(\La^{\!(\a)} _{\mf{b}};\Ups^{(\e{in})}\big) \, = \, \exp\Big\{  -\mc{A}_{\zeta}\big[ \, \wh{z},\, \wh{z} \,\big]  \, - \, \mf{d}\big[\, \wh{z}\,\big]\big( \, \wh{q}_R \big)  \wt{\mc{L}}_{\msc{C}}\big[ \, \wh{z} \,  \big]\big(\, \wh{q}_R + \i\zeta \big)  
\, + \,  \mf{d}\big[\, \wh{z}\,\big] \big( \, \wh{q}_L \big)  \wt{\mc{L}}_{\msc{C}}\big[\,  \wh{z} \, \big]\big( \,\wh{q}_L + \i\zeta \big)  \Big\} \cdot \mc{R}_{ \mc{W} } 
\label{ecriture expression nouvelle pour W}
\enq
where 
\beq
\mc{R}_{ \mc{W} } \; = \; \mc{W}\big(   \big\{ \be_a^{(s)} \big\}_{ 1 }^{  n_{\e{sg}} }   ;    \big\{ \mu_a^{(s)} \big\}_{ 1 }^{  n_{\e{sg}} }   \big) \cdot 
\pl{\a \in \daleth }{} \Big\{  \ex{ -\wt{\mc{L}}_{\msc{C}} [\,  \wh{z} \, ]( \a+\i\zeta )   - \mc{L}_{\msc{C}}[\, \wh{z} \, ]( \a -\i\zeta )   }  \Big\} \;. 
\enq

\end{prop}

\Proof

In order to re-express $\wh{\mc{D}}_{\e{bk}}$, it is convenient to recast the singular product as 
\beq
\mc{D}\big(\La^{\!(\a)} _{\mf{b}};  \Ups^{(\e{in})} \big) 
%
%
%
%
 \; = \; \lim_{\eta \tend 0^+} \Bigg\{  (-\i\eta)^{-|\La^{\!(\a)} _{\mf{b}}|-|\Ups^{(\e{in})}|}
 \ex{  \msc{S}_{\eta}(\La^{\!(\a)} _{\mf{b}};\La^{\!(\a)} _{\mf{b}}) \, + \, \msc{S}_{\eta}(\Ups^{(\e{in})};\Ups^{(\e{in})}) \, - \, \msc{S}_{\eta}( \La^{\!(\a)} _{\mf{b}};\Ups^{(\e{in})} ) 
 \, - \, \msc{S}_{\eta}( \Ups^{(\e{in})};\La^{\!(\a)} _{\mf{b}}) } \Bigg\} 
\nonumber
\enq
leading to 
\bem
\mc{D}\big(\La^{\!(\a)} _{\mf{b}};  \Ups^{(\e{in})} \big)  \; = \; \lim_{\eta \tend 0^+} \Bigg\{ \ex {  \mc{A}_{\eta}[ \, \wh{z},\, \wh{z} \,]  
\, + \, \mf{d}[\, \wh{z}\,]( \, \wh{q}_R )  \wt{\mc{L}}_{\msc{C}}[ \, \wh{z} \,  ]( \, \wh{q}_R+\i\eta )  
\, - \, \mf{d}[\, \wh{z}\,] (  \, \wh{q}_L )  \wt{\mc{L}}_{\msc{C}}[\,  \wh{z} \, ]( \, \wh{q}_L + \i\eta )  }   \\
\times
\pl{\la \in \La^{\!(\a)} _{\mf{b}}}{} \Bigg[  \f{   1-\ex{ - 2\i\pi L \wh{\xi}_{\La^{\!(\a)} }^{\,(\mf{b})} (\la-\i\eta) }      }{  -\i\eta\,  \Big(1-\ex{-2\i\pi L \wh{\xi}_{\Ups}(\la-\i\eta)}\Big)      }  \Bigg]
\pl{\mu \in \Ups^{(\e{in})} }{}  \Bigg[ \f{   1-\ex{ - 2\i\pi L \wh{\xi}_{\Ups}(\mu-\i\eta) }     }{  -\i\eta\,  \Big(1-\ex{-2\i\pi L  \wh{\xi}_{\La^{\!(\a)} }^{\, (\mf{b})}(\mu-\i\eta)} \Big)      }  \Bigg]
\cdot \mc{R}_{\e{bk}}(\eta) \Bigg\} \;. 
\end{multline}
The function $ \mc{R}_{\e{bk}}$ appearing above reads
\beq
 \mc{R}_{\e{bk}}(\eta) \, = \, \f{ \pl{\a \in \daleth }{}  \Big\{ \ex{ \wt{\mc{L}}_{\msc{C}} [\,  \wh{z} \, ]( \a+\i\eta )  + \mc{L}_{\msc{C}}[\, \wh{z} \, ]( \a -\i\eta )   }  \Big\}  }
{ \mc{W}_{\eta}\Big(  \big\{ \be_a^{(s)} \big\}_{ 1 }^{  n_{\e{sg}} }   ;    \big\{ \mu_a^{(s)} \big\}_{ 1 }^{  n_{\e{sg}} }   \Big) }
\pl{\a \in \daleth }{}   \bigg\{ \f{   1-\ex{ - 2\i\pi L \wh{\xi}_{\Ups}(\a+\i\eta) }     }{  1-\ex{-2\i\pi L \wh{\xi}_{\La^{\!(\a)} }^{\, (\mf{b})}(\a+\i\eta)}      }  \bigg\} \;. 
\enq
One can take the $\eta\tend 0^+$ limit of  $\mc{R}_{\e{bk}}^{(s)}(\eta)$ by using the decomposition \eqref{ecriture decomposition fct cptge partie reg et dvgte vers racine sing}. 
More precisely, it holds 
\bem
\lim_{\eta\tend 0^+} \Big\{  \mc{R}_{\e{bk}}(\eta) \Big\} \, = \, \mc{D}\Big(  \big\{ \be_a^{(s)} \big\}_{ 1 }^{  n_{\e{sg}} }   ;    \big\{ \mu_a^{(s)} \big\}_{ 1 }^{  n_{\e{sg}} }   \Big)   
\pl{\a \in \daleth }{}  \Bigg\{ \f{ \ex{  \wt{\mc{L}}_{\msc{C}} [\,  \wh{z} \, ]( \a )   + \mc{L}_{\msc{C}}[\, \wh{z} \, ]( \a  )   } }{ 1-\ex{-2\i\pi L \wh{\xi}_{\La^{\!(\a)} }^{\,(\mf{b})}(\a) }   } \Bigg\}   \\
\times \lim_{\eta\tend 0^+} \pl{a=1}{ n_{\e{sg}}} \bigg\{ (\i\eta)^2 \cdot  \f{   1-\ex{ - 2\i\pi L \wh{\xi}_{\Ups}(\be_a^{(s)}+\i\eta) }     }{   1-\ex{ - 2\i\pi L \wh{\xi}_{\Ups}(\mu_a^{(s)}+\i\eta) }} \bigg\}
\end{multline}
and it remains to invoke that 
\beq
 1-\ex{ - 2\i\pi L \wh{\xi}_{\Ups}(\be_a^{(s)}+\i\eta) } \underset{ \eta \tend 0^+ }{\sim}   -\exp\Big\{-2\i\pi L \wh{\xi}_{\Ups_{\e{reg}}^{\, (a)}}\big( \be_a^{(s)} \big) \Big\}   
\cdot \f{ \sinh\big[ \be_a^{(s)}  -  \big( \be_a^{(s)} \big)^*  \big]  }{   \sinh(\i\eta)   }  \;. 
\enq
The computation relative to $\mc{W}\big(\La^{\!(\a)} _{\mf{b}};\Ups^{(\e{in})}\big)$ is very similar, so that I omit the details. \qed

\vspace{2mm}

The next proposition rewrites $\wh{\mc{D}}_{\e{ex}}\big(\La^{\!(\a)} _{\mf{b}};\Ups\big)$ in a form that is  suited for taking the thermodynamic limit. 
Prior to stating the result, it appears convenient to introduce the set function
\bem
\mc{G}\Big( \Ups^{(p)}\setminus \Ups^{(h)}; \{ \{  c_a^{(p)} \}_{ a=1 }^{ n_p^{(z)} }  \}_{ p = 2 }^{ p_{\e{max}} }  \Big) \; = \; 
\pl{p=2}{ p_{\e{max}} } \pl{ a=1 }{ n_p^{(z)} }\pl{  \substack{ \la \in  \\ \Ups^{(p)} \setminus \Ups^{(h)} } }{}  \Big\{ \Phi_{1,p}\big(\la-c_a^{(p)}\big)  \Phi_{1,p}\big(c_a^{(p)}-\la\big) \Big\} \\
\times \pl{ p,r=2     }{ p_{\e{max}} }    \underset{  (p,a)\not= (r,b) } { \pl{ a=1 }{ n_p^{(z)} }   \pl{ b=1 }{ n_r^{(z)} } }\Phi_{r,p}\big(c_a^{(p)}-c_b^{(r)}\big) 
\cdot \pl{r=2}{  p_{\e{max}} }  \bigg\{ \f{ 1   }{   \sin(r \zeta) } \bigg\}^{ n_r^{(z)} }
\label{definition fonction G cal}
\end{multline}
where $c_a^{(p)}$ are the centres of the strings introduced  \eqref{definition ensemble Ups z}.

\begin{prop}
 
 Let 
\beq
\Ups_{0}^{(z)} \; = \; \Ups^{(z)}_{\e{tot}} \setminus \Big\{ \mu_a^{(s)} \Big\}_{1}^{  n_{\e{sg}} } \qquad
\Ups_{+}^{(z)} \; = \; \Ups^{(z)}_{\e{tot}} \setminus \Big\{ \big( \be_a^{(s)}+\i\zeta \big)^{*} \Big\}_{1}^{  n_{\e{sg}} } \quad and \quad
\Ups_{-}^{(z)} \; = \; \Ups^{(z)}_{\e{tot}} \setminus \Big\{   \be_a^{(s)} +\i\zeta  \Big\}_{1}^{  n_{\e{sg}} }  
\enq
where $\Ups^{(z)}_{\e{tot}}$ is as given in \eqref{definition Ups z tot}. 
Further, for $\eps\in \{0, \pm 1\}$, set 
\beq
\mathbb{Y}\;= \; \Ups^{(z)}_{\e{tot}} \setminus \Ups^{(h)} \qquad 
\mathbb{Y}_{\eps} \;= \; \Ups^{(z)}_{\eps} \setminus \Ups^{(h)} \quad and \quad
\iota_{0}=2 \; , \; \iota_{\pm1}=-1 \;. 
\enq
Assume that  Hypotheses \eqref{propriete espacement ctre corde et particule trou}-\eqref{propriete espacement ctres cordes} on the spacing of the string centres hold. 
Then, within the convention \eqref{ecriture convention produit et somme difference ensemble}, it holds
\bem
\wh{\mc{D}}_{\e{ex}}\big(\La^{\!(\a)} _{\mf{b}};\Ups\big) \, = \, (-1)^{ |\Ups^{(h)}| }\Big( \mc{D} \cdot \mc{W} \Big) \big(\Ups^{(h)};\Ups^{(p)}\big)
\cdot \mc{G}\Big( \Ups^{(p)}\setminus \Ups^{(h)}; \{ \{  c_a^{(p)} \}_{ a=1 }^{ n_p^{(z)} }  \}_{ p = 2 }^{ p_{\e{max}} }  \Big) \\
\times \pl{ \mu \in \mathbb{Y} }{} 
\pl{\eps=0,\pm 1 }{} \Big\{ \ex{ \iota_{\eps}  \mc{L}_{\msc{C}}[ \, \wh{z} \,]( \mu+\i \eps \zeta )  }   \Big\}
\cdot \f{ \pl{\mu \in \Ups^{(h)}}{} \Big(1-\ex{-2\i\pi \wh{F}^{\,(\mf{b})} (\mu) } \Big)^{2} }{ \pl{\mu \in \Ups^{(h)}\cup \Ups^{(p)} }{} \hspace{-4mm}  \big\{ 2\i\pi L \wh{\xi}^{\prime}_{\Ups} (\mu) \big\} }
\cdot   \mc{R}_{\e{ex}}
\end{multline}
where 
\bem
 \mc{R}_{\e{ex}}   \; = \;  \pl{\eps=0,\pm 1 }{} \pl{ \mu \in \mathbb{Y}_{\eps} }{} \pl{ a=1 }{  n_{\e{sg}} }  \bigg\{ \f{ \sinh\big( \mu+\i\eps \zeta-\be_{a}^{(s)} \big) }{ \sinh\big( \mu+\i\eps \zeta-\mu_{a}^{(s)} \big) } \bigg\}^{ \iota_{\eps} } \\
\qquad \times \pl{a\not=b}{n_{\e{sg}}}  \Bigg\{ \f{ \sinh^3\big( \mu_a^{(s)}-\be_b^{(s)}\big) \cdot  \sinh\big( \mu_a^{(s)}-(\be_b^{(s)})^*\big)  }
{  \sinh^2\!\big( \mu_{a}^{(s)}- \mu_{b}^{(s)}\big) \cdot \sinh\big( \be_{a}^{(s)}-\be_{b}^{(s)}\big) \cdot \sinh\big( \be_a^{(s)}-(\be_b^{(s)})^*\big)   }   \Bigg\} \\
\qquad \qquad \times \pl{ a=1 }{  n_{\e{sg}} }  \Bigg\{  \f{ \exp\Big\{ 2\i\pi L \wh{\xi}_{ \Ups_{\e{reg}}^{\, (a)} }\big( \be_a^{(s)} \big) \Big\}  \Big( 2\i\pi L \wh{\xi}_{\Ups}^{\prime}\big( \mu_a^{(s)} \big)  \sinh\big( \mu_a^{(s)} - \be_{a}^{(s)} \big) \Big)^2  }
{ 2\i\pi K\big(\mu_a^{(s)} - \Re(\be_a^{(s)}) \mid \Im(\be_a^{(s)}) \big) \sinh\Big( \be_a^{(s)}  -  \big( \be_a^{(s)} \big)^*  \Big) } \\
\times \f{ \Big( 1-\ex{-2\i\pi L \wh{\xi}_{\La^{\!(\a)} }^{\,(\mf{b})}\big( \be_a^{(s)} \big) }  \Big) 	 \Big( 1-\ex{-2\i\pi L \wh{\xi}_{\La^{\!(\a)} }^{\,(\mf{b})}\big( (\be_a^{(s)})^* \big) }  \Big) }
{ \Big( 1-\ex{-2\i\pi L \wh{\xi}_{\La^{\!(\a)} }^{\,(\mf{b})}\big( \mu_a^{(s)} \big) }  \Big)^{2}  } \Bigg\}  \Big(  1+\e{O}\Big( \f{n_{\e{tot}}^2}{L^{\infty}} \Big) \Big)
\;. 
\label{definition reste R ex}
\end{multline}
The $1+\e{O}\big( \tf{n_{\e{tot}}^2}{L^{\infty}} \big) $ corrections appearing above issue solely from the string deviations. Their expression can be inferred from the content of the proof.  
Finally, $n_{\e{tot}}$ has been defined in \eqref{definition nombre total excitations}. 

\end{prop}

\Proof 

Upon applying the $\eta$-regularisation procedure as in the proof of Proposition \ref{Proposition reecriture integrale D bulk}, one gets
\beq
\pl{\mu \in \Ups^{(z)}_{\e{tot}} }{} V^{2}(\mu) \; = \; \pl{\mu \in \Ups^{(z)}_{\e{tot}} }{} \ex{ 2 \mc{L}_{\msc{C}}[\, \wh{z} \, ](\mu) }  \cdot 
\pl{ a=1 }{  n_{\e{sg}} }  \Bigg\{ \f{ 2\i\pi L \, \wh{\xi}_{\Ups}^{\prime}\big( \mu_a^{(s)} \big)  \sinh\big( \mu_a^{(s)} - \be_{a}^{(s)} \big)  }
			      {  1-\ex{-2\i\pi L \wh{\xi}_{\La^{\!(\a)} }^{\,(\mf{b})}\big( \mu_a^{(s)} \big) }  } \cdot 
\pl{ \mu \in \Ups^{(z)}_{0} }{}  \f{ \sinh\big(\mu-\be_a^{(s)} \big) }{  \sinh\big(\mu-\mu_a^{(s)} \big) }   
    \pl{ a \not=b }{ n_{\e{sg}} }  \f{ \sinh\big(\mu_a^{(s)}-\be_b^{(s)} \big) }{  \sinh\big(\mu_a^{(s)}-\mu_b^{(s)} \big) }   \Bigg\}^{2} 
\nonumber
\enq
and
\beq
\pl{\mu \in \Ups^{(h)} }{} V^{2}_{\mu} (\mu)  \; = \; \pl{\mu \in \Ups^{(h)} }{} \Bigg\{ \ex{  \mc{L}_{\msc{C}}[\, \wh{z} \,](\mu) }   \cdot 
 \f{ 2\i\pi L \, \wh{\xi}^{\prime}_{\Ups} (\mu)  }{ 1-\ex{-2\i\pi \wh{F}^{\,(\mf{b})}(\mu) }  }  \Bigg\}^{2}
\cdot \pl{a=1}{ n_{\e{sg}} }\pl{ \mu \in \Ups^{(h)} }{}  \Bigg( \f{ \sinh\big(\mu-\be_a^{(s)} \big) }{  \sinh\big(\mu-\mu_a^{(s)} \big) }\Bigg)^{2} 
\enq
Further, one has 
\bem
\pl{\mu \in \Ups^{(z)}_{\e{tot}} }{} \pl{\eps=\pm 1 }{}  V^{2} (\mu+\i\eps \zeta) \; = \; \pl{\mu \in \Ups^{(z)}_{\e{tot}} }{} \pl{\eps=\pm 1 }{} \ex{ \mc{L}_{\msc{C}}[\, \wh{z} \,](\mu+\i\eps \zeta) }  \cdot 
\pl{ a=1 }{  n_{\e{sg}} }  \Bigg\{ \f{  \ex{2\i\pi L \wh{\xi}_{\Ups_{\e{reg}}^{\, (a)}}( \be_a^{(s)} )}  \sinh^2\!\big( \be_a^{(s)} - (\be_{a}^{(s)})^* \big)  }
								  {	\sinh\big( \mu_a^{(s)} - \be_{a}^{(s)} \big)	\sinh\big( \mu_a^{(s)} - (\be_{a}^{(s)})^* \big) }  \Bigg\} \\
	\times \pl{a\not=b}{n_{\e{sg}}}  \Bigg\{  \f{   \sinh\big( \be_{a}^{(s)}-\be_{b}^{(s)}\big) \cdot \sinh\big( \be_a^{(s)}-(\be_b^{(s)})^*\big)   }
 { \sinh\big( \mu_a^{(s)}-\be_b^{(s)}\big) \cdot  \sinh\big( \mu_a^{(s)}-(\be_b^{(s)})^*\big)  }  \Bigg\}
	 \cdot \f{  \pl{\eps=\pm 1 }{}  \pl{\mu \in \Ups^{(z)}_{\eps} }{}  \pl{\a \in \Ga}{} \sinh\big( \mu - \a + \i\eps \zeta \big)  }
	{  \pl{ a = 1 }{  n_{\e{sg}} }\Big\{  \Big(1-\ex{-2\i\pi L \wh{\xi}_{\La^{\!(\a)} }^{\,(\mf{b})}\big( \be_a^{(s)} \big) } \Big) \Big(1-\ex{-2\i\pi L \wh{\xi}_{\La^{\!(\a)} }^{\,(\mf{b})}\big( (\be_a^{(s)})^* \big) } \Big) \Big\} } \;. 
\end{multline}
Finally, it holds
\beq
\pl{\mu \in \Ups^{(h)} }{} \pl{\eps=\pm 1 }{}  V^{2} (\mu+\i\eps \zeta) \; = \; \pl{\mu \in \Ups^{(h)} }{} \pl{\eps=\pm 1 }{} \ex{ \mc{L}_{\msc{C}}[\, \wh{z} \,](\mu+\i\eps \zeta) }  
\pl{\eps=\pm 1 }{}  \pl{\mu \in \Ups^{(h)} }{}  \pl{\a \in \daleth}{} \sinh\big( \mu - \a + \i\eps \zeta \big) \;. 
\enq
The obtained expressions can be slightly simplified by using that
\beq
2\i\pi K\big(\mu_a^{(s)} - \Re(\be_a^{(s)}) \mid \Im(\be_a^{(s)}) \big)  \, = \, \f{ \sinh\big[ \be_a^{(s)}  -  \big( \be_a^{(s)} \big)^*  \big] }
{  \sinh\big[ \mu_a^{(s)}  -  \big( \be_a^{(s)} \big)^*  \big] \cdot \sinh\big[ \mu_a^{(s)}  -  \be_a^{(s)}   \big] } \;. 
\enq

It remains to recast the double products as
\beq
 \Big(\mc{D}\cdot \mc{W} \Big)\big(\Ups^{(h)};\Ups^{(z)}_{\e{tot}}\big)\; = \; \mc{P}^{(1)}\Big( \Ups^{(h)};\Ups^{(z)}_{\e{tot}} \Big)\cdot 
\mc{P}^{(2)} \big(\Ups^{(h)} \big) \cdot  \mc{P}^{(2)} \big( \Ups^{(z)}_{\e{tot}} \big)
\enq
where 
\beq
\mc{P}^{(1)}\Big( A;B  \Big) \, = \, \pl{ \substack{ \la \in A \\ \mu \in B}  }{}  \bigg\{ \f{  \sinh\big(\la-\mu-\i \zeta\big)  \sinh\big(\mu-\la-\i \zeta \big) }{  \sinh\big(\la-\mu\big)  \sinh\big(\mu-\la\big)  } \bigg\}
\quad \e{and} \quad
\mc{P}^{(2)} \big( A \big)  \, = \, \f{ \pl{ \la\not= \la^{\prime}   \in A  }{} \sinh\big(\la-\la^{\prime}\big)  }{ \pl{ \substack{ \la, \la^{\prime}  \\ \in A}  }{} \sinh\big(\la-\la^{\prime}-\i \zeta \big) } \;. 
\enq
For further convenience, it is useful to parametrise $ \Ups^{(z)}_{\e{tot}}$ defined in \eqref{definition Ups z tot} in terms of the string centres $\wt{c}_a^{\,(r)}$ as
\beq
  \Ups^{(z)}_{\e{tot}} \; = \; \bigg\{ \Big\{ \big\{\,  \wt{c}_a^{\, (r)} +\i\f{\zeta}{2}\big( r + 1 - 2k \big) \, + \, \de_{a,k}^{(r)} \big\}_{k=1}^{r} \Big\}_{a=1}^{ \wt{n}_z^{(r)}} \bigg\}_{r=1}^{ p_{\e{max}} } \;. 
\enq
Here, the sole difference between $\wt{c}_a^{\, (p)}$ and the string centres $c_a^{(p)}$ introduced in \eqref{definition ensemble Ups z} and \eqref{definition centre cordes pour Ups p off} is for $p=1$ 
where the $\wt{c}_a^{\, (p)}$ also include the elements of $\Ups^{(p)}$ which squeeze down to $\pm q$, \textit{viz}.
the rapidities of the roots $\Ups^{(p)}_{\ups}$. In particular, one has $ \wt{n}_z^{(p)} =  n_p^{(z)} $ for $p\geq 2$ and $ \wt{n}_z^{(1)} = |\Ups^{(p)}| $. 

By using the spacing hypothesis \eqref{propriete espacement ctre corde et particule trou} on central element of odd strings, one readily gets that 
\beq
 \mc{P}^{(1)}\Big( \Ups^{(h)};\Ups^{(z)}_{\e{tot}} \Big) \, = \, \pl{\la \in \Ups^{(h)} }{} \pl{ p=1 }{ p_{\e{max}} }  \pl{ a = 1 }{ \wt{n}_z^{\, (p)} } 
 \f{1}{ \Phi_{p,1}\big( \la-\wt{c}_a^{\, (p)} - \de_{a, \lfloor \frac{p+1}{2} \rfloor }^{(p)} \big) \cdot \Phi_{1,p}\big( \, \wt{c}_a^{\, (p)} + \de_{a, \lfloor \frac{p+1}{2} \rfloor }^{(p)} -\la \big) }
\cdot \Big( 1 +  \e{O}\Big( \f{n_{\e{tot}}^2}{L^{\infty}} \Big)   \Big)
\enq
where $\Phi_{r,p}$ is given by \eqref{formule explicite pour produit double Phi r p sur racines corde} and the $\e{O}\big(  \tf{ n_{\e{tot}}^2 }{ L^{\infty} }  \big) $ remainder issue from neglecting the string deviations. 
One can drop, on the level of the obtained formula, the remaining string deviation. 
Quite similarly, one has
\bem
\mc{P}^{(2)} \big( \Ups^{(z)}_{\e{tot}} \big) \; = \;   \pl{ \substack{ p,r=1 \\ (p,a)\not= (r,b)    }  }{ p_{\e{max}} }   \pl{ a=1 }{ \wt{n}_z^{\,(p)} } \pl{ b=1 }{ \wt{n}_z^{\, (r)} } \Phi_{r,p}\big(\, \wt{c}_a^{\, (p)} - \wt{c}_b^{\, (r)}\big)  
\cdot \pl{ p=2 }{ p_{\e{max}} } \pl{ a = 1 }{ n_p^{(z)} } \f{  \pl{   k \not= s   }{p} \sinh\big[\i\zeta (s-k) \big] }
						      {  \pl{  k,s =1   }{p} \sinh\big[\de_{a,k}^{(p)}-\de_{a,s}^{(p)}+\i\zeta (s-k-1) \big]  } \\
\times \bigg\{ \f{1}{ \sinh(-\i\zeta)  } \bigg\}^{ \wt{n}_z^{(1)} } \cdot \Big( 1+\e{O}\Big( \f{n_{\e{tot}}^2}{L^{\infty}} \Big)  \Big) \;. 
\end{multline}
In the penultimate product one has to keep the string deviations for terms such that $s=k-1$ but, otherwise, these produce as well $ 1+\e{O}\big(  \tf{n_{\e{tot}}^2}{L^{\infty}}  \big) $ contributions. 
Eventually, one gets 
\bem
\mc{P}^{(2)} \big( \Ups^{(z)}_{\e{tot}} \big) \; = \;   \f{ \pl{ a\not=b }{ \wt{n}_z^{\, (1)} } \sinh\big[\,  \wt{c}_a^{\, (1)}-\wt{c}_b^{\, (1)} \big]    }
		{ \pl{ a,b =1 }{ \wt{n}_z^{\, (1)} } \sinh\big[\,  \wt{c}_a^{\, (1)}-\wt{c}_b^{\, (1)}  -\i\zeta \big]    } \cdot 
\f{ \pl{ \substack{ p,r=1 \\ (p,r)\not= (1,1)    }  }{ p_{\e{max}} }  \underset{(p,a)\not= (r,b) }{ \pl{ a=1 }{ n_p^{(z)} } \pl{ b=1 }{ n_r^{(z)} } }\Phi_{r,p}\big(\, \wt{c}_a^{\, (p)} - \wt{c}_b^{\, (r)}\big)    }
{  \pl{ p=2 }{ p_{\e{max}} } \pl{ a = 1 }{ n_p^{(z)} }  \pl{ k =2  }{p}  \sinh\big[\de_{a,k-1}^{(p)}-\de_{a,k}^{(p)}\big]  }
\cdot \f{  1+\e{O}\big(  \tf{n_{\e{tot}}^2}{L^{\infty}}  \big)   }{  \pl{ p=2 }{ p_{\e{max}} } \big\{ (-1)^p\sinh(\i p \zeta) \big\}^{n_p^{(z)} }    } \;. 
\end{multline}
Thus, 
\bem
 (-\i)^{|\Ups^{(z)}| } \Big(\mc{D}\cdot \mc{W} \Big)\big(\Ups^{(h)};\Ups^{(z)}_{\e{tot}}\big) \pl{ p=2 }{ p_{\e{max}} } \pl{ a = 1 }{ n_p^{(z)} }  \pl{ k =2  }{p} \big\{-\i \sinh\big[\de_{a,k-1}^{(p)}-\de_{a,k}^{(p)}\big] \big\} \\ 
\; = \; \Big( \mc{D} \cdot \mc{W} \Big) \big(\Ups^{(h)};\Ups^{(p)}\big) \cdot \mc{G}\Big( \Ups^{(p)}\setminus \Ups^{(h)}; \{ \{  c_a^{(p)} \}_{ a=1 }^{ n_p^{(z)} }  \}_{ p = 2 }^{ p_{\e{max}} }  \Big)
\cdot  \Big(  1+\e{O}\Big( \f{n_{\e{tot}}^2}{L^{\infty}} \Big) \Big) \;. 
\end{multline}

\qed


\subsection{The large-L expansion}

\begin{prop}
\label{Proposition DA Dbk}
  Let $\mf{b}=\varrho+\tf{\de \varrho }{L} \in \mc{D}_{0,r_L}$ with $r_L$ as in Proposition \ref{Proposition rep reguliere pour les FF}. 
Assume that \eqref{propriete espacement ctre corde et particule trou} and  \eqref{propriete espacement ctres cordes} holds. Then, one has  
\bem
\wh{\mc{D}}_{\e{bk}}\big(\La_{\mf{b}}^{\!(\a)};\Ups^{(\e{in})}\big) \cdot \mc{R}_{\e{ex}}\, = \, (-1)^{ |\La_{\mf{b}}^{\!(\a)}|-|\Ups^{(\e{in})}|+\varkappa_{L}(\varkappa_{L}-\varkappa_{R}) }  
\big[ \sinh(2q) \big]^{(\varkappa_{L}-\varkappa_{R})^2} \ex{ \op{H}_0[ F^{(\varrho)} ] }   \\
\times \pl{\ups \in \{L,R\} }{}  \Bigg\{  \exp\bigg[ -2 \sg_{\ups}\varkappa_{\ups} \Int{-q}{q} \f{ F^{(\varrho)}(s)-F^{(\varrho)}(\sg_{\ups}q) }{ \tanh(s-\sg_{\ups} q) } \cdot \dd s \bigg] \\
\times \f{ G\big(1-\op{f}_{\ups}^{\,(\varrho)},1+\op{f}_{\ups}^{\,(\varrho)} \big)  \ex{ -\i \f{\pi}{2} \sg_{\ups} \big( \wh{F}^{(\mf{b})}_{\e{reg}}(\,\wh{q}_{\ups})\big)^2-\i\pi \sg_{\ups} \wh{\op{f}}_{\ups}^{\,(\mf{b})} \varkappa_{\ups} }  }
{  \big[ \tf{ L p^{\prime}(\sg_{\ups}q) \sinh(2q) }{2\pi} \big]^{\big(\op{f}^{\,(\varrho)}_{\ups} \big)^{2}}  }  \Bigg\} 
\cdot \bigg( 1+\e{O}\Big( \f{ \ln L  }{ L }    \Big)  \bigg)
\end{multline}
where the functional $\op{H}_{0}$ has been introduced in \eqref{definition de la fnelle H0}, $G$ is the Barnes function, $\op{f}_{\ups}^{\,(\varrho)}$ is as given by \eqref{definition coeff f ups}, $\mc{R}_{\e{ex}}$ has been defined in \eqref{definition reste R ex}
while 
\beq
\wh{\op{f}}_{\ups}^{\,(\mf{b})}\; = \; \varkappa_{\ups} \, - \, \wh{F}_{\e{reg}}^{\,(\mf{b})}(\, \wh{q}_{\ups} )   
\label{definition hat f ups}
\enq

\end{prop}

Note that in order to have a better given control on the final remainder, one has to keep, at this stage, some of the terms still at their finite-$L$ values. These will subsequently compensate with some other terms in the expansion.

\Proof 

Starting from the rewriting provided by Proposition \ref{Proposition reecriture integrale D bulk}, after invoking equation \eqref{ecriture equivalent dvgt de xi Ups en les mu a sing} of Lemma \ref{Lemme dvgce exp fct cptge proximite des racines sing}
and using the string centres spacing \eqref{propriete espacement ctres cordes} assumption, one obtains that $\mc{R}_{\e{ex}}\cdot \mc{R}_{\e{bk}}=1+\e{O}\big( \tf{n_{\e{tot}}^2}{L^{\infty}} \big)$. 
Further, as a direct consequence of Lemma \ref{Lemme asymptotiques fnelle A0} and of Corollary \ref{Corollaire asymptotiques transformee L et L tilde}, one gets that 
\bem
\ex{ \mc{A}_{0}[\, \wh{z}, \, \wh{z}\, ]  \, - \hspace{-2mm} \sul{\ups \in \{L,R\} }{} \sg_{\ups} \varkappa_{\ups} \wt{\mc{L}}_{\msc{C};-}[\, \wh{z} \,](\,\wh{q}_{\ups})    } \; = \; 
 (-1)^{\varkappa_{L}(\varkappa_{L}-\varkappa_{R}) }  \big[ \sinh(\,\wh{q}_{R}-\wh{q}_{L}) \big]^{(\varkappa_{L}-\varkappa_{R})^2} \ex{ \wh{\op{H}}_{\varkappa_{L};\varkappa_{R}}[ \wh{F}_{\e{reg}}^{\,(\mf{b})} ] }  \\
\times \pl{\ups \in \{L,R\} }{}  \Bigg\{  \f{ G\big(1-\wh{\op{f}}_{\ups}^{\,(\mf{b})},1+\wh{\op{f}}_{\ups}^{\,(\mf{b})} \big) 
\ex{ -\i \f{\pi}{2} \sg_{\ups} \big(\wh{F}_{\e{reg}}^{\,(\mf{b})}\big)^{\,2}\!\!(\,\wh{q}_{\ups})-\i\pi \sg_{\ups} \wh{\op{f}}_{\ups}^{\,(\mf{b})} \varkappa_{\ups} }  }
{  \big[ L \wh{\xi}_{\La}^{\prime}(\,\wh{q}_{\ups}) \sinh(\,\wh{q}_{R}-\wh{q}_{L})  \big]^{ \big(\, \wh{\op{f}}_{\ups}^{\,(\mf{b})} \big)^2} }  \Bigg\}
\cdot \bigg( 1+\e{O}\Big( \f{ \ln L  }{ L } \Big) \bigg)
\end{multline}
with 
\beq
\wh{\op{f}}_{\ups}^{\,(\mf{b})}\; = \; \varkappa_{\ups} \, - \, \wh{F}_{\e{reg}}^{\,(\mf{b})}(\, \wh{q}_{\ups} )   
\label{definition hat f ups}
\enq
and where the functional $\wh{\op{H}}_{\varkappa_{L};\varkappa_{R}}$ takes the explicit form
\bem
\wh{\op{H}}_{\varkappa_{L};\varkappa_{R}}[ \wh{F}_{\e{reg}}^{\,(\mf{b})} ] \; = \; 
\Int{\wh{q}_{L}}{\wh{q}_{R}} \f{ \wh{F}_{\e{reg}}^{\prime}(s)\wh{F}_{\e{reg}}^{\,(\mf{b})}(t)- \wh{F}_{\e{reg}}^{\,(\mf{b})}(s)\wh{F}_{\e{reg}}^{\prime}(t) }{ 2\tanh(s-t)} \dd t \dd s  \\
- \sul{ \ups \in \{L,R\} }{}
\sg_{\ups} \Big( \, \wh{\op{f}}_{\ups}^{\,(\mf{b})}+\varkappa_{\ups} \Big) \Int{\wh{q}_{L}}{\wh{q}_{R}} \f{ \wh{F}_{\e{reg}}^{\,(\mf{b})}(s)-\wh{F}_{\e{reg}}^{\,(\mf{b})}(\,\wh{q}_{\ups}) }{ \tanh(s-\wh{q}_{\ups} ) } \cdot \dd s  \;. 
\end{multline}
A straightforward expansion based on \eqref{ecriture DA racine q har R et L} shows that 
\bem
\wh{\op{H}}_{\varkappa_{L};\varkappa_{R}}[ \wh{F}_{\e{reg}}^{\,(\mf{b})} ] \, + \, \big( \varkappa_R-\varkappa_L\big)^2 \ln \sinh(\,\wh{q}_R  - \wh{q}_L)  \\
\; = \; \wh{\op{H}}_{\varkappa_{L};\varkappa_{R}}[ \wh{F}_{\e{reg}}^{\,(\mf{b})} ]_{\wh{q}_{\ups} \hookrightarrow \sg_{\ups}q}  \, + \, \big( \varkappa_R-\varkappa_L\big)^2 \ln \sinh(2q) 
  \, + \, \e{O}\Big(  \tfrac{ \norm{ \wh{F}_{\e{reg}}^{\,(\mf{b})} }_{ L^{\infty}(\mc{S}_{\de}(\R)) }   }{ L } \Big)  \;. 
\nonumber
\end{multline}
 The claim then follows upon recalling the estimate  \eqref{estimee reste NLIe et deviation Freg a F}. 
 \qed

The next proposition utilises the partitioning \eqref{ecriture partition ens part trou spectre massif et massless} of the set of particle and hole roots into the massless and massive modes
and the shorthand notation for the  discrepancies between the number of particles and holes in the left, right and off-boundary collections of roots 
\eqref{definition notation compactes pour modes massless massif et leur differences sur zone Fermi}.

\begin{prop}
\label{Proposition asymptotiques Dex}
  
 Let $p_{a}^{\ups}, h_a^{\ups}$ be the integers parametrising the particle-hole Bethe roots squeezing on the Fermi zone, \textit{c.f.} the 
decomposition \eqref{ecriture partition ens part trou spectre massif et massless},  \eqref{definition entiers particule et trou bord droit} and \eqref{definition entiers particule et trou bord gauche}.
Under the notations and assumptions of Proposition  \ref{Proposition DA Dbk}, it holds
\bem
\wh{\mc{D}}_{\e{ex}}\big(\La_{\mf{b}}^{\!(\a)};\Ups \big) \cdot \mc{R}^{-1}_{\e{ex}} \, = \,(-1)^{ |\Ups^{(h)}_{\e{off}}| }  \cdot  \f{  \pl{\mu \in \Ups^{(h)}_{\e{off}}}{}  \Big( 1-\ex{-2\i\pi F^{\,(\varrho)}(\mu)} \Big)^2 }
{  \pl{    \mu \in   \Ups^{(h)}_{\e{off}}\cup  \Ups^{(p)}_{\e{off}}    }{} \Big\{L  p^{\prime}(\mu) \Big\}  }
 \cdot \mf{X}_{\e{tot}}\Big( \Ups^{(h)}_{\e{off}}; \mf{C}  ; \{ \ell_{\ups}^{\varkappa} \}  \Big)
\cdot   \bigg|  \f{ \sinh(2q) }{ \sinh(2q+\i\zeta) }  \bigg|^{ 2\varkappa_R \varkappa_L }  \\
\pl{ \ups \in \{L,R\} }{} \Bigg\{ \f{  \mc{R}_{n_{\ups}^{(p)}, n_{\ups}^{(h)} }\big( \{p_a^{\ups}\};\{h_a^{\ups} \} \mid-\sg_{\ups} \op{f}_{\ups}^{\,(\varrho)} \big) }
					{\big[ \tf{ L \sinh(2q) p^{\prime}(\sg_{\ups}q) }{ 2\pi } \big]^{ (\op{f}_{\ups}^{\,(\varrho)} -\sg_{\ups} \ell_{\ups})^2- (\op{f}_{\ups}^{\,(\varrho)})^{\,2} }      }
\cdot  \f{  \ex{ \sg_{\ups}\varkappa_{\ups} \, \aleph^{(\e{bd})}[F^{\,(\varrho)}](\sg_{\ups}q)  }  }{  (2\pi)^{ \varkappa_{\ups}\sg_{\ups}}  }   \bigg( \f{ \sin(\zeta) }{ \sinh(2q) } \bigg)^{ \varkappa_{\ups}^2 }  \Bigg\}  \\
\times \Bigg\{ 1+\e{O}\bigg(   \f{ \descnode+ \descnode_{\ln}  +\ln L }{ L }   \bigg)  \Bigg\}
\label{ecriture asympt dom Dex}
\end{multline}
 where $\descnode$, $ \descnode_{\ln}$ are as defined in \eqref{definition estimee m} and \eqref{definition descnode log},  $\mc{R}_{n^{\ups}_{p}, n^{\ups}_{h} }$ is given by \eqref{definition densite R discrete}, the integral transform
$\aleph^{(\e{bd})}$ has been defined in \eqref{definition transformee aleph bd} while $\mf{X}_{\e{tot}}$ can be found in \eqref{definition fonction G cal} and 
$\op{f}_{\ups}^{\,(\varrho)}$ has been defined in \eqref{definition coeff f ups}.

\end{prop}

\Proof 

It is straightforward to deduce from the large-$L$ expansion \eqref{definition estimee m}  of the $L$ or $R$
 particle/hole roots that 
\bem
\mc{G}\Big( \Ups^{(p)}\setminus \Ups^{(h)}; \{ \{  c_a^{(r)} \}_{ a=1 }^{ n_r^{(z)} }  \}_{ r = 2 }^{ p_{\e{max}} }  \Big) \; = \; 
\mc{G}\Big(\Ups^{(p)}_{\e{off}}\setminus \Ups^{(h)}_{\e{off}} ; \{ \{  c_a^{(r)} \}_{ a=1 }^{ n_r^{(z)} }  \}_{ r = 2 }^{ p_{\e{max}} }   \Big)  \\
\times \pl{  r=2   }{ p_{\e{max}} } \pl{ a=1 }{ n_r^{(z)} } \pl{\ups \in \{L, R \} }{}\Big\{ \Phi_{1,r}\big( \sg_{\ups} q - c_{a}^{(r)} \big) \, \Phi_{1,r}\big(  c_{a}^{(r)} -\sg_{\ups} q \big)\Big\}^{\ell_{\ups}} 
\cdot \Big( 1 + \e{O}\Big(   \f{ \descnode }{ L } \Big)  \Big) \;. 
\end{multline}
Let $ \mathbb{Y}_{\e{off}} = \Ups^{(z)}_{\e{off}}\setminus \Ups^{(h)}_{\e{off}}$.  By virtue of Corollary \ref{Corollaire asymptotiques transformee L et L tilde} and the expansions \eqref{ecriture DA racine q har R et L}, 
 straightforward handlings lead to 
\beq
\pl{ \substack{  \mu \in \mathbb{Y}_{\e{off}} \\  \eps \in \{\pm 1 , 0\} } }{} \ex{\iota_{\eps} \mc{L}_{\msc{C}}[\, \wh{z} \, ](\mu+\i\eps \zeta) } \; = \; 
\f{ \pl{r=1}{ p_{\e{max}} } \pl{a=1}{ n_r^{(z)} }  \Big\{ \ex{\aleph^{(r)}[F^{\,(\varrho)}]( c_a^{(r)} ) } 
\pl{\ups \in \{L, R \} }{}\bigg\{ \Phi_{1,r}\big( \,\wh{q}_{\ups}   - c_{a}^{(r)} \big) \Phi_{1,r}\big( c_{a}^{(r)} - \wh{q}_{\ups}   \big)\Big\}^{-\sg_{\ups}\varkappa_{\ups}}    \bigg\}  }
{  \pl{\mu \in \Ups^{(h)}_{\e{off}} }{  }   \bigg\{ \ex{\aleph^{(1)}_{-}[F^{\,(\varrho)}]( \mu ) } 
\pl{\ups \in \{L, R \} }{}\Big\{ \Phi_{1,1}\big( \, \wh{q}_{\ups}  - \mu \big) \Phi_{1,1}\big( \mu -\wh{q}_{\ups}   \big)\Big\}^{-\sg_{\ups}\varkappa_{\ups}}    \bigg\}    }
\cdot \bigg( 1 + \e{O}\Big( \f{ 1 }{ L } \Big)  \bigg) \;. 
\nonumber
\enq
The integral transforms $\aleph^{(r)}$ have been introduced in  \eqref{definition transformee aleph r} and $\aleph^{(1)}_{-}$ corresponds to the $-$ boundary value of the transform. 

 Corollary \ref{Corollaire asymptotiques transformee L et L tilde} , straightforward expansions based on the 
definition of such roots \eqref{definition entiers particule et trou bord droit}-\eqref{definition entiers particule et trou bord gauche}, their large-$L$ expansion \eqref{definition estimee m}
and the form of the uniform expansion of the Gamma function \eqref{estimee fct Gamma} lead to 
\bem
\pl{\mu \in \Ups^{(p)}_{\ups}\setminus \Ups^{(h)}_{\ups}  }{}\pl{\eps \in \{\pm 1 , 0\} }{} \ex{  \iota_{\eps} \mc{L}_{\msc{C}}[\, \wh{z} \, ](\mu+\i\eps \zeta) } \; = \; 
\f{  \ex{ \ell_{\ups} \aleph^{(\e{bd})}[F^{\,(\varrho)}]( \sg_{\ups} q  ) }     }{   \Big\{ L \sinh(2q) \wh{\xi}_{\La}^{\prime}(\sg_{\ups} q) \Big\}^{ - 2 \sg_{\ups} \ell_{\ups} \op{f}_{\ups}^{\,(\varrho)} }   } 
\pl{ a=1 }{ n_{\ups}^{(p)} }\Ga^2 \left(\ba{c} 1+p_a^{\ups} - \sg_{\ups} \op{f}_{\ups}^{\,(\varrho)}  \\  1 + p_a^{\ups} \ea \right) \\
\times \; \pl{ a=1 }{ n_{\ups}^{(h)} }\Ga^2 \left(\ba{c} 1+h_a^{\ups} + \sg_{\ups} \op{f}_{\ups}^{\,(\varrho)}  \\  1 + h_a^{\ups} \ea \right)
\;  \; \cdot  \pl{   \substack{ \mu \in \Ups^{(p)}_{\ups}\setminus \Ups^{(h)}_{\ups}  \\ \eps= \pm} }{} \bigg\{ 
\bigg( \f{  \sinh\big( \mu-\wh{q}_{\ups}   + \eps \i\zeta\big)    }{ \sinh \big( \mu-\wh{q}_{\ov{\ups}}   \big) } \bigg)^{  \sg_{\ups} \varkappa_{ \ups } } 
\cdot \bigg( \f{ \sinh \big( \mu-\wh{q}_{\ov{\ups}}   \big) }{  \sinh\big( \mu-\wh{q}_{\ov{\ups}}   + \eps \i\zeta\big)    } \bigg)^{  \sg_{\ups} \varkappa_{ \ov{\ups} } }  \bigg\}  \\
\times \pl{ \mu\in \Ups^{(h)}_{\ups} }{}  \Big\{ \ex{2\i\pi \wh{F}^{\,(\mf{b})}_{\e{reg}}(\mu)} \Big\} \cdot \bigg\{ 1 + \e{O}\Big(    \f{ \descnode_{\ln}  + \descnode + \ln L  }{ L } \Big)  \bigg\} \;. 
\end{multline}
Here, $\ov{\ups}=L$ if $\ups=R$ and $\ov{\ups}=R$ if $\ups=L$.

The large-$L$ expansion \eqref{definition estimee m} of the roots belonging to $\Ups^{(p/h)}_{\ups}$ and properties $iii)-iv)$ of Lemma \ref{Lemme structure zeros F} 
ensures that
\bem
\pl{ \mu\in \Ups^{(h)}_{L}\cup \Ups^{(h)}_{R} }{} \hspace{-5mm} \Big\{ \ex{2\i\pi \wh{F}^{\,(\mf{b})}_{\e{reg}}(\mu)} \Big\} \cdot 
\f{ \pl{ \mu \in \Ups^{(h)} }{} \Big(1-\ex{-2\i\pi \wh{F}^{\,(\mf{b})}(\mu)} \Big)^2 }{  \big( 2 \i \pi)^{|\Ups^{(p)}|+|\Ups^{(h)}|}  } 
\; = \; 
\f{ \pl{ \mu \in \Ups^{(h)}_{\e{off}} }{} \Big(1-\ex{-2\i\pi F^{\,(\varrho)}(\mu)} \Big)^2  }{ \big( 2 \i \pi)^{n^{(p)}_{\e{off}}+n^{(h)}_{\e{off}}} }\\ 
\times \pl{ \ups \in \{L,R \} }{  }\Bigg\{ \f{1}{ (2\i\pi)^{\ell_{\ups}} } \cdot \bigg(  \f{\sin\big[\pi F^{\,(\varrho)}(\sg_{\ups}q) \big] }{ \pi    }\bigg)^{2n_{\ups}^{(h)}} \Bigg\}
\cdot \Big( 1 + \Big( \f{   \descnode }{ L } \Big)  \Big) \;. 
\end{multline}

Collecting the $\varkappa_{\ups}$ dependent terms issuing from the various $\mc{L}_{\msc{C}}$ transforms into the function
\bem
\chi\big(\Ups^{(h)};  \Ups^{(p)} \big) \, = \, \pl{\ups \in \{L,R\} }{} \bigg\{  \pl{ \mu \in \Ups^{(p)}_{\e{off}} \setminus \Ups^{(h)}_{\e{off}} }{}  
\Big\{ \Phi_{1,1}\big(\mu-\wh{q}_{\ups}\big) \Phi_{1,1}\big(\,\wh{q}_{\ups}-\mu\big) \Big\}^{-\sg_{\ups} \varkappa_{\ups} }  \bigg\}  \\
\times \pl{\ups \in \{L,R\} }{}  \pl{ \mu \in \Ups^{(p)}_{\ups} \setminus \Ups^{(h)}_{\ups} }{}  \Bigg\{     \pl{\ups^{\prime} \in \{L,R\} }{}
\Big\{ \Phi_{1,1}\big(\mu-\wh{q}_{\ov{\ups}}\big) \Phi_{1,1}\big(\,\wh{q}_{\ov{\ups}}-\mu\big) \Big\}^{-\sg_{\ups^{\prime}} \varkappa_{\ups^{\prime}} }  
 \pl{\eps=\pm }{}\bigg( \f{ \sinh\big(\mu-\wh{q}_{R}+\i\eps \zeta \big)  }{ \sinh\big(\mu-\wh{q}_{L}+\i\eps \zeta \big) } \bigg)^{ \varkappa_{\ups} }
\Bigg\}
\end{multline}
one has that
\bem
\big( \mc{W}\mc{D} \, \chi \big)\big(\Ups^{(h)};  \Ups^{(p)} \big) \, = \, 
\Big\{   \Phi_{1,1}\big(\,\wh{q}_{R}-\wh{q}_{L}\big) \Phi_{1,1}\big(\,\wh{q}_{L}-\wh{q}_{R}\big)   \Big\}^{ \varkappa_{R} \varkappa_{L} } 
\cdot \pl{\ups \in \{L,R,\e{off} \}}{} \hspace{-3mm} \big( \mc{W}\mc{D}\big)\big(\Ups^{(h)}_{\ups};  \Ups^{(p)}_{\ups} \big) \\
\times \pl{ \la \in \Ups_{\e{off}}^{\de} }{}  \pl{ \mu \in \wt{\Ups}_{L}^{\de} \cup \wt{\Ups}_{R}^{\de} }{} \hspace{-3mm} \Big\{ \Phi_{1,1}\big(\la-\mu\big) \Phi_{1,1}\big(\mu-\la\big)  \Big\}
\cdot \pl{ \la \in  \wt{\Ups}_{R}^{\de}  }{}  \pl{ \mu \in \wt{\Ups}_{L}^{\de} }{}  \Big\{   \Phi_{1,1}\big(\la-\mu\big) \Phi_{1,1}\big(\mu-\la\big) \Big\} \\
\times   \pl{ \ups \in \{ L,R  \} }{} \pl{ \mu \in \Ups_{\ups}^{\de}}{} 
\bigg\{   \pl{\eps=\pm }{} \f{ \sinh\big(\mu-\wh{q}_{R}+\i\eps \zeta \big)  }{ \sinh\big(\mu-\wh{q}_{L}+\i\eps \zeta \big) } 
 \Phi_{1,1}^{-\sg_{\ups} }\big(\mu-\wh{q}_{\ov{\ups}}\big) \Phi_{1,1}^{-\sg_{\ups} }\big(\,\wh{q}_{\ov{\ups}}-\mu\big)     \bigg\}^{ \varkappa_{\ups} } \;. 
\end{multline}
I have introduced $\Ups_{\ups}^{\de} \, = \,   \Ups^{(p)}_{\ups} \setminus \Ups^{(h)}_{\ups}  $ and 
$\wt{\Ups}_{\ups}^{\de} \, = \,  \Ups_{\ups}^{\de} \cup \{\, \wh{q}_{\ups} \}^{-\sg_{\ups} \varkappa_{\ups} }  $. There, $\wh{q}_{\ups}$ is repeated 
$|\varkappa_{\ups}|$ times and the set should be added if $-\sg_{\ups} \varkappa_{\ups}\geq 0$ and subtracted otherwise. 
The large-$L$ expansion of most of the individual terms appearing in the above decomposition can be readily accessed. 
Some more care is only needed relatively to $\big( \mc{W}\mc{D}\big)\big(\Ups^{(h)}_{\ups};  \Ups^{(p)}_{\ups} \big)$ when $\ups\in \{L,R\}$.  
One should regularise the singular terms as proposed in \cite{KozKitMailSlaTerEffectiveFormFactorsForXXZ}
leading to 
\beq
\big( \mc{W}\mc{D}\big)\big( \Ups^{(h)}_{ \ups  }; \Ups^{(p)}_{ \ups  } \big) \; = \; \f{ (-1)^{ \ell_{\ups} \tfrac{\ell_{\ups}-1}{2} -n_{\ups}^{(h)} }   }
{ L^{\ell_{\ups}^2}  \hspace{-3mm} \pl{ \la \in  \Ups^{(h)}_{ \ups  } \cup \Ups^{(p)}_{ \ups  }   }{} \hspace{-2mm} \big\{ L \wh{\xi}^{\prime}_{\Ups}(\mu) \big\}^{-1} }
\pl{ \la \in   \Ups_{\ups}^{\de}  }{}  \pl{ \mu \in \Ups_{\ups}^{\de} }{}  \wh{\Psi}_{1,1}\big(\la,\mu)
\f{ \pl{a<b}{n^{(h)}_{\ups}} (h_a^{\ups}-h_b^{\ups} )^2 \cdot  \pl{a<b}{n^{(p)}_{\ups}} (p_a^{\ups} - p_b^{\ups} )^2 }{  \pl{a=1}{ n^{(h)}_{\ups} } \pl{ b=1 }{ n^{(p)}_{\ups} } (h_a^{\ups} + p_b^{\ups} + 1 )^2   }  
\enq
where 
\beq
\wh{\Psi}_{1,1}\big(\la,\mu) \, = \, \f{  \sinh(\la-\mu) }{ \big( \,  \wh{\xi}_{\Ups}(\la) \, - \, \wh{\xi}_{\Ups}(\mu) \big) \sinh(\la-\mu-\i\zeta) } \;. 
\enq
The asympototic expansion \eqref{ecriture DA fct de cptage} and  
spacing properties of the singular roots \eqref{propriete espacement ctre corde et particule trou} and the upper bound \eqref{ecriture deviation des mua au bea} on the spacing between the singular roots $\a_a^{ (s) }$ and  $\be_a^{ (s) }$ 
ensure, owing to Lemma \ref{Lemme borne sur magnitude fct cptge sing}, that that for any
\beq
\la\not= \mu \in   \Ups^{(h)}_{ \ups  } \cup \Ups^{(p)}_{ \ups  }\;\;, \e{it} \; \e{holds}
\qquad 
\wh{\xi}_{\Ups}(\la) \, - \, \wh{\xi}_{\Ups}(\mu) \, = \, \Big( \, \wh{\xi}_{\Ups_{\e{reg}} }(\la) \, - \, \wh{\xi}_{\Ups_{\e{reg}} }(\mu) \Big) \Big( 1+\e{O}\big(n_{\e{sg}}L^{-\infty}\big) \Big)  \;. 
\enq
Straightforward expansions then lead to 
\beq
\big( \mc{W}\mc{D}\big) \big( \Ups^{(h)}_{ \ups  }; \Ups^{(p)}_{ \ups  } \big) \; = \; \f{ (-1)^{\ell_{\ups} \tfrac{\ell_{\ups}-1}{2} -n_{\ups}^{(h)} }  (\i)^{\ell_{\ups}}    }
{ \Big\{ \tfrac{ L \, p^{\,\prime}(\sg_{\ups} q) }{(2\pi)}\Big\}^{\ell_{\ups}^2-n_{\ups}^{(p)}-n_{\ups}^{(h)}} }
\f{ \pl{a<b}{n^{(h)}_{\ups}} (h_a^{\ups}-h_b^{\ups} )^2 \cdot  \pl{a<b}{n^{(p)}_{\ups}} (p_a^{\ups} - p_b^{\ups} )^2 }
{   \sin^{ \ell_{\ups}^2 }(\zeta)  \pl{a=1}{ n^{(h)}_{\ups} } \pl{ b=1 }{ n^{(p)}_{\ups} } (h_a^{\ups} + p_b^{\ups} + 1 )^2   } 
\cdot \bigg\{ 1 + \e{O}\Big( \f{  1 + \descnode }{ L } \Big)  \bigg\} \;. 
\nonumber 
\enq
All-in-all, one gets that 
\bem
\big( \mc{W}\mc{D}\, \chi \big)\big(\Ups^{(h)};  \Ups^{(p)} \big) \, = \, \big( \mc{W}\mc{D}\big) \big( \Ups^{(h)}_{ \e{off} }; \Ups^{(p)}_{ \e{off}  } \big) 
\pl{\mu \in  \Ups^{\de}_{ \e{off} } }{} \pl{\ups \in \{L,R\} }{} \Big\{ \Phi_{1,1}\big(\mu -\sg_{\ups}q  \big) \Phi_{1,1}\big(\sg_{\ups}q - \mu \big) \Big\}^{\ell_{\ups}^{\varkappa} } \\
\times \Big| \f{\sinh(2q) }{ \sinh(2q-\i\zeta) } \Big|^{2 (\varkappa_{L}\varkappa_{R} + \ell_{L}^{\varkappa}\ell_{R}^{\varkappa}  ) }
\pl{\ups \in \{L,R\} }{} \Bigg\{  \bigg(\f{ \sinh(2q) }{ \sin(\zeta) } \bigg)^{ (\ell_{\ups}^{\varkappa})^2-\varkappa_{\ups}^2 }
\f{ \prod_{ a<b }^{ n^{(h)}_{\ups} } (h_a^{\ups}-h_b^{\ups} )^2 \cdot  \prod_{a<b}^{n^{(p)}_{\ups}} (p_a^{\ups} - p_b^{\ups} )^2 }
{     \prod_{a=1}^{ n^{(h)}_{\ups} } \prod_{ b=1 }^{ n^{(p)}_{\ups} } (h_a^{\ups} + p_b^{\ups} + 1 )^2   }  \Bigg\}  \\
\times \pl{\ups \in \{L,R\} }{} \Bigg\{  (\i)^{\ell_{\ups}}(-1)^{n^{(h)}_{\ups} } \bigg(\f{ L p^{\prime}(\sg_{\ups}q) }{ 2\pi } \bigg)^{ n^{(p)}_{\ups} + n^{(h)}_{\ups}}  \cdot 
\bigg( \f{ 2\pi }{ L \sinh(2q) p^{\prime}(\sg_{\ups}q) } \bigg)^{ \ell_{\ups}^2 }   \Bigg\}
\cdot \bigg\{ 1 + \e{O}\Big(   \f{   1 + \descnode }{ L } \Big)  \bigg\}
\nonumber
\end{multline}
Finally, the product $\mc{D}\cdot \mc{W}$ associated to the off-critical particle-hole excitations can be recast as
\beq
\big( \mc{D}\cdot \mc{W} \big)\big( \Ups^{(h)}_{ \e{off} }; \Ups^{(p)}_{  \e{off}  } \big) \; = \; 
 \f{ \pl{  \la \not= \la^{\prime} \\\in \Ups^{(p)}_{ \e{off} }     }{}   \Phi_{1,1}\big(\la-\la^{\prime}\big)  \cdot \pl{  \mu\not=\mu^{\prime}  \in \Ups^{(h)}_{ \e{off} }    }{}   \Phi_{1,1}\big(\mu-\mu^{\prime}\big)  }
 { \Big\{ \sinh(-\i\zeta) \Big\}^{n^{(p)}_{\e{off}}+n^{(h)}_{\e{off}}}   \cdot \pl{   \mu \in \Ups^{(h)}_{ \e{off} }    }{} \pl{ \la \in  \Ups^{(p)}_{  \e{off}  } }{} \Big\{ \Phi_{1,1}(\la-\mu) \, \Phi_{1,1}(\mu-\la) \Big\}  } \;. 
\enq
It solely remains to put the expansions of the various terms together.  \qed

\section{ Analysis of $\mc{A}_{\e{reg}}\big(\La_{\mf{b}}^{\!(\a)}; \Ups \big)$ }
\label{Section Analyse de Areg}

\subsection{Statement of the result}

Define the functional 
\beq
\op{W}_{\varkappa_{L};\varkappa_{R}}\big[f] \; = \;  \f{  \pl{\eps=\pm}{}\big[ \sinh(2q+ \eps \i\zeta)\big]^{\varkappa_{R} \varkappa_{L}}   }{ \big[ \sinh(\i\zeta)\big]^{\varkappa_{R}^2} \big[ \sinh(-\i\zeta)\big]^{\varkappa_{L}^2} } 
\cdot \ex{ \op{H}_1[ f ] } \cdot \pl{\eps=\pm}{} \pl{\ups \in \{L,R\} }{}   \ex{  \sg_{\ups} \varkappa_{\ups} \op{C}_{I_q}[2\i\pi f](\sg_{\ups}q+\eps \i\zeta)  } 
\enq
where the functional $\op{H}_1$ is given by \eqref{definition de la fnelle H1}. 

\begin{prop}
\label{Proposition asymptotiques Areg}
 
 Let $\mf{b}=\varrho+\tf{\de \varrho }{L} \in \mc{D}_{0,r_L}$ with $r_L$ as in Proposition \ref{Proposition rep reguliere pour les FF}. Assume that the string centre hypothesis holds \eqref{propriete espacement ctres cordes}, then 
\beq
\wh{\mc{A}}_{\e{reg}}\big( \La_{\mf{b}}^{\!(\a)}; \Ups \big) \; = \; \mc{A}_{\e{reg}} \big( \Ups^{(h)}_{\e{off}};\mf{C}; \{ \ell_{\ups}^{\varkappa} \} \big) \cdot
\bigg\{ 1\, + \, \e{O}\Big(  \f{ \descnode }{L}   \Big) \bigg\}
\enq
where 
\bem
 \mc{A}_{\e{reg}}\big( \Ups^{(h)}_{\e{off}};\mf{C}; \{ \ell_{\ups}^{\varkappa} \} \big) \; = \; (-1)^{(\varkappa_{R}-\varkappa_{L})\varkappa_{L}} 
\op{W}_{\varkappa_{L};\varkappa_{R}}\big[F^{(\varrho)}\big]  \cdot  \f{  C^{(\ga)}\big( \Ups^{(h)}_{\e{off}};\mf{C}; \{ \ell_{\ups}^{\varkappa} \} \mid \varrho \big) }{ \det^2\big[ \e{id} + \op{K} \big] } \\
 \times \pl{ \ups \in \{ L , R \} }{} \bigg\{  (-1)^{\tfrac{1}{2}\varkappa_{\ups}(\varkappa_{\ups}-1) }   (2\i\pi)^{\sg_{\ups}\varkappa_{\ups}} 
 \f{  \ex{ \i \f{\pi}{2} \sg_{\ups}  \big( F^{(\mf{b})}_{\e{reg}}(\wh{q}_{\ups}) \big)^{2}  + \i \pi \sg_{\ups} \varkappa_{\ups} \wh{\op{f}}_{\ups}^{\,(\mf{b})}  }  } { \big( 2\pi \big)^{ \sg_{\ups}F^{(\varrho)}(\sg_{\ups} q) } }   
\cdot \f{ G (  1-\sg_{\ups}\op{f}_{\ups}^{\,(\varrho)} ) }{ G( 1+ \sg_{\ups}\op{f}_{\ups}^{\,(\varrho)} ) } \bigg\}  
 \;. 
\end{multline}
 where $G$ is the Barnes function and the coefficients $C^{(\ga)}$ have been introduced in \eqref{definition cste Cz non univ}-\eqref{definition cste C+ non univ}. 
 
\end{prop}

\subsection{Analysis}

\begin{prop}

Assume that the string centre separation hypothesis holds \eqref{propriete espacement ctres cordes}, then 
\beq
\mc{W}\big( \La_{\mf{b}}^{\!(\a)}; \Ups^{(\e{in})} \big) \; = \;  (-1)^{(\varkappa_{R}-\varkappa_{L})\varkappa_{L}}  \op{W}_{\varkappa_{L};\varkappa_{R}}\big[ F^{(\varrho)} \big]
\cdot  \Big( 1+\e{O}\Big( \f{1}{L}   \Big) \Big)  \;. 
\enq
\end{prop}

\Proof 

The starting point for the expansion is given by \eqref{ecriture expression nouvelle pour W} appearing in Proposition \ref{Proposition reecriture integrale D bulk}. After an integration by parts, the $\mc{A}_{\zeta}$ transform can be recast as
\beq
\mc{A}_{\zeta}\big[\, \wh{z}, \, \wh{z} \, \big] \, = \, - \sul{\ups \in \{L,R\}}{} \hspace{-2mm} \sg_{\ups} \mf{d}\big[\, \wh{z}\,\big] (\, \wh{q}_{\ups}) \, \op{C}_{\msc{C}}\big[\, \wh{z}  \, \big](\, \wh{q}_{\ups} - \i\zeta) 
\; + \Int{\msc{C} }{}  \f{ \wh{z}\,(s)\,  \wh{z}\,(t)  }{ \sinh^2(s-t+\i\zeta) } \cdot \f{\dd s\, \dd t }{ (2\i\pi)^2 } \;. 
\enq
The double integral  can be estimated starting from the decomposition 
\beq
\wh{z}\, (\om) \, = \, 2 \i \pi  F^{(\varrho)}(\om) \bs{1}_{\msc{C}^{(+)}}(\om) \, + \,  \wh{\mf{z}}(\om)
\label{ecriture decomposition hat z}
\enq
 where 
\beq
 \wh{\mf{z}}\,(\om) \, = \,
	    \left\{  \ba{cc} 2\i\pi \Big( \wh{F}_{\e{reg}}^{\,(\mf{b})} (\om)-F^{(\varrho)}(\om)-L\wh{\xi}_{\Ups_{\e{sing}} }(\om) \Big)   + \wh{u}_{\Ups}^{\, (+)}(\om)-	 \wh{u}_{ \La_{\mf{b}}^{\!(\a)}}^{\, (+)}(\om)  & \om \in \msc{C}^{(+)}  \vspace{2mm}\\ 
					  \wh{u}_{\Ups}^{\, (-)}(\om)-	 \wh{u}_{ \La_{\mf{b}}^{\!(\a)}}^{\, (-)}(\om)      &   \om \in \msc{C}^{(-)}  \ea \right. \;. 
\enq
By Lemmas \ref{Lemme borne sur magnitude fct cptge sing}-\ref{Lemme estimation propriete generales ctg fct} and the bound \eqref{estimee reste NLIe et deviation Freg a F}, one gets
that $\norm{   \, \wh{\mf{z}} \,  }_{ L^{1}(\msc{C}) }=\e{O}(L^{-1})$. The decomposition \eqref{ecriture decomposition hat z} then allows one to recast 
\bem
 \Int{\msc{C} }{}  \f{ \wh{z}(s)\,  \wh{z}(t)  }{ \sinh^2(s-t+\i\zeta) } \cdot \f{\dd s \dd t }{ (2\i\pi)^2 } \, = \, \Int{ \msc{C}^{(+)} }{} \f{ F^{(\varrho)} (t) \, F^{(\varrho)} (s) }{ \sinh^2(s-t+\i\zeta) } \dd s \dd t \\
\, + \,   \Int{ \msc{C}  }{} \f{ \dd s }{ 2\i\pi }  \Int{ \msc{C}^{(+)} }{} \dd t  \bigg\{ \f{ F^{(\varrho)}(t) \,  \wh{\mf{z}}\,(s) + F^{(\varrho)}(s) \,  \wh{\mf{z}}\,(t) }{ \sinh^2(s-t+\i\zeta) }  \bigg\} 
\, + \, \Int{ \msc{C}  }{} \f{ \dd s \dd t  }{ (2\i\pi)^2 } 
\f{  \wh{\mf{z}}\,(t)  \, \wh{\mf{z}}\,(s)  }{ \sinh^2(s-t+\i\zeta) }  \;. 
\end{multline}
The last two integrals produce $\e{O}(    L^{-1})$ corrections owing to direct bounds and estimates on the $L^1$ norm of $\wh{\mf{z}}$ and the bound 
\eqref{borne sur entier varkappa et sur F rho}. 

Further, by invoking the large-$L$ asymptotics of the $\mc{L}_{\msc{C}}$ and $\wt{\mc{L}}_{\msc{C}}$ transforms, the spacing property of the singular  roots \eqref{ecriture deviation des mua au bea} and the 
string centre spacing hypothesis \eqref{propriete espacement ctres cordes}  allow one to infer that\newline  $\mc{R}_{\mc{W}}=1+\e{O}\big( n_{\e{sg}}^2\cdot L^{-\infty} \big)$.

Then, by using the large-$L$ behaviour of the Cauchy transforms given in Proposition \ref{Proposition DA transformee Cauchy} and of the bounds \eqref{estimee reste NLIe et deviation Freg a F}, one gets that 
\bem
\mc{W}\big( \La_{\mf{b}}^{\!(\a)}; \Ups^{(\e{in})} \big) \; = \; \exp\Bigg\{ -\Int{ \wh{q}_L }{ \wh{q}_{R} } \f{ F^{(\varrho)} (t) \, F^{(\varrho)} (s) }{ \sinh^2(s-t+\i\zeta) } \dd s \dd t 
\, + \, \sul{ \eps = \pm }{  } \sul{ \ups\in \{L,R\} }{}  \sg_{\ups} \varkappa_{\ups} \op{C}_{ \intff{ \wh{q}_L }{ \wh{q}_{R} } }\big[2\i\pi  F^{(\varrho)} \big]\big(\wh{q}_{\ups}+\i\eps \zeta \big)  \\
\,-\, \sul{ \substack{\ups, \ups^{\prime} \\  \in \{L,R\} } }{} \sg_{\ups}\sg_{\ups^{\prime}} \varkappa_{\ups} \ln \sinh\big( \wh{q}_{ \ups^{\prime} } - \wh{q}_{\ups} + \i\zeta  \big) 
\; + \; \e{O}\Big(  \f{1}{L}  \Big) \Bigg\} \;. 
\end{multline}
It then remains to invoke the expansion of the endpoints \eqref{ecriture DA racine q har R et L} and then observe that all resulting  correction of first order in $L^{-1}$
cancel out, so that, owing to \eqref{borne sur entier varkappa et sur F rho}, the first corrections are a $\e{O}\Big( \ \norm{ F^{(\varrho)} }_{  L^{\infty}(\mc{S}_{\de}(\R))  }^{2}  \cdot L^{-2} \Big)$. \qed

I now discuss the large-$L$ asymptotics of the other building blocks of $\wh{\mc{A}}_{\e{reg}}^{(\ga)}(\La_{\mf{b}}^{\!(\a)};\Ups)$. This analysis requires the characterisation of the 
poles and zeroes of $1-\ex{2\i\pi \wh{F}^{\,(\mf{b})}}$ inside of $\msc{C}$ which has been achieved in Lemma \ref{Lemme distributions zeros fct comptage}.

\begin{prop}
 Let $\mf{b}=\varrho+\tf{\de \varrho }{L} \in \mc{D}_{0,r_L}\setminus\{ \pm r_{L} \}$ with $r_L$ as in Proposition \ref{Proposition rep reguliere pour les FF} and assume that the string centre separation hypothesis holds \eqref{propriete espacement ctres cordes}. 
Then, 
\beq
\f{ \pl{ \la \in \La^{\!(\a)}_{\mf{b}} }{} \Big(\ex{-2\i\pi \wh{F}^{\,(\mf{b})}(\la)}-1\Big)   }{  \pl{\mu \in \Ups^{(\e{in})} }{} \Big(\ex{-2\i\pi \wh{F}^{\,(\mf{b})}(\mu)}-1\Big)   }  \; = \;  
\exp \bigg\{ \Oint{\msc{C} }{} \wh{z}\,(s) \ln^{\prime}\big[ \ex{-2\i\pi \wh{F}^{\,(\mf{b})}(s)} -1 \big] \cdot \f{ \dd s }{2\i\pi}  \bigg\} \cdot 
\pl{\ups \in \{L,R\} }{} \Big( \ex{-2\i\pi \wh{F}^{\,(\mf{b})}(\,\wh{q}_{\ups})} -1  \Big)^{\sg_{\ups} \varkappa_{\ups}} 
\cdot \mf{R}_{F} 
\label{ecriture rep int L fini exacte pour produit ac fct shift}
\enq
where the remainder takes the form
\beq
 \mf{R}_{F} \; = \; \pl{ a=1 }{  n_{\e{sg}} }  \f{ 1-  \ex{-2\i\pi L \wh{\xi}_{\La^{\!(\a)}}^{\,(\mf{b})}\big( \mu_{a}^{(s)} \big)}  }{ 1-  \ex{-2\i\pi L \wh{\xi}_{\La^{\!(\a)} }^{\,(\mf{b})}\big( (\be_{a}^{(s)} )^*\big)}  }  \;. 
\enq
Furthermore, provided that property \eqref{condition distance racine mua sing au reseau Lambda} holds, one has the large-$L$ expansion  
\beq
\f{ \pl{ \la \in \La^{\!(\a)}_{\mf{b}} }{} \Big(\ex{-2\i\pi \wh{F}^{\,(\mf{b})}(\la)}-1\Big)   }{  \pl{\mu \in \Ups^{(\e{in})} }{} \Big(\ex{-2\i\pi \wh{F}^{\,(\mf{b})}(\mu)}-1\Big)   }  \; = \; 
\pl{ \ups \in \{ L , R \} }{} \bigg\{  (-1)^{\tfrac{1}{2}\varkappa_{\ups}(\varkappa_{\ups}-1) }  
 \f{  \ex{ \i \f{\pi}{2} \sg_{\ups} \big( \wh{F}^{\,(\mf{b}) }_{\e{reg}}(\wh{q}_{\ups}) \big)^{2}  + \i \pi \sg_{\ups} \varkappa_{\ups} \wh{\op{f}}_{\ups}^{\,(\mf{b})}  }  } 
{ \big(2\i\pi\big)^{-\sg_{\ups}\varkappa_{\ups}} \cdot \big( 2\pi \big)^{ \sg_{\ups}F^{\,(\varrho)}(\sg_{\ups} q) } }   
\cdot 
	    \f{ G (  1-\sg_{\ups}\op{f}_{\ups}^{\,(\varrho)} ) }{ G( 1+ \sg_{\ups}\op{f}_{\ups}^{\,(\varrho)} ) }  \bigg\} \cdot \bigg( 1+\e{O}\Big( \f{1 }{ L }   \Big) \bigg)
\enq
with $\op{f}_{\ups}^{\,(\varrho)}$ given by \eqref{definition coeff f ups}. 
\end{prop}

\Proof

It follows from Lemma \ref{Lemme distributions zeros fct comptage} that, for almost all $\eps>0$, $t\mapsto 1-\ex{-2\i\pi \wh{F}^{\,(\mf{b})} (t+\i\eps)}$ has no roots on $\intff{\wh{q}_L}{\wh{q}_R}$. 
Thus, for these $\eps$'s, there exists a small box $V_{\eps} \supset \intff{\wh{q}_L}{\wh{q}_R}$ passing through $\wh{q}_{\ups}$ such that $t\mapsto 1-\ex{-2\i\pi \wh{F}^{\,(\mf{b})}(t+\i\eps)}$ has no zeroes on $V_{\eps}$ 
and is holomorphic there. One can thus define  on this neighbourhood a holomorphic determination for its
logarithm $\ln_{\eps}\big[ \ex{-2\i\pi \wh{F}^{\,(\mf{b})}(t+\i\eps)} -1\big]$. Then, for $\Om\in \{\La^{\!(\a)}_{\mf{b}},\Ups\}$, one gets
\bem
\sul{ \mu \in \Om^{(\e{in})} }{} \ln_{\eps}\big[\ex{-2\i\pi \wh{F}^{\,(\mf{b})}(\mu+\i\eps)}-1 \big] 
%
%
%
 \; = \; - \Oint{ \Dp{}V_{\eps} }{} \wh{u}_{\Om} (s)\ln_{\eps}^{\prime}\big[ \ex{-2\i\pi \wh{F}^{\,(\mf{b})}(s+\i\eps)} -1 \big] \cdot \f{ \dd s }{2\i\pi} \\
\, + \, \sul{ \ups\in \{L,R\} }{} \hspace{-2mm} \sg_{\ups} \,  \mf{d}\big[\, \wh{u}_{\Om}\big] (\,\wh{q}_{\ups}) \ln_{\eps} \big[  \ex{-2\i\pi \wh{F}^{\,(\mf{b})}(\,\wh{q}_{\ups}+\i\eps)} -1 \big] 
\, + \, \de_{\Om,\Ups} \sul{\mu \in \daleth}{} \ln_{\eps}\big[ \ex{-2\i\pi \wh{F}^{\,(\mf{b})}(\mu+\i\eps)}-1 \big] \;. 
\end{multline}
Deforming the contour $\Dp{}V_{\eps}$ to $\msc{C}$ and using that the only singularities of the integrand delimited by these contours are the simple poles of 
$\ln_{\eps}^{\prime}\big[\ex{-2\i\pi \wh{F}^{\,(\mf{b})}(s+\i\eps)} -1\big] $ located at 
\begin{itemize}

\item $ Z_{\eps} \, = \, \{\mf{z}_a^{(s)} -\i \eps \}_1^{ n_{\e{sg}}}$ with residue $1$;

\item $B_{\eps} \, = \,  \{  \big(\be_a^{(s)}\big)^* -\i \eps \}_1^{ n_{\e{sg}}} $ with residue $-1$;

\item $W_{\eps}   \, = \, \big\{  \mf{w}_{a} -\i \eps \big\}_{a=1}^{ \ell_{F} } $ with residue $1$;
 
\end{itemize}
where $\mf{w}_{a}$ a and $\mf{z}_a^{(s)}$  are the simple zeroes introduced in Lemma \ref{Lemme distributions zeros fct comptage}, equation \eqref{ecriture zeros fct shift exponentiee},
one eventually gets 
\bem
\sul{ \mu \in \Om^{(\e{in})} }{} \ln_{\eps}\big[ \ex{-2\i\pi \wh{F}^{\,(\mf{b})}(\mu+\i\eps)}-1 \big] 
 \; = \; - \Oint{ \msc{C} }{} \wh{u}_{\Om} (s)\ln_{\eps}^{\prime}\big[ \ex{-2\i\pi \wh{F}^{\,(\mf{b})}(s+\i\eps)} -1 \big] \cdot \f{ \dd s }{2\i\pi} \\
\, + \hspace{-2mm}  \sul{ \ups\in \{L,R\} }{}\hspace{-2mm} \sg_{\ups} \,  \mf{d}\big[\, \wh{u}_{\Om}\big] (\,\wh{q}_{\ups}) \ln_{\eps} \big[  \ex{-2\i\pi \wh{F}^{\,(\mf{b})}(\,\wh{q}_{\ups}+\i\eps)} -1 \big] 
\, + \, \de_{\Om,\Ups} \sul{\mu \in \daleth}{} \ln_{\eps}\big[ \ex{-2\i\pi \wh{F}^{\,(\mf{b})}(\mu+\i\eps)} -1 \big]  
\, +\hspace{-3mm} \sul{ \substack{ \mu \in Z_{\eps} \\ \cup W_{\eps} \setminus B_{\eps} } }{} \wh{u}_{\Om}(\mu) \;. 
\label{ecriture calcul somme sur racines Omega in de exp shift fct}
\end{multline}
Note that, in the ultimate sum that occurs in \eqref{ecriture calcul somme sur racines Omega in de exp shift fct}, the elements should be summed up according to their respective multiplicities.

The above formula allows one to recast the $\eps$-deformed version of the ratio appearing in the \textit{lhs} of \eqref{ecriture rep int L fini exacte pour produit ac fct shift}. 
The representation holds, in fact, for any $\eps>0$ owing to the continuity in $\eps$ of the original ratio  and its rewriting resulting from \eqref{ecriture calcul somme sur racines Omega in de exp shift fct}. 
One can then compute the  $\eps\tend 0^+$ limit to get representation \eqref{ecriture rep int L fini exacte pour produit ac fct shift}.
In the intermediate calculations, one should use
the decomposition \eqref{ecriture decomposition fct cptge partie reg et dvgte vers racine sing}, the fact that $ \be_a^{(s)}$ is a zero of $\ex{-2\i\pi \wh{F}^{\,(\mf{b})} }$ while 
 $ \big(\be_a^{(s)}\big)^*$ is a zero of $\ex{-2\i\pi L \wh{\xi}_{\Ups} }$  
and that, for any $z\in Z_{0}\cup W_{0}$, one has 
\beq
\lim_{\eps\tend 0^+} \Bigg\{ \f{ 1-\ex{-2\i\pi L \wh{\xi}_{ \La^{\!(\a)} }^{ \,(\mf{b}) }(z-\i\eps) } }{ 1-\ex{-2\i\pi L \wh{\xi}_{\Ups}(z-\i\eps) } }  \Bigg\}
\; = \;  1  \;. 
\enq
The latter limit is a consequence of $\ex{-2\i\pi L \wh{\xi}_{ \La^{\!(\a)}  }^{ \,(\mf{b}) }(z) }\not=1$. Indeed, if this last equation did not hold, then, since $\exp\big\{ -2\i\pi \wh{F}^{\,(\mf{b})}(z) \big\}=1$, 
one would also  have that $\ex{-2\i\pi L \wh{\xi}_{\Ups}(z) } \,  = \,  1 $ . This would imply that \newline 
$z \in  \big\{ \Ups^{(\e{in})}\cup\{ \mu_a^{(s)} \}_{ a = 1 }^{ n_{\e{sg}} } \big\} \cap \La^{\!(\a)}_{ \mf{b} }$. However, the latter set is empty owing to 
\eqref{condition separation ensemble Lambda b et Ups in}.

\vspace{3mm}

It remains to extract the large-$L$ behaviour. Observe that the remainder can be recast as
\beq
\mf{R}^{-1}_{F} \; = \; \pl{ a=1 }{  n_{\e{sg}} }  \Bigg\{ 1-  \ex{-2\i\pi \wh{\xi}_{ \La^{\!(\a)}  }^{ \,(\mf{b}) }(\mu_{a}^{(s)} )}  
\f{   \ex{ - 2\i\pi \big[  \wh{\xi}_{ \La^{\!(\a)}  }^{ \,(\mf{b}) }\big( (\be_{a}^{(s)})^* \big) -\wh{\xi}_{ \La^{\!(\a)}  }^{ \,(\mf{b}) }(\mu_{a}^{(s)} )\big] } -1  }{  1-   \ex{- 2\i\pi \wh{\xi}_{ \La^{\!(\a)}  }^{ \,(\mf{b}) }(\mu_{a}^{(s)} ) }    }       \Bigg\}   \;. 
\enq
The property  \eqref{condition distance racine mua sing au reseau Lambda}, the estimate $\norm{ \wh{\xi}_{ \La^{\!(\a)}  }^{ \,(\mf{b}) } }_{ W^{\infty}_{1}(\mc{S}_{\de}(\R) ) }<C$ for some $\de>0$ and 
the one on the deviation of $(\be_{a}^{(s)} )^*$ in respect to $\mu_a^{(s)}$, \textit{c.f.} \eqref{ecriture deviation des mua au bea}, ensures that 
\beq
 \f{   \ex{-2\i\pi \big[  \wh{\xi}_{ \La^{\!(\a)}  }^{ \,(\mf{b}) }(\mu_{a}^{(s)} )-\wh{\xi}_{ \La^{\!(\a)}  }^{ \,(\mf{b}) }\big( (\be_{a}^{(s)} )^*\big) \big] } -1  }{  1-  \ex{-2\i\pi \wh{\xi}_{ \La^{\!(\a)}  }^{ \,(\mf{b}) }(\mu_{a}^{(s)} )}       } \; = \; \e{O}\Big( L^{-\infty} \Big)  
\qquad \e{so}\, \e{that} \quad  \mf{R}^{-1}_{F} \; = \; 1+ \e{O}\big(n_{\e{sg}} \cdot L^{-\infty}\big) \;. 
\enq

The contour integral can be decomposed as 
\beq
 \Oint{\msc{C} }{} \wh{z}\,(s) \ln^{\prime}\big[ \ex{-2\i\pi \wh{F}^{\,(\mf{b})}(s)} -1 \big] \cdot \f{ \dd s }{2\i\pi} 
 \; = \;  \Int{ \msc{C}^{(+)} }{}  \wh{F}^{\,(\mf{b})}(s)  \ln^{\prime}\big[ \ex{-2\i\pi \wh{F}^{\,(\mf{b})}(s)} -1 \big] \cdot \dd s  
\, + \, \mf{I}_{L} 
\label{estimation de l'integrale contre log prime fct shift}
\enq
where the remainder takes the form
\beq
 \mf{I}_{L}  \, = \, \sul{\eps=\pm}{} \Int{\msc{C}^{(\eps)} }{} \big(\,  \wh{u}_{\Ups}^{\,(\eps)}-\wh{u}_{\La_{ \mf{b} }}^{\,(\eps)} \big)(s) \cdot  \ln^{\prime}\big[ \ex{-2\i\pi \wh{F}^{\,(\mf{b})}(s)} -1 \big] \cdot \f{ \dd s }{2\i\pi} \;. 
\enq
%
%
The remainder term is a $\e{O}\big(  L^{-1} \big)$  as can be seen by invoking the $L^1\big( \msc{C}^{(\eps)} \big)$ bounds \eqref{ecriture estimee hat u Omega reg sur C en L1}
on $\wh{u}_{\Ups}^{\,(\eps)}$, $\wh{u}_{\La_{ \mf{b} }}^{\,(\eps)}$, the lower bound \eqref{bornes sur fct shift upper and lower},  
 the estimates \eqref{estimee reste NLIe et deviation Freg a F}, \eqref{ecriture bornes ctg fct sing sur C} and \eqref{borne sur entier varkappa et sur F rho}.

The first term in the \textit{rhs} of \eqref{estimation de l'integrale contre log prime fct shift} can be computed in closed form . Indeed, one has 
\beq
\Oint{ \msc{C}^{(+)} }{}  \wh{F}^{\,(\mf{b})}(s)  \ln^{\prime}\big[ \ex{-2\i\pi \wh{F}^{\,(\mf{b})}(s)} -1 \big] \cdot \dd s  
\;= \; -\i\f{\pi}{2} \Big( \big(\wh{F}^{\,(\mf{b})}(\,\wh{q}_L) \big)^{\, 2}\, - \, \big(\wh{F}^{\,(\mf{b})}(\,\wh{q}_R) \big)^{\, 2}  \Big) \, + \, \Int{  \wh{F}^{\,(\mf{b})} (\,\wh{q}_R)  }{  \wh{F}^{\,(\mf{b})} (\,\wh{q}_L)  } \pi x \cot(\pi x) \dd x \;. 
\enq
Then, the integral representation  for the ratio of Barnes functions \eqref{ecriture rep int pour fct Barnes} allows one to conclude that
\bem
\exp\bigg\{ \Oint{\msc{C} }{} \wh{z}\,(s) \ln^{\prime}\big[ \ex{-2\i\pi \wh{F}^{\,(\mf{b})}(s)} -1 \big] \cdot \f{ \dd s }{2\i\pi}  \bigg\}  \\ 
 =  \; \pl{ \ups \in \{L,R\} }{}   \Bigg\{    \f{  \ex{ \i \f{\pi}{2} \sg_{\ups}   \big(\wh{F}^{\,(\mf{b})}(\,\wh{q}_{\ups}) \big)^{\, 2} }  } { \big( 2\pi \big)^{ \sg_{\ups}\wh{F}^{\,(\mf{b})}(\, \wh{q}_{\ups}) } }   
\cdot 
	    G\left( \ba{cc}   1+\sg_{\ups}\wh{F}^{\,(\mf{b})}(\,\wh{q}_{\ups} ) \\ 
			  1- \sg_{\ups}\wh{F}^{\,(\mf{b})}(\, \wh{q}_{\ups}) \ea \right)  \Bigg\} \cdot \Big( 1+\e{O}\Big( \f{ 1  }{ L }   \Big) \Big)  \;. 
\end{multline}

Recalling the reflection identity \eqref{ecriture eqn de reflection pour fct Barnes} satisfied by the Barnes function
and upon denoting \newline $\wh{\op{f}}_{\ups,\e{tot}}^{\,(\mf{b})}  =\varkappa_{\ups}-\wh{F}^{\,(\mf{b})}(\,\wh{q}_{\ups})$, one obtains the rewriting
\bem
\pl{ \ups \in \{L,R\} }{}   \Bigg\{ \Big( \ex{-2\i\pi \wh{F}^{\,(\mf{b})}(\, \wh{q}_{\ups})  } -1  \Big)^{ \sg_{\ups} \varkappa_{\ups} }   
		      G\left( \ba{cc}   1+\sg_{\ups}\wh{F}^{\,(\mf{b})}(\, \wh{q}_{\ups})  \\ 
			  1- \sg_{\ups}\wh{F}^{\,(\mf{b})}(\, \wh{q}_{\ups})  \ea \right)  \Bigg\} \\
\; = \; 
\pl{ \ups \in \{ L , R \} }{} \Bigg\{  \f{  (2\i\pi)^{\sg_{\ups}\varkappa_{\ups} } (-1)^{ \tfrac{ \varkappa_{\ups}(\varkappa_{\ups}-1) }{ 2 } } }
 { \ex{ - \i \pi \sg_{\ups} \varkappa_{\ups} \wh{\op{f}}_{\ups,\e{tot}}^{\,(\mf{b})}  }   }
\cdot 
	    G\left( \ba{cc}   1-\sg_{\ups}\wh{\op{f}}_{\ups,\e{tot}}^{\,(\mf{b})}    \\ 
			  1+ \sg_{\ups}\wh{\op{f}}_{\ups,\e{tot}}^{\,(\mf{b})}   \ea \right)  \Bigg\}   \\
\; = \; \pl{ \ups \in \{ L , R \} }{} \Bigg\{  \f{  (2\i\pi)^{\sg_{\ups}\varkappa_{\ups} } (-1)^{ \tfrac{ \varkappa_{\ups}(\varkappa_{\ups}-1) }{ 2 } } }
 { \ex{ - \i \pi \sg_{\ups} \varkappa_{\ups} \wh{\op{f}}_{\ups}^{\,(\mf{b})}  }   }
\cdot 
	    G\left( \ba{cc}   1-\sg_{\ups}\op{f}_{\ups}^{\,(\varrho)}   \\ 
			  1+ \sg_{\ups}\op{f}_{\ups}^{\,(\varrho)}   \ea \right)  \Bigg\} \cdot\Big( 1+\e{O}\Big( \f{ 1 }{L}\Big) \Big) \;. 
\end{multline}
The last line follows as a consequence of \eqref{borne sur entier varkappa et sur F rho} which entails 
$\big| \, \wh{\op{f}}_{\ups,\e{tot}}^{ \,(\mf{b}) } - \op{f}_{\ups}^{\,(\varrho)} \big| \, = \, \e{O}\big(  L^{-1} + n_{\e{sg}}L^{-\infty} \big)$ and, 
by construction, $\big|\op{f}_{\ups}\big|\leq \tf{1}{2}$ one has $\big|\op{f}_{\ups}^{\,(\varrho)}\big|\leq \tf{3}{4}$ 
what ensures that one is far from the poles and zeroes of the Barnes functions. By putting all the estimates together, the claim follows. \qed

\begin{prop}
There exists $r$ such that the conclusions of Lemma \ref{Lemme structure zeros F} hold and such that 
\beq
  \det\Big[ \e{id} + \op{U}_{\a;\th}^{\, (\ga)}\big[ F^{(\varrho)}\big] \, \Big] \, \not= \, 0 \qquad  for\;  any  \quad \varrho \in \Dp{}\mc{D}_{0,r}  \;. 
\enq
Let $r_L=r+\tf{\de r}{L}$ with $\de r$ as given by Proposition \ref{Proposition beta deformation des racines Lambda} and 
$\mf{b}=\varrho+\tf{\de \varrho }{L} \in \mc{D}_{0,r_L}\setminus\{ \pm r_{L} \}$. 
Assume that the string centre separation hypothesis holds \eqref{propriete espacement ctres cordes}. Then one has the asymptotic behaviour 
\beq
\wh{C}^{(\ga)}\big[ \wh{F}^{\,(\mf{b})}\big]\big( \La_{\mf{b}}^{\!(\a)} ; \Ups \big) \, = \, C^{(\ga)}\big( \Ups^{(h)}_{\e{off}};\mf{C}; \{ \ell_{\ups}^{\varkappa} \} \mid \varrho \big)  
\cdot \bigg\{ 1+ \e{O}\bigg(   \f{ \descnode + 1 }{L}     \bigg) \bigg\}  
\enq
where $c\geq 0$,  $\ga \in \{ z, + \}$ and $C^{(\ga)}$ are defined by \eqref{definition cste Cz non univ}-\eqref{definition cste C+ non univ}.  

\end{prop}

\Proof 

Let $r_L=r+\tf{\de r}{L}$ with $r$ such that the conclusions of Lemma \ref{Lemme structure zeros F} hold and 
with the subordinate $\de r$ given by Proposition \ref{Proposition beta deformation des racines Lambda}. 

The form taken by the pre-factors follows from the asymptotic expansion of $\wh{\mc{P}}_{\Ups\setminus\La_{\mf{b}}^{\!(\a)}}$ \eqref{ecriture DA excitation momentum} and from Corollary \ref{Corollaire DA V Ups La}. 
The Fredholm determinant part deserves, however, more attention. 
Lemma \ref{Lemme structure zeros F} ensures the existence of a small open neighbourhood $\msc{V}_{F}$ of the interval $I_q$ such that the only zeroes of 
$s\mapsto 1-\ex{ 2 \i \pi \wh{F}^{ \,(\mf{b})}(s)}$ in  $\msc{V}_{F}$ are given by $Z\cup W$, \textit{c.f.} \eqref{ecriture zeros fct shift exponentiee}. 
Furthermore, there exists $\eps>0$ such that for each $\mf{w}_a\in W$ there exists $z_{w_a}$ so that one has  $w_a\in \mc{D}_{z_w, \eps}$ and no other zero $\mf{w}_{b}$
or element of $Z\cup \big\{ Z^{(s)}-\i\zeta \big\}$ is contained there. Finally, the discs are uniformly away from the endpoints $\pm q$. 
These pieces of information allow one to relate the original Fredholm determinats to ones involving other operators whose thermodynamic limit is, however, easier to cope with. 

By deforming contours in the Fredholm series representation for $\det_{\Ga(\La_{\mf{b}}^{\!(\a)})}\big[ \e{id} + \wh{\op{U}}_{\a;\th}^{(\ga)}\big]$ from the small loop $\Ga(\La_{\mf{b}}^{\!(\a)})$ around the roots $\La_{\mf{b}}^{\!(\a)}$ 
up to  the boundary $\Dp{} \mc{V}_{F}$ one picks up poles at $W\cup Z$. Here, one should note that although the factors $\big( V_{\Ups;\La_{\mf{b}}^{\!(\a)}}\big)^{-1}(\tau+\i\zeta)$ 
contain poles at the elements of $Z^{(s)}-\i\zeta$, this does not generate poles of the integral kernel since these singularities are 
 compensated by the poles of $\ex{ 2\i\pi \wh{F}^{\,(\mf{b})} (\tau) }$. This contour deformation yields 
\beq
\det_{\Ga(\La_{\mf{b}}^{\!(\a)})}\Big[ \e{id} + \wh{\op{U}}_{\a;\th}^{\, (\ga)}\big[ \wh{F}^{\,(\mf{b})}\big]\,  \Big] \, = \,
\det\Big[ \e{id} + \wh{\op{W}}_{\a;\th}^{\, (\ga)}\big[ \wh{F}^{\,(\mf{b})}\big] \, \Big]  
\enq
where $\wh{\op{W}}_{\a;\th}^{\, (\ga)}\big[ \wh{F}^{\,(\mf{b})}\big]$ is the operator on $L^2\big(\Dp{}\msc{V}_F \cup  W \cup Z \big)$:
\beq
\Big( \, \wh{\op{W}}^{(\ga)}_{\a;\th} \cdot f \Big) (\om)  = \Int{  \Dp{}\mc{V}_F  }{} \wh{U}^{(\ga)}_{\a;\th}\big[ \wh{F}^{\,(\mf{b})}\big](\om, \tau) f(\tau) \cdot \dd \tau  
 \, - \,  2\i\pi \hspace{-2mm} \sul{ \mu \in W \cup Z }{  } \hspace{-2mm}f(\mu) \, \underset{\tau=\mu}{\Res}\Big( \,  \wh{U}^{(\ga)}_{\a;\th}\big[ \wh{F}^{\,(\mf{b})}\big](\om, \tau) \dd \tau \Big)  \;.  
\enq

One can then rewrite the functions $V_{\Ups; \La_{\mf{b}}^{\!(\a)} }$ appearing in the integral kernel $\wh{U}^{(\ga)}_{\a;\th}\big[ \wh{F}^{\,(\mf{b})}\big]$ 
by using Corollary \ref{Corollaire DA V Ups La} and the fact that either $\tau$ belongs to the exterior of $\msc{C}$ or that it belongs to the interior but 
coincides with a zero of $1-\ex{ 2 \i \pi \wh{F}^{ \,(\mf{b})}(s)}$. This means that one can replace the integral kernel $\wh{U}^{(\ga)}_{\a;\th}\big[ \wh{F}^{\,(\mf{b})}\big]$
by 
\beq
 \wh{V}^{(\ga)}_{\a;\th}\big[ \wh{F}^{\,(\mf{b})}\big](\om, \tau) \, = \, \f{  \ex{ \mc{L}_{\msc{C}}[\,\wh{z}\, ]( \tau) }  }{  \ex{ \mc{L}_{\msc{C}}[\,\wh{z}\,]( \tau + \i\zeta )  } }
\hspace{-4mm}  \pl{  \mu \in \Ups^{(z)}_{\e{tot}} \cup  \daleth  \setminus \Ups^{(h)}   }{}\hspace{-4mm}  \bigg\{ \f{ \sinh( \tau - \mu) }{  \sinh( \tau - \mu + \i \zeta ) }   \bigg\} \cdot 
\f{ \mc{K}^{(\ga)}_{\a;\th}(\om,\tau) }{ 1 - \ex{ 2\i\pi \wh{F}^{\,(\mf{b})} (\tau) }  }  \;. 
\enq
The resulting kernel is a meromorphic function of $\tau$ outside and inside of $\msc{C}$. The poles inside of $\msc{C}$ are located\symbolfootnote[2]{Here, the poles/zeroes issuing
from the product over $\daleth$ cancel with the contributions issuing from $\Ups^{(z)}$.} at $\Ups^{(h)}\cup W\cup Z $. 
Hence, once that the kernels are replaced, one can reabsorb the residues at $W$ by the contour integrals 
along $\cup_{a=1}^{\ell_F} \Dp{}\mc{D}_{z_w, \eps}$. The above stated properties of these discs ensure that, one does not produce new residues 
when doing so: all the other poles of the integrand are located outside of these discs. Hence, one obtains
\beq
\det_{\Ga(\La_{\mf{b}}^{\!(\a)})}\Big[ \e{id} + \wh{\op{U}}_{\a;\th}^{\, (\ga)}\big[ \wh{F}^{\,(\mf{b})}\big]\,  \Big] \, = \,
\underset{\mf{g} \cup \intn{ 1 }{  n_{\e{sg}} }}{\det}\Big[ \e{id} + \wh{\op{V}}_{\a;\th}^{\, (\ga)}\big[ \wh{F}^{\,(\mf{b})}\big] \, \Big] \;. 
\enq
Here $\mf{g}= \Dp{}\msc{V}_F \setminus \cup_{a=1}^{\ell_F} \mc{D}_{z_w, \eps}$ and $\wh{\op{V}}^{\,(\ga)}_{\a;\th}$ is the integral operator on 
$L^2\big(\mf{g} \cup \intn{ 1 }{  n_{\e{sg}} }\big)$:
\beq
\Big( \, \wh{\op{V}}^{\,(\ga)}_{\a;\th}\big[ \wh{F}^{\,(\mf{b})}\big]\cdot f \Big)(\om) = \Int{ \mf{g} }{} \wh{V}^{(\ga)}_{\a;\th}\big[ \wh{F}^{\,(\mf{b})}\big](\om, \tau) f(\tau) \cdot \dd \tau  
 \, + \, \sul{a=1}{  n_{\e{sg}} }    \wh{\de V}^{(\ga)}_{\a;\th}\big[ \wh{F}^{\,(\mf{b})}\big](\om, \mf{z}^{(s)}_a ) f_{a} \;. 
\enq
where 
\beq
 \wh{\de V}^{(\ga)}_{\a;\th}\big[ \wh{F}^{\,(\mf{b})}\big](\om, \tau ) \, = \, \f{  \ex{ \mc{L}_{\msc{C}}[\, \wh{z} \, ]( \tau) }  }{  \ex{ \mc{L}_{\msc{C}}[\, \wh{z}\, ]( \tau+ \i\zeta )  } }
\hspace{-4mm}  \pl{  \mu \in \Ups^{(z)}_{\e{tot}} \cup  \daleth  \setminus \Ups^{(h)}   }{}\hspace{-4mm}  \bigg\{ \f{ \sinh( \tau - \mu) }{  \sinh( \tau - \mu + \i \zeta ) }   \bigg\} \cdot 
\f{ \mc{K}^{(\ga)}_{\a;\th}(\om,\tau) }{\big( \wh{F}^{\,(\mf{b})}\big)^{\prime}(\tau)  }  \;. 
\enq
It is also useful to introduce a second integral operator on $L^2\big(\mf{g} \cup \intn{ 1 }{  n_{\e{sg}} }\big)$:
\beq
\Big( \op{V}^{(\ga)}_{\a;\th}\big[ F^{(\varrho)}\big]\cdot f \Big)(\om) = \Int{ \mf{g} }{} U^{(\ga)}_{\a;\th}\big[ F^{(\varrho)}\big](\om, \tau) f(\tau) \cdot \dd \tau  
\enq
 whose integral kernel has been defined in  \eqref{definition noyau integral U limite thermo}. 
The operators $\wh{\op{V}}^{(\ga)}_{\a;\th}$ and $\op{V}^{(\ga)}_{\a;\th}$ have $\mc{C}^{1}$ kernel and act on functions supported on compacts. 
They are therefore trace class in virtue of the results of  \cite{DudleyGonzalesBarriosMetricConditionForOpToBeTraceClass}. 
This ensures that they are Hilbert-Schmidt as well.  Also, they admit the matrix representation subordinate to the splitting $\mf{g} \cup \intn{ 1 }{  n_{\e{sg}} }$:
\beq
\wh{\op{V}}^{(\ga)}_{\a;\th} \; = \; \left( \ba{ccc}  \wh{V}^{(\ga)}_{\a;\th}\big[ \wh{F}^{\,(\mf{b})}\big](\om, \tau) & \wh{\de V}^{(\ga)}_{\a;\th}\big[ \wh{F}^{\,(\mf{b})}\big](\om, \mf{z}_b^{(s)})  \vspace{1mm} \\
	 \wh{V}^{(\ga)}_{\a;\th}\big[ \wh{F}^{\,(\mf{b})}\big](\mf{z}_a^{(s)}, \tau)  & \wh{\de V}^{(\ga)}_{\a;\th}\big[ \wh{F}^{\,(\mf{b})}\big](\mf{z}_a^{(s)}, \mf{z}_b^{(s)})      \ea \right)
\label{ecriture rep matricielle pour hat V}
\enq
and
\beq
\op{V}^{(\ga)}_{\a;\th} \; = \; \left( \ba{ccc}  U^{(\ga)}_{\a;\th}[F^{(\varrho)}](\om, \tau) & 0  \\
	 0  & 0      \ea \right) \;. 
\enq

Direct bounds on $\mf{g}\cup Z$ yield 
\beq
  \bigg| \ex{ 2\i\pi \big[ \op{C}_{I_q }[ F^{(\varrho)}](\tau  ) -  2\i\pi  \op{C}_{I_q }[ F^{(\varrho)}](\tau  + \i\zeta) \big] } \cdot G\Big( \Ups^{(h)}_{\e{off}};\mf{C}; \{ \ell_{\ups}^{\varkappa} \}\mid \tau  \Big) \Bigg|
\leq  \ex{ c  }
\enq
for some constant $c \geq 0$. This constant vanishes if the function has modulus lower that one. Then, by using the estimates \eqref{bornes sur fct shift upper and lower}, 
the results of Corollary \ref{Corollaire DA V Ups La} and the fact guaranteed by Lemma \ref{Lemme structure zeros F} that for any root $w_a$, one has  
$d\big(z_{w_a}, \big\{ Z^{(s)}-\i\zeta\big\} \cup \{ \pm q  \} \big) >2\eps$, one can  check that   
\beq
\Norm{ \wh{V}^{(\ga)}_{\a;\th}\big[ \wh{F}^{\,(\mf{b})}\big] - U^{(\ga)}_{\a;\th}[F^{(\varrho)}] }_{ L^{\infty}\big(   \{\mf{g} \cup Z \}  \times \mf{g}  \big) }  \, = \,
\e{O}\Big( \f{\descnode +  1 }{L}   \Big) \;. 
\label{bornes sur norme noyaux hatV moins U}
\enq

In the estimation of the large-$L$ behaviour of the kernel  $ \wh{\de V}^{(\ga)}_{\a;\th}\big[ \wh{F}^{\,(\mf{b})}\big] (\la, \mf{z}_b^{(s)})$ one should pay an extra attention to the
string containing the root $\mu_a^{(s)}$  which approaches $\mf{z}^{(s)}_a$ exponentially fast. The contribution of this string cancel out with the associated singular factor appearing in the product over $\daleth$
while the string deviations appearing in the other only produce exponentially small corrections  owing to the string spacing hypothesis \eqref{propriete espacement ctres cordes}. 
This hypothesis also ensures that 
\beq
\big( \wh{F}^{\,(\mf{b})}\big)^{\prime}(\mf{z}^{(s)}_a) \; = \; \f{-1}{2\pi} \f{ \sin\big[2\Im\big(\be^{(s)}_a\big) \big] }{ \sinh\big(\mf{z}^{(s)}_a -\be^{(s)}_a \big) \sinh\big(\mf{z}^{(s)}_a - \big(\be^{(s)}_a\big)^* \big) } 
\big( 1+\e{O}\big( n_{\e{tot}} L^{-\infty}) \big) \Big)
\enq
so that, for any $\la \in \mf{g}\cup Z$ with $Z=\{ \mf{z}^{(s)}_a \}_{1}^{ n_{\e{sg}} }$, 
\bem
 \wh{\de V}^{(\ga)}_{\a;\th}\big[ \wh{F}^{\,(\mf{b})}\big] (\la, \mf{z}_a^{(s)}) \; = \;  -2\pi 
 \f{  \ex{ 2\i\pi  \op{C}_{I_q^{\ua}}[ F^{(\varrho)}]( \mf{z}_a^{(s)})}   }{  \ex{ 2\i\pi  \op{C}_{I_q }[ F^{(\varrho)}]( \mf{z}_a^{(s)}+\i\zeta)  } } 
G\big(\Ups^{(h)}_{\e{off}}; \mf{C} ; \{\ell^{\varkappa}_{\ups} \} \mid \mf{z}_a^{(s)} \big) \cdot \mc{K}^{(\ga)}_{\a;\th}(\om,\mf{z}_a^{(s)})  \\
\times \f{ \sinh\big(\mf{z}^{(s)}_a -\be^{(s)}_a \big) \sinh\big(\mf{z}^{(s)}_a - \big(\be^{(s)}_a\big)^* \big) }{ \sin\big[2\Im\big(\be^{(s)}_a\big) \big]   } \Big(1+\e{O}\Big( \f{ \descnode +  1}{L}\Big) \Big) \;. 
\end{multline}
Above, $I_q^{\ua}$ is a small deformation of $I_q$ which avoids $\mf{z}_a^{(s)}$ from above. 
Therefore, owing to the estimates \eqref{ecriture bornes zeros za proche de beas},
\beq
\Norm{  \wh{\de V}^{(\ga)}_{\a;\th}\big[ \wh{F}^{\,(\mf{b})}\big]  }_{ L^{\infty}\big(  \{ \mf{g}\cup  Z  \}  \times Z \big) } \, = \,  \e{O}\Big(  L^{-\infty} \Big) \;. 
\label{borne sur noyau deltaV}
\enq

Prior to obtaining the bounds, it remains to establish the statement about the non-vanishing of the determinant. 
The integral kernel $U^{(\ga)}_{\a;\th}[F^{(\varrho)}]$ is an analytic function of $\rho$ on the annulus $r_1>|\rho|>r_2$ with $r_1, r_2$ as given in Lemma \ref{Lemme structure zeros F}. 
Hence, so is $  \underset{ \mf{g}  }{\det}\Big[ \e{id} + \op{V}_{\a;\th}^{\, (\ga)}\big[ F^{(\varrho)}\big] \Big] $. Thus, one can always pick some $r\in \intff{r_2}{r_1}$ 
such that   $ \big| \underset{ \mf{g}  }{\det}\Big[ \e{id} + \op{V}_{\a;\th}^{\, (\ga)}\big[ F^{(\varrho)}\big] | >c > 0 $  on $\Dp{}\mc{D}_{0,r}$.  
This choice of $r$, and the subordinate value of $\de r$, will be assumed in the following.

The Lipschitz bound for $2$-determinants \cite{GohbergGoldbargKrupnikTracesAndDeterminants} of Hilbert-Schmidt operators $\op{A}, \op{B}$
\beq
\Big|  \det_2 \big[ \e{id} + \op{A} \big]  \; - \; \det_2 \big[ \e{id} + \op{B} \big]  \Big| \; \leq \;  
  \norm{\op{A}- \op{B}}_2 \cdot \ex{C \big( \norm{ \op{B} }_2  \, + \,  \norm{\op{A}}_2 \big) }
\enq
yields the below estimate for two trace class operators $\op{A}, \op{B}$
\beq
\Big|  \det \big[ \e{id} + \op{A} \big]  \; - \; \det \big[ \e{id} + \op{B} \big]  \Big| \; \leq \;  
  \bigg\{ \norm{\op{A}- \op{B}}_2 \; + \;  \big| \e{tr} \big( \op{A}- \op{B} \big) \big|  \bigg\} \cdot \ex{C \big( \norm{ \op{B} }_2  \, + \,  \norm{\op{A}}_2 \, + \, |\e{tr}(A)| \, + \, |\e{tr}(B)|+1\big)  }\;. 
\label{ecriture borne determinants}
\enq

One has
\beq
\Big| \e{tr}\Big[ \, \wh{\op{V}}_{\a;\th}^{\, (\ga)}\big[ \wh{F}^{\,(\mf{b})}\big] - \op{V}_{\a;\th}^{\, (\ga)} \big[ F^{(\varrho)}\big] \, \Big] \Big| \, + \, 
\Norm{ \wh{\op{V}}_{\a;\th}^{\, (\ga)}\big[ \wh{F}^{\,(\mf{b})}\big] - \op{V}_{\a;\th}^{\, (\ga)}\big[ F^{(\varrho)}\big]  }_2 
\, =  \, \e{O}\Big(   \f{ \descnode +  1 }{L}   \Big)
\enq
 either directly in virtue of the bounds \eqref{bornes sur norme noyaux hatV moins U}-\eqref{borne sur noyau deltaV} or due to the fact that 
when calculating the Hilbert-Schmidt norm, the last line of the matrix representation \eqref{ecriture rep matricielle pour hat V} for $ \wh{\op{V}}_{\a;\th}^{\, (\ga)}$, which contains \textit{a priori} order 1 terms, 
always appears in pair with the last column in \eqref{ecriture rep matricielle pour hat V} which is exponentially small. Also, I used the upper bound $\ell_{F}\leq C n_{\e{tot}}^{(\e{msv})}$ 
on the number of zeroes of $1-\ex{2\i\pi F^{(\varrho)} } $ inside of $\msc{V}_F$. Similarly, one can check that 
\beq
\Big| \e{tr}\Big[ \op{V}_{\a;\th}^{\, (\ga)} \big[ F^{(\varrho)}\big] \, \Big] \Big| \, + \, 
\Norm{\op{V}_{\a;\th}^{\, (\ga)}\big[ F^{(\varrho)}\big]  }_2 
\, =  \, \e{O}\big(   1  \big) \;. 
\enq
Therefore, by using the lower bound on the Ferdholm determinant, one gets that 
\beq
 \underset{\mf{g} \cup \intn{ 1 }{  n_{\e{sg}} }}{\det}\Big[ \e{id} + \wh{\op{V}}_{\a;\th}^{\, (\ga)}\big[ \wh{F}^{\,(\mf{b})}\big] \, \Big]  \; = \; 
  \underset{ \mf{g}  }{\det}\Big[ \e{id} + \op{V}_{\a;\th}^{\, (\ga)}\big[ F^{(\varrho)}\big] \, \Big] 
 \bigg\{ 1 \, + \,  \e{O}\bigg(   \f{ \descnode + 1 }{L}   \bigg)    \bigg\} \;. 
\enq
It remains to gather the bounds.  \qed

\section{Analysis of $\wh{\mc{A}}_{\e{sing}}\big( \Ups \big)$}
\label{Section analyse de Asing}

This section discusses the large-$L$ expansion of $\wh{\mc{A}}_{\e{sing}}\big( \Ups \big)$. The most delicate part consists in extracting the 
large-$L$ behaviour out of the "norm"-determinants, this while keeping a satisfactory control on the remainder.

\begin{prop}
\label{Proposition asymptotiques Asing}
 The below large-$L$ asymptotic expansion holds 
\beq
\wh{\mc{A}}_{\e{sing}}\big(  \Ups \big) \, = \, \f{1}{ \pl{r=2}{ p_{\e{max}} } \pl{ a=1 }{ n_r^{(z)} } \Big\{ L p_r^{\prime}\big( c_a^{(r)} \big) \Big\} } \cdot \Big( 1 \, + \, \e{O}\Big( \f{ \descnode  +1 }{L}\Big)  \Big) 
\enq

\end{prop}

\Proof  
The claim follows by straightforward handlings of the results obtained in Lemma \ref{Lemme DA det Ups}.  \qed

\subsection{An auxiliary determinant identity}

 One of the key results which allows one to carry out the large-$L$ analysis of $\wh{\mc{A}}_{\e{sing}}\big(  \Ups \big)$ is an auxiliary determinant identity
 which allows one to recast the "norm"-determinants in a form that allows one to extract their large-$L$ asymptotics 
 in the presence of string solutions, this while providing a control on the corrections.

\begin{lemme}
\label{Lemme reecriture totale det norme}
 Let 
\beq
\De\Big( P_a ; X_{ab} \Big) \; = \; \det_{N}\Big[ \de_{ab}\Big( P_a - \sul{k=1}{ N }X_{ak}\Big)+X_{ab}   \Big]
\enq
be a determinant defined in terms of an $N \times N $ symmetric matrix $X_{ab}$ having a vanishing diagonal $X_{aa}=0$.
Then it holds
\beq
\De\Big( P_a ; X_{ab} \Big) \; = \; \det_N\bigg[ \sul{a = \e{max}(k,\ell) }{N}\hspace{-3mm}  P_a  - \sul{s=1}{ \e{min}(k,\ell)-1 } \hspace{-2mm}  \sul{t= \e{max}(k,\ell) }{N} \hspace{-2mm}  X_{st}     \bigg]
\enq

\end{lemme}

 \Proof 
 
This rewriting of the original determinant can be obtained by doing the chain of linear combinations on the lines $L_a$ and columns $C_a$: $L_{a}\hookrightarrow \sum_{k=a}^{N}L_k$
and $C_{a}\hookrightarrow \sum_{k=a}^{N}C_k$ with $a=1,2,\dots, N-1$. The claim is proven by induction, where the induction hypothesis is 
$\De\Big( P_a ; X_{ab} \Big) \; = \;  \det_N\big[ M^{(j)} \big]$ for any $j$ with 
\beq
M^{(j)}   \; = \; \left( \ba{cc} 
\sul{a = \e{max}(k,\ell) }{N}\hspace{-3mm}  P_a \; \;  - \sul{s=1}{ \e{min}(k,\ell)-1 } \hspace{-2mm}  \sul{t= \e{max}(k,\ell) }{N} \hspace{-2mm}  X_{st}     &     P_{\ell}  - \sul{s=1}{k-1 }   X_{k s }    \\ 
					 P_k  - \sul{s=1}{\ell-1 }    X_{s\ell } 									 &  	  \de_{k\ell}\Big( P_k - \sul{s=1}{ N }X_{ks}\Big)+X_{k\ell}  
					 \ea    	\right) \;. 
\enq
Above, the matrix is written in a block decomposition subordinate to $k,\ell \leq j$ for the first diagonal and $k,\ell \geq j+1$ for the second diagonal. 
The details are left to the reader.

 \begin{prop}
 \label{Proposition asymptotiques determinant des normes}
  Let 
\beq
\De\Big( P_a ; X_{ab} \Big) \; = \; \det_{ n_{\Ups} }\Big[ \de_{ab}\Big( P_a - \sul{k=1}{ N }X_{ak}\Big)+X_{ab}   \Big]
\enq
be a determinant defined in terms of an $n_{\Ups} \times n_{\Ups} $ symmetric matrix $X_{ab}$ having a vanishing diagonal $X_{aa}=0$ and let $n_{\Ups}=\sul{r=2}{p_{\e{max}} } r n_r^{(z)} $. 
Let 
\beq
v_{p,a,k}\; = \;  \sul{r=2}{p-1} r n_r^{(z)} \, + \, (a-1)p \, + \, k
\enq
be a linear index labelling a block matrix decomposition into blocks 
labelled by the length $r$,  ranging over $2,\dots,p_{\e{max}}$,  sub-blocks of fixed length $r$ but having indices $a=1,\dots, n_r^{(z)}$ and, finally, 
$r\times r$ sub-blocks labelled by $k=1,\dots, r$. Let 
\beq
\mf{P}^{(p,a)}_k\; = \; P_{ v_{p,a,k} }\; - \hspace{-2mm} \sul{   \substack{  t\not = v_{p,a,k} \\ k=1,..,p }   }{ n_{\Ups} } \hspace{-1mm} X_{t,v_{p,a,k}}
\quad, \quad  \mf{X}^{(p,a)}_{k,\ell} \; = \; X_{v_{p,a,k},v_{p,a,\ell}} \quad  and  \quad 
\quad \mf{X}^{(p,a),(r,b) }_{k,\ell} \; = \; X_{v_{p,a,k},v_{r,b,\ell}} \;. 
\enq
Then one has the block matrix decomposition 
\beq
\De\Big( P_a ; X_{ab} \Big) \; = \; \det\left(\ba{cc} A & B \\ C & D   \ea  \right)\;. 
\label{ecriture rep matric block ABCD pour determinant normes generique}
\enq
The block matrix $A$ has its entries labelled by the $2$-ple $(p,a)$ 
\beq
A_{(p,a),(r,b)} \; = \; \ov{\mf{P}}^{(p,a)}_1 \cdot \de_{(p,a),(r,b)} \, - \, \big( 1\, - \, \de_{(p,a),(r,b)} \big)\cdot \ov{\mf{X}}_{1,1}^{(p,a),(r,b)}
\quad  with  \quad
\left\{  \ba{cc} \ov{\mf{P}}^{(p,a)}_k \, = \, \sul{s=k}{p} \mf{P}^{(p,a)}_s   \\ 
		  \ov{\mf{X}}_{k,\ell}^{(p,a),(r,b)} \, = \, \sul{s=k}{p} \sul{t=\ell}{r}\mf{X}_{s,t}^{(p,a),(r,b)}   \ea   \right.  \;. 
\enq
The Kronecker symbol $\de_{(p,a),(r,b)}$ appearing above is such that $ \de_{(p,a),(r,b)}=1$ if the $2$-uples coincide and zero otherwise. The off-diagonal blocks $B$ and $C$ read
\beq
B_{(p,a),(r,b,k)} \; = \; - \ov{\mf{X}}_{1,k}^{(p,a),(r,b)} \quad , \quad 
C_{(p,a,k),(r,b)} \; = \; - \ov{\mf{X}}_{k,1}^{(p,a),(r,b)}
\enq
where $(p,a)$ and $(r,b)$ run through $\Big\{ (w,t) \, :   \, w=2,\dots, p_{\e{max}} \;\; t=1,\dots,n_w^{(z)} \Big\}$ and $k$ in  $(p,a,k)$, resp.  $(r,b,k)$, 
$k=2,\dots,p$, resp. $k=2,\dots,r$.   
Finally, the $D$ block takes the form 
\beq
D_{(p,a,k),(r,b,\ell)} \; = \; \bigg[ \ov{\mf{P}}^{(p,a)}_1    - \sul{s=1}{\min(\ell,k)-1} \sul{t=\max(\ell,k) }{p}\mf{X}_{s,t}^{(p,a)}   \bigg]
\cdot \de_{(p,a),(r,b)} \, - \, \big( 1\, - \, \de_{(p,a),(r,b)} \big) \cdot \ov{\mf{X}}_{k,\ell}^{(p,a),(r,b)} \;. 
\enq

In particular, if the matrix $A$ is invertible, then one has 
\beq
\De\Big( P_a ; X_{ab} \Big) \; = \; \det A \cdot \pl{r=2}{ p_{\e{max}} } \pl{a=1}{n_r^{(z)}} \pl{k=1}{r-1} \Big\{- \mf{X}^{(p,a)}_{k,k+1} \Big\}
\cdot \bigg\{ 1+ \e{O}\bigg( n_{\Ups}   \f{  \underset{(r,b,\ell) \not= (r,b,k\pm 1)  }{ \max}{  X_{k,\ell  }^{(p,a),(r,b)}  } }{  \min \mf{X}^{(p,a)}_{k,k+1} } \cdot \max (A^{-1})_{(p,a),(r,b)} \bigg) \bigg\} \;. 
\label{ecriture asymptotiques determinant a la norme}
\enq

 \end{prop}

 A similar result, but without the explicit control of the remainder was obtained in \cite{KirillovKorepinNormsStateswithStrings}. However, 
 the technique developed in \cite{KirillovKorepinNormsStateswithStrings} does not allow one for any estimate on the remainder. This control is important
 since the matrix $\de_{ab} \Big( P_a - \sum_{k=1}^{ N }X_{ak}\Big) \,+ \,X_{ab}$ of interest to the analysis will have, in principle, various entries that will diverge. 
 Some at exponential speed and some at algebraic. The control in \eqref{ecriture asymptotiques determinant a la norme} allows one to neglect the contribution of the entries 
 diverging only algebraically in $L$. 
 
\Proof 

The block matrix decomposition \eqref{ecriture rep matric block ABCD pour determinant normes generique} is obtained, first, by doing linear combinations of lines and columns
$L_{v_{a,p,k}}\hookrightarrow \sum_{s=k}^{p} L_{v_{a,p,s}}$ and $C_{v_{a,p,k}}\hookrightarrow \sum_{s=k}^{p} C_{v_{a,p,s}}$ for $k=1,\dots,p-1$ and $(p,a)$ fixed and this for each $(a,p)$ index. 
The form of the diagonal $(a,p)$ block is taken care of by Lemma \ref{Lemme reecriture totale det norme} whereas the form of the off-diagonal blocs is given by a simple sum. 
It then solely remains to exchange appropriate lines and columns of the determinant so that the $\Big((p,a,1), (r,b,1) \Big)$ entry of each diagonal bloc moves to the 
$\Big((p,a ), (r,b ) \Big)$ entry of the resulting matrix, hence giving rise to $A$.
The estimate on the remainder follows from \eqref{ecriture borne determinants}.  \qed

\subsection{The large-$L$ analysis}

In order to recast $\wh{\mc{A}}_{\e{sing}}\big( \Ups \big)$ in a form allowing one to take the thermodynamic limit readily, the main point is to rewrite $\det\big[ \Xi_{\Ups}\big]$
in a different form. Such manipulations, eventually, lead to

\begin{lemme}
 \label{Lemme DA det Ups}
 Let $ \Ups_{\e{red}} = \Ups \setminus \Ups^{(z)}$. 
 It holds for $L$ large enough:
\beq
\det \big[ \Xi_{\Ups} \big] \; = \; \det \Big[ \Xi_{ \Ups_{\e{red}} } \Big]\cdot 
\f{  \pl{r=2}{ p_{\e{max}} } \pl{ a=1 }{ n_r^{(z)} } \Big\{ \tfrac{L}{2\pi}   p_r^{\prime}\big( c_a^{(r)} \big) \Big\}  }{    \pl{  \mu \in \Ups^{(z)}  }{} \Big\{ L \wh{\xi}^{\prime}_{\Ups}(\mu) \Big\}  }    
\cdot \pl{r=2}{ p_{\e{max}} } \pl{ a=1 }{ n_r^{(z)} } \pl{k=1}{r-1}\Big\{ -K\big(\de_{a,k}^{(r)} \, - \, \de_{a,k+1}^{(r)} +\i\zeta \big) \Big\} 
\cdot \bigg\{  1  \, + \, \e{O}\bigg( \f{ 1  }{ L }    n_{\e{tot}}^{(z)} \bigg) \bigg\} 
\enq
 where 
\beq
\big[ \Xi_{  \Ups_{\e{red}}  } \big]_{ab}=\de_{ab} \, + \,    \f{ K(\nu_a-\nu_b) }{  L \, \wh{\xi}^{ \, \prime }_{ \Ups }(\nu_b)  } \;. 
\label{definition matrice Xi Ups red}
\enq

\end{lemme}

\Proof

Consider the parametrisation 
\beq
\Ups \setminus \Ups^{(z)} \, \equiv \, \Ups_{\e{red}}  \; = \; \big\{  \nu_a \big\}_1^{ | \Ups_{\e{red}}  | } \quad ,  \quad 
\Ups^{(z)} \, = \, \big\{ z_a \big\}_1^{ |\Ups^{(z)}| }  \;. 
\enq
Clearly, one has  $\det\big[ \Xi_{\Ups} \big]=\det\big[ \wt{\Xi} \big]$, where $\wt{\Xi}$ is an $ |\Ups| \times |\Ups|$ matrix with 
 which takes the form 
\beq
\wt{\Xi} \; = \; \left( \ba{cccc}    \big[ \Xi_{  \Ups_{\e{red}}   }\big]_{ab}    &  \wt{K}\big(\nu_a, z_b \big) \vspace{1mm} \\ 
					\wt{K}\big( z_a ,\nu_b \big)  &   \de_{ab} + \wt{K}\big(  z_a ,z_b\big)    \ea \right) 
\quad \e{with} \quad 	\wt{K}(\la,\mu) \; = \; \f{ K(\la-\mu) }{   L \, \wh{\xi}^{ \, \prime }_{ \Ups }(\mu)  } \;. 
\enq
It follows from Lemma \ref{Lemme Invertibilite Xi Ups in} that 
the matrix $\Xi_{ \Ups_{\e{red}}  }$ is invertible. Its inverse can be represented in the form 
\beq
\big[ \Xi_{ \Ups_{\e{red}} }^{-1}\big]_{ab} \; = \; \de_{ab} \, - \, \f{ \wh{R}(\nu_a,\nu_b)  }{ L \wh{\xi}^{\prime}_{\Ups}(\nu_b) } \;. 
\label{ecriture resolvent discret sur Ups red}
\enq
The so-called discrete resolvent $\wh{R}$ is a function on the discrete set $ \Ups_{\e{red}}  \times \Ups_{\e{red}}$. 
One can extend the definition of this function almost everywhere to  $\Cx^2$, first  for $\tau \in \Ups_{\e{red}} $ and  $z$ a generic complex number by the formula:
\beq
\wh{R}\big(\tau,z) \; = \; K(\tau-z) \, - \, \sul{ \a \in  \Ups_{\e{red}}   }{} \f{ \wh{R}\big(\tau,\a)K(\a-z)  }{ L \wh{\xi}^{\prime}_{\Ups}(\a) }  \;, 
\label{ecriture resolvent discret en une variable continue}
\enq
 and then, for $z,z^{\prime}$ generic, by
\beq
\wh{R}\big(z,z^{\prime}) \; = \; K\big( z - z^{\prime} \big) \, - \, \sul{ \a \in  \Ups_{\e{red}}  }{} \f{ K\big( z-\a \big)\wh{R}\big(\a, z^{\prime})  }{ L \wh{\xi}^{\prime}_{\Ups}(\a) } \;.  
\label{ecriture resolvent discret en deux variables continues}
\enq
$\wh{R}\big(z,z^{\prime})$ constitutes the discrete version of the continuous resolvent operator $\op{R}$, see Lemma \ref{Lemme convergence resolvent discrete vers resolvent continu} for a precise statement. 

Owing to the factorisation for block determinants
\beq
\det\left( \ba{cc} A & B \\ C & D  \ea \right) \; = \; \det[A] \cdot \det\big[ D\, -\, C A^{-1} B \big] 
\enq
 valid whenever the block matrix  $A$ is invertible, one gets the determinant factorisation 
\beq
\det \big[ \Xi_{\Ups} \big] \; = \; \det \big[ \Xi_{  \Ups_{\e{red}}  } \big] \f{ \det_{n_{\Ups}}\big[ \, \wh{Y} \, \big]   }
{  \pl{  \mu \in   \Ups^{(z)}    }{} \Big\{ L \wh{\xi}^{\prime}_{\Ups}(\mu) \Big\}  }
\qquad \e{with} \quad n_{\Ups} \, = \, |\Ups^{(z)}|
\enq
where the matrix $\wh{Y}_{ab}$ has entries $\wh{Y}_{ab}= L\,\wh{\xi}^{\prime}_{\Ups}\big( z_a\big) \de_{ab}  + \wh{R}\big( z_a, z_b \big)$. 
The function 
\beq
\wh{\rho}(\om) \, = \, K\big( \om\mid \tf{\zeta}{2}\big)  \, - \, \sul{ \a \in \Ups_{\e{red}} }{ } \wh{R}(\om,\a) K\big( \a \mid \tf{\zeta}{2}\big) 
\enq
allows one to recast $\wh{\xi}^{\, \prime}_{\Ups}$ in the form 
\beq
L\, \wh{\xi}^{\, \prime}_{\Ups}\big(  \om \big)  \, = \, L \, \wh{\rho}(\om) \, - \, \sul{ \a \in \Ups^{(z)} }{} \wh{R}(\om,\a)  \;. 
\enq
The above representation for the counting function thus ensures that the matrix $\wh{Y}$ takes the generic form 
\beq
\wh{Y}_{ab}=\de_{ab}\Big( P_a + \sul{k=1}{n_{\Ups}}X_{ak}\Big)-X_{ab}  \qquad \e{for} \quad a,b=1,\dots, n_{\Ups}
\label{ecriture forme matrix Y pour AA}
\enq
for an appropriate choice of the symmetric matrix $X_{ab}$ and the $P_a$'s. 
The matrix $\wh{Y}$ has thus precisely  the form taken care of by Proposition \ref{Proposition asymptotiques determinant des normes}. In the notations of that proposition, one has
\beq
\ov{\mf{P}}_1^{(p,a)} \; = \;  L \big(\,  \wh{\xi}^{\, (r)}_{\Ups;a} \, \big)^{\prime}\big( c_a^{(r)} \big) \quad \e{and} \quad \ov{\mf{X}}_{1,1}^{(p,a)(r,b)} \, = \, \wh{\mc{R}}_{ab}^{(pr)}\big( c_a^{(p)}, c_b^{(r)} \big)
\enq
where the diagonal entries are expressed in terms of 
\beq
\Big( \wh{\xi}^{\, (r)}_{\Ups;a}\Big)^{\prime} \big( \om \big) \; = \; \sul{ k=1 }{ r } \wh{\xi}^{\, \prime}_{\Ups \setminus \mc{S}^{(r)}_a } \Big( \om \, + \; \i \f{\zeta}{2}\big(r+1-2k \big)+\de_{a,k}^{(r)} \Big) 
\enq
where 
\beq
\mc{S}^{(r)}_a \, = \, \Big\{ c_a^{(r)} +  \i \f{\zeta}{2}\big(r+1-2k \big)+\de_{a,k}^{(r)} \Big\}_{ k=1 }^{ r }
\enq
and
\beq
\wh{\xi}^{\, \prime}_{\Ups \setminus \mc{S}^{(r)}_a } \big( \om \big) \, = \, K\big( \om \mid \tfrac{\zeta}{2} \big) 
\, - \, \f{1}{L} \sul{ \a \in \Ups\setminus \mc{S}^{(r)}_a  }{} K\big( \om - \a \big) \;. 
\enq
The off-diagonal entries are expressed in terms of the function 
\beq
\wh{R}_{ab}^{\, (rq)}\big( \la, \mu \big)  \; = \; 
\sul{k=1}{r}\sul{\ell=1}{q}\wh{R}\Big( \la \, + \; \i \f{\zeta}{2}\big(r+1-2k \big)+\de_{a,k}^{(r)} \, , \mu \, + \; \i \f{\zeta}{2}\big(q+1-2\ell \big)+\de_{b,\ell}^{(q)} \Big) \;. 
\enq
It follows from Lemma \ref{Lemme convergence fct cptge centre cordres} and Lemma \ref{Lemme convergence resolvent discrete vers resolvent continu} that the matrix 
$\ov{\mf{P}}_1^{(p,a)}\de_{(p,a),(r,b)} \, - \, \ov{\mf{X}}_{1,1}^{(p,a)(r,b)} \big(1-\de_{(p,a),(r,b)} \big) $ is invertible. Hence, 
after pulling out from the determinant the diagonal terms, one gets 
\bem
\det\big[ \wh{Y} \big] \; = \; \pl{r=2}{ p_{\e{max}} } \pl{ a=1 }{ n_r^{(z)} } \pl{k=1}{r-1}\Big\{ -K\big(\de_{a,k}^{(r)} \, - \, \de_{a,k+1}^{(r)} +\i\zeta \big) \Big\}  \\
\times \pl{r=2}{ p_{\e{max}} } \pl{ a=1 }{ n_r^{(z)} } \Big\{ L \big(\,  \wh{\xi}^{\, (r)}_{\Ups;a} \, \big)^{\prime}\big( c_a^{(r)} \big) \Big\} \cdot 
 \det\big[\wh{Y}_{\e{red}} \big] \cdot \Big( 1 \, + \, \e{O}\Big(   n^{ (z) }_{ \e{tot} }  \cdot  L^{-\infty} \Big) \,  \Big) \;. 
\end{multline}
The reduced matrix takes the form
\beq
\big[ \wh{Y}_{\e{red}} \big]_{(a,r),(b,q)} \; = \;   \de_{(a,r),(b,q)}  +  \f{ \wh{\mc{R}}_{ab}^{(rq)}\big( c_a^{(r)}, c_b^{(q)} \big)  }{  L \big(\,  \wh{\xi}^{\, (r)}_{\Ups;a} \, \big)^{\prime}\big( c_a^{(r)} \big)   } \;. 
\enq
In order to proceed further and obtain the leading asymptotics of the determinant, one needs to invoke the simplified expression for the discrete resolvent 
obtained in Lemma \ref{Lemme convergence resolvent discrete vers resolvent continu} and the one for the re-summed counting functions obtained in Lemma \ref{Lemme convergence fct cptge centre cordres}. 
Then formula \eqref{ecriture borne determinants} yields $\det\big[ \wh{Y}_{\e{red}} \big] \, = \, 1 + \big(   n^{(z)}_{\e{tot}}   \cdot  L^{-1} \big)$.  \qed

\begin{lemme}
 \label{Lemme convergence fct cptge centre cordres}
 
 Assume that the Bethe roots satisfy to the string centre spacing hypotheses \eqref{hypothese sur borne inf sur espacement des diverses racines} and \eqref{propriete espacement ctres cordes}.
Then, it holds that 
\beq
 \Big( \wh{\xi}^{\, (r)}_{\Ups;a}\Big)^{\prime} \big(  c_a^{(r)} \big) \; = \; \frac{1}{2\pi} \cdot  p^{\prime}_{r}\big(  c_a^{(r)} \big) \; + \; \e{O}\bigg( \f{\descnode + 1  }{ L }  \bigg) 
\label{fct cptge r-sommee comme densite r-corde}
\enq
where $p_r$ is the dressed momentum of an $r$-string introduced in \eqref{defintion moment habille r-corde}. 
The remainder in \eqref{fct cptge r-sommee comme densite r-corde} is uniform in respect to the various parameters at play.

\end{lemme}

\Proof 

Let 
\beq
\mf{s}_{a,k}^{(r)} \; = \; c_{a}^{(r)}  \, + \,  \i\f{\zeta}{2}\big( r + 1 - 2k \big) \, + \, \de_{a,k}^{(r)} 
\qquad \e{and} \qquad 
\mf{c}_{a,k}^{(r)} \; = \; c_{a}^{(r)}  \, + \,  \i\f{\zeta}{2}\big( r + 1 - 2k \big) \, + \, \de_{a,\lfloor \f{r+1}{2} \rfloor }^{(r)} 
\enq
denote components of the various strings depending on whether one kept -or not- the exponentially small string deviations. 

It follows from the Taylor-integral expansion and from hypothesis \eqref{propriete espacement ctres cordes}  that for any $(a,r)\not=(b,s)$, 
\beq
\big| K\big( \mf{s}_{a,k}^{(r)} -  \mf{s}_{b,\ell}^{(s)} \big) \, - \, K\big( \mf{c}_{a,k}^{(r)} - \mf{c}_{b,\ell}^{(s)} \big)   \big|   \; \leq \; 
\big|  \de_{a,k}^{(r)}  - \de_{b,\ell}^{(s)}  \Big| \sup_{t\in \intff{0}{1}}  \big| K^{\prime}\Big( \mf{c}_{a,k}^{(r)} -  \mf{c}_{b,\ell}^{(s)}  + t \big[ \de_{a,k}^{(r)}  - \de_{b,\ell}^{(s)} \big]  \Big) \Big| 
\; \leq \; C^{\prime}  \big|  \de_{a,k}^{(r)}  - \de_{b,\ell}^{(s)}  \big| L^{2\kappa} \;. 
\enq
Similarly, the hypothesis on the lower bound  \eqref{propriete espacement ctre corde et particule trou} on roots spacings ensures that, for any $\a \in \Ups^{(\e{in})}\cup \Ups^{(p)} $, one has 
\beq
\big| K\big( \mf{s}_{a,k}^{(r)} - \a \big) \, - \, K\big( \mf{c}_{a,k}^{(r)} - \a \big)   \big| 
%
%
%
\; \leq \;  C \cdot \left\{\ba{ccc}  \big|  \de_{a,k}^{(r)}  - \de_{a,\lfloor \f{r+1}{2} \rfloor }^{(r)}  \big|    & , &k\not= \tfrac{r\pm1}{2}, \tfrac{r-3}{2}  \vspace{3mm}  \\ 
		  \big| \Im\big( \de_{a, \f{r-1}{2}  }^{(r)} \big) \big|^{1-\ups}             &, & k = \tfrac{r\pm1}{2}   \ea  \right.  
\label{ecriture bornes sur distance noyau cordes avec et sans deviation}
\enq
where $\ups$ is as given in \eqref{propriete espacement ctre corde et particule trou} and I made use of \eqref{ecriture deviation des mua au bea} in the intermediate bounds. 
Finally, hypothesis \eqref{hypothese sur borne inf sur espacement des diverses racines} one also has that 
\beq
\big|  K\big( \mf{s}_{a,k}^{(r)}  \mid \tf{\zeta}{2} \big) \, - \, K\big( \mf{c}_{a,k}^{(r)}  \mid \tf{\zeta}{2} \big)    \big|  
		\; \leq \; C^{\prime}  \cdot \left\{  \ba{ccc }    \big|  \de_{a,k}^{(r)}  - \de_{a,\lfloor \f{r+1}{2} \rfloor }^{(r)}    \big| \; L^{2\kappa}   & k \not= \tfrac{r}{2}, \tfrac{r}{2} +1  \vspace{3mm}  \\ 
									\big|  \de_{a, \f{r}{2} + 1 }^{(r)}   - \de_{a, \f{r}{2} }^{(r)}    \big|^{\tf{1}{2}}   & k=	\tfrac{r}{2} +1	      \ea \right.  \;.
\enq
This allows one to infer that 
\beq
\wh{\xi}^{\, \prime}_{\Ups \setminus \mc{S}^{(r)}_a } \big(\mf{s}_{a,k}^{(r)} \big)  \, - \, \wh{\xi}^{\, \prime}_{\Ups \setminus \mc{S}^{(r)}_a } \big(\mf{c}_{a,k}^{(r)} \big) 
\; =  \;  \e{O}\Big( n_{\e{tot}} L^{-\infty} \Big)  \;. 
\enq
Thus using the summation identity \eqref{expression explicite noyau K r s}
one gets that 
\beq
 \Big( \wh{\xi}^{\, (r)}_{\Ups;a}\Big)^{\prime} \big(  c_a^{(r)} \big)  \; = \; \chi_{a}^{(r)}\Big( \mf{c}_{a, \lfloor \f{r+1}{2} \rfloor }^{(r)}  \Big)   \; + \;  \e{O}\Big( n_{\e{tot}}^2 L^{-\infty} \Big)
\; = \; \chi_{a}^{(r)}\big( c_a^{(r)} \big)   \; + \;  \e{O}\Big(  L^{-\infty} \Big) 
\enq
where 
\beq
\chi_{a}^{(r)}(\om) \; = \; K\Big(\om  \mid r \tfrac{\zeta}{2}\Big)
 \, -\, \f{1}{L} \hspace{-3mm} \sul{ \substack{ \a \in \Ups^{(\e{in})} \\ \cup \Ups^{(p)} \setminus \Ups^{(h)} } }{} \hspace{-3mm} K_{r,1}\big( \om - \a \big)
  \, -\, \f{1}{L} \sul{ s=2 }{p_{\e{max}} } \sul{ \substack{ b=1 \\ (s,b) \not= (r,a) } }{n_s^{(z)} } K_{r,s}\big(\om - c_b^{(s)} \big) 
\enq
For real $\om$ it holds
\beq
 \, -\, \f{1}{L} \sul{   \a \in \Ups^{(\e{in})}  }{} K_{r,1}\big( \om - \a \big)  \;  =  \; 
 %
%
%
%
 -\Int{ - q }{ q }  K_{r,1}\big( \om - s \big)  p^{\prime}\big( s  \big) \cdot \f{ \dd s  }{ 2\pi }\, + \, \mf{R}_r(\om)
\enq
 and
\bem
\mf{R}_r(\om) \, = \,  -\bigg\{ \Int{ q}{ \wh{q}_{R} } +\Int{  \wh{q}_L }{ -q } \bigg\}  K_{r,1}\big( \om - s \big)p^{\prime}\big( s  \big) \cdot \f{ \dd s  }{ 2\pi }
\; + \, \Int{ \wh{q}_{L} }{ \wh{q}_{R} } K_{r,1}\big( \om - s \big)\Big[ \f{ p^{\prime}( s ) }{ 2\pi } \, - \,  \wh{\xi}^{\prime}_{\Ups_{\e{reg}} } (s) \Big]\cdot \dd s \\
\, - \, \sul{\eps = \pm }{} \eps \Int{ \msc{C}^{(\eps)} }{} K_{r,1}\big( \om - s \big) 
	  \Bigg\{   \big( \, \wh{u}^{\,(\eps)}_{\Ups}\big)^{\, \prime}(s)   \, - \, \wh{\xi}^{\, \prime}_{\Ups_{\e{sing}} } (s)  \cdot \de_{\eps,+} \Bigg\}  \cdot \f{\dd s }{2\i\pi L} 
\, - \, \f{1}{L}  \sul{  \a \in \daleth  }{    }  K_{r,1}\big( \om - \a \big)  \;. 
\end{multline}
Owing to the bounds \eqref{ecriture DA racine q har R et L}, \eqref{ecriture bornes ctg fct sing sur C}, the estimates \eqref{ecriture deviation des mua au bea}, 
the asymptotic expansion of the counting function \eqref{ecriture DA fct de cptage} given in Proposition \ref{Proposition DA ctg fct},
and the bounds \eqref{ecriture brone sur U Ups moins U Ups reg}, \eqref{ecriture estimee hat u Omega reg sur C en L1}
one has that $\mf{R}_{r}(\om) \, = \, \e{O}\big( L^{-1} \big)$ on $\mc{S}_{\de}(\R)$. Thence,
\beq
 \Big( \, \wh{\xi}^{\, (r)}_{\Ups;a}\Big)^{\prime} \big(  c_a^{(r)} \big)  \; = \; K\Big(c_a^{(r)}  \mid \tfrac{r \zeta}{2}\Big)
\, - \, \Int{ - q }{ q }  K_{r,1}\big(  c_a^{(r)} - s \big) \cdot p^{\prime}\big( s \big) \cdot \dd s
\, + \, \mc{R}_{a}^{(r)}\big(  c_a^{(r)}  \big)
\enq
with 
\beq
\mc{R}_{a}^{(r)}(\om)  \, = \, \mf{R}_{r}(\om)  \, -\, \f{1}{L} \sul{ \substack{ \a \in \Ups^{(p)} \\ \setminus \Ups^{(h)} } }{} K_{r,1}\big( \om - \a \big)
  \, -\, \f{1}{L} \sul{ s=2 }{p_{\e{max}} } \sul{ \substack{ b=1 \\ (s,b) \not= (r,a) }  }{ n_z^{(s)} } K_{r,s}\big( \om - c_b^{(s)} \big) \; = \;  
  \e{O}\Big(\f{1}{L} \big( 1  + \tfrac{\descnode}{L}\big) \Big)
\enq
again, uniformly in $\om \in \R$ and owing to \eqref{definition estimee m}. The leading term can be identified with the derivative of the dressed momentum of $r$-strings owing to \eqref{ecriture eqn lin pour densite r-corde}.  \qed

\begin{lemme}
\label{Lemme Invertibilite Xi Ups in}

Let $\Ups_{\e{red}}=\Ups\setminus \Ups^{(z)}$ and assume that hypothesis \eqref{propriete espacement ctre corde et particule trou} holds. 
The matrix $\Xi_{ \Ups_{\e{red}} }$ defined in \eqref{definition matrice Xi Ups red} is invertible provided that $L$ is large enough. Furthermore, one has 
\beq
\det\big[ \Xi_{ \Ups_{\e{red}} } \big] \; = \; \det\big[ \e{id} + \op{K} \big] \cdot \bigg( 1+\e{O}\Big( \f{ 1   }{L} \Big) \, \bigg)
\label{ecriture DA determinant XiUpsRed}
\enq
and 
\beq
\det\big[ \Xi_{ \La_{\mf{b}}^{\!(\a)}} \big] \; = \; \det\big[ \e{id} + \op{K} \big] \cdot \Big( 1+\e{O}\Big( \f{1}{L} \Big) \Big) \;. 
\label{ecriture DA det Xi lambda}
\enq

\end{lemme}

\Proof 

 Let $r\geq 1$. Then, for any function  $f$ holomorphic in a neighbourhood of $\R$  it holds
\beq
  \sul{ \a \in \Ups_{\e{red}}  }{} \f{ K_{r,1}(\la-\a) f(\a) }{ L \wh{\xi}_{\Ups}^{\prime}(\a) } \; = \;   \Int{ - q }{q } f(s) K_{r,1}(\la-s) \cdot \dd s 
\; + \; \op{O}_{r}[f](\la)
\label{ecriture somme sur Ups in vers integrale}
\enq
with 
\bem
\op{O}_{r}[f](\la) \; = \;  \sul{\eps=\pm }{}  \Int{ \msc{C}^{(\eps)} }{}  \f{ K_{r,1}(\la-s) \cdot \big( \,\wh{u}_{\Ups}^{\,(\eps)} \big)^{\prime}(s) }{ L \,  \wh{\xi}_{\Ups}^{\prime}(s) } \cdot f(s)  \cdot \f{ \dd s }{2\i\pi}
  \; -   \sul{ \substack{ \a \in V \cup \Ups^{(p)} \\ \setminus \Ups^{(h)} } }{}    \f{  K_{r,1}( \la - \a ) f(\a) }{L \, \wh{\xi}_{\Ups}^{\prime}(\a) }   \\ 
\;+\; \Int{ \R }{}  K_{r,1}(\la-s) f(s) \big[ \bs{1}_{ \intff{ \wh{q}_L }{ \wh{q}_R } }(s) \, -\,  \bs{1}_{ I_{q} }(s)  \big] \cdot \dd s
\label{definition operateur Or}
\end{multline}
and where  $V=\big\{ \mu_a^{(s)} \big\}_1^{ n_{\e{sg}}}$. 
For $\de>0$ small enough, $\op{O}_{r}$ is a continuous linear operator on $L^{\infty}(\mc{S}_{\de}(\R))$ and
\beq
\Norm{ \op{O}_{r}[f] }_{L^{\infty}\big( \mc{S}_{\de}(\R) \big) } \; \leq \; C \f{ n_{\e{sg}}+ 1 + \tf{\descnode}{L} }{L}  \norm{ f }_{L^{\infty}\big( \mc{S}_{\de}(\R) \big) }  \;. 
\label{ecriture borne cte de Or}
\enq
Such estimates can be obtained as follows. 
\begin{itemize}

\item The estimates \eqref{ecriture bornes ctg fct sing sur C} and \eqref{ecriture DA fct de cptage} ensure that $ \wh{\xi}_{\Ups}^{\prime}>c >0$ for some constant $c$ on the compact $\msc{C}$. Then,  
\eqref{ecriture estimee hat u Omega reg sur C en L1} allows one to bound the first term in \eqref{definition operateur Or}. 

\item The integral appearing in the second line of \eqref{definition operateur Or} can be bounded thanks to \eqref{ecriture DA racine q har R et L}. 

\item The discrete sum over $\Ups^{(p)}\setminus \Ups^{(h)}$ can be bounded by first, invoking \eqref{propriete espacement ctre corde et particule trou} and then 
using \eqref{ecriture estimee norme Wk ctg fct sing loin des sings}, the expansion \eqref{ecriture DA fct de cptage}
and the fact that $\Ups^{(p)}$ and $\Ups^{(h)}$ are both bounded in $L$. The occurrence of $\tf{(1+\descnode)}{L}$ then appears due to further simplifications for the massless modes. 

\item Finally,  the sum over $V$ is bounded by using the estimates  \eqref{ecriture equivalent dvgt de xi Ups en les mu a sing} on $\wh{\xi}_{\Ups}^{\prime}(\a)$.

\end{itemize}

Expanding $\det\big[ \Xi_{ \Ups_{\e{red}} } \big] $ into a discrete Fredholm series and then replacing the discrete sums over the elements of $\Ups^{(\e{in})}$ one gets  
\beq
\det\big[ \Xi_{ \Ups_{\e{red}} } \big] \; = \; \sul{ n \geq 0 }{} \f{1}{n!} \pl{a=1}{n} \Bigg\{ \Int{ \msc{C} }{} \f{ \dd s_a }{ \ex{2\i\pi L \wh{\xi}_{\Ups}(s_a)}-1 }
\, -\, \sul{ \substack{ s_a\in V \cup \Ups^{(p)} \\ \setminus \Ups^{(h)} } }{} \f{1}{L \wh{\xi}_{\Ups}^{\prime}(s_a) } \Bigg\}
\cdot \det_{n}\big[ K(s_a-s_b) \big]
\enq
with $V=\big\{ \mu_1^{(s)},\dots, \mu_{ n_{\e{sg}}}^{(s)} \big\}$. Upon a slight rewriting of the integrations, and upon interpreting the 
the operator $\op{O}_{1}$ as an 
operator on $L^2\Big(\intff{ \min(-q,\wh{q}_L) }{ \e{max}(q,\wh{q}_R) } \cup \msc{C}^{(+)}\cup \msc{C}^{(-)}\cup V  \cup  \Ups^{(p)} \cup  \Ups^{(h)}\Big)$
one gets 
\beq
\det\big[ \Xi_{ \Ups_{\e{red}} } \big] \; = \; \det\big[ \e{id} + \op{K} + \op{O}_1 \big] \;. 
\enq
The operator $\op{K}$ appearing in the \textit{rhs} of this equality corresponds to the injection of the operator $\op{K}$ into the $L^{2}$-space defined earlier.  
The determinant is well defined since, according to the criterion established in \cite{DudleyGonzalesBarriosMetricConditionForOpToBeTraceClass}, the operators $K$ and $\op{O}_1$ are trace class: 
they acts on functions supported on a compact and have integral kernels that are of class $\mc{C}^1$. They are as well
Hilbert-Schmidt. It is easily seen on the basis of \eqref{ecriture borne cte de Or} that $| \e{tr}\big[ \op{O}_1  \big] | +  || \op{O}_1  ||_{2}=\e{O}\big(  L^{-1} \big)$ where  $||\cdot ||_{2}$ is the Hilbert-Schmidt norm.
Then, owing to the bound \eqref{ecriture borne determinants}, 
one readily gets that 
\beq
\big| \det\big[ \e{id}+ \op{K} + \op{O}_1 \big]\, - \, \det\big[ \e{id} + \op{K} \big] \big| \, \leq \, C \cdot \f{   1  }{L}
\enq
so that \eqref{ecriture DA determinant XiUpsRed} follows. The expansion \eqref{ecriture DA det Xi lambda}  is obtained by similar handlings. \qed

\begin{lemme}
\label{Lemme convergence resolvent discrete vers resolvent continu} 
 Under the hypotheses \eqref{propriete espacement ctre corde et particule trou} and \eqref{propriete espacement ctres cordes}, it holds
\beq
\wh{R}(\la,\mu) \; = \; R(\la,\mu) \; + \; \e{O}\Big( \f{  1 }{L} \Big) \quad uniformly\; in \; \; \la,\mu \in \R
\enq
and
\bem
\wh{R}_{ab}^{\,(r s)}\big( c_a^{(r)}, c_b^{(s)} \big) \; = \; K_{r,s}\big( c_a^{(r)} - c_b^{(s)} \big) \, - \, \Int{-q }{ q }K_{r,1}\big( c_a^{(r)}-t) K_{1,s}\big(t- c_b^{(s)} \big) \cdot \dd t \\
\, + \, \Int{-q }{ q }K_{r,1}\big( c_a^{(r)}-t)R(t,v)K_{1,s}\big(v- c_b^{(s)} \big) \cdot \dd  t \dd v   \; + \; \e{O}\bigg( \f{   1 }{L} \bigg) 
\end{multline}
\end{lemme}

\proof

The first step consists in characterising the discrete resolvent $\wh{R}(\la,\mu)$ for $\la,\mu \in \R$. Since $\Xi_{\Ups_{\e{reg}}}$ is invertible for $L$ large enough, the latter is well defined
by means of \eqref{ecriture resolvent discret sur Ups red}, \eqref{ecriture resolvent discret en une variable continue} and \eqref{ecriture resolvent discret en deux variables continues} 
In fact, starting form \eqref{ecriture resolvent discret en deux variables continues}  and by using the summation identity \eqref{ecriture somme sur Ups in vers integrale},
the discrete resolvent can be recast as 
\beq
\wh{R}(\la,\mu) \,  =\, K(\la-\mu) \, - \,  \Int{ - q }{ q }  K(\la-s)\wh{R}(s,\mu) \cdot \dd s 
\;-\; \op{O}_{1}\big[ \wh{R}(*,\mu) \big](\la)
\enq
leading to 
\beq
\wh{R}(\la,\mu) \,  = \, R\big( \la , \mu \big) \; - \; \Big( \e{id}-\op{R}\Big) \circ \op{O}_{1}\big[ \wh{R}(*,\mu) \big](\la) \;. 
\enq
Owing to \eqref{ecriture borne cte de Or} and to the continuity of $ \e{id}-\op{R}$ one has the operator norm bound
\beq
\Norm{\Big( \e{id}-\op{R}\Big) \circ \op{O}_{1} }_{  \mc{L}\big( L^{\infty}\big( \mc{S}_{\de}(\R) \big) \big) } \; \leq  \; \f{ C }{ L } 
\enq
and hence  $\e{id}+\Big( \e{id}-\op{R}\Big) \circ \op{O}_{1}$ is invertible on $L^{\infty}\big( \mc{S}_{\de}(\R) \big)$. The Neumann series representation for this
inverse then ensures that, in fact, $ \wh{R}(\la,\mu)$ is holomorphic on $\mc{S}_{\de}(\R)\times\mc{S}_{\de}(\R)$ for some $\de>0$ small enough and that 
\beq
\wh{R}(\la,\mu) \,  = \, R\big( \la , \mu \big) \; + \; \e{O}\bigg( \f{ 1 }{L} \bigg)
\label{ecriture DA dominant resolvent discret}
\enq
with a remainder that is uniform on $\mc{S}_{\de}(\R)\times\mc{S}_{\de}(\R)$.  

Starting from the expression \eqref{ecriture resolvent discret en deux variables continues} and then using the bounds \eqref{ecriture bornes sur distance noyau cordes avec et sans deviation}, the fact that $\wh{R}$
is bounded on $\Ups \setminus \Ups^{(z)}$ by virtue of \eqref{ecriture DA dominant resolvent discret} and also the fact that  $\wh{\xi}_{\Ups}^{\prime}(\a)>C>0$  for $\a \in \Ups \setminus \Ups^{(z)}$, one gets
\beq
\Big| \wh{R}_{ab}^{\,(r s)}\big( c_a^{(r)}, c_b^{(s)} \big) \, - \, \wh{R}_{r,s}\big( c_a^{(r)}, c_b^{(s)} \big) \Big| \; \leq \; C \big| \Ups\setminus \Ups^{(z)} \big|^2\cdot \max\big\{ \de_{a,k}^{(r)} \big\} 
L^{2\kappa} \, = \, \e{O}\big( L^{-\infty} \big)
\qquad \e{for}  \quad (r,a)\not= (s,b) \;. 
\label{ecriture controle R discrete vs R discrete sans deviation cordes}
\enq
The function $\wh{R}_{r,s}$ appearing above is expressed, owing to the summation identity \eqref{expression explicite noyau K r s}, as
\beq
\wh{R}_{r,s}(\la,\mu) \; = \; K_{r,s}(\la-\mu) \; - \hspace{-2mm}  \sul{ \a \in \Ups \setminus \Ups^{(z)} }{} \hspace{-2mm}  \f{ K_{r,1}(\la-\a) K_{1,s}(\a-\mu) }{ L\, \wh{\xi}_{\Ups}^{\prime}(\a) }
\; + \hspace{-3mm} \sul{ \a, \, \be \in \Ups \setminus \Ups^{(z)} }{} \hspace{-2mm}  \f{ K_{r,1}(\la-\a) \wh{R}(\a,\be) K_{1,s}(\be-\mu) }{ L \, \wh{\xi}_{\Ups}^{\prime}(\a) \cdot L \, \wh{\xi}_{\Ups}^{\prime}(\be) } \;. 
\enq
By using \eqref{ecriture somme sur Ups in vers integrale}, one can recast $\wh{R}_{r,s}(\la,\mu)$ in the form 
\bem
\wh{R}_{r,s}(\la,\mu) \; = \; K_{r,s}(\la-\mu) \, - \, \Int{-q }{ q }K_{r,1}(\la-t)K_{1,s}(t-\mu) \cdot \dd t  \\
\, + \, \Int{-q }{ q }K_{r,1}(\la-t)R(t,v)K_{1,s}(v-\mu) \cdot \dd  t \dd v   \; + \; \mf{R}_{r,s}[\, \wh{R}\, ](\la,\mu)
\end{multline}
where 
\bem
 \mf{R}_{r,s}[\, \wh{R}\,](\la,\mu) \; = \; 
 \Int{-q }{ q } \Big\{  K_{r,1}(\la-t)\op{O}_{s}\big[\wh{R}(t,*)\big](\mu) \, + \,   \op{O}_{r}\big[\wh{R}(*,t)\big](\la) K_{1,s}(t-\mu)   \Big\} \cdot  \dd t \\
- \, \op{O}_{r}\big[K_{1,s}(*-\mu) \big](\la)  \, + \, \op{O}_{r}\otimes \op{O}_{s}\big[\wh{R}(*,*)\big](\la,\mu) 
\, + \, \, + \, \Int{-q }{ q }K_{r,1}(\la-t) \, \Big\{ \wh{R}(t,v) - R(t,v) \Big\} K_{1,s}(v-\mu) \cdot \dd  t \dd v  \;. 
\end{multline}
The estimates \eqref{ecriture DA dominant resolvent discret} on $\wh{R}$ and the continuity \eqref{ecriture borne cte de Or} of the operators $\op{O}_{s}$ then allow one to conclude. \qed

\section{Conclusion}

This paper addressed the problem of extracting the large-volume asymptotic behaviour of the form factors of local operators in the massless regime of the XXZ spin-$1/2$ chain in the case where 
the expectation value is taken between the ground state at finite magnetic field below the critical one and an excited state containing both, particle-hole excitations and bound states. 
The asymptotic expansion were obtained on rigorous grounds and provide an explicit control of the remainder uniformly in respect to a large-class of excited states built by the Bethe Ansatz. 
Such a precise control opens the possibility to extract rigorously, under mild assumptions, 
\begin{itemize}
 \item the large-distance and long-time asymptotic expansion of two-point functions of the model;
 
 \item the edge exponents and the associated non-universal prefactors characterising the singular behaviour of the space and time Fourier transforms of two-point function. 
\end{itemize}
The results obtained in this paper thus provide one with a fundamental tool allowing one to deal with the dynamical properties of a model containing bound states and particle-hole excitations. 
I refer to the forthcoming paper for a better exposition of the problematic.

\section*{Acknowledgements}

K.K.K. acknowledges support from a CNRS-installation grant. The author is indebted to F. Göhmann, N. Kitanine, J.M. Maillet, G. Niccoli for stimulating discussions.




\appendix

\section{Asymptotic expansions of auxiliary integral transforms}
\label{Appendix Asymptotics auxiliary integrals}

\subsection{Special functions of interest and auxiliary results}

The Euler Gamma function satisfies to the uniform in $\De\geq 0$ and $z \in \intoo{\eps}{+\infty}$, $\eps>0$, estimate
\beq
\f{ \Ga\big(z+\De) }{ \Ga(z) } \, = \, z^{\De} \Big\{ 1\, + \, \e{O}\Big( \f{\De}{z} \Big)\Big\}
\label{estimee fct Gamma}
\enq

The Barnes function is a generalisation of the Gamma function. Its ratio admits the integral representation

\beq
(2\pi)^z \cdot \f{ G(1-z) }{ G(1+z) } \; = \; \exp\Bigg\{ \Int{0}{z} \pi x \cot(\pi x) \dd x \Bigg\}
\label{ecriture rep int pour fct Barnes}
\enq
and it satisfies to the reflection identity
\beq
 (-1)^{\tfrac{1}{2}\ell(\ell+1) } \f{ G(1-z-\ell) G(1+z)}{ G(1+z+\ell) G(1-z) }\; = \;  \bigg( \f{ \sin(\pi z) }{ \pi } \bigg)^{\ell}
\label{ecriture eqn de reflection pour fct Barnes}
\enq

\vspace{2mm}

\begin{lemme}\cite{IlYashenkoandYakovenkoCtgZeroesofHoloFctsSatisfyingLinearDiffEqns}
\label{Lemme borne sup zeros fct holomorphe}
Let U be a simply connected domain in $\Cx$ and $K$ be a compact in $U$. Given a   holomorphic function $f$ on $U$, 
denote by $N_{K}(f)$ the number of its zeroes, counted with their multiplicities and lying inside of $K$.  
Then, there exists $C>0$ such that, for any $f$ holomorphic on $U$, continuous on $\Dp{}U$ and having no zeroes on $\Dp{}U$, 
one has the bound 
\beq
N_{K}(f) \; \leq \; C \f{  \ln \norm{f}_{L^{\infty}(U)} }{   \ln \norm{f}_{L^{\infty}(K)} } \;. 
\enq
\end{lemme}

\subsection{Uniform asymptotics of the Cauchy transform}

For $\ups\in \{L,R\}$, let $\op{M}_{\ups}$ be the integral transform 
\beq
\wh{\op{M}}_{\ups}[f](\om) \, = \, - \Int{ \wh{q}_L }{ \wh{q}_R } \f{ f(s)- f(\om) }{ \tanh(s-\om) } \cdot \dd s  \; + \; \i \pi \bs{1}_{\e{Int}(\msc{C}) }(\om) f(\om) 
-f(\om) \ln \Bigg( \f{ \sg_{\ups} \sinh\big( \om - \wh{q}_R\big) }{ \sinh\big( \om - \wh{q}_L\big)  \big[ \, \wh{\xi}_{\La}(\om)-\wh{\xi}_{\La}(\, \wh{q}_{\ups})\big]^{\sg_{\ups}}  }  \Bigg) 
\enq
where $\sg_{\ups}$ is as given in \eqref{definition parametres sg L et R}. 

For the purpose of this appendix, given $\Om \in \{\Ups, \La_{\mf{b}}^{\!(\a)} \}$, I adopt the notations
\beq
\vsg_{\ups}^{(\Om)}(\om) \; = \; \tau_{\ups} + \de_{\Om; \Ups}\Big(\varkappa_{\ups} - \wh{F}_{\e{reg}}(\om)\Big)_{\mid \a_{\La}=0}    + \;  \de_{\Om;\La_{\mf{b}}^{\!(\a)} }\wh{\mf{f}}(\om) \qquad \e{and}, \; \e{for} \, \e{short},  \qquad 
\vsg_{\ups}^{(\Om)}  \; = \;\vsg_{\ups}^{(\Om)}(\, \wh{q}_{\ups})
\label{definition vsg ups Omega}
\enq
where the regular part $\wh{F}_{\e{reg}}$ of the shift function has been defined in \eqref{definition partie reg fct shift}, 
\beq
\wh{\mf{f}}(\om) \, = \, L \Big( \, \wh{\xi}^{\,(\mf{b})}_{ \La^{\!(\a)} }(\om)  \, - \, \wh{\xi}_{ \La }(\om) \Big)  \qquad \e{and} \qquad 
\wh{\mf{f}}_{\ups} \, = \, \mf{f}(\, \wh{q}_{\ups}) \;. 
\enq

\begin{prop}
\label{Proposition DA transformee Cauchy}

Given any $\de>0$  small enough, the Cauchy transform $\op{C}_{\msc{C}}[\,\wh{z}\,](\om)$ admits the asymptotic expansion 
\beq
\op{C}_{\msc{C}}[\,\wh{z}\,](\om) \, = \,  2\i \pi \op{C}_{ \msc{C}^{(+)} }\big[  \wh{F}_{\e{reg}}^{\,(\mf{b})} \big](\om) \; + \; \e{O}\Big( \f{1}{L} \Big)  \; \qquad for \quad 
\om \in \Big\{ \Cx \setminus \mc{D}_{ \, \wh{q}_L, \de}\cup \mc{D}_{ \, \wh{q}_R, \de} \Big\} / \big\{ \i\pi \mathbb{Z} \big\} \;. 
\label{ecriture DA Cauchy transfom z hat fct far from bdry}
\enq
When $\om \in  \mc{D}_{ \, \wh{q}_{\ups}, \de}$ with $\ups \in \{L,R\}$, the large-$L$ asymptotic expansion takes the form
\bem
\op{C}_{\msc{C}}[\,\wh{z}\,](\om) \, = \, \wh{\op{M}}_{\ups}\big[ \wh{F}_{\e{reg}}^{\,(\mf{b})} \big](\om) 
\; + \; \sg_{\ups} \Big( \wh{F}_{\e{reg}}^{\,(\mf{b})}( \, \wh{q}_{\ups})-\varkappa_{\ups}- \wh{F}_{\e{reg}}^{\,(\mf{b})} (\om) \Big) 
\cdot \ln\Big[ \pm  \sg_{\ups} L \cdot \Big( \, \wh{\xi}_{\La}(\, \wh{q}_{\ups}) - \wh{\xi}_{\La}(\om) \Big)  \Big]  \\
\, + \, \sg_{\ups}   \wh{F}_{\e{reg}}^{\,(\mf{b})}( \om) \ln L   \;  \mp \; 
\ln \Ga \Bigg( \ba{c} \tfrac{1}{2}  \pm  \sg_{\ups} \Big( L \cdot \big[ \, \wh{\xi}_{\La}(\, \wh{q}_{\ups}) - \wh{\xi}_{\La}(\om) \big] - \vsg_{\ups}^{(\Ups)} \Big)  \vspace{1mm} \\ 
			 \tfrac{1}{2}   \pm  \sg_{\ups} \Big( L \cdot \big[ \,  \wh{\xi}_{\La}(\, \wh{q}_{\ups}) - \wh{\xi}_{\La}(\om) \big] - \vsg_{\ups}^{(\La_{\mf{b}}^{\!(\a)})} \Big) 	 \ea \Bigg) 
			 \, + \; \e{O}\bigg(    \f{ \ln L }{ L }\bigg) \;. 
\end{multline}
In each of these local asymptotics, the remainder is uniform in the whole region where the expansion holds. Also, the $+$ sign corresponds to $\om \in \e{Int}(\msc{C})/\{\i\pi \mathbb{Z}\}$
while the $-$ sign corresponds to $\om \in \e{Ext}(\msc{C})/\{\i\pi \mathbb{Z}\}$.

\end{prop}

\Proof

By using the notation \eqref{definition Uomega sing et reg}, one can decompose the Cauchy transform as
\beq
\op{C}_{\msc{C}}\big[ \, \wh{u}_{\Om} \big](\om) \; = \; - 2\i\pi L \op{C}_{ \msc{C}^{(+)} }\big[ \, \wh{\xi}_{\Om_{\e{reg}}} \big](\om) 
\, + \, \sul{\eps=\pm}{} \op{C}_{\msc{C}^{(\eps)}}\big[ \, \wh{u}_{\Om_{\e{reg}} }^{\, (\eps)} \big](\om)
\, + \, \sul{\eps=\pm}{} \op{C}_{\msc{C}^{(\eps)}}\big[ \mf{u}^{(\eps)}_{\Om} \big](\om)
\label{ecriture decomposition transfo Cauchy}
\enq
with 
\beq
\mf{u}^{(\eps)}_{\Om} \, = \, \wh{u}_{\Om}^{\, (\eps)} - \wh{u}_{\Om_{\e{reg}} }^{\, (\eps)} - 2\i\pi L \de_{\eps;+} \, \wh{\xi}_{\Om_{\e{sing}}} \;. 
\enq
Note that $\mf{u}^{(\eps)}_{\La_{ \mf{b}}^{\!(\a) } }=0$, so that the last term in \eqref{ecriture decomposition transfo Cauchy} is only relevant when $\Om=\Ups$.  
Then, however, this term only generates uniform $\e{O}(L^{-\infty})$ corrections. Indeed, let $\mf{ln}_{\eps}$ be some determination 
of the logarithm such that 
\begin{itemize}
\item[$i)$] $\mf{ln}_{\eps}(s-\om)$ is holomorphic on $\ov{\mathbb{H}}^{\, \eps}$; 
\item[$ii)$]$ \mf{ln}_{+}\big( \wh{q}_{\ups}-\om\big)- \mf{ln}_{-}\big( \wh{q}_{\ups}-\om\big)=2\i\pi n_{\ups}(\om)$
with $n_{\ups}(\om)$ bounded in $L$, this uniformly in \newline
$\om \in \big\{ \Cx \setminus \mc{D}_{ \, \wh{q}_L, \de}\cup \mc{D}_{ \, \wh{q}_R, \de} \big\} / \big\{ \i\pi \mathbb{Z} \big\}$;
\item[$iii)$] $\norm{  \mf{ln}_{\eps}\big( * - \om\big) }_{L^{1}(\msc{C}^{(\eps)}) } \leq C$.
\end{itemize}
Then, it holds
\bem
\sul{\eps=\pm}{} \op{C}_{\msc{C}^{(\eps)}}\big[ \mf{u}^{(\eps)}_{\Ups} \big](\om) \, = \;
\sul{\eps =\pm }{} \f{\eps}{2\i\pi} \Big\{ \mf{u}^{(\eps)}_{\Ups} (\, \wh{q}_L) \cdot \mf{ln}_{\eps}\big( \, \wh{q}_L-\om\big) \, -\,  \mf{u}^{(\eps)}_{\Ups} (\, \wh{q}_R) \cdot \mf{ln}_{\eps}\big(\, \wh{q}_R-\om\big)    \Big\} \\
\; - \; \sul{\eps=\pm}{}\Int{ \msc{C}^{(\eps)} }{} \big(  \mf{u}^{(\eps)}_{\Ups} \big)^{\prime}(s)\,  \mf{ln}_{\eps}\big( s - \om\big) \cdot \f{\dd s }{ 2 \i \pi } \;. 
\end{multline}
It is readily checked, owing to \eqref{ecriture bornes ctg fct sing sur C} and \eqref{condition saut log neper}, that $\mf{u}^{(+)}_{\Ups} (\, \wh{q}_{\ups}) = \mf{u}^{(-)}_{\Ups} (\, \wh{q}_{\ups})$.
This ensures that the logarithmic singularities at the boundaries cancel out and, due to the other properties of $\mf{ln}_{\eps}\big( s - \om\big)$,  one gets 
\beq
\Norm{ \sul{\eps=\pm}{} \op{C}_{\msc{C}^{(\eps)}}\big[ \mf{u}^{(\eps)}_{\Ups} \big]  }_{ L^{\infty} (K  ) }  \; \leq \; 
      C^{\prime} \sul{\eps=\pm}{}  \,  \norm{ \mf{u}^{(\eps)}_{\Ups}  }_{ W_{1}^{\infty} ( \msc{C}^{(\eps)}  ) }  = \e{O}\big( L^{-\infty} \big) 
\label{ecriture borne sur Cauchy transform de mathfrac u Ups}
\enq
due to \eqref{ecriture bornes ctg fct sing sur C} and \eqref{ecriture brone sur U Ups moins U Ups reg}, where $K$ is some sufficiently narrow compact neighbourhood of $\msc{C}$ in $\Cx$. 
One obtains similar bounds on $\big\{ \Cx\setminus K \big\}/ \{ \i \pi \mathbb{Z} \}$ this time by straightforward bounds since $\tanh|s-z|$ is uniformly bounded from below for $s\in \msc{C}/ \{ \i \pi \mathbb{Z} \}$ and 
$z\in \big\{ \Cx\setminus K \big\}/ \{ \i \pi \mathbb{Z} \}$. 

If  $\om \in \big\{ \Cx \setminus \mc{D}_{ \wh{q}_L, \de}\cup \mc{D}_{ \wh{q}_R, \de} \big\} / \{ \i \pi \mathbb{Z} \}$, then, if necessary, deforming the part of $\msc{C}$ 
that is uniformly away from $\wh{q}_{\ups}$ so as to make the distance of $\om$ to $\msc{C}$ finite, the bound \eqref{ecriture estimee hat u Omega reg sur C en L1} leads to 
\beq
\Big| \op{C}_{\msc{C}^{(\eps)}}\big[ \, \wh{u}_{\Om_{\e{reg}} }^{\, (\eps)} \big](\om) \Big|  \; = \; \e{O}(L^{-1})\;. 
\enq
All together, these bounds yield \eqref{ecriture DA Cauchy transfom z hat fct far from bdry}.

\vspace{3mm}

The treatment of $\om \in  \mc{D}_{ \, \wh{q}_{\ups}, \de}$ needs more care. The bound \eqref{ecriture borne sur Cauchy transform de mathfrac u Ups} ensures that the last term in 
\eqref{ecriture decomposition transfo Cauchy} still produces $\e{O}\big( L^{-\infty} \big) $ corrections in that case.  The first term in \eqref{ecriture decomposition transfo Cauchy} can be decomposed as
\bem
- 2\i\pi L \op{C}_{ \msc{C}^{(+)} }\big[ \wh{\xi}_{\Om_{\e{reg}}} \big](\om) \; = \; 
L \Int{ \wh{q}_L }{ \wh{q}_R } \f{ \wh{\xi}_{\Om_{\e{reg}}}(s)  - \wh{\xi}_{\Om_{\e{reg}}}(\om)  }{ \tanh(s-\om) } \cdot \dd s  \\
+ \;  L \,  \wh{\xi}_{\Om_{\e{reg}}}(\om)  \bigg[  \ln\bigg( \f{ \sinh(\, \wh{q}_R-\om) }{ \sinh(\, \wh{q}_L-\om) } \bigg)  -2\i\pi  \bs{1}_{\e{Int}(\msc{C})\cap\mathbb{H}^{+}}(\om) \bigg] \;. 
\end{multline}

In order to estimate the behaviour of $\op{C}_{\msc{C}^{(\eps)}}\big[ \,  \wh{u}_{\Om_{\e{reg}}}^{\, (\eps)}\big](\om) $,  it is useful 
to decompose the contour $\msc{C}$ into portions neighbouring the endpoints $\wh{q}_{\ups}$ and portions that are uniformly away from $\R$. For this purpose, one introduces the 
intervals
\beq
J_{\de}^{(L)}=  \f{ \tf{1}{2} + \tau_{L} }{ L } + \intff{\i\de}{-\i\de} \qquad \e{and} \qquad 
J_{\de}^{(R)}=  \f{ \tf{1}{2}+  |\La| + \tau_{R} }{ L } + \intff{-\i\de}{\i\de} 
\label{definition intervalles locaux J delta Left et Right}
\enq
which then allow one to define the sub-sets of $\msc{C}$
\beq
\msc{C}^{(\e{out})}\, = \, \wh{\xi}_{\La}^{-1}\Big(\,  \wh{\Ga} \setminus \big\{ J_{\de}^{(L)} \cup J_{\de}^{(R)} \big\} \Big) \;,
\hspace{1cm} \msc{C}^{(\e{out};\pm)}\, = \,\msc{C}^{(\e{out})} \cap \mathbb{H}^{\pm}  
\label{definition contours C out et C out pm}
\enq
and
\beq
\msc{C}^{(\ups)}\, = \, \wh{\xi}_{\La}^{-1}\Big( J_{\de}^{(\ups)} \Big)  \;\;, \qquad  \msc{C}^{(\ups;\eps)}\, = \, \msc{C}^{(\ups)} \cap \mathbb{H}^{\eps}\;. 
\label{definition contours locaux C ups et C ups pm}
\enq
Let $\ov{\ups}=L$ if $\ups=R$ and $\ov{\ups}=R$ if $\ups=L$. Then one decomposes the Cauchy transforms as
\bem
\sul{\eps=\pm}{} \op{C}_{ \msc{C}^{(\eps)} } \big[\,  \wh{u}_{\Om_{\e{reg}}}^{\, (\eps)} \big](\om) \; = \; \sul{\eps=\pm}{} \op{C}_{ \msc{C}^{(\e{out};\eps)} } \big[\,  \wh{u}_{ \Om_{\e{reg}} }^{\, (\eps)} \big](\om) 
\, + \,  \sul{\eps=\pm}{} \op{C}_{ \msc{C}^{(\ov{\ups};\eps)} } \big[ \, \wh{u}_{ \Om_{\e{reg}} }^{\, (\eps)} \big](\om) \\
\, + \, \sul{\eps= \pm }{} \Int{ \msc{C}^{(\ups;\eps)} }{}  \wh{u}_{ \Om_{\e{reg}} }^{\, (\eps)}(s) 
			  \bigg\{ \coth(s-\om) \, - \, \f{ \wh{\xi}^{\prime}_{\La}(s) }{\wh{\xi}_{\La}(s) - \wh{\xi}_{\La}(\om)}  \bigg\} \cdot \f{ \dd s }{ 2\i\pi }
\, +\, \sul{\eps= \pm }{} \Int{ \msc{C}^{(\ups;\eps)}  }{}  \f{ \wh{u}_{\Om_{\e{reg}}}^{\, (\eps)}(s)  \cdot \wh{\xi}^{\prime}_{ \La }(s) }{\wh{\xi}_{  \La }(s) - \wh{\xi}_{ \La }(\om)}    \cdot \f{ \dd s }{ 2\i\pi } \; . 
\nonumber
\end{multline}
 Since the $\wh{u}_{\Om_{\e{reg}}}^{\, (\eps)} $ independent part of the integrand is bounded in the case of the first three terms, the estimate \eqref{ecriture estimee hat u Omega reg sur C en L1}
 ensures that these are a $\e{O}(L^{-1})$. In its turn, upon the change of variables 
\beq
s_{t}\, = \, \wh{\xi}^{-1}_{\La} \Big( \, \wh{\xi}_{\La}\big(\, \wh{q}_{\ups}\big)  - \tfrac{\sg_{\ups} t}{2\i\pi L } \Big) 
\qquad \e{and} \; \e{after} \; \e{setting} \qquad
a \; = \; 2\i\pi \sg_{\ups} L \Big( \, \wh{\xi}_{\La}\big(\, \wh{q}_{\ups}\big) \, - \,  \wh{\xi}_{\La}\big( \om \big)  \Big)
\label{ecriture chgmt vars loc voisinage hat q ups}
\enq
the last term can be recast in the form 
\beq
\sul{\eps= \pm }{} \Int{ \msc{C}^{(\ups;\eps)}  }{}  \f{ \wh{u}_{\Om_{\e{reg}}}^{\, (\eps)}(s)  \cdot \wh{\xi}^{\prime}_{ \La }(s) }{\wh{\xi}_{  \La }(s) - \wh{\xi}_{ \La }(\om)}  \f{ \dd s}{2\i\pi}
\; = \;   \Int{- 2\pi \de L }{  2\pi \de L }   \f{ \ln \Big(1+\ex{- |t|+2\i\pi \vsg_{\ups}^{(\Om)}(s_t) \sg_{\ups}\e{sgn}(t)} \Big)  }{   t - a }  \cdot \f{ \dd t }{ 2\i\pi} \;. 
\enq
Note that $a\in \mathbb{H}^{+}$ if $\om \in \e{Int}(\msc{C})/\{ \i\pi\mathbb{Z}\}$ and $a\in \mathbb{H}^{-}$ if $\om \in \e{Ext}(\msc{C})/\{ \i\pi\mathbb{Z}\}$. 
One can expand the $s_t$ dependent part into a series in $t$. By using bounds as in \eqref{ecriture borne inf sur exp decroissante} and integrating by parts, 
since $\norm{ \wh{F}_{\e{reg}} }_{ L^{\infty}\big(\mc{S}_{\de}(\R)\big) }< C n_{\e{tot}}^{(\e{msv})}   $ is bounded in virtue of \eqref{estimee reste NLIe et deviation Freg a F},  
one gets that 
 the $t$-dependent part of the expansion produces $\e{O}\big(\tf{\ln L}{L}\big) $ corrections. 
Once the replacement $s_t \hookrightarrow s_0$ is made one can extend the integration to $\R$, this for the price of $\e{O}(L^{-\infty})$ corrections. The resulting integral can then be computed
by means of Lemma \ref{Lemme integral pole simple vers fct Gamma}. All of this yields:
\bem
\sul{\eps=\pm}{} \op{C}_{ \msc{C}^{(\eps)} } \big[ \wh{u}_{\Om_{\e{reg}}}^{\, (\eps)} \big](\om) \; = \; 
\mp \,  \ln \Ga \bigg( \f{1}{2} \pm \sg_{\ups} \Big[ L \Big( \, \wh{\xi}_{\La}\big(\, \wh{q}_{\ups}\big) \, - \,  \wh{\xi}_{\La}\big( \om \big)  \Big)  - \vsg_{\ups}^{(\Om)} \Big]\bigg) 
 \\
\hspace{2cm} + \, \sg_{\ups}\Big[    L \Big( \, \wh{\xi}_{\La}\big(\, \wh{q}_{\ups}\big) \, - \,  \wh{\xi}_{\La}\big( \om \big)  \Big)  - \vsg_{\ups}^{(\Om)} \Big] 
\cdot  \ln \Big[\pm \sg_{\ups} L  \Big( \, \wh{\xi}_{\La}\big(\, \wh{q}_{\ups}\big) \, - \,  \wh{\xi}_{\La}\big( \om \big)  \Big) \Big]  \\
 \pm\,  \ln \sqrt{2\pi} \, - \, \sg_{\ups}L \Big( \, \wh{\xi}_{\La}\big(\, \wh{q}_{\ups}\big) \, - \,  \wh{\xi}_{\La}\big( \om \big)  \Big)  \; + \; \e{O}\Big(   \f{\ln L }{L }\Big) \;. 
\label{estimation somme transfos Cauchy de hau u Omega}
\end{multline}
In order to obtain the claimed form for the Cauchy transform, one needs to observe that 
\beq
\ln\bigg( \f{ \sinh(\om-\wh{q}_R) }{ \sinh(\om-\wh{q}_L) } \bigg) \; = \; 
\ln\bigg( \f{ \sg_{\ups} \sinh(\om-\wh{q}_R) }{ \sinh(\om-\wh{q}_L)  \big[ \,  \wh{\xi}_{\La}\big( \om \big) \,-\, \wh{\xi}_{\La}\big(\, \wh{q}_{\ups}\big)  \big]^{\sg_{\ups}}   } \bigg)
+ \sg_{\ups} \ln\Big( \sg_{\ups} \big[ \,  \wh{\xi}_{\La}\big( \om \big) \,-\, \wh{\xi}_{\La}\big(\, \wh{q}_{\ups}\big)  \big] \Big)
\enq
and that, for $\om \in \e{Int}(\msc{C})$, 
\beq
 \ln\Big( \sg_{\ups} \big[ \, \wh{\xi}_{\La}\big( \om \big) \, - \,  \wh{\xi}_{\La}\big(\, \wh{q}_{\ups}\big)  \big] \Big) \; = \; 
\ln\Big( \sg_{\ups} \big[ \, \wh{\xi}_{\La}\big(\, \wh{q}_{\ups}\big) \, - \,  \wh{\xi}_{\La}\big( \om \big)  \big] \Big) 
\, + \,  \i\pi \sg_{\ups}  \e{sgn}\big( \Im(\om) \big) \;. 
\enq
The rest is straightforward algebra. \qed

\vspace{3mm}

Proposition \ref{Proposition DA transformee Cauchy} provides one with the large-$L$ asymptotics of the $\mc{L}_{\msc{C}}$ and $\wt{\mc{L}}_{\msc{C}}$ transforms.
\begin{cor}
 \label{Corollaire asymptotiques transformee L et L tilde}

 Given any $\de>0$  small enough, given $\om \in \Big\{ \Cx \setminus \mc{D}_{ \, \wh{q}_L, \de}\cup \mc{D}_{ \, \wh{q}_R, \de} \Big\} / \big\{ \i\pi \mathbb{Z} \big\}$ it holds 
\beq
\ex{ \mc{L}_{\msc{C}}[\,\wh{z}\,](\om) } \, = \,  \ex{ - 2\i \pi \op{C}_{ \msc{C}^{(+)} }[  F^{(\varrho)} ](\om) }  \cdot \f{ \sinh^{\varkappa_L}(\om - \wh{q}_L) }{  \sinh^{\varkappa_R}(\om-\wh{q}_R) } 
\Big( 1 \; + \; \e{O}\Big( \f{1  }{L} \Big)  \, \Big) \;. 
\enq
 For $\om \in \mc{D}_{\wh{q}_{\ups},\eta}$ with $\ups\in \{L,R\}$ one has 
\bem
\ex{ \mc{L}_{\msc{C}}[\,\wh{z}\,](\om) } \; = \; \f{  \ex{ - \wh{\op{M}}_{\ups}\big[ \wh{F}_{\e{reg}}^{\,(\mf{b})}-\varkappa_{\ups} \big](\om)    } }{ L^{ \sg_{\ups} \big( \wh{F}_{\e{reg}}^{\,(\mf{b})} (\om) -\varkappa_{\ups} \big) } }
\Big\{ \pm  \sg_{\ups} L \cdot \big[ \, \wh{\xi}_{\La}(\, \wh{q}_{\ups}) - \wh{\xi}_{\La}(\om) \big]  \Big\}^{ - \sg_{\ups} \big( \wh{F}_{\e{reg}}^{\,(\mf{b})}( \, \wh{q}_{\ups}) - \wh{F}_{\e{reg}}^{\,(\mf{b})} (\om) \big) } 
 \\
\times \big[ \sinh(\om-\wh{q}_{\ov{\ups}})\big]^{\varkappa_{L}-\varkappa_{R}}  \cdot \Ga^{\pm 1} \Bigg( \ba{c} \tfrac{1}{2}  \pm  \sg_{\ups} \Big( L \cdot \big[ \, \wh{\xi}_{\La}(\, \wh{q}_{\ups}) - \wh{\xi}_{\La}(\om) \big] - \vsg_{\ups}^{(\Ups)} \Big)  \vspace{1mm} \\ 
			 \tfrac{1}{2}   \pm  \sg_{\ups} \Big( L \cdot \big[ \,  \wh{\xi}_{\La}(\, \wh{q}_{\ups}) - \wh{\xi}_{\La}(\om) \big] - \vsg_{\ups}^{(\La_{\mf{b}}^{\!(\a)})} \Big) 	 \ea \Bigg) 
			 \cdot \bigg( 1 \, + \; \e{O}\Big(   \f{ \ln L }{ L }\Big) \bigg) \;. 
\end{multline}
Above, $\ov{\ups}=L$ if $\ups=R$ and viceversa and one should take $+$ if $\om \in \e{Int}(\msc{C})/\big\{\i\pi\mathbb{Z} \big\}$ and $-$ otherwise. Further, one has
\beq
\ex{ \wt{\mc{L}}_{\msc{C}}[\,\wh{z}\,](\om) } \; = \; (-1)^{\varkappa_{L}-\varkappa_{R}} \cdot \ex{ \mc{L}_{\msc{C}}[\,\wh{z}\,](\om) } \; .
\enq

\end{cor}

\subsection{Asymptotics of the $\mc{A}_0$ transform}

The uniform asymptotics of the Cauchy transform $\op{C}_{\msc{C}}\big[ \, \wh{z} \,  \big](\om)$ allow one to determine the ones of $\mc{A}_0\big[ \, \wh{z}, \wh{z} \,  \big]$
which was introduced in \eqref{definition transfor integrale A eta}. Prior to stating the result, 
I introduce the functional
\beq
\wh{\mc{T}}[f] \; = \; \Int{ \wh{q}_{L} }{ \wh{q}_{R} }   \f{ f^{\prime}(s)f(t)-f^{\prime}(t)f(s)  }{ 2 \tanh(s-t) } \dd t \dd s \; + \; 
\sul{\ups \in \{L,R\} }{} \sg_{\ups} \Big(f(\, \wh{q}_{\ups})  -\varkappa_{\ups} \Big)  \Int{ \wh{q}_{L} }{ \wh{q}_{R} }   \f{ f(s) \, - \, f(\, \wh{q}_{\ups}) }{ \tanh(s-\wh{q}_{\ups}) }  \dd s \;. 
\enq

\begin{lemme}
 \label{Lemme asymptotiques fnelle A0}
Let $\wh{\op{f}}_{\ups}$ be as defined in \eqref{definition hat f ups}. Then, it holds 
\bem
\mc{A}_{0}\big[ \, \wh{z}, \wh{z} \,  \big] \, = \, \wh{\mc{T}}\big[ \wh{F}_{\e{reg}}^{\,(\mf{b})} \big] \, + \, \sul{\ups \in \{L,R\} }{}  \Bigg\{ -\i \sg_{\ups} \f{\pi}{2} \Big(\wh{F}_{\e{reg}}^{\,(\mf{b})} \big(\, \wh{q}_{\ups} \big) \Big)^2
\, + \, \wh{\op{f}}_{\ups}^{\,(\mf{b})}\wh{F}_{\e{reg}}^{\,(\mf{b})} \big(\, \wh{q}_{\ups}\big) \ln \Big(  L \sinh \big(\, \wh{q}_{R}- \wh{q}_{L} \big)\,  \wh{\xi}^{\prime}_{\La}\big(\, \wh{q}_{\ups}\big) \Big)  \\
\, + \, \i\pi \sg_{\ups} \varkappa_{\ups}  \wh{F}_{\e{reg}}^{\,(\mf{b})} \big(\, \wh{q}_{\ups}\big)  \, + \, \ln G\Big( 1- \wh{\op{f}}_{\ups}^{\,(\mf{b})} ,  1+  \wh{\op{f}}_{\ups}^{\,(\mf{b})}\Big)  
\; + \; \sg_{\ups} \varkappa_{\ups} \ln \Ga\Bigg( \ba{c}  \tfrac{1}{2} + \sg_{\ups} \big( \tau_{\ups} + \wh{\mf{f}}_{\ups} \big)  \vspace{1mm} \\ 
						      \tfrac{1}{2} + \sg_{\ups} \big(\tau_{\ups} + \wh{\op{f}}_{\ups\mid \a_{\La}=0} \big)    \ea   \Bigg) \Bigg\}  \; + \; \e{O}\bigg(   \f{ \ln L }{ L } \bigg)
\end{multline}

\end{lemme}

\Proof

One has the decomposition $\mc{A}_{0}\big[ \, \wh{z}, \, \wh{z} \,  \big]=\sul{a=1}{3} \mc{A}_{0}^{(a)}\big[ \, \wh{z}, \, \wh{z} \,  \big]$ with 
\beqa
\mc{A}_{0}^{(1)}\big[ \, \wh{z}, \, \wh{z} \,  \big] & = & - \Int{\msc{C}^{(+)} }{} \big(\wh{F}_{\e{reg}}^{\,(\mf{b})}(t) \big)^{\prime} \cdot  \op{C}_{\msc{C};+}\big[ \, \wh{z} \,  \big](t) \cdot  \dd t 
\; = \; \Int{\msc{C} }{}  \wh{z}\,(s) \cdot \op{C}_{\msc{C}^{(+)};-}\big[ \, \big(\wh{F}_{\e{reg}}^{\,(\mf{b})} \big)^{\prime}  \,  \big](s) \dd s  \\
\mc{A}_{0}^{(2)}\big[ \, \wh{z},\, \wh{z} \,  \big] & = & 
- \sul{\eps=\pm }{}\Int{\msc{C}^{(\eps)} }{}  \Big( \wh{u}_{\Ups_{\e{reg}}}^{\, (\eps)} \, - \, \wh{u}_{\La_{\mf{b}}^{\!(\a)} }^{\, (\eps)} \Big)^{\prime}(t) \cdot  \op{C}_{\msc{C};+}\big[ \, \wh{z} \,  \big](t) \cdot \f{\dd t }{2\i\pi} \\
\mc{A}_{0}^{(3)}\big[ \, \wh{z},\, \wh{z} \,  \big] & = & 
- \sul{\eps=\pm }{}\Int{\msc{C}^{(\eps)} }{}  \Big( \wh{u}_{\Ups}^{\, (\eps)} \, - \, \wh{u}_{\Ups_{\e{reg}}}^{\, (\eps)} -2\i\pi \de_{\eps;+}L \wh{\xi}_{\Ups_{\e{sing}}}\Big)^{\prime}(t) 
\cdot  \op{C}_{\msc{C};+}\big[ \, \wh{z} \,  \big](t) \cdot \f{\dd t }{2\i\pi} \;. 
\eeqa
We remind that $\op{C}_{\msc{C};+}$ refers to the $+$ boundary value of the Cauchy transform on $\msc{C}$. 
The estimate $\mc{A}_{0}^{(3)}\big[ \, \wh{z},\, \wh{z} \,  \big]=\e{O}\big( L^{-\infty}\big)$   can be inferred from the bound
\beq
\norm{ \op{C}_{\msc{C};+}\big[ \, \wh{z} \,  \big] }_{ L^{1}(\msc{C}) } \leq C \ln L 
\enq
which is a consequence of Proposition \ref{Proposition DA transformee Cauchy} and of the bounds \eqref{ecriture brone sur U Ups moins U Ups reg}-\eqref{ecriture bornes ctg fct sing sur C}.

Next, one recasts  $\mc{A}_{0}^{(1)}\big[ \, \wh{z}, \wh{z} \,  \big]$ as
\bem
\mc{A}_{0}^{(1)}\big[ \, \wh{z}, \wh{z} \,  \big] \, = \, 2\i\pi \Int{ \msc{C}^{(+)} }{}  \wh{F}_{\e{reg}}^{\,(\mf{b})}(s) \cdot \op{C}_{\msc{C}^{(+)};-}\big[ \,  \big(\wh{F}_{\e{reg}}^{\,(\mf{b})} \big)^{\prime} \;   \big](s) \cdot \dd s \\
\, + \, \sul{\eps=\pm}{}\Int{ \msc{C}^{(\eps)} }{}   \Big( \wh{u}_{\Ups}^{\, (\eps)}(s) \, - \, \wh{u}_{\La^{\!(\a)}_{\mf{b}}}^{\, (\eps)}(s)  -2\i\pi \de_{\eps;+}L \wh{\xi}_{\Ups_{\e{sing}}}(s) \Big) \cdot 
\op{C}_{\msc{C}^{(+)};-}\big[ \,  \big(\wh{F}_{\e{reg}}^{\,(\mf{b})} \big)^{\prime} \;   \big](s) \cdot \dd s \;. 
\end{multline}
%
%
Upon observing that the Cauchy transform has logarithmic singularities at the edges $\wh{q}_{\ups}$ and invoking the bounds \eqref{ecriture bornes ctg fct sing sur C}, \eqref{estimee norme L1 u plus sing log}
one infers that the second line produces at most $\e{O}\big(  L^{-1} \cdot \ln L \big)$ contributions. The first line can be estimated by deforming the contour $\msc{C}^{(+)}$ to $\intff{ \wh{q}_{R} }{ \wh{q}_L }  $
and then symmetrising the integral. All-in-all, one obtains
\beq
\mc{A}_{0}^{(1)}\big[ \, \wh{z}, \wh{z} \,  \big] \, = \,
\Int{ \wh{q}_{L} }{ \wh{q}_{R} }   \f{  \big(\wh{F}_{\e{reg}}^{\,(\mf{b})} \big)^{\prime}(s)\wh{F}_{\e{reg}}^{\,(\mf{b})}(t)- \big(\wh{F}_{\e{reg}}^{\,(\mf{b})} \big)^{\prime}(t)\wh{F}_{\e{reg}}^{\,(\mf{b})}(s)  }{ 2 \tanh(s-t) } \dd t \dd s
 \, + \,\i \f{\pi}{2}  \sul{\ups \in \{L,R\} }{}   \sg_{\ups} \Big( \wh{F}_{\e{reg}}^{\,(\mf{b})} \Big)^2 \big(\, \wh{q}_{\ups} \big)  \; + \; \e{O}\Big( \f{ \ln L }{ L } \Big)\;. 
\enq
Finally, one has 
\beq
\mc{A}_{0}^{(2)}\big[ \, \wh{z}, \wh{z} \,  \big] \, = \, \sul{\ups \in \{L,R\} }{}   \mc{A}_{0;\ups}^{(2)}\big[ \, \wh{z}, \wh{z} \,  \big]
\; - \;  \sul{\eps=\pm }{}\Int{\msc{C}^{(\e{out};\eps)} }{}  \Big( \wh{u}_{\Ups_{\e{reg}}}^{\, (\eps)} \, - \, \wh{u}_{\La^{\!(\a)}_{\mf{b}}}^{\, (\eps)} \Big)^{\prime}(t) \cdot  \op{C}_{\msc{C};+}\big[ \, \wh{z} \,  \big](t) \cdot \f{\dd t }{2\i\pi} \;. 
\enq
The last term produces $\e{O}\big(L^{-\infty}\big)$ corrections while, upon implementing the change of variables \eqref{ecriture chgmt vars loc voisinage hat q ups} and inserting the local asymptotics
of the Cauchy transform, one gets 
%
%
\bem
\mc{A}_{0;\ups}^{(2)}\big[ \, \wh{z}, \wh{z} \,  \big] \; = \; -  \Int{ -2\pi L \de }{ 2\pi L \de } \bigg\{
\f{ \e{sgn}(t)  \Big(1\, -\, \tf{ \big(\wh{F}_{\e{reg}}^{\,(\mf{b})} \big)^{\prime}(s_t) }{ \big(L \wh{\xi}^{\prime}_{\La}(s_t) \big) } \Big)     }{  1+\ex{|t|-2\i\pi \e{sgn}(t)\sg_{\ups} \vsg_{\ups}^{(\Ups)} (s_{t}) }   } 
\, - \, \f{ \e{sgn}(t) }{  1+\ex{|t|-2\i\pi \e{sgn}(t)\sg_{\ups} \vsg_{\ups}^{(\La^{\!(\a)}_{\mf{b}})} (s_{t})}   }  \bigg\} \\
 \times \bigg\{ \wh{\op{M}}_{\ups}\big[ \wh{F}_{\e{reg}}^{\,(\mf{b})} \big](s_{t}) \;  + \;  \sg_{\ups}   \wh{F}_{\e{reg}}^{\,(\mf{b})}(s_{t}) \ln L 
\, + \,  \ln \Ga \Bigg( \ba{c} \tfrac{1}{2}  + \tfrac{t-a}{2\i\pi}  \vspace{2mm} \\
\tfrac{1}{2}  + \tfrac{t-a-b}{2\i\pi} 	 \ea \Bigg)   \\ 
 \, + \, \sg_{\ups} \Big( \wh{F}_{\e{reg}}^{\,(\mf{b})}( \, \wh{q}_{\ups})-\varkappa_{\ups}- \wh{F}_{\e{reg}}^{\,(\mf{b})} (s_{t}) \Big)  \cdot \ln \big[ -\i(\tf{t}{2\pi}+\i0^+)\big]
\bigg\}  \cdot  \f{\dd t }{2\i\pi} 
 \, + \; \e{O}\bigg(    \f{ \ln L }{ L }\bigg) \;. 
\end{multline}
The form of the remainder follows from \eqref{ecriture estimee hat u Omega reg sur C en L1} and I have set $a=2\i\pi \sg_{\ups} (\tau_{\ups}+ \wh{\mf{f}}_{\ups})$ and $b=2\i\pi \sg_{\ups} \wh{\op{f}}_{\ups }^{\,(\mf{b})}$. 
The $s_t$ dependent part can, again, be replaced by $s_0$ up to $\e{O}\big(    L^{-1} \ln L \big)$ correction and then one can extend the integration to $\R$ up to exponentially small corrections. 
One gets 
\bem
\mc{A}_{0;\ups}^{(2)}\big[ \, \wh{z}, \wh{z} \,  \big] \; = \; -  \Int{ \R  }{  } \bigg\{  \f{ \e{sgn}(t) }{  1+\ex{|t|-\e{sgn}(t)a}   } \, - \, \f{ \e{sgn}(t) }{  1+\ex{|t|-\e{sgn}(t)(a+b)}   }  \bigg\}
 \Bigg\{ \wh{\op{M}}_{\ups}\big[ \wh{F}_{\e{reg}}^{\,(\mf{b})} \big](\, \wh{q}_{\ups} ) \;  + \;  \sg_{\ups}   \wh{F}_{\e{reg}}^{\,(\mf{b})}( \, \wh{q}_{\ups}) \ln L  \\ 
 \, - \, \sg_{\ups} \varkappa_{\ups}  \cdot \ln \big[ -\i(\tf{t}{2\pi}+\i0^+)\big]
+ \ln \Ga \Bigg( \ba{c} \tfrac{1}{2}  + \tfrac{t-a}{2\i\pi}  \vspace{2mm} \\
\tfrac{1}{2}  + \tfrac{t-a-b}{2\i\pi} 	 \ea \Bigg) \Bigg\}  \cdot  \f{\dd t }{2\i\pi} 
 \, + \; \e{O}\bigg( \f{ \ln L }{ L }\bigg) \;. 
\end{multline}
The resulting integrals can be computed by means of Lemma \ref{Lemme integrale fct Gamma et autres}, hence leading to 
\bem
\mc{A}_{0;\ups}^{(2)}\big[ \, \wh{z}, \wh{z} \,  \big] \; = \; \sg_{\ups} \wh{\op{f}}_{\ups}^{\,(\mf{b})} \, \wh{\op{M}}_{\ups}\big[ \wh{F}_{\e{reg}}^{\,(\mf{b})} \big](\,\wh{q}_{\ups})
\,+\, \wh{\op{f}}_{\ups}^{\,(\mf{b})}  \wh{F}_{\e{reg}}^{\,(\mf{b})} (\,\wh{q}_{\ups}) \ln L  \\ 
+ \, \ln G\Big( 1- \wh{\op{f}}_{\ups}^{\,(\mf{b})} ,  1+  \wh{\op{f}}_{\ups}^{\,(\mf{b})} \Big)  
\; + \; \sg_{\ups} \varkappa_{\ups} \ln \Ga\Bigg( \ba{c}  \tfrac{1}{2} + \sg_{\ups} (\tau_{\ups}+\wh{\mf{f}}_{\ups} )  \vspace{1mm} \\ 
						      \tfrac{1}{2} + \sg_{\ups} (\tau_{\ups}+\wh{\op{f}}_{\ups\mid \a_{\La}=0} )   \ea   \Bigg)  
 \, + \; \e{O}\bigg(  \f{ \ln L }{ L }\bigg) \;. 
\end{multline}
Upon straightforward algebra, one gets the claim. \qed





%
\section{A few auxiliary bounds}
\label{Appendix auxiliary bounds}

\subsection{The singular counting function and singular roots}

\begin{lemme}
\label{Lemme dvgce exp fct cptge proximite des racines sing}
 Assume that the algebraic spacing between string centres holds \eqref{propriete espacement ctres cordes} and also the real/singular root spacing \eqref{propriete espacement ctre corde et particule trou}. Then, one has 
\beq
\big|  \Big( \wh{\xi}_{\Ups_{\e{reg}}^{\, (a)} } \Big)^{\prime} ( \mu_a^{(s)} ) \big| \; =  \; \e{O}(1) 
\enq
and 
\beq
\wh{\xi}^{\, \prime}_{\Ups}( \mu_a^{(s)} )   \;  =  \; \f{1}{L}   K\Big( \mu_a^{(s)}-\Re(\be_a^{(s)})  \mid \Im(\be_a^{(s)}) \Big) \cdot  \big( 1 + \e{O}(L^{-\infty})\big) 
\quad so \, that\quad \; \big| \wh{\xi}^{\, \prime}_{\Ups}( \mu_a^{(s)} )  \big| \; \geq \; C L^{k}
\label{ecriture equivalent dvgt de xi Ups en les mu a sing}
\enq
for any $k \in \mathbb{N}$ and $C>0$. 
 
\end{lemme}

\Proof 

One has, by construction, that 
\beq
\Big( \wh{\xi}_{ \Ups_{\e{reg}}^{\, (a)} }  \Big)^{\prime} ( \mu_a^{(s)} )\;= \;  \wh{ \xi }_{ \Ups_{\e{reg}} }^{\, \prime}( \mu_a^{(s)} )
\; + \; \f{1}{L} \hspace{-5mm} \sul{ \be+\i\zeta \in Z^{(s)} \! \!  , \;  \be \not= \be_a^{(s)}  }{} \hspace{-3mm} K\Big( \mu_a^{(s)}-\Re(\be)  \mid \Im(\be) \Big)\;. 
\enq
The first term is clearly an $\e{O}(1)$. As for the terms arising in the sum, since $\be\not= \be_a^{(s)}$, one has, due to \eqref{propriete espacement ctres cordes}, that 
\beq
\min_{\be\not= \be_a^{(s)}} \Big( |\mu_a^{(s)}-\be|, |\mu_a^{(s)}-\be^*| \Big) \, > \,  C \cdot L^{-\kappa}
\qquad \e{so}\, \e{that} \qquad 
  K\Big( \mu_a^{(s)}-\Re(\be)  \mid \Im(\be) \Big)  \; = \; \e{O}\Big( L^{-\infty} \Big)
\enq
where the bound also follows from the estimate $\Im(\be)=\e{O}(L^{-\infty})$. This result will then entail \eqref{ecriture equivalent dvgt de xi Ups en les mu a sing}
as soon as one has the lower bound on $\big|  K\big( \mu_a^{(s)}-\Re(\be_a^{(s)})  \mid \Im(\be_a^{(s)}) \big) \big|$. 
Recall that $\mu_a^{(s)}$ belongs to a disk of radius $C |\Im(\be_a^{(s)}) |^{1-\ups}$ around $\be_a^{(s)}$ for some $L$ independent $C>0$, \textit{c.f.} \eqref{ecriture deviation des mua au bea}. Thus, one has that 
\beq
\Big| K\Big( \mu_a^{(s)}-\Re(\be_a^{(s)})  \mid \Im(\be_a^{(s)}) \Big) \Big| \; \geq \;  \f{ C^{\prime}  |\Im(\be_a^{(s)}) | }{ \big| \mu_a^{(s)}-\be_a^{(s)} \big|^2  }
\geq \f{ C^{\prime \prime} }{   \big| \Im(\be_a^{(s)}) \big|^{1-2\ups} }
\enq
and the claim follows since $\ups < \tf{1}{2}$. \qed




\begin{lemme}
 \label{Lemme borne sur magnitude fct cptge sing}
Assume \eqref{propriete espacement ctres cordes}-\eqref{propriete espacement ctre corde et particule trou}. Then, given $k\in \mathbb{N}$, one has 
\beq
 \norm{ \wh{\xi}_{\Ups_{\e{sing}}} }_{ L^{\infty}\big(\msc{Z}^{(s)} \big) }  \; = \; \e{O}\big(n_{\e{sg}} L^{-\infty}\big) 
\label{ecriture estimee norme Wk ctg fct sing loin des sings}
 \enq
with
\beq
 \msc{Z}^{(s)}  \, = \, \bigg\{ z\in \mc{S}_{\de}(\R) \quad : \quad  d\Big(z, c_a^{(r)}+\de^{(r)}_{a, \f{r+1}{2}} \Big)  > C \big| \Im\big(\de^{(r)}_{a, \f{r-1}{2}} \big)  \big|^{\ups}  \quad%
r \, odd \,, \ba{c} r=3,\dots, p_{\e{max}} \\ a=1,\dots, n_r^{(z)}  \ea \bigg\}  \;. 
\nonumber
\enq
  In particular, one has  
\beq
 \norm{\,  \wh{\xi}_{\Ups_{\e{sing}}}  }_{W_k^{\infty}(\msc{C}) }  \; =\; \e{O}\Big(  n_{\e{sg}} L^{-\infty}\Big) \;. 
\label{ecriture bornes ctg fct sing sur C}
\enq

Furthermore, given $ \wh{\xi}_{ \Ups_{\e{sing}}^{(a)} }$ as in \eqref{definition xi Ups sing local regulier en a} and $R>0$, one has
\beq
 \norm{ \wh{\xi}_{\Ups_{\e{sing}}^{(a)} } }_{ L^{\infty}\big( \mc{D}_a \big) }  \; = \; \e{O}\big(n_{\e{sg}} L^{-\infty}\big) 
\qquad  with  \quad \mc{D}_a\,=\,\mc{D}_{ \be_a^{(s)}, R | \Im( \be^{(s)}_a ) | }
\label{ecriture estimee norme Wk ctg fct sing regularisee en a}
\enq
where $\msc{Z}^{(s)}(\be_a^{(s)})$ is defined analogously to $\msc{Z}^{(s)}$  with the exception that the central roots $\mu^{(s)}_a$ associated with $\be^{(s)}_a$
is absent from the constraint on the lower bound on the distances.

\end{lemme}

\Proof

The representation 
\beq
 \wh{\xi}_{\Ups_{\e{sing}}}(\om)  \; = \;  \f{1}{2\i\pi L } \sul{  \substack{ \be+\i\zeta \\ \in Z^{(s)} }  }{} \ln \Big[ \cosh(\be-\be^{*}) - \coth(\be^{*}-\om)  \sinh(\be-\be^*) \Big] 
\label{ecriture representation xi Ups Sing}
\enq
which follows readily from \eqref{definition partie sing fct cptge} leads to the bounds
\beq
 \Big| \, \wh{\xi}_{\Ups_{\e{sing}}}(\om) \Big|  \; \leq \; C\cdot  \f{ \# Z^{(s)} }{L}  
 \max_{  \substack{ \be+\i\zeta \\ \in Z^{(s)} } } \bigg| \f{ \be - \be^* }{ \om-\be^* } \bigg|  \;. 
\label{ecriture estimee xi Ups sing}
\enq
Then, by using \eqref{ecriture deviation des mua au bea}, \eqref{ecriture estimee norme Wk ctg fct sing loin des sings} follows, just as \eqref{ecriture bornes ctg fct sing sur C}
 for $k=0$ since $\e{d}\big( \be^{(s)}_a, \msc{C} \big)\geq \tf{C}{L}$. Further, with $\sinh\!\e{c}(x)=\sinh(x)/x$, for $k\geq 1$, one has 
\beq
 \wh{\xi}_{\Ups_{\e{sing}}}^{\, (k)}(\om) \; = \; \f{1}{2\i\pi L } \sul{  \substack{ \be+\i\zeta \\ \in Z^{(s)} }  }{} \ln^{(k)} \bigg( \f{ \sinh\!\e{c}(\be-\om) }{  \sinh\!\e{c}(\be^{*}-\om) }\bigg)
\, + \,   \f{(-1)^{k-1} (k-1)!}{2\i\pi L } \sul{  \substack{ \be+\i\zeta \\ \in Z^{(s)} }  }{} \bigg\{  \f{ 1 }{ (\om-\be)^{k} } \, -\, \f{ 1 }{ (\om-\be^*)^{k} } \bigg\}
\enq
The first summand is bounded by $\big| \be-\be^* \big| $ while the second is bounded by 
\beq
 \bigg| \f{ 1 }{ (\om-\be)^{k} } \, -\, \f{ 1 }{ (\om-\be^*)^{k} } \bigg| \; \leq \; \f{ C \cdot \big| \be-\be^* \big| }{ |\om-\be^*|^{k} \cdot |\om-\be|^{k} } \;. 
\enq
These allow one to conclude relatively to the $W_k^{\infty}\big( \msc{C} \big)$, $k\geq 1$, bounds on $ \wh{\xi}_{\Ups_{\e{sing}}} $. 
Finally, \eqref{ecriture estimee norme Wk ctg fct sing regularisee en a} follows from the string centre spacing hypothesis  \eqref{propriete espacement ctres cordes} 
and bounds analogous to \eqref{ecriture estimee xi Ups sing} with the root $\be_{a}^{(s)}$ removed.

\subsection{The functions $\wh{u}_{\Om}^{\,(\eps)}$ on $\msc{C}$}

\begin{lemme}
\label{Lemme estimation propriete generales ctg fct} 
 For any $k\in \mathbb{N}$, it holds
\beq
 \norm{\,  \wh{u}_{\Ups}^{\, (\eps)} \, - \, \wh{u}_{\Ups_{\e{reg}}}^{ \, (\eps)}  }_{W_k^{\infty}(\msc{C}^{(\eps)}) } \;  \leq \; 
 C \cdot L^{k+1}  \norm{\,  \wh{\xi}_{\Ups_{\e{sing}}} }_{W_k^{\infty}(\msc{C}) }  \;. 
\label{ecriture brone sur U Ups moins U Ups reg}
\enq

Finally, for $\Om \in \{\Ups, \La_{\mf{b}}^{\!(\a)}\}$, one has
\beq
 \Norm{\,  \wh{u}_{\Om_{\e{reg}}}^{ \, (\eps)}  }_{L^{1}(\msc{C}^{(\eps)}) } \;  =\; \e{O}\Big( L^{-1} \Big) \qquad and  \qquad 
  \Norm{\,  \Big( \wh{u}_{\Om_{\e{reg}}}^{ \, (\eps)} \Big)^{\prime}  }_{L^{1}(\msc{C}^{(\eps)}) } \;  =\; \e{O}\big( 1 \big)  
\label{ecriture estimee hat u Omega reg sur C en L1}
\enq
as well as
\beq
 \Norm{\,  \ln|\,  \wh{q}_{\ups}-* | \cdot \wh{u}_{\Om_{\e{reg}}}^{ \, (\eps)}  }_{L^{1}(\msc{C}^{(\eps)}) } \;  =\; \e{O}\bigg( \f{\ln L }{ L} \bigg) 
\qquad for \quad \ups \in \{L, R\} \;. 
\label{estimee norme L1 u plus sing log}
\enq
\end{lemme}

\Proof

One has 
\beq
\wh{u}_{\Ups}^{\, (\eps)}(s) \, - \, \wh{u}_{\Ups_{\e{reg}}}^{ \, (\eps)}(s) \; = \; \Int{0}{1} \f{ 2\i\pi \eps L   \wh{\xi}_{\Ups_{\e{sing}}}(s)  }{ 1 \, - \,  \ex{-2\i\pi L \eps \wh{\xi}_{\Ups_{t}}(s)  }     }  \dd t 
\qquad \e{with} \quad \wh{\xi}_{\Ups_{t}}(s) \; = \;  \wh{\xi}_{\Ups_{\e{reg}}}(s) \, + \,  t\wh{\xi}_{\Ups_{\e{sing}}}(s) \;. 
\label{ecriture rep int pour u Ups moins u Ups reg}
\enq
Then, given some $\de>0$ small enough, for $\ups \in \{L,R\}$, define the contours $ \msc{C}^{(\e{out})},  \msc{C}^{(\e{out};\eps)}, \msc{C}^{(\ups)}$ and 
$\msc{C}^{(\ups;\eps)}$ as in \eqref{definition contours C out et C out pm}-\eqref{definition contours locaux C ups et C ups pm}. Then, on $\msc{C}^{(\ups;\eps)}$
one has the parametrisation $s_x=\wh{\xi}_{\La}^{-1}\big(\tf{n_{\ups}}{L} +\i\eps x \big)$ with $n_{R}= |\La| + \tau_R + \tf{1}{2}$,  $n_L = \tau_L + \tf{1}{2} $ and $x\in \intff{0}{\de}$. 
This leads to 
\beq
\ex{-2\i\pi L \eps \wh{\xi}_{\Ups_{t}}(s_x)  } -1 \; = \; - \Big( \ex{2\pi L x} \ex{2\i\pi \eps \wh{\ga}_{\ups}(s_x) } +1  \Big)
\quad \e{with} \quad 
\wh{\ga}_{\ups}(s) \; = \; \wh{F}_{\e{reg}}(s)-\tau_{\ups}-\varkappa_{\ups}-t L \wh{\xi}_{\Ups_{\e{sing}}}(s) \;. 
\enq

The bounds \eqref{bornes sur fct shift en q hat left right}, \eqref{ecriture bornes ctg fct sing sur C} and the estimate $\norm{ \wh{F}_{\e{reg}} }_{ W_1^{\infty}\big( \mc{S}_{\de}(\R) \big) } < C n_{\e{tot}}^{(\e{msv})}$ 
then ensure that, 
\beq
\wh{\ga}_{\ups}(s_x) \; = \; \Re\big(\wh{\ga}_{\ups}(s_0)\big) \, + \,   x \, \big(v_{\ups}(s_x) \, + \, \i   w_{\ups}(s_x) \big) \quad \e{with} \quad%
\left\{ \ba{c } \Big| \Re\big(\wh{\ga}_{\ups}(s_0)\big) \Big| \,  < \, \tfrac{ 1 }{ 2 } \, - \,  \tfrac{ 3  }{ 4 } \eps_{\Ups}   \vspace{3mm} \\
	|v_{\ups}(s_x)|+|w_{\ups}(s_x)|  \, \leq  \, C n_{\e{tot}}^{(\e{msv})}	  \ea \right. \;. 
\enq
One has 
\bem
\Big| \ex{-2\i\pi L \eps \wh{\xi}_{\Ups_{t}}(s_x)  } -1 \Big| \geq \ex{2\pi x(L-\eps w_{\ups}(s_x)) } \cdot  \Big|  1 + \ex{-2\pi x(L-\eps v_{ \ups }(s_x)) } \ex{ - 2\i\pi x v_{\ups}(s_x)  } \Big|  \\
\geq 
\left\{  \ba{cc} 1-\cos(\pi \eps_{\Ups} )  & ,  x \in \Big[ 0 ;  \tfrac{ \eps_{\Ups} }{  4Cn_{\e{tot}}^{(\e{msv})} } \Big] \vspace{2mm}  \\ 
			1/2 			& , x \in \Big[  \tfrac{ \eps_{\Ups} }{  4Cn_{\e{tot}}^{(\e{msv})}  }  ; \de \Big]	\ea \right. >c>0 \;. 
\label{ecriture borne inf sur exp decroissante} 
\end{multline}
Note that the lower bound on $ \Big[  \tfrac{ \eps_{\Ups} }{  4Cn_{\e{tot}}^{(\e{msv})}  }  ; \de \Big]$ follows from  $\ex{-2\pi x(L-\eps v_{R/L}(s_x)) }<\tf{1}{2}$ for $L$ large enough. 

Thus, the above lower bound and \eqref{ecriture rep int pour u Ups moins u Ups reg} imply that, for $s\in \msc{C}^{(\ups;\eps)} $, one has
\beq
 \Big|\, \wh{u}_{\Ups}^{\, (\eps)}(s) \, - \, \wh{u}_{\Ups_{\e{reg}}}^{ \, (\eps)}(s) \Big|  \leq C  \big| \, \wh{\xi}_{\Ups_{\e{sing}}}(s) \big| \;. 
\label{ecriture bornes locales sur u Ups et u Ups reg}
\enq
Next,  for $s\in \msc{C}^{(\e{out};\eps)} $, one has the direct bounds
\beq
\Big| \wh{F}_{\e{reg}}(s)+t L \wh{\xi}_{ \Ups_{\e{sing}} }(s) \Big| \, \leq \, C \qquad \e{and} \quad 
\eps \Im \big( \wh{\xi}_{\La}(s) \big) > \de 
\enq
so that, for $L$ large enough,
\beq
\Big| \ex{-2\i\pi L \eps \wh{\xi}_{\Ups_{t}}(s_x)  } -1 \Big| \geq  \ex{2\pi (L\de - C)}-1 >1 \;. 
\enq
Hence, the bounds \eqref{ecriture bornes locales sur u Ups et u Ups reg} also holds for $s\in \msc{C}^{(\e{out};\eps)}$. The above entails \eqref{ecriture brone sur U Ups moins U Ups reg} 
for $k=0$. The result for general $k$ can be obtained by taking derivatives of \eqref{ecriture rep int pour u Ups moins u Ups reg} and using similar types of bounds as described above.

 It remains to establish the last set of bounds \eqref{ecriture estimee hat u Omega reg sur C en L1}-\eqref{estimee norme L1 u plus sing log}. By using the contours introduced in 
\eqref{definition contours C out et C out pm}-\eqref{definition contours locaux C ups et C ups pm} and the fact that $C>| (\, \wh{q}_{\ups}-s)/\big(\, \wh{\xi}_{\La}(\,\wh{q}_{\ups})- \wh{\xi}_{\La}(s)\big) |>C^{-1}>0$
on $\msc{C}^{(\ups)}$, and that $\big|\ln|\, \wh{q}_{\ups}-s| \big|$ is bounded on $\msc{C} \setminus \msc{C}^{(\ups)}$ one gets 
\bem
\Int{ \msc{C}^{(\eps)} }{} \Big( 1+ \big|\ln|\, \wh{q}_{\ups}-s| \big| \Big) \big| \, \wh{u}_{ \Om_{\e{reg}} }^{\,(\eps)}(s)\big| \cdot |\dd s | \, \leq \, 
C_1 \Int{ \msc{C}^{(\e{out};\eps)} }{}\big| \, \wh{u}_{ \Om_{\e{reg}} }^{\, (\eps)}(s)\big| \cdot |\dd s | \; + \; C_2 \Int{ \msc{C}^{(\ov{\ups};\eps)} }{}\big| \, \wh{u}_{ \Om_{\e{reg}} }^{\, (\eps)}(s)\big| \cdot |\dd s |  \\
\; + \; \Int{ \msc{C}^{(\ups; \eps)} }{} \Big( C_3 + \big|\ln|\,  \wh{\xi}_{\La}(\,\wh{q}_{\ups})- \wh{\xi}_{\La}(s) | \big| \Big) \big| \, \wh{u}_{ \Om_{\e{reg}} }^{\, (\eps)}(s)\big| \cdot |\dd s | \;. 
\end{multline}
Here, $\ov{\ups}=L$ if $\ups=R$ and \textit{vice versa}. 
Since $\big| \ex{2\i\pi \eps L \wh{\xi}_{ \Om_{\e{reg}} } } \big| \leq C \ex{-c^{\prime} L }$ with $c^{\prime}>0$ on $\msc{C}^{(\e{out};\eps)}$, the first term will only generate exponentially small corrections. 
The second term is similar to the third one, so that it remains to focus on the last line. There, implementing the change of variable $s_{\eps t}$ with $s_{t}$ as defined in \eqref{ecriture chgmt vars loc voisinage hat q ups}, 
leads to  
\beq
\Int{ \msc{C}^{(\ups; \eps)} }{} \Big( 1+ \big|\ln|\,  \wh{\xi}_{\La}(\,\wh{q}_{\ups})- \wh{\xi}_{\La}(s) | \big| \Big) \big| \, \wh{u}_{ \Om_{\e{reg}} }^{\, (\eps)}(s)\big| \cdot |\dd s |   \; \leq  \; 
\Int{ 0 }{ 2\pi L \de } \Big( C+ \big|\ln|\f{t}{2\pi L } | \big| \Big) \big| \, \ln \big[ 1\, + \, \ex{-|t|+2\i\pi \eps \sg_{\ups} \vsg^{(\Om)}_{\ups}(s_{\eps t})  } \big] \big| \cdot \f{ \dd t }{ 2\pi L  }
\enq
with $\vsg^{(\Om)}_{\ups}$ as defined in \eqref{definition vsg ups Omega}. It then remains to observe that bounds similar to \eqref{ecriture bornes sur fonction hat gamma ups} allow
one to use the bounds $|\ln(1+z) | \leq C |z|$ for $\e{arg}(z) \in \intoo{ \eta - \pi}{ \pi- \eta}$ with $\eta>0$ and fixed. The latter bound entails the claim. 
Finally, the bounds relative to  $ \big(\, \wh{u}_{ \Om_{\e{reg}} }^{\,(\eps)} \big)^{\prime}$ can be obtained through similar handlings.

\qed




\section{Auxiliary integrals and special functions}
\label{Appendix integrales auxiliaires}

\subsection{A few auxiliary integrals}

\begin{lemme}
\label{Lemme integral pole simple vers fct Gamma}
Let $a_{\pm} \in \mathbb{H}_{\pm}$ and $\Im(\a) \in \intoo{ - \tf{\pi}{2} }{ \tf{\pi}{2} }$\, , then
\beq
\Int{\R}{}  \f{  \ln\big(1+\ex{-|t| +\a \e{sgn}(t) }\big) }{t-a_{\pm}} \f{\dd t}{2\i\pi} \; = \;  \mp  \ln \Ga\bigg( \f{1}{2} \pm \f{a_{\pm}-\a }{2 \i \pi }  \bigg)
\pm \f{1}{2} \ln(2\pi)   \, + \, \f{a_{\pm}-\a }{2 \i \pi } \ln \Big( \f{ a_{\pm} }{ \pm 2 \i \pi  } \Big) \,  - \,  \f{ a_{\pm} }{ 2\i\pi } \;.
\enq
\end{lemme}

\Proof
We set $f(t) \, = \, (t-a_{\pm})\ln(t-a_{\pm})-(t-a_{\pm})$ and integrate twice by parts
\beq
\Int{\R}{}  \f{   \ln\big(1+\ex{-|t| +\a \e{sgn}(t) }\big) }{t-a_{\pm}} \f{\dd t}{2\i\pi} \; = \;
\Int{\R}{} \f{ f(t)  }{ \big( \ex{\f{t-\a}{2}} +\ex{-\f{t-\a}{2}} \big)^2 } \f{\dd t}{2\i\pi}  \; - \;  \f{f(0)}{2\i\pi} \; - \, \a \f{ f^{\prime}(0) }{ 2\i\pi } \;.
\enq
As $f$ is holomorphic in the lower/upper half plane (depending whether $a_{\pm}$ belongs to the upper/lower half plane),
one can deform the integration contour to $\R \mp 2 \i\pi N$, $N\in \mathbb{N}$, for the price of picking the poles at $t=\mp \i\pi (2p+1) $, with
$p=1, \dots, N-1$. Then
\beq
\Int{\R}{}  \f{   \ln\big(1+\ex{-|t| +\a \e{sgn}(t) }\big) }{t-a_{\pm}} \f{\dd t}{2\i\pi}  \, = \,  \pm  \sul{p=0}{N-1} f^{\prime}\big( \mp \i \pi (2p+1) +\a \big)
 +\Int{\R \mp \i 2 \pi N }{} \f{ f(t)  }{ \big( \ex{\f{t-\a}{2}} +\ex{-\f{t-\a}{2}} \big)^2 } \f{\dd t}{2\i\pi}  \; - \;  \f{f(0)}{2\i\pi} \; - \, \a \f{ f^{\prime}(0) }{ 2\i\pi } \;.
\enq
Now, one integrates twice by parts  and compute the sum over the crossed poles in terms of $\Ga$ functions
\bem
\Int{\R}{}  \f{   \ln\big(1+\ex{-|t| +\a \e{sgn}(t) }\big) }{t-a_{\pm}}  \f{\dd t}{2\i\pi} \,  = \;  \f{ f(\mp 2 \i \pi N) - f(0) }{2\i\pi} \; + \, \a \f{ f^{\prime} (\mp 2 \i \pi N)-f^{\prime}(0) }{ 2\i\pi }
\pm  N \ln (\mp 2 \i \pi) \\
\pm  \ln \Ga\Big(N+\tfrac{1}{2} \pm \tfrac{a_{\pm}-\a }{2 \i \pi}   \Big) \mp \ln \Ga\Big(\tfrac{1}{2} \pm \tfrac{a_{\pm}-\a }{2 \i\pi }  \Big)
\; + \; \Int{\R}{}  \f{\ln\big(1+\ex{-|t| +\a \e{sgn}(t) }\big)  }{ \mp 2 \i\pi N + t -a_{\pm} }  \f{\dd t}{2\i\pi}  \;.
\end{multline}
The last integral is a $\e{O}(N^{-1})$ by the dominated convergence theorem and
\bem
\f{ f(\mp 2\i\pi N)  }{2\i\pi} \; + \, \a \f{ f^{\prime} (\mp 2\i\pi N) }{ 2\i\pi }
\pm  N \ln ( \mp 2 \i \pi )
\; \pm \;  \ln \Ga\Big( N+\tfrac{1}{2} \pm \tfrac{a_{\pm}-\a }{2 \i \pi}    \Big) \,  = \, 
\pm  \f{1}{2} \ln 2\pi  \, + \, \f{ \a - a_{\pm} }{ 2\i\pi } \ln( \mp 2 \i \pi )  \, + \, \e{O}\Big( \f{1}{N}\Big)\; .
\end{multline}
Putting all these results together and then using that 
\beq
- \f{a_{\pm}-\a}{2\i \pi}\ln \Big( \f{ \mp 1  }{ 2 \i\pi} \Big) + \f{a_{\pm}-\a}{2\i \pi}\ln\pa{-a_{\pm}} = \f{a_{\pm}-\a}{2\i\pi} \ln \Big( \f{ a_{\pm} }{ \pm 2 \i \pi  } \Big)
\enq
yields the value of the integral. \qed

\begin{lemme}
\label{Lemme integrale fct Gamma et autres}
Let $a,b \in \Cx$ be small enough, then
\beq
\Int{\R}{ }   \bigg\{  \f{ \e{sgn}(t) }{ 1+\ex{|t| - a \e{sgn}(t)  }  } - \f{ \e{sgn}(t) }{ 1 + \ex{|t| - (a+b) \e{sgn}(t)} }   \bigg\}  
\ln \Ga\Bigg( \ba{c} \tfrac{1}{2} + \tfrac{t-a}{ 2 \i \pi }   \vspace{2mm} \\
	      \tfrac{1}{2} + \tfrac{t-a-b}{2 \i \pi}  \ea \Bigg) \cdot \f{\dd t}{2 \i \pi}
\; = \;   -  \ln G\Big( 1-\tfrac{ b }{ 2 \i \pi },  1 + \tfrac{ b }{2 \i \pi}  \Big)
\label{formule pour integrale avec Gamma}
\enq
and
\beq
\Int{\R}{ }   \bigg\{  \f{ \e{sgn}(t) }{ 1+\ex{|t| - a \e{sgn}(t)  }  } - \f{ \e{sgn}(t) }{ 1 + \ex{|t| - (a+b) \e{sgn}(t)} }   \bigg\}   \cdot \f{\dd t}{2 \i \pi}
\; = \; - \f{b}{2 \i \pi} \;.
\label{formule pour integrale difference exponentielles}
\enq
Here, we agree upon $G(x,y)=G(x)G(y)$ and upon similar notations for products of $\Ga$ functions. Also, one has that 
\beq
 \Int{\R}{ }   \bigg\{  \f{ \e{sgn}(t) }{ 1+\ex{|t| - a \e{sgn}(t)  }  } - \f{ \e{sgn}(t) }{ 1 + \ex{|t| - (a+b) \e{sgn}(t)} }   \bigg\} \ln \big[ -\i(t+i0^+) \big] \cdot \f{\dd t }{ 2 \i \pi }
\; = \;   \ln \Ga \bigg( \ba{c}  \tfrac{1}{2} +  \tfrac{a}{2i\pi}  \vspace{1mm} \\   \tfrac{1}{2} +  \tfrac{a+b}{2i\pi}  \ea \bigg)  \, - \, \f{b}{2\i\pi} \ln(2\pi) \;.
\label{formule pour integrale avec ln}
\enq
\end{lemme}

\Proof

The integrals can be computed either by means of direct integration or owing to a residue calculation in the spirit of Lemma \ref{Lemme integral pole simple vers fct Gamma}
after observing that  
\beq
 \f{ \e{sgn}(t) }{ 1+\ex{|t| - a \e{sgn}(t)  }  } - \f{ \e{sgn}(t) }{ 1 + \ex{|t| - (a+b) \e{sgn}(t)} }  \; = \; 
 \f{ 1 }{ 1+\ex{t - a   }  } - \f{ 1 }{ 1 + \ex{t - (a+b) } } \;. 
\enq
\qed

\end{document}